\documentclass[pdflatex,twocolumn,epjc3]{svjour3}          
\RequirePackage[T1]{fontenc}
\usepackage[dvipsnames]{xcolor}
\smartqed  

\RequirePackage{graphicx}
\RequirePackage{mathptmx}      
\RequirePackage{flushend}
\RequirePackage[numbers,sort&compress]{natbib}
\RequirePackage[colorlinks,citecolor=blue,urlcolor=blue,linkcolor=blue]{hyperref}

\def\be{\begin{equation}}
\def\ee{\end{equation}}
\def\bea{\begin{eqnarray}}
\def\eea{\end{eqnarray}}

\newcommand{\gsim}{\lower.7ex\hbox{$\;\stackrel{\textstyle>}{\sim}\;$}}
\newcommand{\lsim}{\lower.7ex\hbox{$\;\stackrel{\textstyle<}{\sim}\;$}}
\newcommand{\ovl}[1]{\overline{#1}}
\usepackage{slashed}
\usepackage{graphicx}
\usepackage{epstopdf}
\usepackage{mathrsfs}
\usepackage{amssymb}
\usepackage{verbatim}
\usepackage{color}
\usepackage{multirow}
\usepackage{soul,xcolor}
\usepackage{amsmath}
\usepackage{subfig}
\usepackage{soul,xcolor}
\usepackage[utf8]{inputenc}
\usepackage[normalem]{ulem}
\journalname{Eur. Phys. J. C}
\usepackage{eso-pic}
\newcommand{\reportnum}[2]{
  \AddToShipoutPictureBG*{%
    \AtPageUpperLeft{%
      \hspace{0.75\paperwidth}%
      \raisebox{#1\baselineskip}{%
        \makebox[0pt][l]{\textnormal{#2}}
  }}}%
}
\begin{document}
\reportnum{-6}{DESY-24-042}
\setstcolor{red}
\title{The Waning of the WIMP: Endgame?}


\author{Giorgio Arcadi\thanksref{e1,addr8,addr9}
        \and
        David Cabo-Almeida\thanksref{e2,addr8,addr9,addr15}
        \and
        Ma\'ira Dutra\thanksref{e3, addr6,addr7} 
\and
        Pradipta Ghosh\thanksref{e4, addr2}
\and
        Manfred Lindner\thanksref{e5, addr1}  
  \and
        Yann Mambrini\thanksref{e6, addr3} 
        \and
              Jacinto P. Neto\thanksref{e7, addr8,addr10,addr11} 
   \and
        Mathias Pierre\thanksref{e8, addr31}  
     \and
        Stefano Profumo\thanksref{e9, addr4, addr5}  
   \and
        Farinaldo S. Queiroz\thanksref{e10, addr10, addr11,addr12} 
}

\thankstext{e1}{e-mail: giorgio.arcadi@unime.it}
\thankstext{e2}{e-mail: david.cabo@ct.infn.it}
\thankstext{e3}{e-mail: mdutrava@mail.nasa.gov}
\thankstext{e4}{e-mail: tphyspg@physics.iitd.ac.in}
\thankstext{e5}{e-mail: lindner@mpi-hd.mpg.de}
\thankstext{e6}{e-mail: yann.mambrini@th.u-psud.fr}
\thankstext{e7}{e-mail: jacinto.neto.100@ufrn.edu.br}
\thankstext{e8}{e-mail: mathias.pierre@desy.de}
\thankstext{e9}{e-mail: profumo@ucsc.edu}
\thankstext{e10}{e-mail:farinaldo.queiroz@ufrn.br}

\institute{
Dipartimento di Scienze Matematiche e Informatiche, Scienze Fisiche e Scienze della Terra, \\ Universita degli Studi di Messina, Via Ferdinando Stagno d'Alcontres 31, I-98166 Messina, Italy\label{addr8}
\and 
INFN Sezione di Catania, Via Santa Sofia 64, I-95123 Catania, Italy \label{addr9}
\and
Departament de Física Quàntica i Astrofísica, Universitat de Barcelona,\\
Martí i Franquès 1, E08028 Barcelona, Spain\label{addr15}
\and Astroparticle Physics Laboratory, NASA Goddard Space Flight Center, Greenbelt, MD 20771, United States of America \label{addr6}
\and NASA Postdoctoral Program Fellow \label{addr7}
          \and
          Department of Physics, Indian Institute of Technology Delhi,
Hauz Khas,  New Delhi 110016, India \label{addr2}
\and 
Max Planck Institut f\"ur Kernphysik, Saupfercheckweg 1, D-69117 Heidelberg, Germany\label{addr1}
          \and
        Universit\'e Paris-Saclay, CNRS/IN2P3, IJCLab, 91405 Orsay, France\label{addr3}    
             \and    
Departamento de Física, Universidade Federal do Rio Grande do Norte, 59078-970, Natal,
RN, Brasil\label{addr10}
        \and    
International Institute of Physics, Universidade Federal do Rio Grande do Norte, Campus Universitario, Lagoa Nova, Natal-RN 59078-970, Brazil\label{addr11}
        \and
        Deutsches Elektronen-Synchrotron DESY, Notkestr. 85, 22607 Hamburg, Germany \label{addr31}
          \and
          Department of Physics, University of California, Santa Cruz, 
1156 High St, Santa Cruz, CA 95060, United States of America\label{addr4}
          \and
          Santa  Cruz  Institute  for  Particle  Physics,  Santa  Cruz,  
1156  High  St,  Santa  Cruz,  CA 95060, United States of America\label{addr5}
    \and    
Millennium Institute for Subatomic Physics at the High-Energy Frontier (SAPHIR) of
ANID, Fernández Concha 700, Santiago, Chile\label{addr12}
}

\date{Received: date / Accepted: date}


\maketitle

\begin{abstract}
Weakly Interacting Massive Particles (WIMPs) continue to be considered some of the best-motivated Dark Matter (DM) candidates. No conclusive signal, despite an extensive search program that combines, 
often in a complementary way, direct, indirect, and collider probes, has been however detected so far. This situation 
might change in the near future with the advent of even larger, multi-ton Direct Detection experiments. We provide here an updated review of the WIMP paradigm, with a focus on selected models that can be probed with upcoming facilities, all relying on the standard freeze-out paradigm for the relic density. We also discuss Collider and Indirect Searches when they provide complementary experimental information. 
\end{abstract}



\section{Introduction}
\label{sec:Introduction}
Cold Dark Matter (CDM) is a pillar of the Standard Cosmological Model, which represents the best fit of a broad variety of cosmological and astrophysical observations \cite{Planck:2018vyg}, covering states of the history of the Universe from the primordial Big Bang Nucleosynthesis (BBN) to the Cosmic Microwave Background (CMB) and more recent times. Furthermore, the presence of a DM component in the Early Universe is a fundamental requirement to achieve a mechanism for structure formation, in agreement with experimental observations \cite{Blumenthal:1984bp,Bullock:1999he}. 

While there is broad consensus about the hypothesis that the DM is made by one or more new particle states beyond the spectrum of the Standard Model (SM) of particle physics, the latter has not yet received definitive experimental confirmation. Experimental hints, moreover, do not provide unequivocal guidelines for particle model building; consequently a very broad plethora of theoretical proposals are available in the literature. 
Nevertheless, there is a set of general requirements that any particle model should fulfil, to provide a viable DM candidate:
\begin{enumerate}
    \item Being stable, at least on cosmological scales. While decaying DM candidates are not strictly excluded,  strong constraints force their  lifetime to be over 10 orders of magnitude longer than the lifetime of the Universe, see e.g. \cite{Queiroz:2014yna,Audren:2014bca,Giesen:2015ufa,Mambrini:2015sia,Baring:2015sza,Lu:2015pta,Slatyer:2016qyl,Jin:2017iwg}. 
    \item Having weak enough interactions with the ordinary matter to justify the absence of non-gravitational detection so far. In particular, the DM should be electrically neutral, or at most millicharged, to comply with null searches for stable, charged particles \cite{SanchezSalcedo:2010ev,McDermott:2010pa}. 
    \item Account for a production mechanism in the Early Universe, leading to the experimentally determined, via CMB observations, value of the the DM relic density \cite{Planck:2018vyg}.
    \item To comply with structure formation, the DM should be in large part non-relativistic at  matter-radiation equality. How such requirement is translated into the parameters of a particle model depends on the DM phase space distribution in the Early Universe which depends
    in turn on its interactions. In this paper we will always consider the DM as a thermal relic, i.e. it was as some early stage of the history of the Universe in thermal equilibrium with the primordial plasma. In such a case, a lower bound of the order of a few keVs \cite{Benson:2012su,Lovell:2013ola,Kennedy:2013uta} can be put on the DM mass.
    \item Have a rate of self-interaction non conflicting with the observations e.g. of cluster collisions, such as the Bullet Cluster \cite{Clowe:2006eq}.
    \item Comply with a broad variety of null dedicated DM searches at Earth-scale experiments. The bounds depend on the specific class of DM candidates under consideration, which will be spelled out in detail later.
\end{enumerate}

In this work we will focus on the popular class of DM candidates represented by the WIMPs. We refer to particle states which were existing in thermal equilibrium in the very early stages of the history of the Universe and, at later times, decoupled ({\it freeze-out}) from the primordial plasma. In such a setup, and assuming a standard cosmological history for the Universe, the DM relic density is in one-to-one correspondence with a single particle physics input, the so-called thermally averaged pair annihilation cross-section. Such particle input can be related to a series of complementary observables probed by two dedicated search strategies, dubbed Direct Detection (DD) and Indirect Detection (ID) as well as broader perspective New Physics searches at particle accelerators and low energy physics experiments.

This review aims to provide an overview of the status of  the aforementioned category of DM candidates in light of the recent updates in experimental searches, with a particular focus on DD updating and augmenting our previous review \cite{Arcadi:2017kky}: Besides the relevant update of the experimental results, the present work will  include a broader and different selection of particle physics models under scrutiny.  Ref. \cite{Arcadi:2017kky}, indeed, mainly focused on the so-called portal models where very useful benchmarks were obtained with only a few free parameters. The present work, besides portals, will also investigate more realistic particle physics frameworks. The latter
not only assures a straightforward correlation between the requirement of the correct DM relic density and experimental outcomes but also overpowers the potential theoretical loopholes affecting simplified models. 

The paper is organised as follows. In section 2 we will provide a brief review of the freeze-out paradigm. Section 3 and 4 will be devoted to the most salient features of DM Direct Detection (DD) and Indirect Detection (ID), respectively. Section 5 contains some general remarks which will be useful to guide a reader throughout the paper. The review of the WIMP model will start in section 6 with the "Simplified Models": s-channel portals, t-channel portals and models with the DM interacting via the $SU(2) \times U(1)$ gauge interactions. In this last case we will focus essentially on the features related to DD. The following section will be devoted to increasingly refined models based on the idea that the DM interacts with the SM via the 125 GeV Higgs boson. In section 8 the interaction between the DM and the Higgs sector will be again considered, but this time the latter will be extended with a further $SU(2)$ doublet and possibly a $SU(2)$ singlet. Before stating our conclusion, we will consider in section 9 some realistic realization of spin-1 portals.



\section{The WIMP Paradigm}
\label{sec:wimpparam}
Any DM model has to account for a dynamical production mechanism at the early stages of the Universe, before the BBN, in which the DM relic abundance agrees with the observed value inferred by CMB experiments, such as the Planck satellite (see \cite{Planck:2018vyg} for the most recent results). For the particle physics scenarios discussed in this work, we will consider the so-called thermal freeze-out mechanism. It arises from the application of the principles of particle physics and statistical mechanics to an expanding Universe. As will be stressed below, the most appealing feature of this scenario consists of the fact that, if the Standard Cosmological Model is considered, the DM relic density is determined by a single particle physics input. Moreover, it requires sizeable couplings between the SM and DM particles, making such DM models testable at the current experiments.

The freeze-out paradigm arises from a statistical description of the Early Universe in which each particle species is described by a distribution function $f(p, T)$, where $p$ stands for the modulus of the momentum (this is due to the assumption of homogeneity and isotropy at large scales of the Early Universe) while the temperature $T$ is a measure of the time. Macroscopic observables, such as number density and energy density, are obtained as integrals, over the phase space of such distribution functions. For example, the number density, $n_\chi$, of a particle species $\chi$ is given by:

\begin{equation}
  n_\chi=g_\chi \int \frac{d^3 p}{(2\pi)^3}f_\chi(p,T),
\end{equation}
with $g_\chi$ stemming from the "internal" degrees of freedom ($dof$), like the number of spin states. $f_\chi(p,T)$ depicts distribution function for the particle species $\chi$.
The time evolution of the distribution function of a particle species can be tracked according to the rate of interactions with the other species, via the so-called Boltzmann equation. In the case of WIMPs, one can actually rely on an integrated Boltzmann equation, describing the time evolution of the number density. 
Considering a DM candidate $\chi$ interacting with a pair of SM states via a $2 \rightarrow 2$ annihilation processes, its Boltzmann equation, assuming the aforesaid processes to be in thermal equilibrium during the DM production process, is given by:
\begin{equation}
\label{eq:bolbase}
\frac{dn_{\chi}}{dt} +3 H(T) n_\chi = -\langle \sigma v \rangle (n_{\chi}^2 -n_{\chi, eq}^2),
\end{equation}
where $n_{\chi, eq}$ represents the DM matter number density at equilibrium, $H(T)$ represents the Hubble expansion rate, and $\langle \sigma v \rangle$ is the thermally averaged pair annihilation cross-section of the DM, which can be written as:
\bea
    && \langle \sigma v \rangle=\frac{\int d^3 p_1 d^3 p_2 \sigma v f_{\chi,eq}(p_1,T) f_{\chi,eq}(p_2,T)}{\int d^3 p_1 d^3 p_2  f_{\chi,eq}(p_1,T) f_{\chi,eq}(p_2,T)}\\
    && =\frac{1}{8 m_\chi^4 T K_2\left(\frac{m_\chi}{T}\right)^2}\int_{4 m_\chi^2}^{\infty}ds \sigma(s) \sqrt{s}(s-4 m_\chi^2)K_1\left(\frac{\sqrt{s}}{T}\right),\nonumber
    \eea

where $m_\chi$ denotes the DM mass and $s$ represents the center-of-mass energy for the aforesaid $2 \rightarrow 2$ annihilation processes. The functions $K_1, K_2$ depict modified Bessel functions.
$\sigma$(s) is the annihilation cross-section computed with conventional field theory techniques. The equilibrium distribution function $f_{\chi,eq}=\exp\left[-E/T\right]$ is the Maxwell-Boltzmann distribution leading to:
\begin{equation}
    n_{\chi,eq}=g_\chi \int \frac{d^3 p}{(2\pi)^3}f_{\chi,eq}=\frac{g_\chi m_\chi^2 T}{2\pi^2}K_2\left(\frac{m_\chi}{T}\right).
\end{equation}

The Boltzmann equation can be solved semi-analytically by introducing the comoving number density:
\begin{equation}
Y_\chi=\frac{n_\chi}{s},\,\,\,\,\,s=\frac{2\pi^2}{45}h_{\rm eff}(T)T^3,
\end{equation}
where $s$ is the entropy density of the Universe and $h_{\rm eff}(T)$ is the effective number of entropy degrees of 
freedom at the temperature $T$. This change of variables gauges out the term on the left-handed side, depending on the Hubble expansion rate:

\begin{equation}
\label{eq:bol_real}
\frac{dY_\chi}{dt}=\frac{ds}{dt}\frac{\langle \sigma v \rangle}{3H}Y_\chi^2 \left(1-\frac{Y_{\chi,eq}^2}{Y_\chi^2}\right),
\end{equation}
with $Y_{\chi,eq}$ denoting the comoving number density at equilibrium.
Using the entropy conservation, $\frac{ds}{dt}=-3Hs$, it is possible to use the temperature $T$ to replace the time as an independent variable. The former can be then, in turn, possibly replaced with $x=m_\chi/T$. The solution of Eq.~\eqref{eq:bol_real} can be written as: 
\begin{equation}
Y(T_0)\equiv Y_0 \simeq \sqrt{\frac{\pi}{45}}M_{\rm Pl}{\left[\int_{T_0}^{T_f} g_{*}^{1/2} \langle \sigma v \rangle dT \right]}^{-1},
\end{equation}
where $T_0$ denotes the present time temperature of the Universe, $M_{\rm Pl}$ represents the Planck mass, and:
\begin{equation}
g_{*}^{1/2}=\frac{h_{\rm eff}}{g^{1/2}_{\rm eff}}\left(1+\frac{1}{3}\frac{T}{h_{\rm eff}}\frac{dh_{\rm eff}}{dT}\right),
\end{equation}
where $g_{*}^{1/2}$ depicts relativistic $dof$ of the
primordial thermal bath, $T_f$ is dubbed freeze-out temperature and corresponds to the time at which the DM number density deviates from thermal equilibrium. For WIMP models $T_f \sim \frac{m_\chi}{20}\div\frac{m_\chi}{30}$.
The relative energy density of relic dark matter particles normalized by the critical energy density of the Universe, $\Omega_{\rm DM}$,  can be determined from the solution of the Boltzmann equation as: 
\bea
&&\Omega_{\rm DM}=\rho_{\rm DM}/\rho_{\rm cr}(T_0), \,\,\rho_{\rm DM}=m_\chi s_0 Y_0, \nonumber\\ 
&&\rho_{\rm cr}(T)=3 H(T)^2 M_{\rm PL}^2/8 \pi,\,\,\,\rho_{\rm cr}(T_0) \simeq 10^{-5}~ \mathrm{GeV ~cm^{-3}},\,\,\,
\eea
where , $\rho_{\rm cr}(T)$ denotes the critical energy density at a temperature $T$, and $s_0 \equiv s(T_0)$ is the entropy density at present times.
Replacing the numerical values for $s_0$ and $\rho_{\rm cr},$ we can arrive at the following compact expression for the DM relic density:
\begin{equation}
\label{eq:relic_final}
\Omega_{\rm DM}h^2 \approx 8.76 \times 10^{-11}\, {\mbox{GeV}}^{-2} {\left[\int_{T_0}^{T_f} 
g_{*}^{1/2} \langle \sigma v \rangle \frac{dT}{m_\chi} \right]}^{-1}.
\end{equation}
As well known, experimental determination of $\Omega_{\rm DM}h^2 \approx 0.12$~\cite{Ade:2015xua} is matched by a value of the cross-section of the order of $10^{-9} {\mbox{GeV}}^{-2}$ corresponding to $\langle \sigma v \rangle \sim 10^{-26}\,{\mbox{~cm} }^3 
{\mbox{~s}}^{-1}$.

The DM relic density in the standard freeze-out mechanism described above is in one-to-one correspondence with a single particle physics input, i.e., the thermally averaged cross-section $\langle \sigma v \rangle$. Firstly, this kind of solution relies on the assumption of a standard cosmological evolution during the DM production. The second important remark is that, looking at the extrema of the integral as shown in Eq.\eqref{eq:relic_final}, the DM abundance is determined by the values of the DM annihilation cross-section at temperatures below the one of freeze-out,and thus below the DM masses. Even if in principle, the solution of the Boltzmann equation would require an integral over a wide range of temperatures, and hence, a particle physics framework possibly valid up to an arbitrary high energy scale, the thermal freeze-out is actually an "infrared" mechanism as the low energy behaviour of the DM interaction rate is also relevant. Consequently, effective or simplified models are viable benchmarks to test WIMP scenarios.

The DM relic density is determined with great precision for arbitrary particle physics models by publicly available numerical packages as micrOMEGAs \cite{Belanger:2006is,Belanger:2008sj,Alguero:2023zol}, DARKSUSY \cite{Gondolo:2004sc,Bringmann:2018lay} or MadDM \cite{Backovic:2013dpa,Arina:2023msd}. All the results, about DM relic density, shown in this work, are based on the package micrOMEGAs. Nevertheless, it is anyway useful to dispose of an analytical approximation for a better understanding of the underlying dynamics. Such approximation is provided by the so-called velocity expansion \cite{Gondolo:1990dk}. 
\begin{equation}
    \langle \sigma v \rangle \simeq a+\frac{3}{2}b \frac{1}{x} \equiv a+ b v^2,
\end{equation}
The velocity expansion is essentially a non relativistic expansion of the cross-section as in the Standard freeze-out paradigm the relic density is determined at times corresponding to $1/x=T/m_\chi \ll 1$.  The velocity expansion can be reliably adopted for WIMP models with some relevant exceptions: the DM annihilation cross-section has a s-channel resonance, coannihilations (see below), the center-of-mass energy of the annihilation processes is in vicinity of the opening threshold of a final state \cite{Griest:1990kh}. The coefficients of the expansion  $a,\,b$ are determined by the content (e.g. masses and couplings) of the underlying particle theory. 
As evident, the thermally averaged cross-section features a temperature (and hence, time) independent term, described by the coefficient $a$, dubbed s-wave term given the analogy of the velocity expansion with the partial wave analysis in  quantum mechanics. Note that according to the spin assignments of the DM and the mediator, the s-wave term might vanish. The leading velocity (temperature) dependent term is dubbed p-wave contribution. The majority of WIMP models have an s-wave or p-wave-dominated cross-section. Some examples with d-wave ($v^4$) dominated cross-section nevertheless exist (see later on in the text).  Via the velocity expansion, one can obtain the following approximate expression for the relic density:
\begin{equation}
    \Omega_\chi h^2 \simeq 1.07\times 10^{9}{\mbox{GeV}}^{-1}\frac{x_f}{g_{*}^{1/2}M_{\rm Pl}\left(a+3b/x_f\right)},
\end{equation}
where $x_f=m_\chi/T_f$ is the freeze-out ``time''.

Note that the results presented until now rely on the assumption that the DM particle is the only particle added to the SM spectrum (or at least the only relevant for phenomenology). In most realistic scenarios, the DM is part of a larger new sector. In general, one should then replace a single simple Boltzmann's equation written before by a system of equations of the following form \cite{Edsjo:1997bg}:
\bea
    && \frac{dn_i}{dt}+3Hn_i=-\sum_{j=1}^N \langle \sigma_{ij}v_{ij} \rangle \left(n_{\chi_i} n_{\chi_j}-n_{\chi_i,eq}n_{\chi_j,eq}\right)\nonumber\\
    && -\sum_{j \neq i}^N\langle \sigma_{Xij} v_{ij} \rangle \left(n_i n_X-n_{i,eq}n_{X, eq}\right)-\langle \sigma_{Xji} v_{ji} \rangle \left(n_j n_X-n_{j,eq}n_{X, eq}\right)\nonumber\\
    && -\sum_{j \neq i}^N\langle\Gamma_{ij}\rangle\left(n_{\chi_i}-\frac{n_{\chi_j}}{n_{\chi_j,eq}}n_{\chi_i,eq}\right)-\langle \Gamma_{ji} \rangle \left(n_{\chi_j}-\frac{n_{\chi_i}}{n_{\chi_i,eq}}n_{j,eq}\right)\nonumber\\
    && - \sum_{j\neq i}^N \langle \sigma_{iijj} v_{iijj} \rangle \left(n_{\chi_i}^2-n_{\chi_j}^2 \frac{n_{\chi_i,eq}}{n_{\chi_j,eq}}\right)\nonumber\\
    && -\frac{1}{2}\langle \sigma v_{iiiX} \rangle \left(n_{\chi_i}^2-n_{\chi_i} n_{\chi_i,eq}\right)-\frac{1}{2}\langle \sigma v_{iiij} \rangle \left(n_{\chi_i}^2-n_{\chi_i} n_{\chi_j} \frac{n_{\chi_i,eq}}{n_{\chi_j,eq}}\right)\nonumber\\
    && -\frac{1}{2}\langle \sigma v_{ijjj} \rangle \left(n_{\chi_i} n_{\chi_j}-n_{\chi_j}^2 \frac{n_{\chi_i,eq}}{n_{\chi_j,eq}}\right) \nonumber \\ 
    &&-\frac{1}{2}\langle \sigma v_{ijjX} \rangle \left(n_{\chi_i}n_{\chi_j}-n_{\chi_i,eq}n_{\chi_j}\right)\nonumber\\
    && +\frac{1}{2}\langle \sigma v_{jjiX} \rangle \left(n_{\chi_j}^2-n_{\chi_i}\frac{n_{\chi_j,eq}^2}{n_{\chi_i,eq}}\right),
\eea

with $i,j$ being labels for beyond the SM (BSM) particles while $X$ generically indicates SM states. The first term on the right-hand side describes the general annihilation processes of the BSM $ij$ states into pairs of SM states. The second line describes $\chi_i X \leftrightarrow \chi_j X, (i \leftrightarrow j)$ processes while the third line accounts for the plausible $\chi_i \leftrightarrow \chi_j+X+(i \leftrightarrow j)$  decays.
The last three lines contain terms which arise when the BSM sector contains more stable particle species.  $\sigma v_{iijj}$ describes $2 \rightarrow 2$ conversion processes between the different stable species. The other processes, which feature an odd number of the BSM particles between the initial and final states are dubbed semi-annihilations \cite{DEramo:2010keq}. These kinds of processes arise in models in which the DM is stabilized by larger complex symmetries, as $Z_N$ with $2 < N \leq 10$ \cite{Belanger:2012vp,Belanger:2014bga, Yaguna:2019cvp}.

 The case when the BSM sector contains only a single stable particle  i.e. a single DM candidate, it is possible to sum the equations of the system. Defining $n_\chi =\sum \limits_i n_i$ we can recover the original Boltzmann's equation(see Eq. (\ref{eq:bolbase})) as:

\begin{equation}
    \frac{dn_\chi}{dt}+3Hn_\chi=-\langle \sigma_{\rm eff} v \rangle \left(n_\chi^2-n_{\chi,eq}^2\right),
\end{equation}
with $\langle \sigma_{\rm eff} v \rangle$ representing an effective cross-section:
\begin{equation}
    \langle \sigma_{\rm eff} v \rangle=\langle \sigma_{ij}v_{ij}\rangle \frac{n_{\chi_i,eq}}{n_{\chi,eq}}\frac{n_{\chi_j,eq}}{n_{\chi,eq}},
\end{equation}
encoding in the DM annihilation processes involving other particles in the initial state is dubbed coannihilations.

In the case of multiple stable particles, i.e.,  multi-components DM, the experimental value of the DM relic density should be reproduced by the sum of the contributions of the single states, $\Omega_\chi h^2=\sum \limits_i \Omega_{\chi_i}h^2$.

\section{Direct Detection}
\label{sec:ddDM}

The DM detection strategy dubbed direct detection (DD) is based on the possibility that DM particles, belonging to the halo surrounding our Galaxy, might interact with the ordinary matter present in the suitable detectors while flowing through the Earth.

In the case of WIMPs, the main process, responsible for a feasible detection, is the elastic scattering between the DM particle and the nuclei or electrons of the chemical elements composing the detector. The scattering process implies a small transfer of kinetic energy between the DM and the ordinary matter which, in turn, releases it as recoil energy. Different DD experiments are distinguished by the kind of detector material used, by the strategy of measuring the recoil energy (e.g., phonons rather than scintillation light), and by different background mitigation techniques. 
For the models of concern in this review, DD relies essentially on DM elastic scattering leading to nuclear recoils. In the following we will then review the basics aspects related to the detection of such process. Elastic scattering on electrons is nevertheless an interesting possibility gathering increasing attention from the experimental community. The interested reader could refer, for example, to \cite{Essig:2011nj,Essig:2012yx,Graham:2012su,Essig:2015cda,Hochberg:2015pha,Hochberg:2015fth,Derenzo:2016fse,Essig:2017kqs} 
The main observable extracted from the experimental outcome is the differential scattering rate:
\begin{equation}
\frac{dR(E,t)}{dE}  = \frac{N_T \rho_{\chi} }{m_{\chi}\, m_T} \int_{v_{\rm min}}^{v_{\rm max}}  
v f_E(\vec{v},t) \frac{d\sigma (v,E)}{dE} d^3\vec{v},
\label{scatteringrate}
\end{equation}
with $E$ being the recoil energy associated with the scattering events and $\frac{d\sigma (v,E)}{dE}\equiv \frac{d\sigma}{dE}$ denoting the differential scattering cross-section. Three kinds of inputs determine the DM scattering rate. First of all, we have the information about the target material contained in the parameters $N_T$ (number of target nuclei per kilogram of the detector) and $m_T$ (mass of the target nucleus). 
Secondly, we have an astrophysical input represented by $\rho_\chi$, i.e., the local DM density  and, thirdly, $f_E(\vec{v},t)$, i.e., the velocity distribution of the DM particles flowing through the Earth. Only the DM particles with velocity in the interval $[v_{\rm min},v_{\rm max}]$ contribute to the scattering rate. $v_{\rm min}$ is the minimal velocity, for which a scattering event with recoil energy $E$ can occur. It is determined by kinematics considerations to be $v_{\rm min}=\sqrt{m_T E/(2 \mu^2_{\chi T})}$ with $\mu_{\chi T}=m_\chi m_T/(m_\chi+m_T)$ being the reduced mass of the concerned system. $m_\chi,\,m_T$ represent the mass of DM  particle $\chi$ and the target nuclei $T$, respectively. $v_{\rm max}$ is instead, the maximal velocity for which a DM particle is still gravitationally bound to our Galaxy.

Experimental results are given considering a fixed choice of the astrophysical parameters. A common choice for the latter is the so-called Standard Halo Model (SHM)~\cite{Drukier:1986tm}. In the SHM, the DM is described by an isotropic velocity distribution in the Galactic frame:
\begin{equation}  
f_{\rm gal}(v)=\left \{ \begin{array}{cc} N \exp\left(-|v|^2/v_c^2 \right)  & |v| \leq v_{\rm max}  \\ 0     &  |v| \geq v_{\rm max} \end{array} \right. \, , 
\end{equation}  
This function describes an isothermal sphere. $N$ is a normalization factor ensuring that $\int
f_{\rm gal}(v)dv=1$  while $v_c$ is a circular velocity.  There is a sharp cut for velocities above $v_{\rm max}$, i.e., the escape velocity of the DM particle from the Galaxy. The velocity distribution $f_E$ in the differential rate is obtained as $f_{E}(v)=f_{\rm
gal}(|\vec{v}+\vec{v}_s+\vec{v}_{e}(t)|)$ with $\vec{v}_s$ and $\vec{v}_e$
being, respectively, the Sun's velocity with respect to the centre of the Galaxy
and the Earth's velocity with respect to the Sun. The local DM density is
determined from astrophysical observations either through local methods, i.e.,
using kinematical data from the nearby population of stars, or through global methods, i.e., modelling the DM and baryon content of the Milky Way and using kinematical data from the whole Galaxy;  see
Refs.~\cite{Read:2014qva,Catena:2009mf,Weber:2009pt,Salucci:2010qr,McMillan:2011wd,Garbari:2011dh,Iocco:2011jz,Bovy:2012tw,Zhang:2012rsb,Bovy:2013raa}
for more details. 

The SHM adopts fiducial values for these three parameters, namely, $\rho_{\chi}=0.3\,\mbox{GeV}/{\mbox{cm}}^3$, $v_c=220$ (or 230) km/s and $v_{\rm esc}=544$ km/s. These parameters are, nevertheless, subject to uncertainties which in turn affect the limits on the DM scattering, see e.g., ~\cite{Calore:2015oya,Bernal:2016guq,Benito:2016kyp} for some examples. Note that proposals to supersede the SHM, which is anyway an approximate model  of the DM distribution, are present in Ref. \cite{Bozorgnia:2016ogo,Bozorgnia:2017brl,Necib:2018iwb,Necib:2018igl,Evans:2018bqy}. Alternatively, one might encompass astrophysical uncertainties via the so-called halo independent methods~\cite{Fox:2010bz,McCabe:2011sr,DelNobile:2013cva,Ibarra:2017mzt,Gondolo:2017jro,Catena:2018ywo,Kahlhoefer:2018knc}. 

The particle physics inputs of the DM scattering rate are contained in the differential cross-section, as depicted in Eq. (\ref{scatteringrate}). Fixing the astrophysical input and the detector properties, the experimental limits (or hypothetical experimental signals) are directly translated to $\frac{d\sigma}{dE}$ and then to the particle content of the specific particle physics model under scrutiny. In WIMP models it is possible to relate the differential cross-section to the scattering cross-section of the DM $\chi$ over the target $(T)$ nuclei. This is typically done via the following decomposition:   
\begin{align}
\label{eq:sigma_diff}
 \frac{d\sigma}{dE} &= {\left(\frac{d\sigma}{dE}\right)}_{\rm SI}+  {\left(\frac{d\sigma}{dE}\right)}_{\rm SD}\nonumber\\ 
&= \frac{m_T}{2\mu^2_{\chi T} v^2} \left(\sigma^{SI}_{\chi T,0} \left \vert F_{\rm SI}(q) \right \vert^2 +  \sigma^{SD}_{\chi T,0}\left \vert F_{\rm SD}(q) \right \vert^2\right),
\end{align}

where we identify two plausible classes of interactions between the DM and nuclei dubbed Spin Independent (SI) and Spin Dependent (SD). From the second line, we see that the differential cross-section can be expressed as the product of a cross-section $\sigma_{\chi T,0}^{SI,SD}$, with the  subscript $``_0"$ stemming from the fact that it is computed in the limit of zero momentum transfer, and a concerned form factor $(F_{\rm SI}$ or $F_{\rm SD})$.  The latter accounts for the extended structure of the nuclei which depends on the momentum transfer $q$, related in turn to the recoil energy $E$ as $q=\sqrt{2 \mu_T E}$. For details on the determination of the SI/SD form factors we refer to Refs. \cite{Jungman:1995df,Duda:2006uk,Bednyakov:2006ux,Schnee:2011ooa}. The form factors are input parameters for the analysis of the experimental signal and, are not related to specific particle physics models. Experimental bounds can subsequently be translated into bounds on the microscopic cross-section. It is now possible to perform a further step and convert the scattering cross-sections over nuclei into scattering cross-section over nucleons. The detailed relation actually requires information on the particle physics input (see Eq. ({\ref{nucleitonucleon}}) ). A good insight is already provided by the following relations:
\begin{align}\label{nucleitonucleon}
    & \sigma_{\chi T,0}^{\rm SI}\approx \left[\frac{\mu_{\chi T}}{\mu_{\chi p}}Z \sqrt{\sigma_{\chi p}^{\rm SI}}+\frac{\mu_{\chi T}}{\mu_{\chi n}}(A-Z) \sqrt{\sigma_{\chi n}^{\rm SI}}\right]^2,\nonumber\\
    & \sigma_{\chi T,0}^{\rm SD}\approx \left[\frac{\mu_{\chi T}}{\mu_{\chi p}}S_p \sqrt{\sigma_{\chi p}^{\rm SD}}+\frac{\mu_{\chi T}}{\mu_{\chi n}}S_n \sqrt{\sigma_{\chi n}^{\rm SD}}\right]^2,
 \end{align}
 here $\sigma_{\chi N=p,n}^{\rm SI, SD}$ represent the scattering cross-sections of the DM over protons and neutrons, respectively. $Z, A$ are the atomic and mass number of the target nuclei while $S_{p,n}$ are parameters associated with the contributions of protons and neutrons to the nuclear spin. $\mu_{\chi N=p,\,n}$ denotes the reduced mass of the DM and nucleon (proton, neutron) system. The relation above highlights an important property of SI interactions, i.e., they are coherent; in simpler words, the interaction rate of the DM with a nucleus is obtained by summing the contributions of the individual nucleons. Assuming that the DM interacts in the same way with protons and neutrons, the scattering cross-section over a nucleus with mass number $A$ is enhanced by a factor $A^2$ with respect to the scattering cross-section over protons. This motivates the use of heavy elements, i.e., with high mass number, like the Xenon, as target material in detectors. As will be seen, the great sensitivity ensured by the coherent nature of SI interactions, combined with the great volumes achieved by the current and near future generation of detectors, will allow also to probe DM interactions originating at the loop level. The SD interactions have, instead, no coherent character as the contributions of the different nucleon spin tend to average out so that the scattering cross-section over nuclei is essentially accounted for an eventual unpaired nucleon. The cross-section over nucleon and nucleus differ by a $O(1)$ factor. In light of this, not all the detectors are suitable to probe SD interactions. Xenon, having two isotopes with odd $A (129, 131)$, can be used to probe SD interactions over neutrons. 
Using the relations illustrated above, the experimental outcome (exclusion limit or signal) can be interpreted directly in terms of the DM scattering cross-section over nucleons. For this reason, experimental papers show limits of the latter quantity as a function of the DM mass.

To test a particle physics model of DM against the DD one essentially has to determine the scattering cross-section of the DM over nucleons. For this, we must have in mind that the DM particles flowing through the Earth are fairly non-relativistic, and hence, the momentum transfer in the scattering processes is small. Besides, the typical energy scale for the DM DD is $O(\mbox{GeV})$. Starting from the full Lagrangian of the model under scrutiny, defined at some high-energy (NP) scale $\Lambda_{\rm NP}$, one constructs an effective Lagrangian containing interaction among the DM particle (more precisely of the slowly varying components of the DM field \cite{Cirelli:2013ufw,Bishara:2016hek} taken as static source) and the residual dynamical $dof$ available at an energy scale $<< \Lambda_{\rm NP}$, i.e., the light quark flavours, $u,\,d,\,s$, and the gluon. The heavy degrees of freedom of the theory, i.e., heavy mediators, heavy quark flavours and high-frequency modes of the DM field are integrated out and their collective effects are encoded in the Wilson coefficients of the effective Lagrangian. The effective interactions between the DM and quarks/gluons should then be translated into effective interactions between the DM and nucleons. These kinds of relations will be illustrated afterwards once specific models are introduced. 

Some relevant remarks are in order. An important feature of the analysis discussed so far is the factorization of the DM scattering rate into a term encoding the microscopic interactions of the DM, independent on the momentum transfer, and a form factor determined by the nuclear physics. As already pointed, this relies on the assumption that the interactions between the DM and nucleons can be written in terms of momentum-independent contact operators. This assumption is valid, for example, if the mediator of the interactions between the DM and the SM states is always heavy compared to the scale of typical momentum transfer in the elastic scattering processes. Another relevant assumption relies on the fact that the effective interactions between the DM and nucleon/nuclei are only either SI or SD. While the vast majority of WIMP models indeed fall within these two categories, interesting models exist with momentum-dependent or "long-range" DM nucleon interactions possibly corresponding to different form factors with respect to the conventional SI and SD ones.

A more general description of the DM DD can be achieved, for example, as proposed in Refs.~\cite{Fitzpatrick:2012ix,Fitzpatrick:2012ib,Anand:2013yka}. A generic BSM Lagrangian can be mapped into a basis of non-relativistic operators as,
\begin{equation}
    \mathcal{L}_{\rm BSM}\rightarrow \mathcal{L}_{\rm NR}=\sum_i c^N_i O_i^{\rm NR},
\end{equation}
where $c^N_i, \,N=p,\,n$ are coefficients depending on the specific particle physics model under scrutiny while $\{O_i^{\rm NR}\}$ is a set of linearly independent operators dependant on the momentum transfer $(q)$, the DM spin ($\vec{\sigma}_\chi$) and the nucleon spin $(\vec{\sigma}_N)$:
\begin{align}\label{eq:NRoperators}
    & O_1^{\rm NR}=I, \,\,\,\,O_3^{\rm NR}=i \vec{\sigma}_N \cdot \left(\vec{q} \times \vec{v^\bot}\right),\,\,\,\,O_4^{\rm NR}=\vec{\sigma}_N \cdot \vec{\sigma}_\chi,\nonumber\\
    & O_5^{\rm NR}=i \vec{\sigma}_N \cdot \left(\vec{q} \times \vec{v^\bot}\right),\,\,\,\,O_6^{\rm NR}=\left(\vec{\sigma}_N \cdot \vec{q}\right)\left(\vec{\sigma}_\chi \cdot \vec{q}\right),\nonumber\\
    & O_7^{\rm NR}=\vec{\sigma}_N \cdot \vec{v^\bot},\,\,\,\,O_8^{\rm NR}=\vec{\sigma}_\chi \cdot \vec{v^\bot},\,\,\,\,O_9^{\rm NR}=i \vec{\sigma}_\chi \cdot \left(\vec{\sigma}_N \times \vec{q}\right), \nonumber\\
    & O_{11}^{\rm NR}=i \vec{\sigma}_\chi \cdot \vec{q},\,\,\,\,\,O_{12}^{\rm NR}=\vec{v^\bot}\cdot \left(\vec{\sigma}_N \times \vec{\sigma}_\chi\right).   
\end{align}
where $\vec{v^{\bot}}=\vec{v}+\frac{\vec{q}}{2\mu_{\chi }}$
Note that the operator $O_{2}^{\rm NR}$ has been omitted on purpose as no relativistic invariant operator can be mapped into it \cite{DelNobile:2013cva}. The conventional SI and SD interactions are automatically incorporated in this formalism as they correspond, respectively, to the $O_1^{\rm NR}$ and $O_4^{\rm NR}$ operators. 
With the decomposition in the non-relativistic basis at hand, one can write a general differential rate as:
\begin{align}\label{eq:dRdE}
    & \frac{d R}{dE}\propto \sum_{i,j=1}^{12} \sum_{N,N^{'}=p,n}c_i^N c_j^{N^{'}} \mathcal{F}_{i,j}^{N,N^{'}},\nonumber\\
    & \mathcal{F}_{i,j}^{N,N^{'}}=\int d^3 v \frac{1}{v}f_E(v)F_{i,j}(v,E),
\end{align}
with $F_{i,j}$ being combinations of a set of five fundamental form factors, including the conventional SI and SD ones. In this formalism, experimental limits are expressed in terms of parameters contained in the $c_i^N$ coefficients, considering the NR operators individually. For example, through $\Lambda_{\rm NP}$  which encodes the mass of the mediator of DM interactions and the corresponding couplings. See Refs.~\cite{XENON:2017fdd,LUX:2020oan,LZ:2023lvz} for some examples. While it will not be discussed explicitly here, we mention, for completeness, that an alternative formulation of a general Effective Field Theory (EFT) for the DM DD is presented in Refs.~ \cite{Bishara:2016hek,Brod:2017bsw}. Interpretations of DD limits in this framework have also been addressed in literature, e.g., in Ref. \cite{XENON:2022avm}.

\section{Indirect Detection}
\label{sec:idDM}

The ID of DM particles is based on the detection of gamma rays, cosmic rays and neutrinos stemming from either DM annihilation or decay that appear as excess over the expected 
background. The detection of DM signals occurs from Earth-based telescopes such as H.E.S.S. (The High Energy Stereoscopic System) and CTA (Cherenkov Telescope Array), or satellites 
like the AMS (Alpha Magnetic Spectrometer) and Fermi-LAT (Fermi Gamma Ray Space Telescope) \cite{Abdo:2010ex,Abramowski:2011hc,Ackermann:2011wa,Ackermann:2012nb,Abramowski:2012au,
Ackermann:2013uma,Ackermann:2013yva,Abramowski:2014tra,Ackermann:2015tah,HESS:2015cda,
Ackermann:2015zua,Ackermann:2015lka,Abdallah:2016ygi,Abdalla:2016olq,Fermi-LAT:2016uux}.

The ID search strategy offers an exciting possibility of the DM 
detection as it allows us to search for heavier DM candidates, compared to the DD and collider searches and provides an orthogonal insight into the DM properties. In this work, we will focus on gamma-rays. The gamma ray flux arising from the DM annihilation depends on:
\begin{itemize}
\item The squared number density of particles, i.e., $n_{\chi}^2=\rho_
{\chi}^2/m_{\chi}^2$; 
\item The WIMP annihilation cross-section today, $\sigma$;
\item The mean WIMP velocity $v$ within the target region;
\item The volume of the sky observed within a solid angle $\Omega$;
\item The number of gamma rays produced per annihilation at a given energy, known as the energy spectrum ($dN/dE$).
\end{itemize}

In summary, it is found to be:
%
\bea
\overbrace{\frac{d\Phi}{d\Omega dE}}^{\rm Diff.~ Flux} &=& {\color{blue} \frac{ \overbrace{ \sigma v }^{\rm Annihilation\, 
Cross-Section}}{8\pi m_{\chi}^2}} \times {\color{OliveGreen} \underbrace{\frac{dN}{dE}}_{\rm Energy\, Spectrum}} \nonumber\\ 
&&\times {\color{red} \overbrace{\int_{\rm l.o.s} ds}^{\rm Line\, of\, Sight\, Integral}} \times {\color{magenta}  
\underbrace{\rho_
\chi^2 (\overrightarrow{r}(s,\Omega))}_{\rm DM\, Distribution}}.
\label{eq:flux}
\eea

Hence, the differential gamma-ray flux in Eq.~(\ref{eq:flux}) is sourced by three different inputs namely, the
DM annihilation cross-section, the energy spectrum that is computed once a specific annihilation final state is given, and the integral over the line of sight (l.o.s) of the DM 
density distribution which is subject to large uncertainties, especially in high-density regions such as the Galactic Center.  As far as particle physics is concerned, the key quantity is the WIMP annihilation cross-section. The DM is known to be non-relativistic today, thus if the DM annihilation cross-section today depends on the relative velocity, the corresponding DM signal will be highly suppressed.

Regarding the integral over the line of sight, we emphasize that it is carried out from the observer to the source, and it does depend on the DM density profile. We point out that the DM density is not tightly constrained, and several DM density profiles have been considered in the 
literature leading to either spike or core DM densities toward the center of galaxies \cite{Burkert:1995yz,
Navarro:1995iw,Salucci:2000ps,Graham:2005xx,Salucci:2007tm,Navarro:2008kc}. A commonly adopted profile is the Navarro-Frenk-White (NFW) \cite{Navarro:1995iw} which reads,
%
\begin{equation}
\rho(r)  = \frac{r_s}{r} \frac{\rho_s}{[1+ r/r_s]^2},
\end{equation}
where $r_s=24.42$~kpc is the scale radius of the halo, as used by Fermi-LAT collaboration in Ref.~\cite{Ackermann:2015zua}, and $\rho_s=0.184$ is a normalization constant to guarantee that the 
DM density at the location of the Sun is $0.3{\rm ~GeV/cm^3}$. It is important to emphasize that the NFW profile is known to be as steep profile, as it leads to a large DM density toward the inner regions of the galactic center. Alternative density profiles with a core-like behavior \cite{Benito:2016kyp} yield much weaker limits \cite{CTA:2020qlo,NFortes:2022dkj,CTAConsortium:2023yak,Abe:2024cfj}.

From Eq.~(\ref{eq:flux}) it is clear that the ID probes complementary properties of the DM particles. It is sensitive to how the DM is distributed, to the annihilation cross-section today, which 
might be different than the annihilation cross-section relevant for the relic density, and to the WIMP mass.
Therefore, after measuring the flux of gamma-rays from a given source, we compare it with the background 
expectations. If no excess is observed, we can choose a DM density profile and select an annihilation 
final state  needed for $dN/dE$, and then derive a limit on the ratio $\sigma v/m_\chi^2$ according to 
Eq.~(\ref{eq:flux}). This is the basic idea behind experimental limits. Although, more sophisticated 
statistical methods have been conducted such as likelihood analysis \cite{Chiappo:2016xfs,CTA:2020qlo}. 

An interesting aspect of the indirect DM detection, when it comes to probing WIMP models, is the fact that if the annihilation cross-section, $\sigma v$, is not velocity dependent, bounds on $\sigma v$ today are 
directly connected to the DM relic density. In particular, the observation of gamma-rays in the direction of dwarf spheroidal galaxies (dSphs) results in stringent limits on the plane of annihilation cross-section vs WIMP mass \cite{Fermi-LAT:2011vow,Bonnivard:2014kza,Geringer-Sameth:2014qqa, Fermi-LAT:2015att,MAGIC:2016xys,Hayashi:2016kcy}. If 
for a given channel the annihilation cross-section of $10^{-26}~{\rm cm}^3~{\rm s}^{-1}$ is excluded for DM masses 
below $100$~GeV, it also means that one cannot reproduce the right relic density for WIMP masses below 
$100$~GeV \footnote{There are still some exceptions to this direct relation between non-velocity dependent 
annihilation cross-section and relic density as discussed in detail in Ref.~\cite{Griest:1990kh}.}. In other 
words, in this particular case, ID limits should trace the relic density curve.

%

\section{General remarks}
\label{sec:GeneralRemarks}

In this section, we aim to convey the content of this work to a common reader. Thus, we  summarise here a few general assumptions, that are common to all the models discussed in this work, and the general conventions adopted to present our results.

First of all,  except one case, we have considered only charge-parity (CP)-preserving extensions of the SM. This assumption is mostly dictated by simplicity. 

Concerning the presentation and discussion of the results, this review is mostly focused on the DM phenomenology. For each of the chosen model, we will primarily focus on a comparative study of the parameter space compatible with the correct DM relic density with the exclusion regions coming from the dedicated direct and indirect searches. Complementary constraints from more general NP searches, like collider ones or of theoretical origin, whenever relevant, might be considered case-by-case.

In this work, models with different grades of refinement and complexity will be presented, from simplified models with 2-3 free parameters to more realistic scenarios very close to Ultra-Violet (UV) complete models. In most cases, it will be possible to identify pairs of parameters playing an important role in characterising the DM phenomenology. We will then illustrate collective effects originating from the relevant constraints, e.g., relic density, DD, ID, etc., for a given model in a bidimensional plane. In this setup, the correct relic density will be represented by a (narrow) isocontour; the points of the line corresponds to the assignations giving the value of the DM relic density determined by the Planck satellite, $\Omega_{DM} h^2 = 0.120 \pm 0.001$ \cite{Planck:2018vyg}. 
Subsequently, in each plot, we will show the regions excluded by dedicated DM searches with highlighted coloured areas in the aforesaid bidimensional plane.
In the context of DM searches, firstly, we will show the current and projected limits for the SI interactions. For the former, we will combine the exclusion limit given by LZ \cite{LZ:2022lsv}  which is relevant for DM masses above $10$ GeV and the one from the search of XENON1T \cite{XENON:2019gfn} of ionization signals, which might be used to constrain DM candidates with masses between $1-10$ GeV. Notice that XENONnT has determined limits \cite{XENON:2023cxc}, very close to the ones from LZ. We will however, often skip  XENONnT results simply to avoid over-filled plots that are in general hard to comprehend. We will also consider the projected sensitivities of the next generation 
DD experiments, using the DARWIN experiment as a reference \cite{Aalbers:2016jon}. Further, constraints arising from the SD interactions will also be considered for a given model. For the latter we will adopt the update limits coming from the ID searches, whenever relevant, will also be considered. For ID probes, the models considered in this work can be tested mostly via $\gamma$-ray signals.  We have considered the most recent limits due to the non-observation of a DM signal from a set of 30 dSphs with 14.3 years of Fermi-LAT data \cite{McDaniel:2023bju}. We will also show how this limit would improve with 15 years of Fermi-LAT data observing 60 dSphs \cite{Fermi-LAT:2016afa}. Both limits are for a $\gamma$-ray signal due to a DM annihilation into $b\bar b$. Regarding the future prospects, we consider the expected limits on the DM annihilation into $b\bar b$ and $\tau^+ \tau^-$ states by future CTA observation of the Galactic center \cite{CTA:2020qlo}, assuming the Einasto DM profile. Apart from the DD and ID probes, as pointed out before, when ever appropriate, further exclusion bounds from theoretical arguments or complementary searches for the NP have also been considered.

\begin{figure*}
    \centering
    \subfloat{\includegraphics[width=0.4\linewidth]{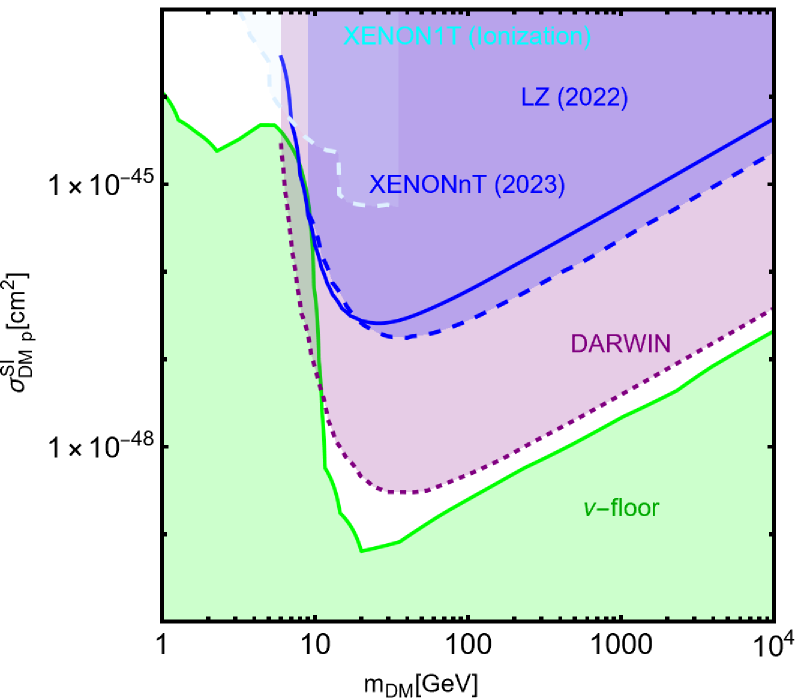}}
    \subfloat{\includegraphics[width=0.37\linewidth]{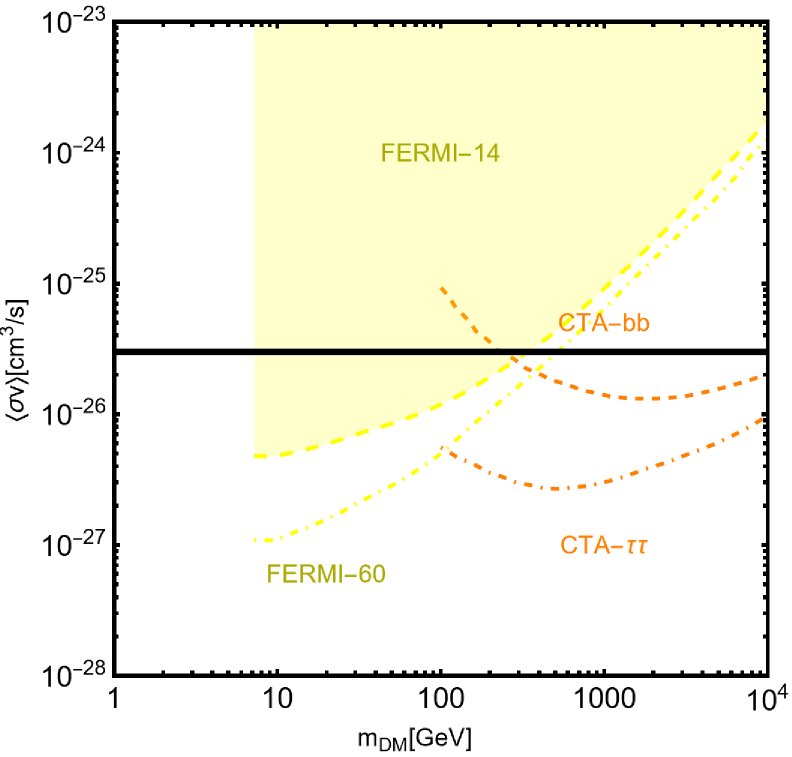}}
    \caption{Most relevant constraints (both current and projected) applied in this work. The left panel shows the DD limits in the $(m_{\rm DM}.\sigma_{DM p}^{\rm SI})$ plane while the right panel refers to the ID constraints in the $(m_{\rm DM},\langle \sigma v \rangle)$ plane. See text body for the detailed description.}
    \label{fig:limits}
\end{figure*}

A summary of the most important current and near future bounds, from the dedicated DM searches, is shown by Fig. \ref{fig:limits}. The left panel shows the exclusion bounds, in a generic $(m_{\rm DM},\sigma_{DM p}^{\rm SI})$ bidimensional plane. The cyan-colored and blue-coloured regions, above the dashed lines of the same color, represent the exclusions by XENON1T and LZ respectively. For reference, we also show the exclusion line (solid blue) by the XENONnT experiment.  The purple coloured region represents the expected sensitivity reach from the DARWIN experiment. As can be seen that it is very close and overlaps, at small DM masses ($m_{\rm DM} \lesssim 20$ GeV), with the so-called $\nu$-floor \cite{Billard:2013qya}. This is the sensitivity region to the coherent scattering of the solar and the atmospheric neutrinos, mediated by the SM Z-boson. Coherent neutrino scattering can mimic a WIMP signal, hence representing an irreducible background, at least for the current design of the DD experiments. The right panel of Fig. \ref{fig:limits} relies on the DM ID. The yellow coloured region, dubbed FERMI-14, corresponds to the portion of the $(m_{\rm DM},\langle \sigma v \rangle)$ plane currently excluded by searches of $\gamma$-ray signal by FERMI. The dot-dashed yellow coloured line corresponds to the near future sensitivity reach of the same experiment, dubbed FERMI-60. The two orange coloured contours represent the projected sensitivities of the CTA experiments to the DM annihilation processes into $\tau^+ \tau^-$ and $b \bar b$ final states. The right panel of Fig. \ref{fig:limits} also shows a horizontal black coloured line corresponding to the thermally favoured value of the DM annihilation cross-section. In case the DM features a s-wave dominated cross-section, ID can probe values of the DM mass up to around $200$ GeV. Negative signals by CTA might exclude DM masses in the multi-TeV range.

In models with a higher number of free parameters, the picture depicted above will be complemented by a scan over all the relevant parameters. These scans will identify the sets of model assignations complying with all the constraints applied to a given model. The model assignations are still viable after an eventual negative signal by the DARWIN will be highlighted as well.

As final remark, we point out that the strong correlation among DM relic density and experimental outcome, shown by the models illustrated below, is due to the fact the DM relic density is mostly accounted for $2 \rightarrow 2$ annihilation processes into SM final states. We will evidence via some relevant example that the picture changes substantially when this assumption does not hold.



%
\section{Simplified models}
Simplified models are minimal extensions of the SM including just the minimal content, in terms of particles and couplings, to accommodate DM phenomenology. The study of these models played a relevant role in our previous review \cite{Arcadi:2017kky}. As already pointed out, these models have been mostly superseded by more refined benchmarks. Nevertheless, it is worth examining the updated constraints on these models. Indeed their simplicity, allows us to interpret the results via analytical expressions which will prove useful for the more complicated models discussed in the second part of this work.

\subsection{s-channel portals}
One of the simplest realizations of the WIMP models is represented by the so-called s-channel portals~\cite{DiFranzo:2013vra,
Berlin:2014tja,Abdallah:2014hon,Buckley:2014fba,Godbole:2015gma,Abdallah:2015ter,Duerr:2015wfa,Baek:2015lna,
Carpenter:2016thc,Bauer:2016gys,Sandick:2016zut,Bell:2016uhg,Bell:2016ekl,
Khoze:2017ixx,ElHedri:2017nny}. In these models the SM is extended by two extra particle states: a cosmologically stable DM candidate and a "mediator" state coupled to the DM pairs as well as the SM fermions $(f)$. The absence of interactions involving an odd-number of DM particles is ensured by an ad-hoc discrete $Z_2$ or global $U(1)$ symmetry. This is connected to the fact whether the DM belongs to a real or a complex representation of the Lorentz group, respectively. Both the DM and the mediator field are typically assumed to be singlets under the SM gauge groups. These kinds of models show a very strong complementarity between the relic density and the dedicated DM search strategies. Furthermore, they represented a first generation of benchmarks for collider studies, see e.g., Refs.~\cite{Jacques:2015zha,Xiang:2015lfa,Backovic:2015soa,Bell:2015rdw,Brennan:2016xjh,
Boveia:2016mrp,Englert:2016joy,Goncalves:2016iyg,DeSimone:2016fbz,Liew:2016oon,Kraml:2017atm,Bauer:2017ota,Albert:2017onk}. 
One can conceive several variants of s-channel portals, according to the possible spin assignations for the DM and the s-channel mediator.


\subsubsection{Spin-0 mediator -- CP-even}

To start with we consider cases when spin-$0$ ($\chi$), spin-$1/2$ ($\psi$), and spin-$1$ ($V^\mu$) DM coupled with a spin-$0$, CP-even state ($S$) as:
\bea
\label{eq:scalarMed}
\mathcal{L}&&=\xi \mu_\chi^S \lambda_\chi^S \chi \chi S 
+\xi (\lambda_\chi^S)^2 \chi^2 S^2+\frac{c_S}{\sqrt{2}}\frac{m_f}{v_h}\ovl f f S\nonumber\\
\mathcal{L}&&=\xi g_\psi \ovl \psi \psi S+\frac{c_S}{\sqrt{2}}\frac{m_f}{v_h}\ovl f f S,\nonumber\\
\mathcal{L}&&=\mu_V^S \eta_V^S V^\mu V_\mu S+\frac{1}{2}(\eta_V^S)^2 V^\mu V_\mu SS+\frac{c_S}{\sqrt{2}}\frac{m_f}{v_h}\ovl f f S,
\eea
$\xi=1$ for a real scalar or a Dirac fermion and $\xi=1/2$ for a complex scalar or Majorana fermion. Terms proportional to $S^3$ would be in general present in the lagrangians above. We neglect them for simplicity.
Given the simplicity of the models, we can provide analytical expressions for an elucidated illustration of our analysis. Starting from the relic density, it is accounted for the DM annihilation proccesses into the SM fermion pairs final states and, if kinematically accessible, $SS$ final states.  

The corresponding cross-sections for the three spin assignations of the DM can be approximated by:

\begin{align}
    &  \langle \sigma v \rangle (\chi \chi \rightarrow \ovl f f)
\approx \sum _f \frac{3 n_c^f}{16 \pi} {(\lambda_\chi^S)}^2 c_S^2 
\frac{m_f^2}{v_h^2}\frac{m_\chi^2}{(4 m_\chi^2-m_S^2)^2},\nonumber\\
& \langle \sigma v \rangle(\ovl \psi \psi \rightarrow \ovl f f)~
\approx \sum_f \frac{3 n_c^f v^2}{4\pi}g_\psi^2 c_S^2 \frac{m_f^2}{v_h^2}\frac{m_\psi^2}{(4 m_\psi^2-m_S^2)^2},\nonumber\\
& \langle \sigma v \rangle (V V \rightarrow \ovl f f)\approx
\sum \frac{n_c^f}{4\pi} {(\eta_{V}^S)}^2 c_S^2 \frac{m_f^2}{v_h^2}\frac{m_V^2}{(4 m_V^2-m_S^2)^2},
\end{align}
where $n_c^f$ is a colour factor. $n_c^f=3$ for quark final states while $n_c^f=1$ in the other cases.
\begin{align}\label{eq:dmpairtoss}
    & \langle \sigma v \rangle (\chi \chi \rightarrow S S) \approx \frac{{(\lambda_\chi^S)}^4}{64 \pi m_\chi^2},\nonumber\\
   & \langle \sigma v \rangle(\ovl \psi \psi \rightarrow S S) \approx 
 \frac{3}{64\pi}g_\psi^4 \frac{1}{m_\psi^2}v^2,\nonumber\\
 & \langle \sigma v \rangle(V V \rightarrow S S) \approx \frac{11}{2304\pi}{(\eta_V^S)}^4 \frac{1}{m_V^2},
\end{align}
for the $SS$ final state (the cross-section are evaluated in the limit $m_\chi,m_\psi,m_V \gg m_S$). To obtain the previous expressions we have assumed $\mu_\chi^S=m_S$ and $\mu_V^S=m_V$ for, respectively, scalar and vector DM (see Ref. \cite{Arcadi:2017kky} for more details).
While the different cross-sections have analogous parametric dependence, we notice that the ones of scalar and vector DM are s-wave dominated (velocity independent). In contrast, the fermionic DM has instead a p-wave (velocity dependent) cross-section. This implies that, for the same values of the DM and mediator masses, fermionic DM requires stronger couplings to get the thermally favoured value for its annihilation cross-section. Furthermore, the velocity dependence implies that the parameter region corresponding to the correct relic density cannot be probed via ID, contrary to the cases of scalar and vector DM.
Moving to DD, we have:

\begin{align}
\label{eq:DDsimplifiedS0}
& \sigma_{\chi p}^{\rm SI}=\frac{\mu_{\chi p}^2}{4 \pi}\frac{{(\lambda_\chi^S)}^2 c_S^2}{m_\chi^2 m_S^2} 
\frac{m_{p}^2}{v_h^2}{\left[f_p \frac{Z}{A}+f_n \left(1-\frac{Z}{A}\right)\right]}^2,\nonumber\\
& \sigma^{\rm SI}_{\psi p}=\frac{\mu_{\psi p}^2}{\pi}g_\psi^2 c_S^2 \frac{m_p^2}{v_h^2}{\left[f_p \frac{Z}{A}+f_n \left(1-\frac{Z}{A}\right)\right]}^2 \frac{1}{m_S^4},\nonumber\\
& \sigma_{Vp}^{\rm SI}=\frac{\mu_{Vp}^2}{4\pi} {(\eta_V^S)}^2 c_S^2 \frac{m_p^2}{v_h^2}{\left[f_p \frac{Z}{A}+f_n \left(1-\frac{Z}{A}\right)\right]}^2 \frac{1}{m_S^4},
\end{align}
where $A$ and $Z$ represent the atomic and proton number of the chemical element constituting the detector, respectively. $\mu_{\rm DM p}=m_{\rm DM} m_p/(m_{\rm DM}+m_p)$, with ${\rm DM}=\chi,\,\Psi,\,V$, denotes the reduced mass of the WIMP-proton system with $m_p$ representing 
the mass of the latter. $f_p$ and $f_n$ represent the effective couplings of the DM with protons and neutrons.
The simultaneous presence of the effective couplings of the DM with protons and neutrons, $f_{p,\,n}$, and the explicit dependence on $Z$ and $A$, as 
depicted Eq.~\eqref{eq:DDsimplifiedS0}, is not arising due to computation from the first principles but coming from an ad-hoc rescaling accounting for the fact 
that the conventional experimental analysis assumes equal interactions of the DM with protons and neutrons. 
In the case of a scalar mediator, we have:
\begin{eqnarray}
    f_N=\sum_{q=u,d,s} f_q^N+\frac{6}{27}f^N_{\rm TG}
    \label{eq:formfac1}
\end{eqnarray}
with
\begin{equation}
     f^N_{\rm TG}=1-\sum_{q=u,d,s} f_q^N,\,\,\,\,N=p,n.
\end{equation}
$f_{q=u,d,s}^N,f^N_{\rm TG}$ are form factors defined from the expectation value of the $\bar q q $ bilinear between initial and final nucleon states. More precisely we have
\begin{equation}
    \langle N| m_q \bar q q | \rangle N= m_N f_q^N 
\end{equation}
for $q=u,d,s$ and with $m_N$ being the nucleon mass. In the case of the heavy quarks $Q=c,b,t$ we have instead used \cite{Shifman:1978zn}:
\begin{align}
    & m_Q \bar Q Q=-\frac{\alpha_s}{12 \pi}G_{\mu \nu}G^{\mu \nu}\\
    & \langle N | G_{\mu \nu}G^{\mu \nu}| N \rangle=-m_N\frac{8\pi}{9 \alpha_s}f_{TG}
\end{align}
This is due to the fact that, at the energy scale relevant for DD, the heavy quark flavor, namely $c,b,t$, are integrated out leading to EFT operators between DM and gluon pairs, i.e. $|\chi|^2 G_{\mu \nu}G^{\mu \nu},\bar \psi \psi G_{\mu \nu} G^{\mu \nu}$, $V^\rho V_\rho G_{\mu \nu}G^{\mu \nu}$.
The form factors $f^N_{u,d,s}$ can be determined from pion--nucleon scattering~\cite{Alarcon:2011zs,Crivellin:2013ipa,Hoferichter:2015dsa,Hoferichter:2017olk}. For our DD computation we have used the central values of the following determinations:
\begin{align}
    & f_u^p=(20.8 \pm 1.5) \times 10^{-3},\,\,\,\,\,\,f_u^n=(18.9 \pm 1.4) \times 10^{-3}, \nonumber\\
    & f_d^p=(41.1 \pm 2.8) \times 10^{-3},\,\,\,\,\,\,f_u^n=(45.1 \pm 2.7) \times 10^{-3}, \nonumber\\
    & f_s^p=f_s^n=0.043 \pm 0.011 \,,
\end{align}
which lead to $f_{TG}\approx 0.894$. From these values it is evident that in the case of a CP-even spin-0 mediator, one has $f_p \simeq f_n$. For such a reason, we will neglect, in analogous models presented in the next sections, the scaling factors used in eq. \ref{eq:DDsimplifiedS0}.
Given the small number of free parameters, the main features of the DM phenomenology of simplified DM models with a spin-$0$ CP-even mediator can be visualized via simple bidimensional plots, in the concerned $(m_S,m_{\rm DM})$-planes  for some fixed assignations of the concerned couplings, as mentioned in Eq.~(\ref{eq:scalarMed}).
\begin{figure*}
\begin{center}
\subfloat{\includegraphics[width=0.33\textwidth]{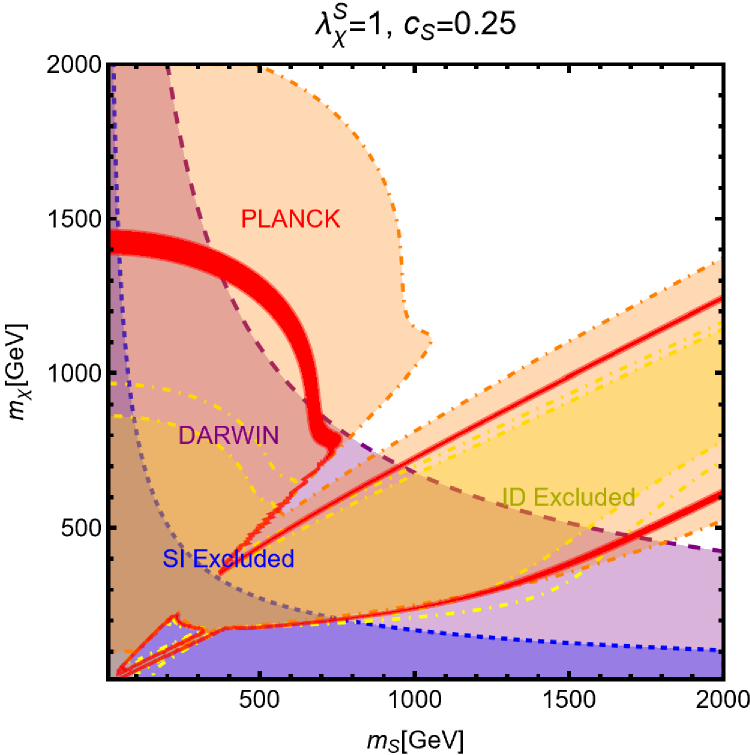}}
\subfloat{\includegraphics[width=0.33\textwidth]{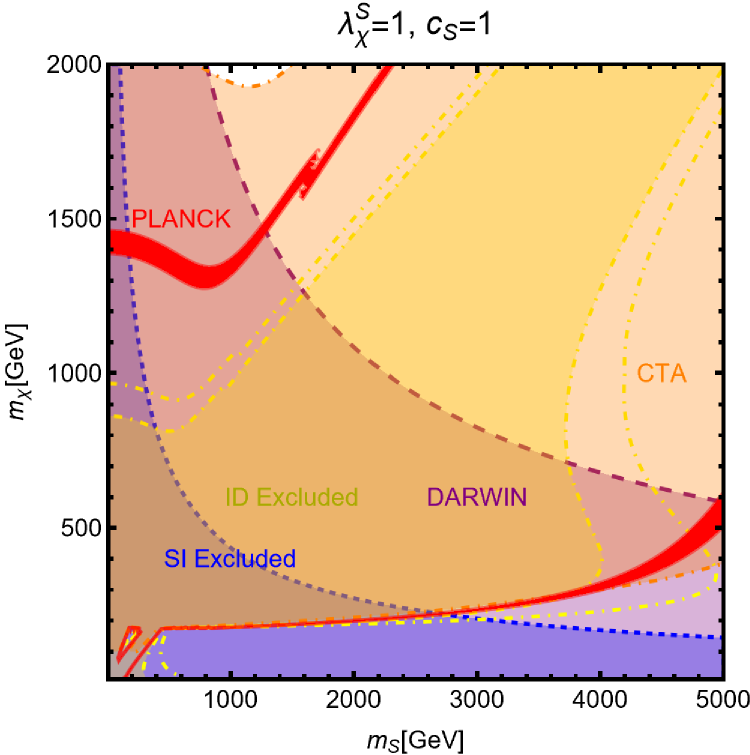}}
\subfloat{\includegraphics[width=0.33\textwidth]{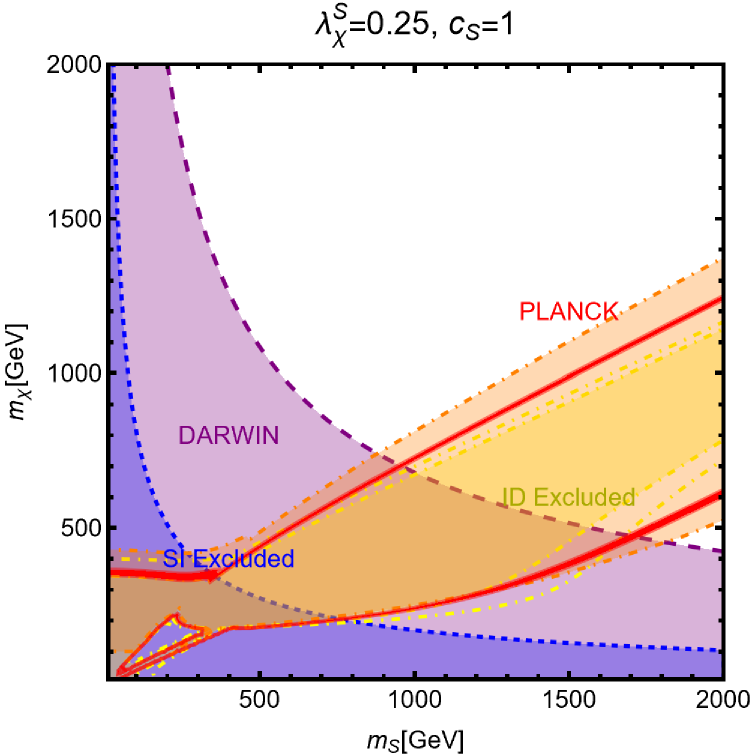}}\\
\subfloat{\includegraphics[width=0.33\textwidth]{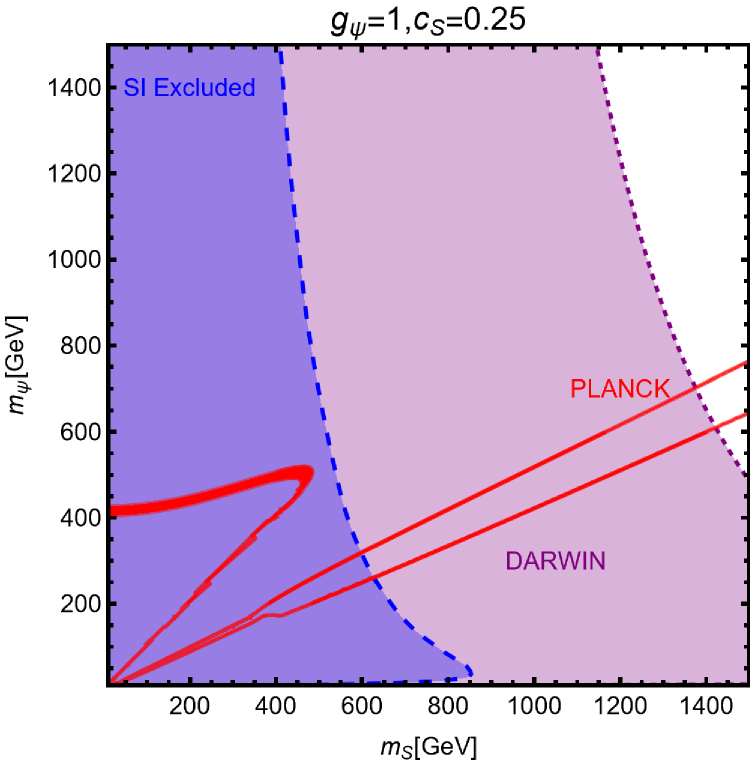}}
\subfloat{\includegraphics[width=0.33\textwidth]{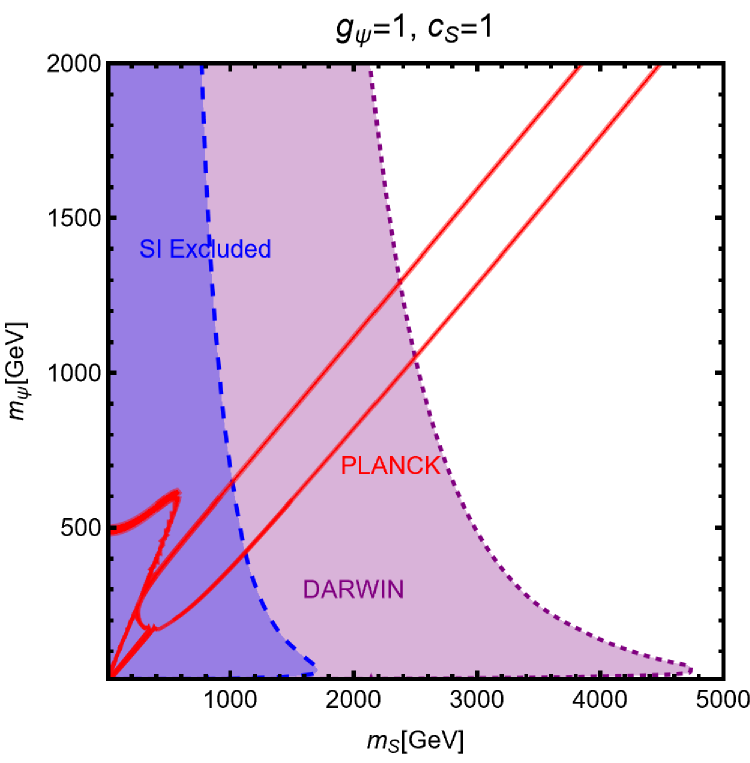}}
\subfloat{\includegraphics[width=0.33\textwidth]{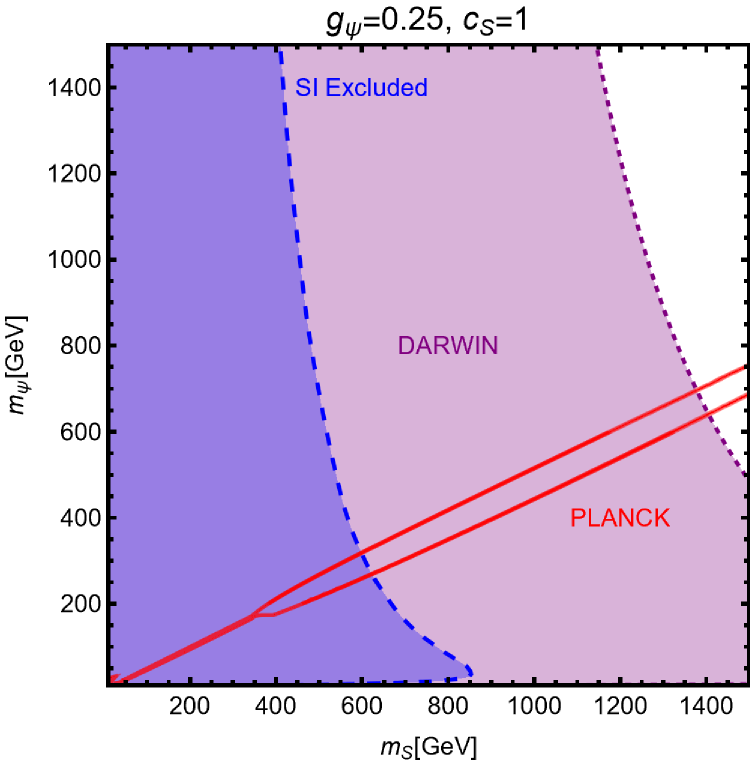}}\\
\subfloat{\includegraphics[width=0.33\textwidth]{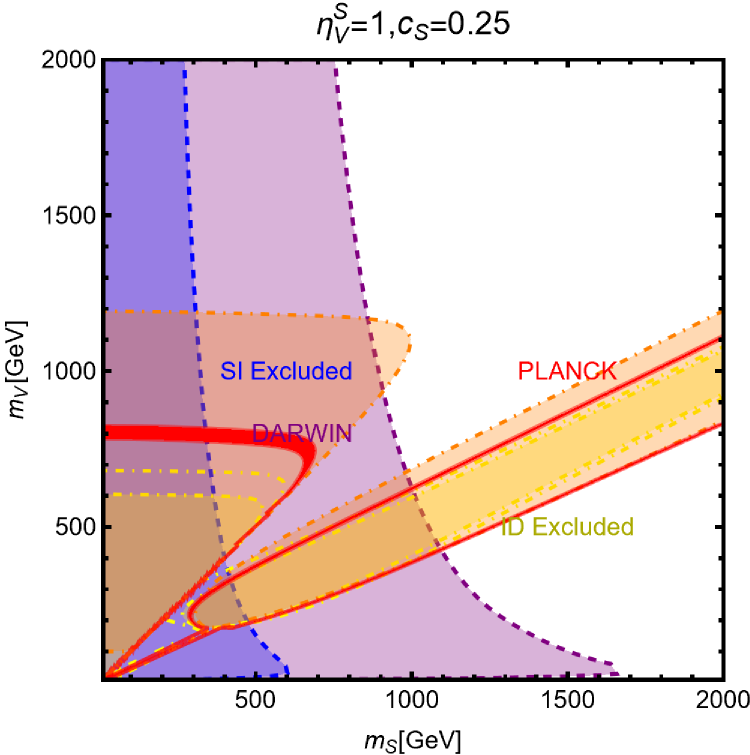}}
\subfloat{\includegraphics[width=0.33\textwidth]{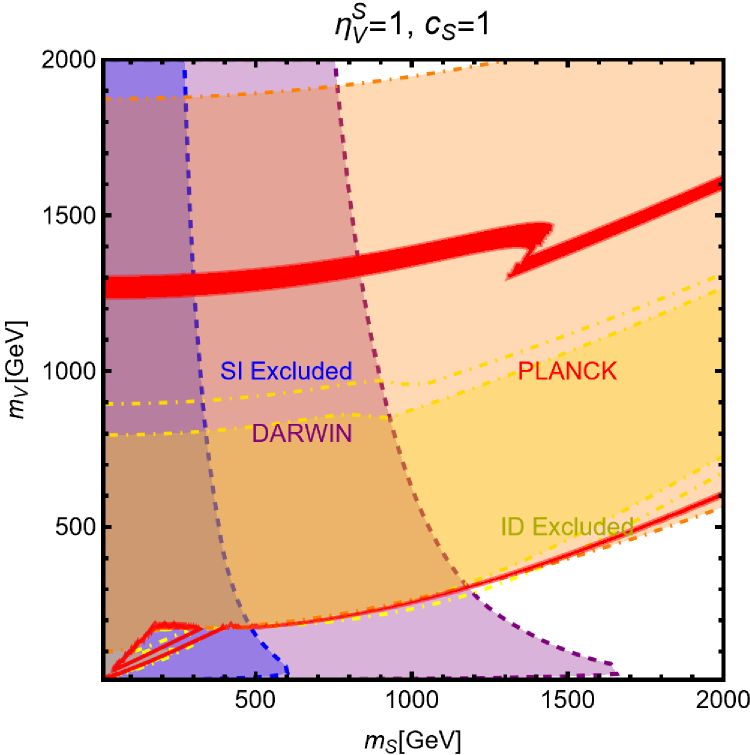}}
\subfloat{\includegraphics[width=0.33\textwidth]{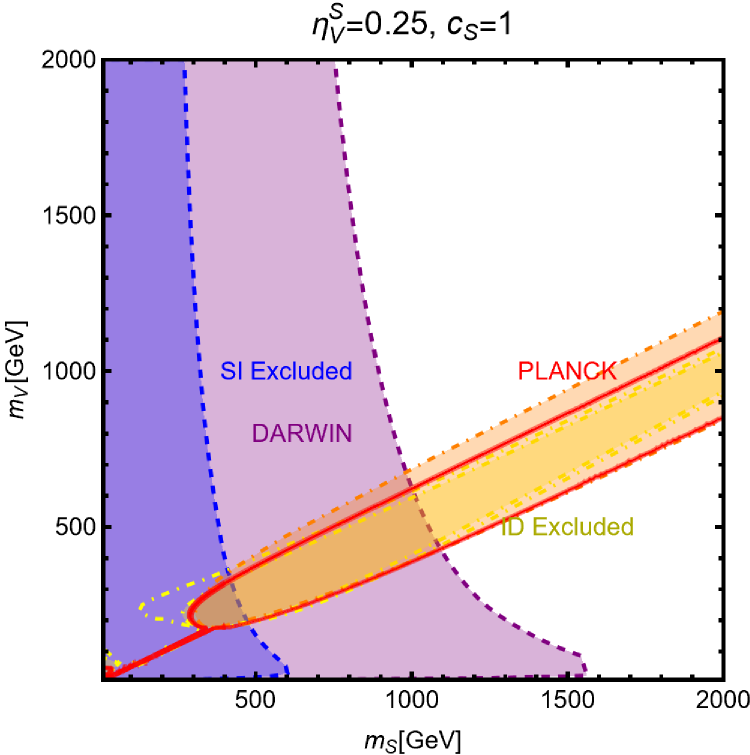}}
\end{center}
\caption{Summary of plausible constraints for the simplified portals with a CP-even scalar mediator. The top row shows results for a scalar DM $\chi$.The middle row depicts the same for a fermionic DM $\Psi$ while the bottom row refers to the case of a vector DM $V$. Each row contains three panels corresponding to different assignations of the pertinent couplings (see Eq. (\ref{eq:scalarMed})), as reported at the top of each plot. In each plot, the red coloured contour corresponds to the correct DM relic density. The blue (purple) coloured region corresponds to the current (projected) exclusions related to the missing experimental signatures coming from the SI interactions. The yellow coloured regions described the excluded parameters space from the absence of ID signals from the DM annihilation processes.}
\label{fig:Sportal}
\end{figure*}

Such plots are shown in Fig.~\ref{fig:Sportal}. In each panel, the isocontours corresponding to the correct relic density are shown in red coloured while the region currently excluded by the DD (ID) experiments have been marked with blue (yellow) colour. The purple (orange) coloured region will be ruled out if the next generation experiment DARWIN (CTA) does not detect any DM signals. Fig.~\ref{fig:Sportal} is an update, with the latest experimental results, of an analysis already discussed in Ref.~\cite{Arcadi:2017kky}. We refer to the original reference for a discussion of the shape of the contours.

\subsubsection{Spin-$0$ mediator -- Pseudoscalar}
The next simplified model that we will review now contains again a spin-$0$ mediator, but this time it will be a pseudoscalar ($a$). Thus, with our assumption of the CP-conservation, only a fermionic DM $\psi$ will be considered in this case. The relevant Lagrangian is: 

\begin{equation}\label{eq:Lagpsmediator}
\mathcal{L}=-i \lambda_\psi^a \ovl \psi \gamma_5 \psi a-i \sum_f \frac{c_a}{\sqrt{2}} \frac{m_f}{v_h}\ovl f \gamma_5 f a.
\end{equation}
The change in the parity of the mediator has a crucial impact on the DM phenomenology. Looking at the analytical expression of the DM annihilation cross-section:
\begin{align}
& \langle \sigma v \rangle{(\ovl\psi\psi \to \ovl ff)} \approx \sum_f 
\frac{n_c^f c_a^2 {(\lambda_\psi^a)}^2}{2 \pi}\frac{m_f^2}{v_h^2}\frac{m_\psi^2}{{\left(4 m_\psi^2-m_a^2\right)}^2}, \nonumber\\
& \langle \sigma v \rangle{(\ovl \psi \psi \to aa)} \approx \frac{{(\lambda_\psi^a)}^4}{192 \pi m_\psi^2}v^2,
\end{align}
we see that unlike Eq.~(\ref{eq:dmpairtoss}), the velocity dependence of the annihilation cross-section into the $ \ovl f f$ final states is lifted so that the latter becomes s-wave dominated. Given the more efficient annihilations, the DM can get the correct relic density in a larger parameter space, at the price of ID signals which should be compared against experiments. The most important result is relative to DD though. The operator $(i \bar \psi \gamma^5 \psi) (i \bar q \gamma^5 q)$ does not correspond neither to the conventional SI nor to the conventional SD interactions. Adopting the more general formalism of NR operators,  $(i \bar \psi \gamma^5 \psi) (i \bar q \gamma^5 q)$ is associated to $\mathcal{O}_6^{\rm NR}$, leading to a DM recoil rate suppressed with the momentum exchange in DM scattering processes, given by~\cite{Arina:2014yna,Dolan:2014ska}: 

\bea
\label{eq:coyDD}
 \frac{d \sigma_T}{dE_R}&&=\frac{|\lambda_\psi^a|^2 c_a^2}{128 \pi}\frac{q^4}{m_a^4}
 \frac{m_T^2}{m_\psi m_N}\frac{1}{v_E^2}\sum_{N,N'=p,n} g_N g_{N'} F_{\Sigma^{''}}^{NN'}(q^2), \nonumber\\
 g_N &&= \sum_{q=u,d,s} \frac{m_N}{v}\left[1-\frac{\overline{m}}{m_q}\right]\Delta_q^{N},\nonumber\\
 && \overline{m}={\left(1/m_u+1/m_d+1/m_s\right)}^{-1},
\eea 
where $m_T$ is the mass of the target nucleus, $v_E$ represents the DM speed in the Earth frame, $\Delta^N_q={u,d,s}$ are form factors denoting the quark spin content of the nucleon (see next subsection for more details),
$E_R$ is the recoil energy and $q$ the momentum transfer. Finally, $F^{NN'}_{\Sigma^{''}}$ are (squared) form factors whose (approximate) analytical expressions are given in Ref.~\cite{Fitzpatrick:2012ix}. Given its strong dependence on the momentum transfer, very small in WIMP elastic scattering processes, the scattering rate eq. \ref{eq:coyDD} is very suppressed. However, as the SI interaction involving a pseudoscalar mediator arises at the loop level, a mere tree-level analysis
appears insufficient to access the concerned detection prospects. It will be shown subsequently that no momentum suppression arises for the CP-odd mediator case.  Further, the coherence of the DM pseudoscalar interaction can compensate for the loop suppression and put the considered framework, at least for some assignation of the relevant parameters, in the reach of current and near future detectors.

The most refined computation of the loop-induced (an example of loop diagram is shown in fig.\ref{fig:pseudo1} cross-section can be found in Refs.~\cite{Abe:2018emu,Ertas:2019dew} (see also Refs.~\cite{Ipek:2014gua,Arcadi:2017wqi,Bell:2018zra} for earlier attempts): 

\begin{figure}
    \centering
    \includegraphics{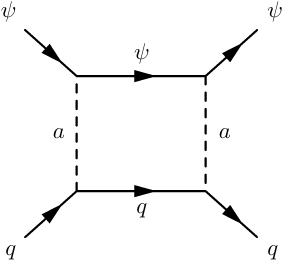}
    \caption{{\it Example of Feynman diagram responsible for the loop induced SI scattering cross-section of DM in the simplified model with pseudoscalar mediator.}}
    \label{fig:pseudo1}
\end{figure}

\begin{equation}
    \sigma_{\psi p}^{\rm SI}=\frac{\mu_{\psi p}^2}{\pi}|C_N|^2,
\end{equation}
where: 
\begin{align}
    & C_N=m_N \left[\sum_{q=u,d,s} f_q^N C_q+ C_G f^N_{TG}\right.\nonumber\\
    & \left. +\frac{3}{4}\sum_{q=u,d,s,c,b} \left(m_\psi C_{1q}+m_\psi^2 C_{2q}\right)\left(q^N (2)+\bar{q}^N(2)\right)\right],
\end{align}
with:
\bea
     C_q&&= -\frac{m_\psi}{(4\pi)^2}|c_a|^2 {\left(\frac{m_q}{v_h}\right)}^2\frac{(\lambda_\psi^a)^2}{m_a^2}   \nonumber \\
     &&\times \Big[G\left( m_\psi^2,0,m_a^2\right)
      -G\left(m_\psi^2,m_a^2,0\right)\Big],
     \eea
where: 
\begin{align}\label{eq:Gfn}
     G(m_\psi^2,m_1^2,m_2^2)&=6 X_{001}(m_\psi^2,m_\psi^2,m_1^2,m_2^2)\nonumber\\
    & + m_\psi^2 X_{111}(m_\psi^2,m_\psi^2,m_1^2,m_2^2),
\end{align}

\bea
     C_{1q}&&=-\frac{8}{(4\pi)^2}|c_a|^2 {\left(\frac{m_q}{v_h}\right)}^2\frac{(\lambda_\psi^a)^2}{m_a^2}\nonumber\\
     &&\times \Big[ X_{001}(p^2,m_\psi^2,0,m_a^2)
     -X_{001}(p^2,m_\psi^2,m_a^2,0)\Big],
     \eea

\bea
     C_{2q}&&=-\frac{4 m_\psi}{(4\pi)^2}|c_a|^2 {\left(\frac{m_q}{v_h}\right)}^2\frac{(\lambda_\psi^a)^2}{m_a^2}\nonumber\\
     && \times\Big[X_{111}(p^2,m_\psi^2,0,m_a^2)
     -X_{111}(p^2,m_\psi^2,m_a^2,0)\Big],
\eea

\begin{align}\label{eq:X001fn}
   & X_{001}(p^2,M^2,m_1^2,m_2^2)=\nonumber\\
   & \int_0^1 dx \int_{0}^{1-x}dy \frac{\frac{1}{2} (1-x-y)}{M^2 x+m_1^2 y+m_2^2 (1-x-y)-p^2 x (1-x)},
\end{align}

\begin{align}\label{eq:X111fn}
    & X_{111}(p^2,M^2,m_1^2,m_2^2)=\nonumber\\
    & \int_0^1 dx \int_{0}^{1-x}dy \frac{-x^3 (1-x-y)}{(M^2 x+m_1^2 y+m_2^2 (1-x-y)-p^2 x (1-x))^2}.
\end{align}

Finally,
\begin{align}
    & C_G=-\frac{m_\psi}{432\pi^2}|c_a|^2 |\lambda_\psi^a|^2 \sum_{Q=c,b,t}{\left(\frac{m_Q}{v_h}\right)}^2 \frac{\partial F(m_a^2)}{\partial m_a^2},
\end{align}
with
\begin{align}\label{eq:Ffn}
     F(m_a^2)&=\int_{0}^1 dx  \Big[ 3 Y_1 (p^2,m_\psi^2,m_a^2,m_Q^2)\nonumber\\
    & -m_Q^2 \frac{2+5x-5x^2}{x^2 (1-x)^2} Y_2 (p^2,m_\psi^2,m_a^2,m_Q^2)\nonumber\\
    &  -2 m_Q^4 \frac{1-2x+2x^2}{x^3 (1-x)^3} Y_3 (p^2,m_\psi^2, m_a^2,m_Q^2) \Big].
\end{align}

In $Y_1,\,Y_2,\,Y_3$ functions $p$ represents the momentum of the external DM particle. Thus, we set $p^2=m^2_\Psi$ for 
numerical analysis and use the same for
$Y_1,\,Y_2,\,Y_3$, namely:

\begin{align}
    & Y_1 (m_\psi^2,m_\psi^2,m_a^2,m_q^2)=\nonumber\\
    & \int_0^1 dy \int_{0}^{1-y}dz \frac{-2y}{m_\psi^2 y^2+\frac{m_q^2}{x (1-x)}z+m_a^2 (1-y-z) },
\end{align}

\begin{align}
    & Y_2 (m_\psi^2,m_\psi^2,m_a^2,m_q^2)=\nonumber\\
    & \int_0^1 dy \int_{0}^{1-y}dz \frac{2xy}{\left[m_\psi^2 y^2+\frac{m_q^2}{x (1-x)}z+m_a^2 (1-y-z)\right]^2},
\end{align}
and
\begin{align}
    & Y_3 (m_\psi^2,m_\psi^2,m_a^2,m_q^2)=\nonumber\\
    & \int_0^1 dy \int_{0}^{1-y}dz \frac{-4yz^2}{\left[m_\psi^2 y^2+\frac{m_q^2}{x (1-x)}z+m_a^2 (1-y-z)\right]^3}.
\end{align}


\begin{figure*}
    \centering
\subfloat{\includegraphics[width=0.33\linewidth]{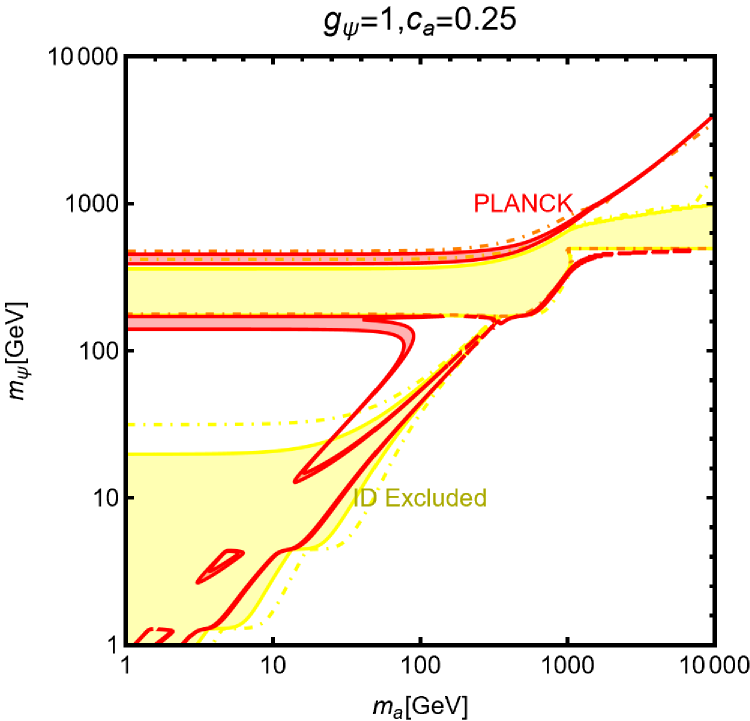}}
\subfloat{\includegraphics[width=0.33\linewidth]{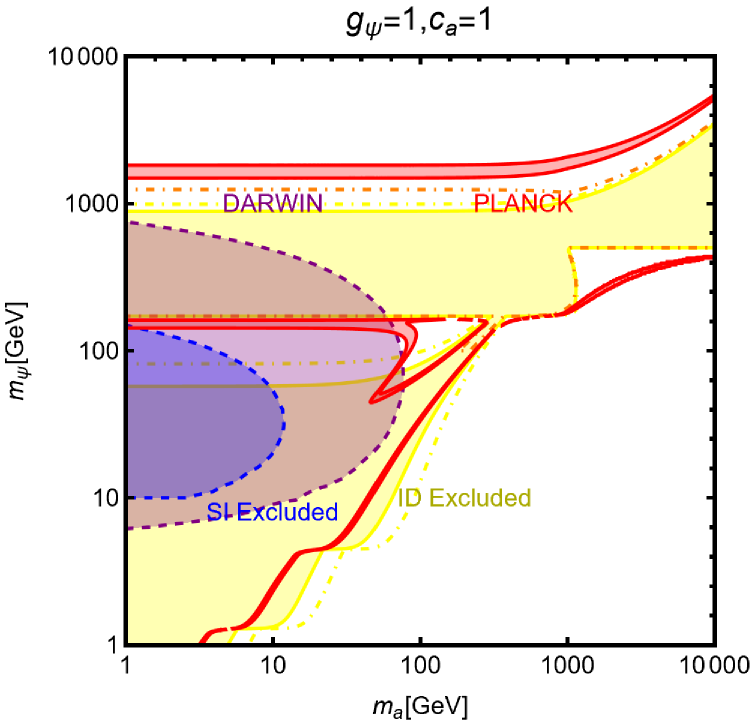}}
\subfloat{\includegraphics[width=0.33\linewidth]{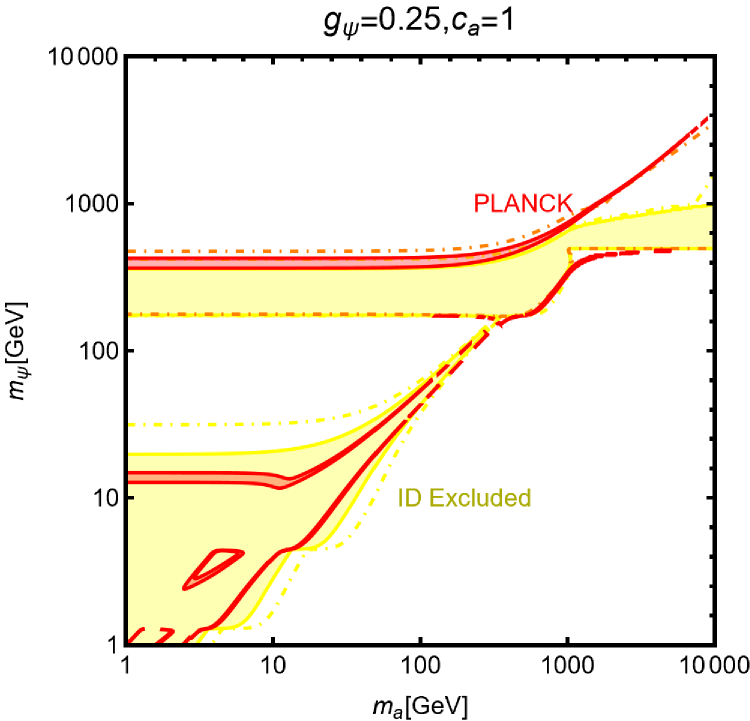}}
\caption{Summary of constraints, in the $(m_a,m_\psi)$ plane for a simplified s-channel portal with a fermionic DM $\Psi$ and a CP-odd mediator $a$. The three panels refer to different assignations of the relevant couplings (see Eq.~(\ref{eq:Lagpsmediator})), reported at the top of each plot. The colour code is the same as of the ones used for Fig.~\ref{fig:Sportal}.}
\label{fig:PSportal}
\end{figure*}
The DM constraints on the model are summarized in Fig.~\ref{fig:PSportal}, with the same colour codes as Fig \ref{fig:Sportal}. Similarly to the case of the CP-even spin-0 mediator, we consider three assignations for the $(g_\psi,c_a)$ pair, namely (from left to right in the figure) $(1,0.25)$, $(1,1)$ and $(0.25,1)$. As evident, the most effective experimental probe is represented by ID. DD is relevant for $g_\psi=c_a=1$ and $m_a \lesssim 100\,\mbox{GeV}$. In the rest of the parameter space of the model one expects a DM cross-section below the $\nu$-floor \cite{Arcadi:2017wqi}.

\subsubsection{Spin-1 mediator}
We will now consider the case of a spin-$1$ mediator.  We can define the following simplified models for a complex scalar DM $\chi$ and a fermion (Dirac and Majorana) DM $\Psi$:

\begin{align}
\label{eq:vectorMed}
\mathcal{L}&=i g_\chi \left(\chi^{*} \partial_\mu \chi -\chi \partial_\mu \chi^{*}\right)Z^{'\,\mu} \nonumber\\
 &+g_\chi^2  |\chi|^2 Z^{'\mu}Z^{'}_\mu 
+g_\chi \ovl f \gamma^\mu (V_f^{Z^{'}}-A _f^{Z^{'}} \gamma_5) Z_\mu^{'} f, \nonumber\\
 \mathcal{L}&=g_\psi \xi \ovl \psi \gamma^\mu (V_\psi^{Z^{'}}-A _\psi^{Z^{'}} \gamma_5)\psi Z_\mu^{'} \nonumber\\
&+g_\psi \ovl f \gamma^\mu (V_f^{Z^{'}}-A _f^{Z^{'}} \gamma_5)f Z_\mu^{'}.
\end{align}
We start our discussion for the case of a scalar DM. Concerning the relic density for this case, one has to consider DM annihilation processes into $\ovl f f$ and $Z'Z'$ final states, whose cross-sections can be written as:
\begin{align}
    \langle \sigma v \rangle (\chi \chi^* \rightarrow \ovl f f)&\approx \frac{g_\chi^4 m_\chi^2 v^2}{3\pi \left(4 m_\chi^2-m_{Z'}^2\right)^2}\nonumber\\
    & \times \sum_f n_c^f \left(|V_f^{Z'}|^2+|A_f^{Z'}|^2\right),\nonumber\\
    \langle \sigma v \rangle (\chi \chi^* \rightarrow Z' Z')&=\frac{g_\chi^4}{8\pi m_\chi^2},
\end{align}
in the limit $m_f, m_{Z'}\rightarrow 0$.
Moving to the DD, for a more effective illustration of the feasible phenomenological prospects, we will consider various possibilities of couplings depicted in Eq. (\ref{eq:vectorMed}) individually:
\begin{itemize}
    \item {\bf Only vectorial couplings among the SM fermions $f$ and the $Z'$ for a complex scalar DM , i.e., $A^{Z'}_f=0 \, \forall \, f$}: 
    the combination of the $\left(\chi^{*} \partial_\mu \chi -\chi \partial_\mu \chi^{*}\right)$ operator with a vectorial quark current would lead, in the NR limit, to a SI operator that corresponds the following cross-section of the DM over protons:
\bea
\label{eq:scalarZp}
\sigma_{\chi p}^{\rm SI}&&= \frac{\mu_{\chi p}^2}{\pi}\frac{g_\chi^4}{m_{Z'}^4}\frac{{\left[ Z f_p+(A-Z) f_n \right]}^2}{A^2}, ~~{\rm with}
\nonumber\\
&& f_p=2 V_u^{Z'}+V_d^{Z'},f_n=V_u^{Z'}+2 V_d^{Z'}.
\eea
We see that although we are discussing the SI interactions, the translation of the microscopic interaction between the DM and quarks into interactions between the DM and nucleon is not the same as the case of a spin--$0$ mediator (see Eq.~({\ref{eq:DDsimplifiedS0}})). Indeed, the bilinear operator $\ovl q \gamma^\mu q$  once evaluated among the initial and final nucleon states, is related to the electric charge of the nucleon. The associated bilinear operator $\ovl N \gamma^\mu N$, with $N$ being the nucleon's field, will be then determined only by the valence quarks. The effective couplings $f_{p,\,n}$ will be then just linear combinations of the couplings of the $Z'$ with the up and down quarks. Unless the $Z'$ has the same couplings with the up and the down quarks, the DM will couple differently to protons and neutrons; it is then essential to account for the scaling factor related to the detector material to perform a consistent comparison with the experimental outcome. 
Since, contrary to the case of the spin--0 mediator, no small form factors are present, we expect comparatively stronger limits. 

\item {\bf Only axial vector couplings among $f$ and the $Z'$ for a complex scalar DM, i.e., $V^{Z'}_f=0 \, \forall \, f$:} Here, integrating out the $Z'$ mediator in the NR limit, one would obtain an operator that can be mapped in $O_7^{\rm NR}$ (see Eq.~(\ref{eq:NRoperators})). This operator depends on the nucleon's spin (hence no coherent enhancement) and would be suppressed by the DM velocity. This picture, however, does not take into account a relevant fact. As already pointed out, once determining the interactions relevant for the DD, one should take into account their low characteristic scale. Besides integrating out the heavy $dof$, the running of the BSM couplings from the initial high NP scale to $1$ GeV should also be accounted for. In this process, operator mixing occurs in general, so that couplings which are set to zero at some initial high energy scale, might re-appear again at a lower energy by the renormalization group (RG) evolution. As pointed out in Ref.~\cite{DEramo:2016gos}, the RG running of the axial couplings of the $Z'$ will generate vectorial couplings at the scale $\mu_N=1$ GeV, whose approximate expression is given  by ( the mass of $Z'$, $m_{Z'}$, has been taken as the initial scale):
\begin{align}
     \widetilde{V}_u^{Z'}=(3-8& s_W^2)\Big[\frac{\alpha_t}{2\pi} A_u^{Z'} 
 \log\left(\frac{m_{Z'}}{M_Z}\right)\nonumber\\
 &\left.- \left[\frac{\alpha_b}{2 \pi} 
 A_d^{Z'}+\frac{\alpha_\tau}{6\pi}A_e^{Z'}\right] \log\left(\frac{m_{Z'}}{\mu_N}\right)\right], \nonumber\\
 \widetilde{V}_d^{Z'}=(3-4& s_W^2)\Big[-\frac{\alpha_t}{2\pi} 
 A_u^{Z'} \log\left(\frac{m_{Z'}}{M_Z}\right) \nonumber\\
 & \left. + 
 \left[\frac{\alpha_b}{2 \pi} A_d^{Z'}+\frac{\alpha_\tau}{6\pi}A_e^{Z'}\right] 
 \log\left(\frac{m_{Z'}}{\mu_N}\right)\right],
\end{align}
where $s_W\equiv \sin \theta_W$, $\theta_W$ being the weak mixing angle.
Thus, for the DM DD, we can adopt the same expression as the previous case, just with the replacement $V_{u,d}^{Z'} \rightarrow \widetilde{V}_{u,d}^{Z'}$.
\end{itemize}

\begin{figure*}
    \centering
    \subfloat{\includegraphics[width=0.33\linewidth]{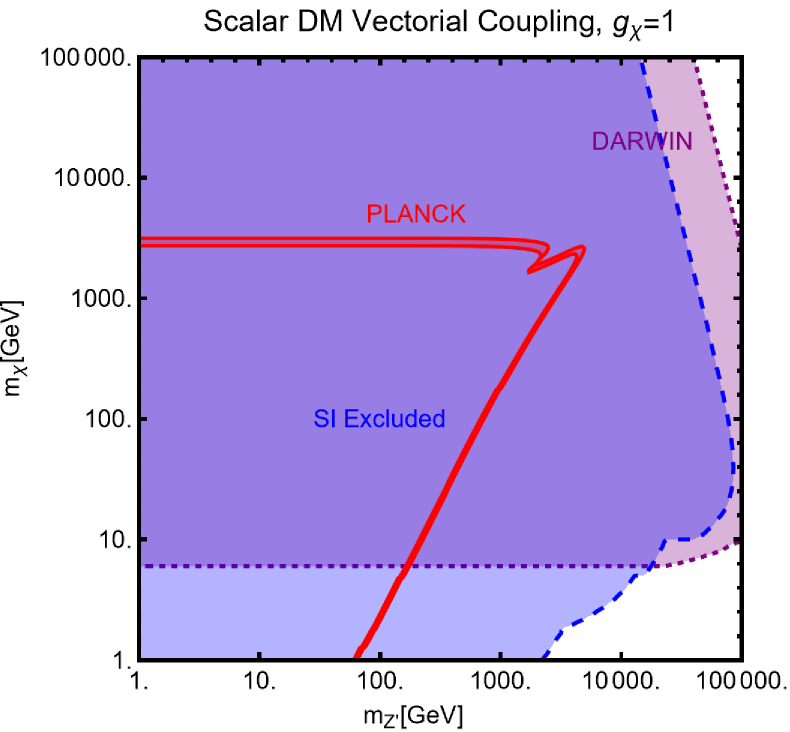}}
    \subfloat{\includegraphics[width=0.32\linewidth]{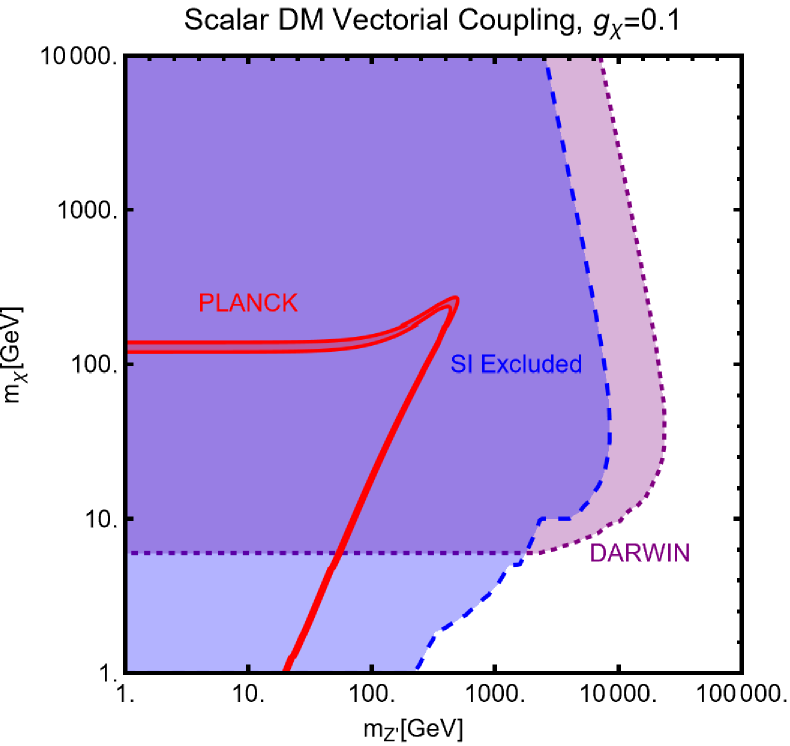}}
    \subfloat{\includegraphics[width=0.32\linewidth]{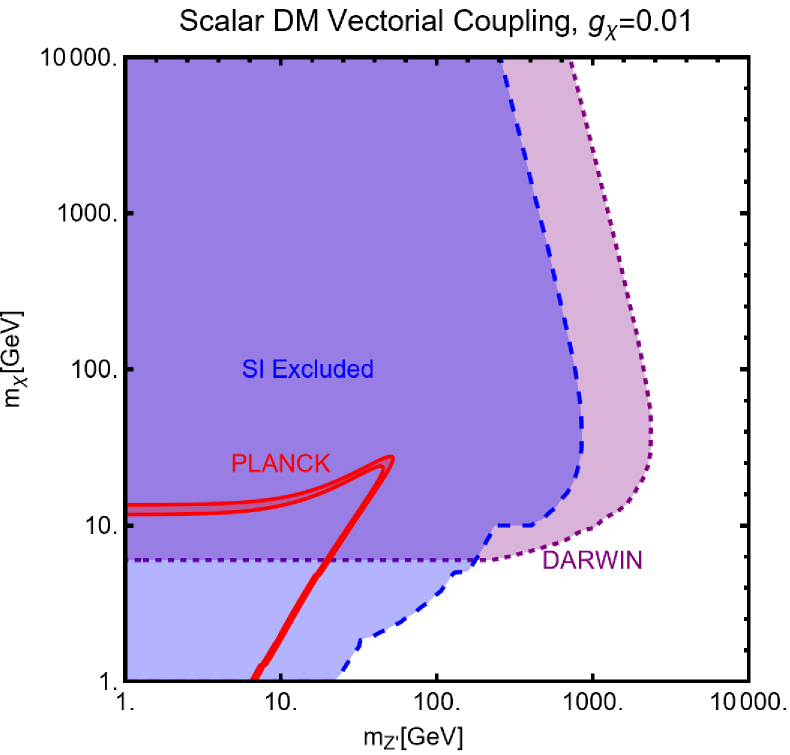}}\\
    \subfloat{\includegraphics[width=0.33\linewidth]{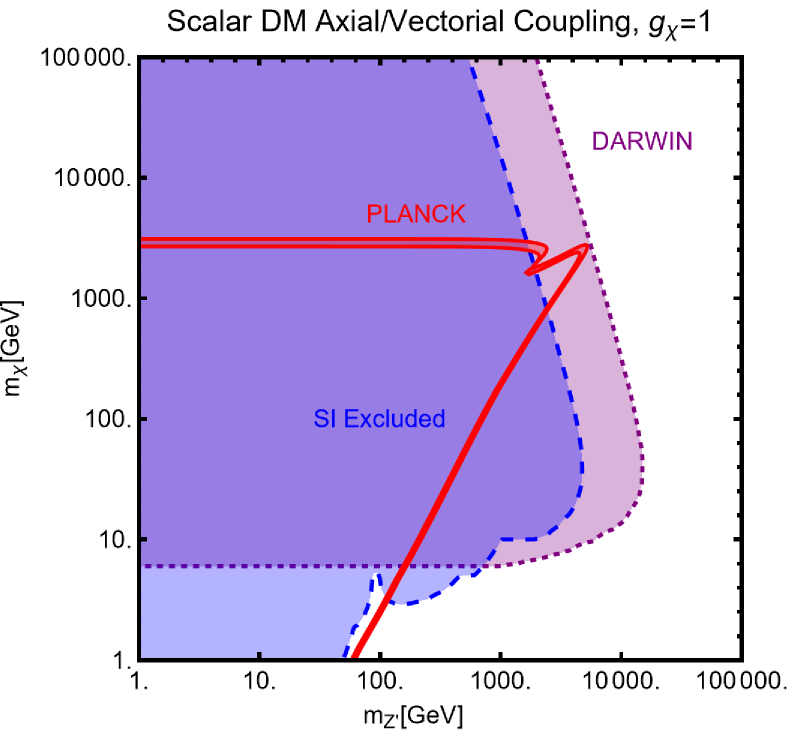}}
    \subfloat{\includegraphics[width=0.32\linewidth]{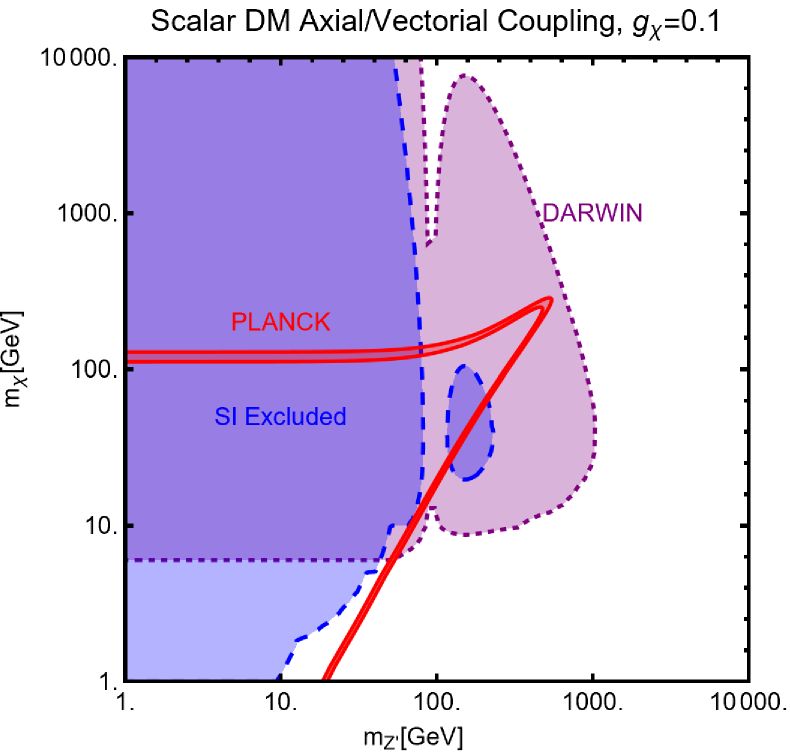}}
    \subfloat{\includegraphics[width=0.31\linewidth]{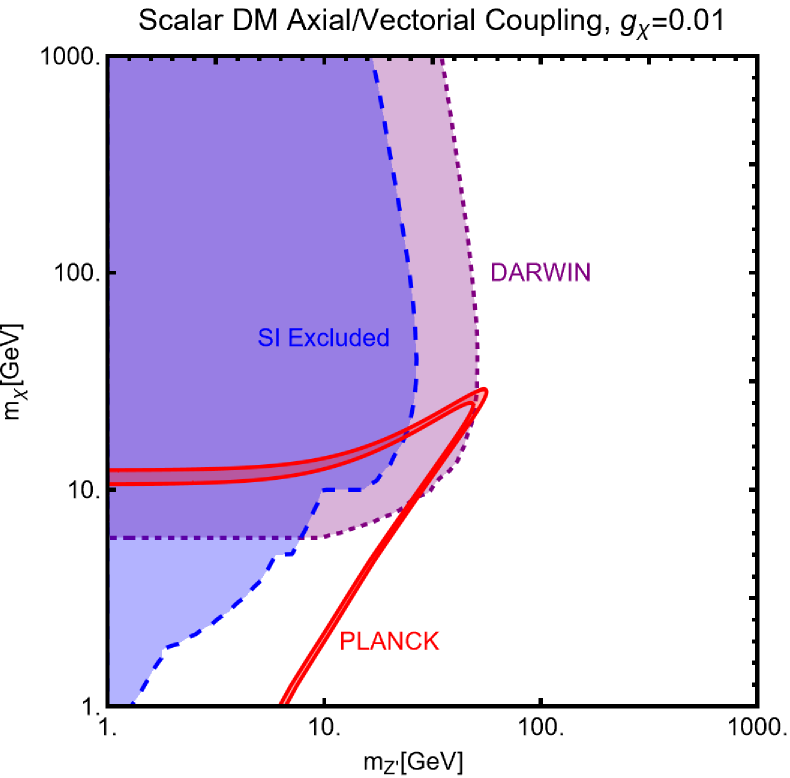}}
    \caption{Summary of constraints for a simplified model with a complex scalar DM interacting via an s-channel spin-$1$ mediator  $Z'$. The constraints are shown in the $(m_{Z'},m_\chi)$ plane. For each plot, the viable parameters space is the area where the red coloured isocontours, representing the correct DM relic density, lie outside the blue and purple coloured regions. The colour code is the same as of the ones used for Fig.~\ref{fig:Sportal}. The phrases "Scalar DM Vectorial Coupling" (top row) and "Scalar DM Axial Vectorial Coupling" (bottom row) refer to $A_f^{Z'}=0\,\,\forall\,\,f$ and   $V_f^{Z'}=0\,\,\forall\,\,f$, respectively. We refer to the main text for details on the assignations of the couplings.}
    \label{fig:ScalarZpportal}
\end{figure*}

Again, to characterize the model, it is sufficient to study the $(m_{Z'},m_\chi)$ bidimensional plane as shown in Fig.~\ref{fig:ScalarZpportal}.
For simplicity, we have adopted only $g_\chi$  as varying coupling and fixed $V_{f}^{Z'}=1,\, A_{f}^{Z'}=0$ in the first row and $V_{f}^{Z'}=0,\, A_{f}^{Z'}=1$ in the second.

Now we move to the case of a fermionic DM $\Psi$, as depicted in Eq. (\ref{eq:vectorMed}). Like the case of a complex scalar DM, in this case also, the presence or absence of vectorial couplings rather than the axial ones among the $Z'$ with the DM and the SM fermions strongly impact the DM phenomenology. For instance, if both vector and axial-vector couplings are present, strong limits from atomic parity violation arise (see for instance \cite{Cosme:2021baj}). For this reason, we will again consider different possible cases:
\begin{itemize}

    \item {\bf $Z'$ couplings to the DM and the SM fermions are only vectorial}: 
    The cross-sections accounting for the DM relic density can be described via the following analytical approximations:
    \begin{align}
        \langle \sigma v \rangle (\ovl \psi \psi \rightarrow \bar f f)&=\frac{g_\psi^4 |V_\psi^{Z'}|^2}{\pi}\nonumber\\
        &\times \frac{m_\psi^2}{(4 m_\psi^2-m_{Z'}^2)^2}\sum_f n_f^c |V_f^{Z'}|^2,\nonumber\\
        \langle \sigma v \rangle (\ovl \psi \psi \rightarrow Z' Z')&=\frac{g_\psi^4 |V_\psi^{Z'}|^4}{16\pi m_\psi^2}.
    \end{align}
    where, again, the limit of null final state masses has been taken.
    In both cases, we have s-wave-dominated cross-sections. Hence, we expect that the ID can also probe the parameter space corresponding to the correct DM relic density.
    Moving to the DD, the DM current $\ovl \psi \gamma^\mu \psi$ behaves in the same way as the derivative interactions of the complex scalar DM. Hence, we obtain the exact same SI cross-section as previously defined for a complex scalar DM (see Eq.~(\ref{eq:scalarZp})).
    
    \item {\bf $Z'$ couplings with the $\Psi$ are only vectorial while the same with $f$s are purely axial}:
    Neglecting the final state fermion masses, the annihilation cross-sections of the DM, at the leading order in the velocity expansion, coincide with the same of the previous case, just by replacing $V_f^{Z'}\rightarrow A_{f}^{Z'}$ for the $\ovl f f$ final states. Also, the scattering cross-section over the proton retains the same analytical expression as the one with vectorial couplings of the $Z'$ with quarks, generated at the scale of the DD processes through the RG running.
    
    \item {\bf  $Z'$ couples with $\Psi$ only axially while couples with $f$s purely vectorially}: Changing the $Z'$-DM couplings from vectorial to axial  has a significant impact on the relic density and ID constraints as the DM annihilation cross-section into fermions becomes velocity dependent:
    \begin{equation}
        \langle \sigma v \rangle (\ovl \psi \psi \rightarrow \bar f f)=\sum_f n_f^c \frac{g_\psi^4 |A_\psi^{Z'}|^2 |V_f^{Z'}|^2 m_\psi^2 v^2}{3\pi (4 m_\psi^2-m_{Z'}^2)^2},
    \end{equation}
    while the one into $Z'Z'$ can be approximated as:
    \begin{equation}\label{eq:psipsiZpZpAxial}
        \langle \sigma v \rangle (\ovl \psi \psi \rightarrow Z' Z')=\frac{g_\psi^4 |A_\psi^{Z'}|^4}{\pi}\left(\frac{1}{m_\psi^2}+\frac{m_\psi^2}{m_{Z'}^4}v^2\right).
    \end{equation}
    It is now evident that we have included the next-to-leading order term in the velocity expansion as it features a $m_\psi^2/m_{Z'}^2$ enhancement. The importance of this term will be clarified below. Concerning the DD, the interaction Lagrangian would lead to operators which would be mapped into a combination of the $O_8^{\rm NR}$ and $O_9^{\rm NR}$ operators (see Eq.~(\ref{eq:NRoperators})). Again, these operators are suppressed by the highly NR velocity of the DM and by the absence of coherent enhancement, as they contain the spins of the DM and the nucleon. Similarly, based on what we already observed in the case of a pseudoscalar mediator, one might wonder whether SI interactions could arise at the loop-level overcoming the "tree-level" ones. This indeed happens \cite{Hisano:2011cs,Haisch:2013uaa,Belyaev:2022qnf}, thanks to the enhancement in heavy nuclei and the absence of velocity/momentum transfer suppression. 
    Diagrams with box-shaped topology induce (as the ones show in fig. \ref{fig:Zploop}) an SI scattering cross-section over protons of the following form:
    \begin{align}
         \sigma_{\psi p}^{\rm SI}&=\frac{\mu_{\psi p}^2}{\pi} \bigg\vert  \sum_{q=u,d,s}f_q f_q^p-\frac{8\pi}{9 \alpha_s}f_G f_{TG} \nonumber\\
        &  +\frac{3}{4}m_p \sum_{q=u,d,s,c,b}\left(q(2)+\bar q(2)\right)\left(g_q^{(1)}+g_q^{(2)}\right)  \bigg\vert^2,
    \end{align}
with $f_q,g_q^{(1,2)},f_G$ being Wilson coefficients given by:   
    \begin{align}
        & f_q=\frac{g_\psi^4}{m_{Z'}^3}\left(A_q^{Z'2}-V_q^{Z'2}\right) g_s\left(\frac{m_{Z'}^2}{m_\psi^2}\right),\nonumber \\
        & g_q^{(1)}=\frac{2 g_\psi^4}{m_{Z'}^3}\left(A_q^{Z'2}+V_q^{Z'2}\right) g_{T1}\left(\frac{m_{Z'}^2}{m_\psi^2}\right),\nonumber\\
        & g_q^{(2)}=\frac{2 g_\psi^4}{m_{Z'}^3}\left(A_q^{Z'2}+V_q^{Z'2}\right) g_{T1}\left(\frac{m_{Z'}^2}{m_\psi^2}\right),\nonumber\\
        & f_G=\frac{\alpha_s}{4\pi}\frac{g_\psi^4}{4m_{Z'}^3}g_Z\left(\frac{m_t^2}{m_\psi^2},\frac{m_{Z'}^2}{m_\psi^2}\right).
    \end{align}
     The loop functions are written as:
\begin{align}
     g_s(x)=&-\frac{2}{b_x}(2+2x-x^2){\tan}^{-1}\left(\frac{2b_x}{\sqrt{x}}\right) \nonumber \\ &
+\frac{1}{4}\sqrt{x}(2-x\log x),\nonumber \\
     g_{T_1}(x)=&\frac{1}{3}b_x(2+x^2){\tan}^{-1}\left(\frac{2b_x}{\sqrt{x}}\right)\nonumber\\
    & +\frac{1}{12}\sqrt{x}(1-2x-(2-x)\log x),\nonumber\\
     g_{T_2}(x)=&\frac{1}{4b_x}x(2-4x+x^2){\tan}^{-1}\left(\frac{2b_x}{\sqrt{x}}\right)\nonumber\\
    & -\frac{1}{4}\sqrt{x}(1-2x-x(2-x)\log x),
\end{align}
with $b_x=\sqrt{1-\frac{x}{4}}$. The function $g_Z$ can be evaluated only numerically. We refer to Ref.~\cite{Hisano:2011cs} for details.
\begin{figure*}
    \centering
    \subfloat{\includegraphics[width=0.33\linewidth]{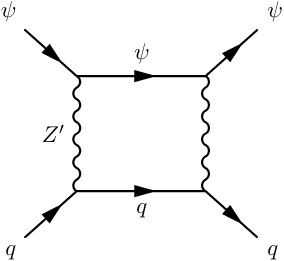}}
    \subfloat{\includegraphics[width=0.33\linewidth]{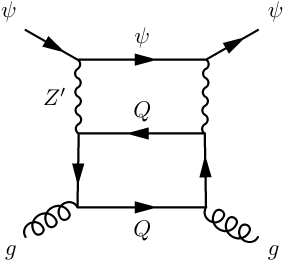}}
    \caption{{\it Examples of loop diagrams inducing, at one loop, SI cross-section via exchange of $Z'$ in the internal lines}}
    \label{fig:Zploop}
\end{figure*}

As pointed out, e.g., in Ref.~\cite{Kahlhoefer:2015bea}, the annihilation cross-section for the DM into a pair of $Z'$, in the presence of only axial couplings, has a pathological behaviour triggered by the longitudinal $dof$ of the spin--$1$ mediator. This is evidenced by the presence of a $p$-wave term in Eq. (\ref{eq:psipsiZpZpAxial}) which increases with the mass hierarchy between the fermionic DM and the $Z'$. Since the cross-section is computed in the NR limit, the increase in the DM mass is actually a symptom of the increase of the cross-section, before the thermal average, with the center-of-mass energy. Hence,to avoid the unitarity violation the following condition should be satisfied \cite{Shu:2007wg,Hosch:1996wu,Babu:2011sd}:
\begin{equation}
    \sqrt{s}< \frac{\pi m_{Z'}^2}{g_\psi^2 |A_\psi^{Z'}|^2 m_\psi}.
\end{equation}
Since $s \sim 4 m_\psi^2$, the latter implies that there cannot be a too strong mass hierarchy between the masses of $\psi$ and $Z'$.

\item {\bf Only axial couplings of the $Z'$ with both the DM and the SM fermions}: In this case the thermally averaged cross-section is again p-wave dominated:
\begin{equation}
    \langle \sigma v \rangle (\ovl \psi \psi \rightarrow \ovl {f} f)=\sum_f n_f^c \frac{g_\psi^4 |A_\psi^{Z'}|^2 |A_f^{Z'}|^2 v^2}{3\pi}\frac{m_\psi^2}{(4 m_\psi^2-m_{Z'}^2)^2}.
\end{equation}
In the presence of only axial couplings,
i.e., $V_{\psi,f}^{Z'}, A_{\psi,f}^{Z'}$, a $s$-wave term actually appears in the annihilation cross-section. The same, however, suffers a helicity suppression $m_f^2/m_\psi^2$, leaving
the $p$-wave contribution to be the dominant one.

The combination of the $\ovl \psi \gamma^\mu \gamma_5 \psi$ and $\ovl f \gamma^\mu \gamma_5 f$ operators leads to the conventional SD interaction which can be described by the following cross-section:
\begin{equation}
    \sigma_{\psi p}^{\rm SD}=\frac{3 \mu_{\psi\,p}^2}{\pi}\frac{g_\psi^4}{m_{Z'}^4} |A_\psi^{Z'}|^2
{\left[A_u^{Z'}\Delta_u^p+A_d^{Z'}\left(\Delta_d^p+\Delta^p_s\right)\right]}^2.
\end{equation} 
the form factors $\Delta_q={u,d,s}^{N=p,n}$ describe the contributions from the quarks to the nucleon spin and are defined by:
\begin{equation}
    \langle N | \bar q \gamma_\mu \gamma_5 q |N\rangle=2 s_\mu \Delta_q^N
\end{equation}
with $s_\mu$ being the nucleon's spin. We adopted, for our study, the same as implemented in the package micrOMEGAs \cite{Belanger:2008sj}.

\end{itemize}

\begin{figure*}
    \centering    
    \subfloat{\includegraphics[width=0.25\linewidth]{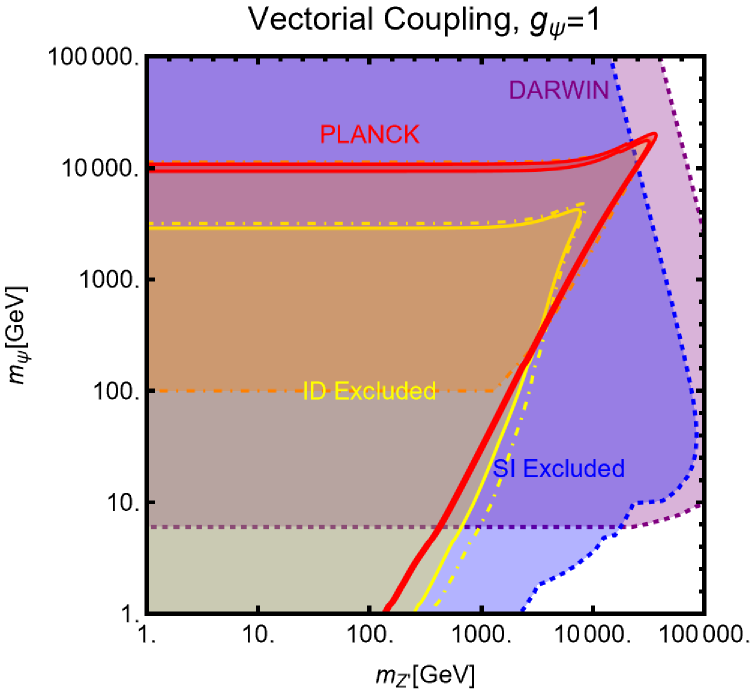}}    \subfloat{\includegraphics[width=0.25\linewidth]{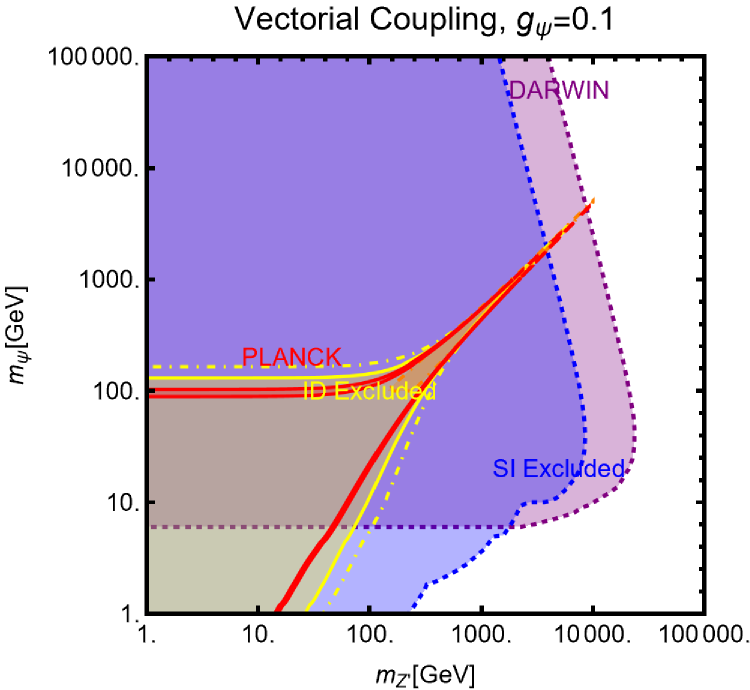}}    \subfloat{\includegraphics[width=0.25\linewidth]{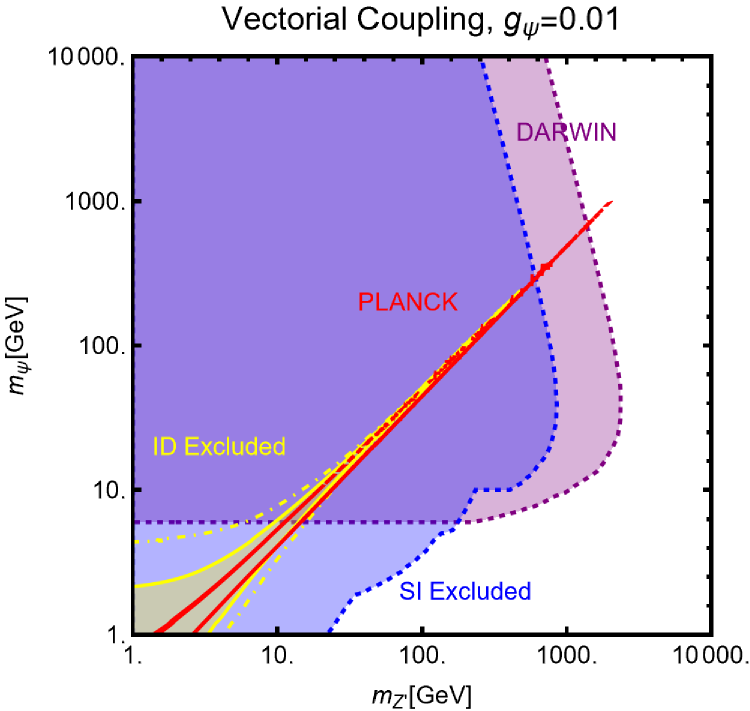}}\\
    \subfloat{\includegraphics[width=0.25\linewidth]{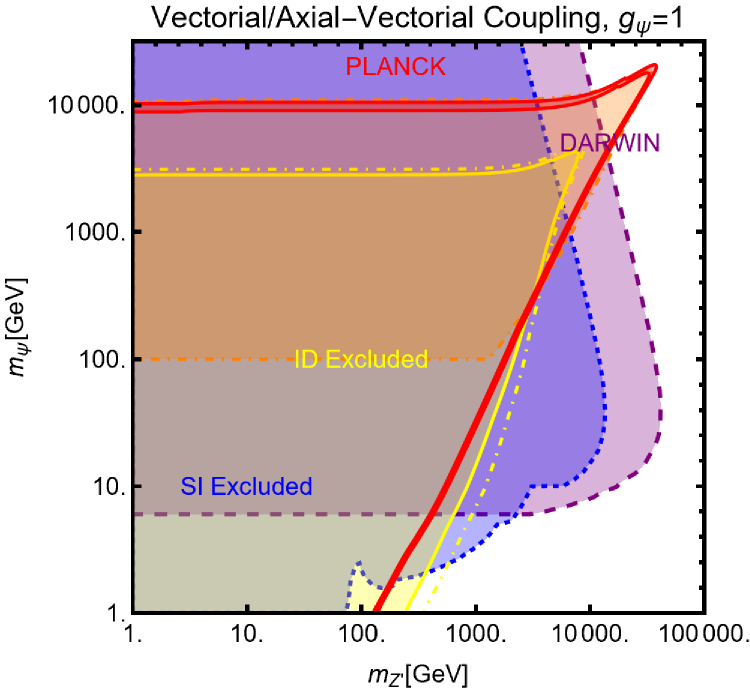}}    \subfloat{\includegraphics[width=0.25\linewidth]{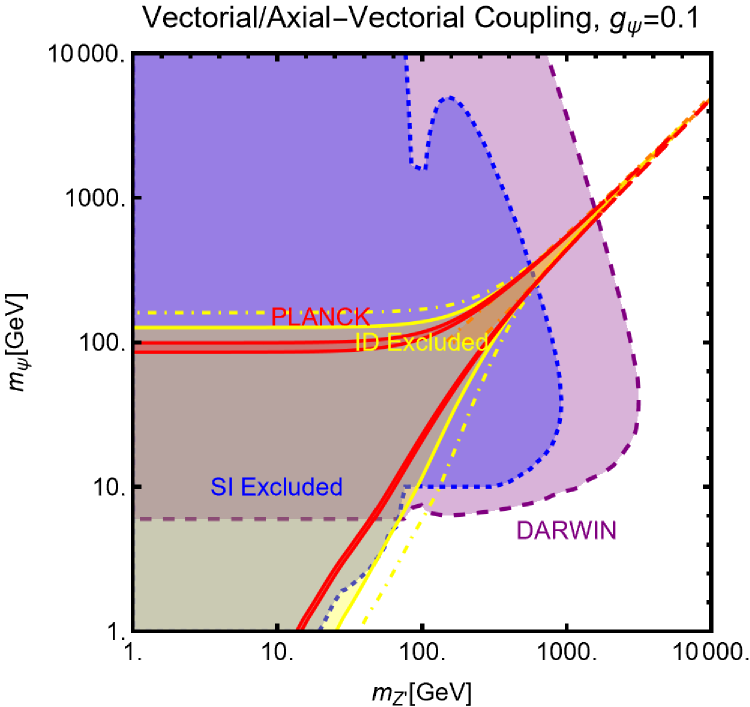}}    \subfloat{\includegraphics[width=0.25\linewidth]{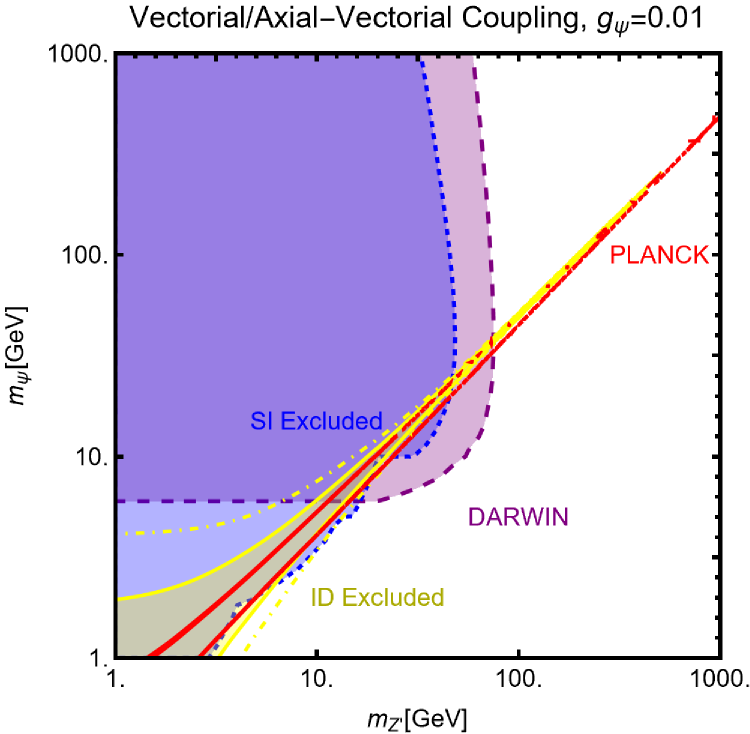}}\\
    \subfloat{\includegraphics[width=0.25\linewidth]{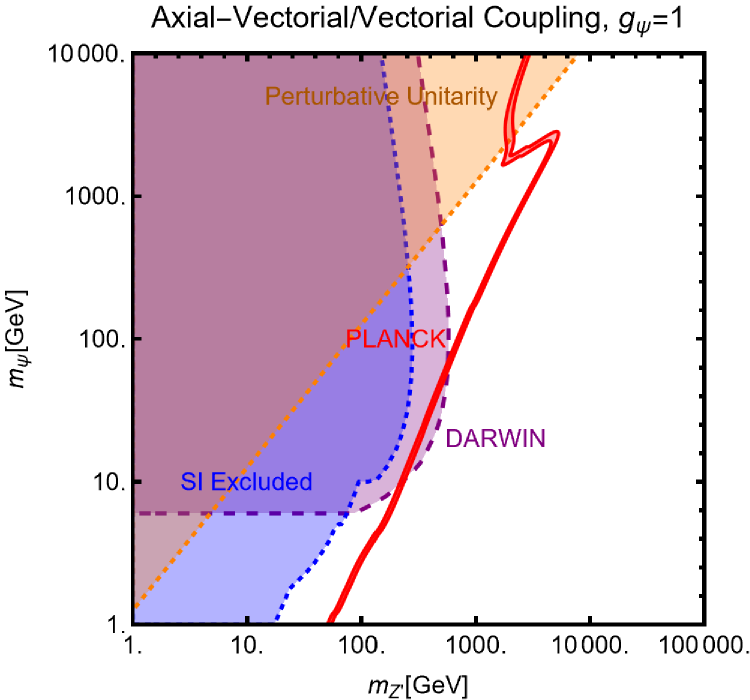}}    \subfloat{\includegraphics[width=0.25\linewidth]{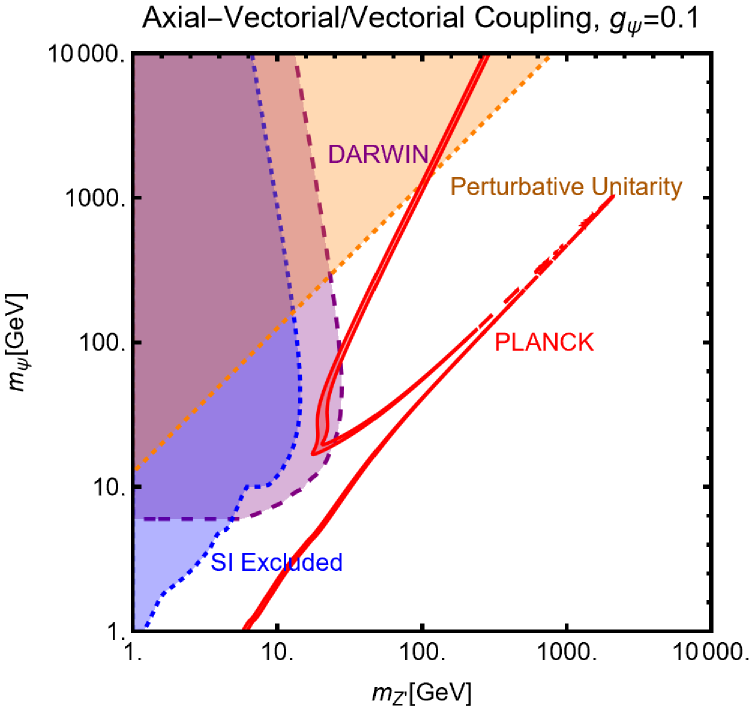}}    \subfloat{\includegraphics[width=0.25\linewidth]{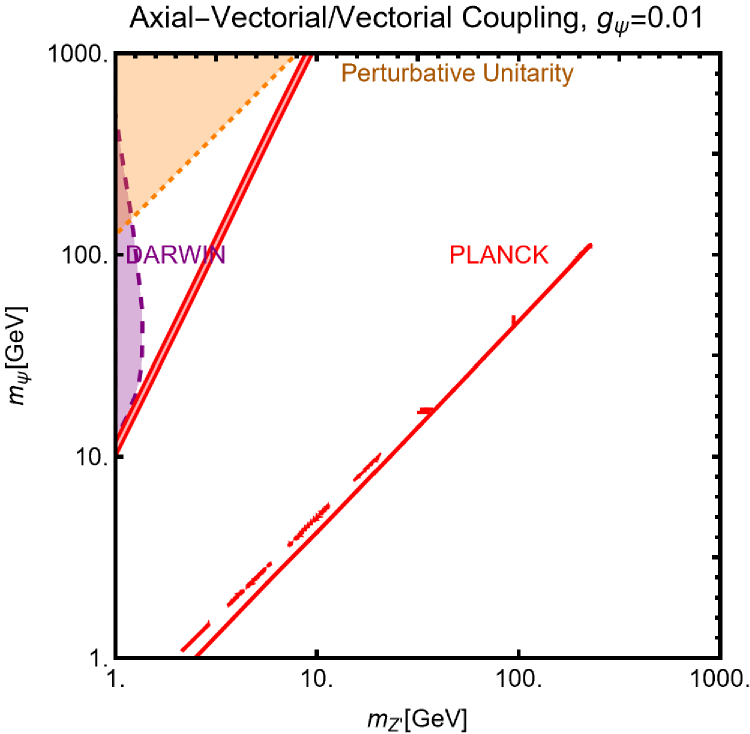}}\\
     \subfloat{\includegraphics[width=0.25\linewidth]{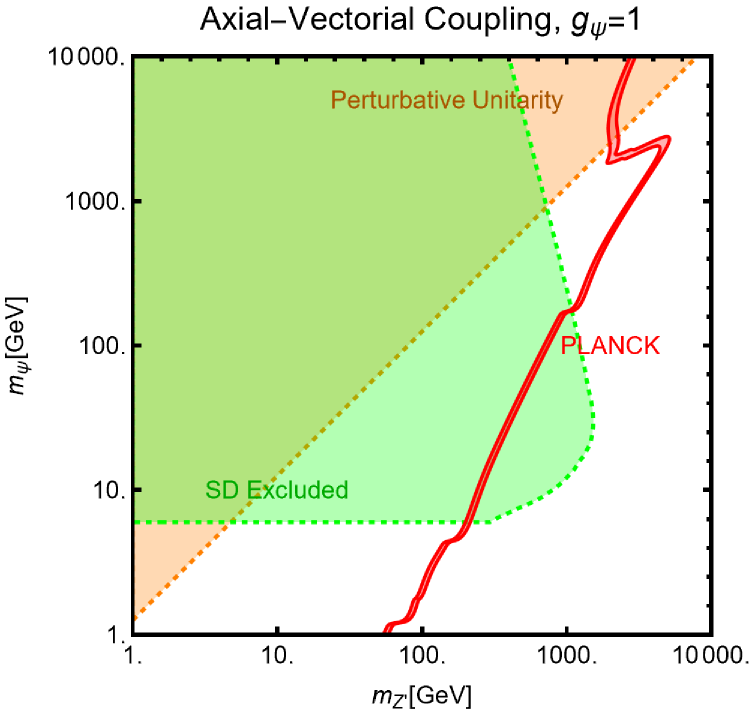}}    \subfloat{\includegraphics[width=0.25\linewidth]{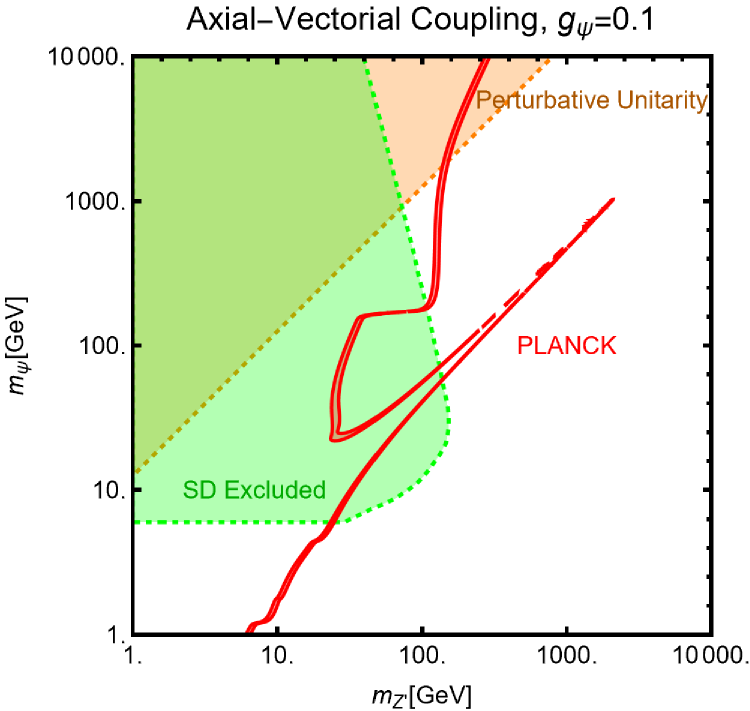}}    \subfloat{\includegraphics[width=0.25\linewidth]{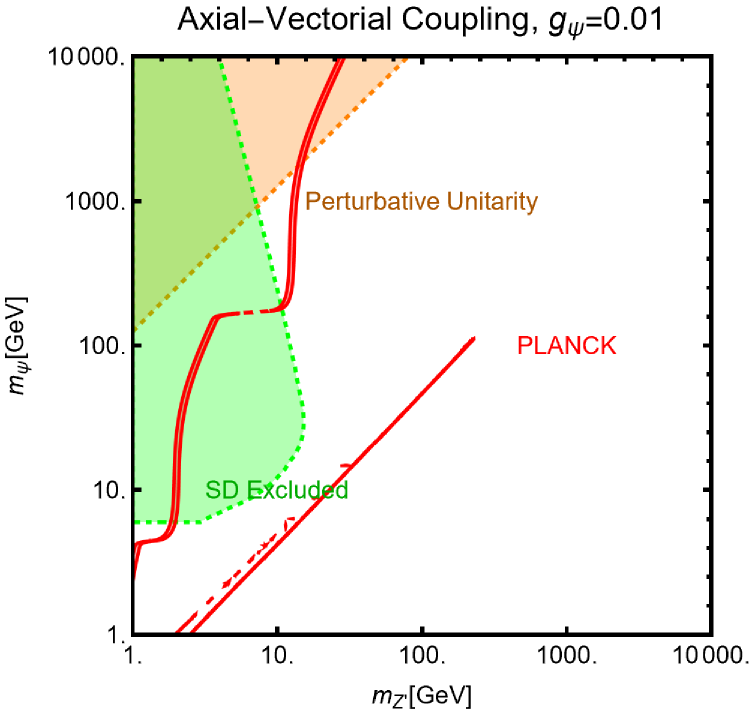}}
    \caption{Summary of constraints for a simplified model with a fermionic DM interacting via an s-channel spin-$1$ mediator $Z'$. The constrains are shown in the $(m_{Z'},m_\psi)$ plane. For each  plot, the viable parameters space is the area where the red coloured isocontours, representing the correct DM relic density, lie outside the blue and purple coloured regions.  The colour code is the same as of the ones used for Fig.~\ref{fig:Sportal}. The 
    phrases used, "Vectorial Coupling" (top row), "Vectorial/Axial-Vectorial Coupling" (second row), "Axial-Vectorial/Vectorial Coupling" (third row) and "Axial-Vectorial Coupling" (bottom row) refer to $(A_\psi^{Z'}=0,\, A_f^{Z'}=0\,\,\forall\,\,f)$,
    $(A_\psi^{Z'}=0,\, V_f^{Z'}=0\,\,\forall\,\,f)$, $(V_\psi^{Z'}=0,\, A_f^{Z'}=0\,\,\forall\,\,f)$, $(V_\psi^{Z'}=0,\, V_f^{Z'}=0\,\,\forall\,\,f)$, respectively. We refer to the main text for details on the assignations of the couplings.}
    \label{fig:Zpportal}
\end{figure*}

Similar to the case of a complex scalar DM, we combine the DM constraints in Fig.~\ref{fig:Zpportal}, in the $(m_{Z'},m_\psi)$ bidimensional plane, for three assignations of $g_\psi$ and fixing the couplings $V_{\psi,f}^{Z'},A_{\psi,f}^{Z'}$ to 1 or 0 according to the four cases previously illustrated. 

\subsection{$t$-channel portals for SM singlet DM}
Another viable class of simplified DM model relies on Yukawa interactions involving a scalar (fermion) DM $\Phi_{\rm DM}~(\Psi_{\rm DM})$, a suitable\footnote{Based on the concerned SM gauge charges, such that the overall term remains gauge invariant.} SM fermion $f_i$ and, another BSM fermion (scalar) state $\Psi_{f_i}~(\Phi_{f_i})$.  The most general interaction Lagrangians are given by \cite{Arcadi:2023imv}:
\begin{align}
 \mathcal{L}_\text{scalar}
&={\Gamma_L^{f_i}\ovl {f_i} }{P_R}{\Psi_{f_i}}{\Phi_{\rm DM}}+  {\Gamma_R^{f_i}\ovl {f_i} }{P_L}{\Psi_{f_i}}{\Phi_{\rm DM}} +
{\rm{H}}{\rm{.c.}}
\nonumber\\
& + \lambda_{1H\Phi}  (\Phi_{\rm DM}^\dagger \Phi_{\rm DM})  (H^\dagger H)\nonumber\\
& + \lambda_{2H\Phi}  (\Phi_{\rm DM}^\dagger T^a_\Phi \Phi_{\rm DM} )  (H^\dagger \dfrac{\sigma^a}{2} H),
\label{eq:tchannel_scalar_lagrangian}
\end{align}
and
\begin{align}
\mathcal{L}_\text{fermion}&=
{\Gamma_L^{f_i}\ovl {f_i} }{P_R}{\Phi_{f_i}}{\Psi_{\rm DM}}+  {\Gamma_R^{f_i}\ovl {f_i} }{P_L}{\Phi_{f_i}}{\Psi_{DM}} +
{\rm{H}}{\rm{.c.}}
\nonumber\\
& + \lambda_{1H\Phi}  (\Phi_{f_i}^\dagger \Phi_{f_i})  (H^\dagger H) \nonumber\\
& + \lambda_{2H\Phi}  (\Phi_{f_i}^\dagger T^a_\Phi \Phi_{f_i} )  (H^\dagger \dfrac{\sigma^a}{2} H), \ 
\label{eq:tchannel_fermion_lagrangian}
\end{align}
for a scalar and a fermionic DM, respectively, assuming that the concerned interactions are parametrized in the same way for both of these two scenarios. 
For the aforesaid Lagrangians. the main DM annihilation processes responsible for the relic density calculation occur via the $t$-channel exchange of $\Psi_{fi},\, \Phi_{f_i}$ which justify why these are named $t$-channel portals. Invariance of $\mathcal{L}_\text{scalar},\,\mathcal{L}_\text{fermion}$, as shown in Eqs.~\eqref{eq:tchannel_scalar_lagrangian} and (\ref{eq:tchannel_fermion_lagrangian}), under the SM gauge group suggests that $\Psi_{fi},\, \Phi_{f_i}$ should be charged at least under some of the components of the SM group, depending on how they couple to $f_i$. Even the DM itself might not be a pure SM gauge singlet but, according to the quantum numbers of $\Phi_{f_i}~{\rm or}~\Psi_{f_i}$ and the SM fermion $f_i$ to which it couples, might be just the lightest electrically neutral component of a $SU(2)_L$ multiplet. We will not consider this possibility in this subsection. Similarly, we will assume that the DM has a zero hypercharge, otherwise, its phenomenology would be dominated by the unavoidable couplings with the $Z$-boson. Classification of the possible assignations of the SM gauge quantum numbers of the BSM fields have been discussed, for example, in Refs. \cite{Arcadi:2021glq, Arcadi:2021cwg}. Contrary to the case of $s$-channel portals, discussed in the previous section, in the minimal realizations of a $t$-channel portal model the DM is coupled only with a specific quark or lepton species. Of course one could overcome this issue by introducing more mediators with different quantum numbers under the SM gauge group. As a final remark, note that we have also included, in Eqs.~\eqref{eq:tchannel_scalar_lagrangian} and (\ref{eq:tchannel_fermion_lagrangian}), the quadrilinear coupling between the BSM scalars $\Psi_{f_i},\,\Phi_{f_i}$, and the SM Higgs doublet $H$, as these renormalizable interaction terms are allowed by the SM gauge symmetry. As we will clarify below, such quadrilinear interaction plays a crucial role in the case of scalar DM $\Phi_{\rm DM}$. 

Similar to what was done in the previous subsection, we start illustrating the DM phenomenology from the relic density calculation. Contrary to the case of $s$-channel simplified models, one should adopt an effective thermally averaged cross-section as coannihilation processes associated with the $t$-channel mediator are present and might be important for the relic density. The effective annihilation cross-section of the DM can be schematically written as \cite{Bai:2013iqa}:
%
\begin{align}
\langle \sigma v \rangle_{\rm eff}~ &=\frac{1}{2}\langle \sigma v \rangle_{\rm DM\, DM}\frac{g_{\rm DM}^2}{g_{\rm eff}^2}\nonumber\\
& +\langle \sigma v \rangle_{\rm DM\, M}\frac{g_{\rm DM} g_{\rm M}}{g^2_{\rm eff}}{\left(1+\tilde{\Delta}\right)}^{3/2} \exp\left[-x \tilde{\Delta} \right] 
\nonumber\\
& + \frac{1}{2}\langle \sigma v \rangle_{\rm M^{\dagger}M}\frac{g_{\rm M}^2}{g_{\rm eff}^2}{\left(1+\tilde{\Delta}\right)}^3 \exp\left[-2 x \tilde{\Delta} \right]\,,
\end{align}
in the case of complex scalar or Dirac fermionic DM. In the case of real scalar or Majorana fermion, we have a slightly different expression:

\begin{align}
\langle \sigma v \rangle_{\rm eff} &= \langle \sigma v \rangle_{\rm DM \,DM}\frac{g_{\rm DM}^2}{g_{\rm eff}^2}\nonumber\\
& +\langle \sigma v \rangle_{\rm DM\, M}\frac{g_{\rm DM} g_{\rm M}}{g^2_{\rm eff}}{\left(1+\tilde{\Delta}\right)}^{3/2} \exp\left[-x \tilde{\Delta} \right]
\nonumber
\\
& + \left(\langle \sigma v \rangle_{\rm M^{\dagger}M}+\langle \sigma v \rangle_{\rm M\,M}\right)\nonumber\\
&\times\frac{g_{\rm M}^2}{g_{\rm eff}^2}{\left(1+\tilde{\Delta}\right)}^3 \exp\left[-2 x \tilde{\Delta} \right]\,.
\end{align}
In the above equations $\tilde{\Delta}={(M_{\rm M}-M_{\rm DM})}/{M_{\rm DM}}$ denotes relative splitting between the 
DM mass $M_{\rm DM}$ and the mediator mass $M_{\rm M}$ with respect to $M_{\rm DM}$ while:
\begin{equation}
g_{\rm eff}=g_{\rm DM}+g_{\rm M}{\left(1+\tilde{\Delta}\right)}^{3/2}\exp\left[-x \tilde{\Delta} \right]\,,
\end{equation}
with $g_{\rm M}$ and $g_{\rm DM}$ denoting the internal $dof$ of the mediator and the DM. $x$ is the temperature parameter $\sim {\rm Mass}/T$.
Given the exponential suppression, the contribution from coannihilations is relevant only for small $\tilde{\Delta}$, typically remaining below $20\%$. A rigorous numerical treatment is required to solidly account for connaihilations though. Assuming a sufficiently large mass splitting, coannihilations are not relevant, and the relic density due to DM pair annihilations into SM fermions are found to be \cite{Bai:2013iqa,Bai:2014osa,Giacchino:2015hvk,Arcadi:2017kky}:
\begin{align}
 \langle \sigma v \rangle_{{\rm DM}\, {\rm DM}}^{\text{Complex}}
&=\frac{3 |\Gamma^f_{L,R}|^4 m_f^2}{2 \pi {\left(M_{\Psi_f}^2+M_{\Phi_{\rm DM}}^2-m_f^2\right)}^2} \nonumber\\
&\times \left(1-\frac{m_f^2}{M_{\Phi_{\rm DM}}^2}\right)\nonumber\\
& + n_c^f\frac{|\Gamma^f_{L,R}|^4 M_{\Phi_{\rm DM}}^2 v^2}{48 \pi {\left(M_{\Phi_{\rm DM}}^2+M_{\Psi_f}^2\right)}^2},\nonumber
\end{align}
\begin{align}
    & \langle \sigma v \rangle_{{\rm DM}\, {\rm DM}}^{\text{Dirac}}=
 n_c^f\frac{|\Gamma^f_{L,R}|^4 M_{\Psi_{\rm DM}}^2}{32 \pi {\left(M_{\Psi_{\rm DM}}^2+M_{\Phi_f}^2\right)}^2},\nonumber
\end{align}
\begin{align}
    \langle \sigma v \rangle_{{\rm DM}\, {\rm DM}}^{\text{Real}}&=\frac{12 |\Gamma^t_{L,R}|^4 }{\pi}{\left(1-\frac{m_f^2}{M_{\Phi_{\rm DM}}^2}\right)}^{3/2}
\nonumber\\
& \times \frac{m_f^2}{{\left(M_{\Phi_{\rm DM}}^2+M_{\Psi_f}^2-m_f^2\right)}^2} \nonumber\\
&+ n_c^f\frac{|\Gamma^f_{L,R}|^4 M_{\Phi_{\rm DM}}^6 v^4}{60 \pi {\left(M_{\Phi_{\rm DM}}^2+M_{\Psi_f}^2\right)}^4},\nonumber
\end{align}
\begin{align}
  \langle \sigma v \rangle_{{\rm DM}\,{\rm DM}}^{\text{Majorana}}&=\frac{3 |\Gamma_{L,R}^f|^4}{2\pi}\frac{m_f^2}{{\left(M_{\Phi_f}^2+M_{\Psi_{\rm DM}}^2-m_f^2\right)}^2}\nonumber\\
  &\times \sqrt{1-\frac{m_f^2}{M_{\Psi_{\rm DM}}^2}}\nonumber\\
&+ n_c^f\frac{|\Gamma^f_{L,R}|^4 M_{\Psi_{\rm DM}}^2 \left(M_{\Psi_{\rm DM}}^4+M_{\Phi_f}^4\right) v^2}{48 \pi {\left(M_{\Psi_{\rm DM}}^2+M_{\Phi_f}^2\right)}^4}\ .
\end{align}

%
To achieve this analytical approximation we have considered the leading order term in the velocity expansion in the limit $m_f \rightarrow 0$ and then added to it the leading helicity suppressed contribution. In the previous expressions, the sum is carried out over the kinematically accessible final states.

In the limit $m_f\rightarrow 0$, only in the case of a Dirac fermionic DM, we have an $s$-wave-dominated DM annihilation cross-section. In the cases of Majorana fermion and complex scalar DM, we have a velocity suppression. In contrast,for the case of a real scalar DM, we have the very peculiar scenario of a $d$-wave, i.e., $v^4$ suppressed, cross-section. The velocity suppression is, however, lifted in the case when the DM mass is not too far from the one of a fermionic final state. As can be easily argued, this kind of scenario mostly occurs when the DM can annihilate into top-quark pairs.

The study of the DD is more complicated for the $t$-channel portals, compared to the case of $s$-channel portals. Let us first consider the case of a scalar (real/complex) DM $\Phi_{\rm DM}$. The low-energy effective Lagrangian for the DD is given by the following expression:
\begin{align}    
L_{\rm eff}^{{\rm Scalar},q}
&=  \sum_{q=u,d} c^q
\left( \Phi_{\rm DM}^\dagger i\overset{\leftrightarrow}{\partial_\mu} \Phi_{\rm DM}\right) \bar q \gamma^\mu q \nonumber\\
& + \sum_{q=u,d,s} d^q m_q \Phi_{\rm DM}^\dagger \Phi_{\rm DM}\, \bar q q \nonumber\\
&+ d^g \frac{\alpha_s}{\pi}\Phi_{\rm DM}^\dagger \Phi_{\rm DM} \, G^{a\mu \nu}G^a_{\mu \nu}
\nonumber\\ 
& + \sum_{q=u,d,s} \frac{g_1^q}{M_{\Phi_{\rm DM}}^2}
\Phi_{\rm DM}^\dagger (i \partial^\mu)(i \partial^\nu ) \Phi_{\rm DM}\, \mathcal{O}^{q}_{\mu \nu}\nonumber\\
& + \frac{g_1^g}{M_{\Phi_{\rm DM}}^2}
\Phi_{\rm DM}^\dagger (i \partial^\mu)(i \partial^\nu) \Phi_{\rm DM}\, \mathcal{O}^{g}_{\mu \nu}
\ ,
\label{Scalar:leff}
\end{align}
where $\mathcal{O}^{q}_{\mu \nu}$ and $\mathcal{O}^{g}_{\mu \nu}$ are the twist-2 components:
\begin{align}
& \mathcal{O}^q_{\mu \nu}=\bar q \left(\dfrac{iD_\mu \gamma_\nu+iD_\nu \gamma_\mu}{2}-\frac{1}{4}g_{\mu \nu}i\slashed{D}\right)q, \nonumber\\
& \mathcal{O}^g_{\mu \nu}=G_\mu^{a\rho} G^a_{\nu \rho}-\frac{1}{4}g_{\mu \nu}G^a_{\rho \sigma}G^{a\rho \sigma}\,.
\end{align}
The Lagrangian $L_{\rm eff}^{{\rm Scalar},q}$ leads to the following SI cross-section:

\begin{equation}
\sigma_{\Phi_{\rm DM}}^{{\rm SI},\, p}= \frac{\mu_{\Phi_{DM}\,p}^2}{\pi}\, \frac{\left[Z f_p +(A-Z)f_n\right]^2}{A^2}\,,
\end{equation}
with:
\begin{align}\label{eq:ffsDM}
f_{N=p,n}
&=c^N +M_N\sum_{q=u,d,s}\left(f_q^N d_q  +\frac{3}{4}g_1^q \left(q(2)+\bar{q}(2)\right)\right)
\nonumber\\
& +\frac{3}{4}m_N\sum_{q=c,b,t}g_1^g G(2) -\frac{8}{9}f_{TG}f_G.
\end{align}
The form factors $\bar q(2),q(2),G(2)$ are defined by:
\begin{align}
& \langle N |\mathcal{O}^q_{\mu \nu} |N \rangle
=
\frac{1}{m_N}\left(p_\mu p_\nu-\frac{1}{4}m_N^2 g_{\mu \nu}\right) \left(\bar q(2)+q(2)\right)\ ,
\nonumber\\
& \langle N |\mathcal{O}^g_{\mu \nu} |N \rangle
=
\frac{1}{m_N}\left(p_\mu p_\nu-\frac{1}{4}m_N^2 g_{\mu \nu}\right) G(2)\
\end{align}
Again, we have adopted the micrOMEGAs defaults for their values.

Let's now illustrate the Wilson coefficients entering the cross-section. For more details, we refer to Ref. \cite{Arcadi:2023imv}. 

\begin{figure*}
\begin{center}
\begin{tabular}{ccc}
\subfloat[]{\includegraphics[width=0.3\textwidth]{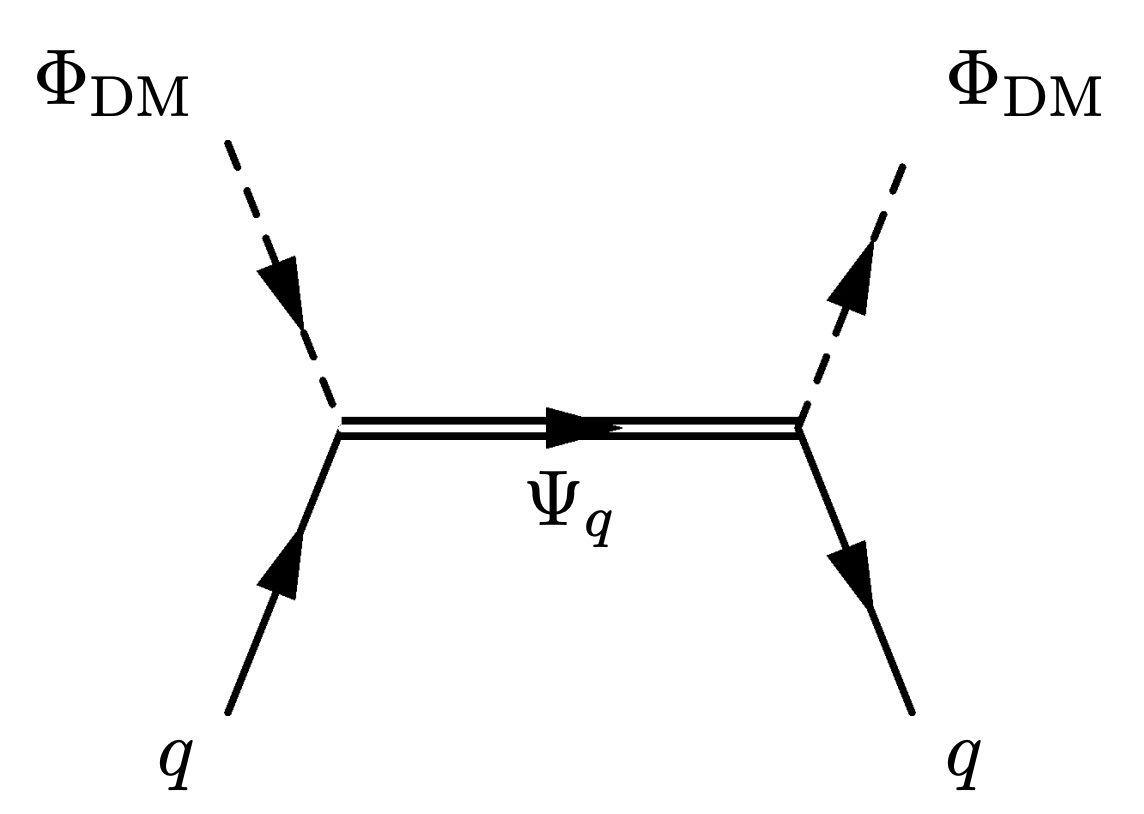}}
&
\subfloat[]{\includegraphics[width=0.3\textwidth]{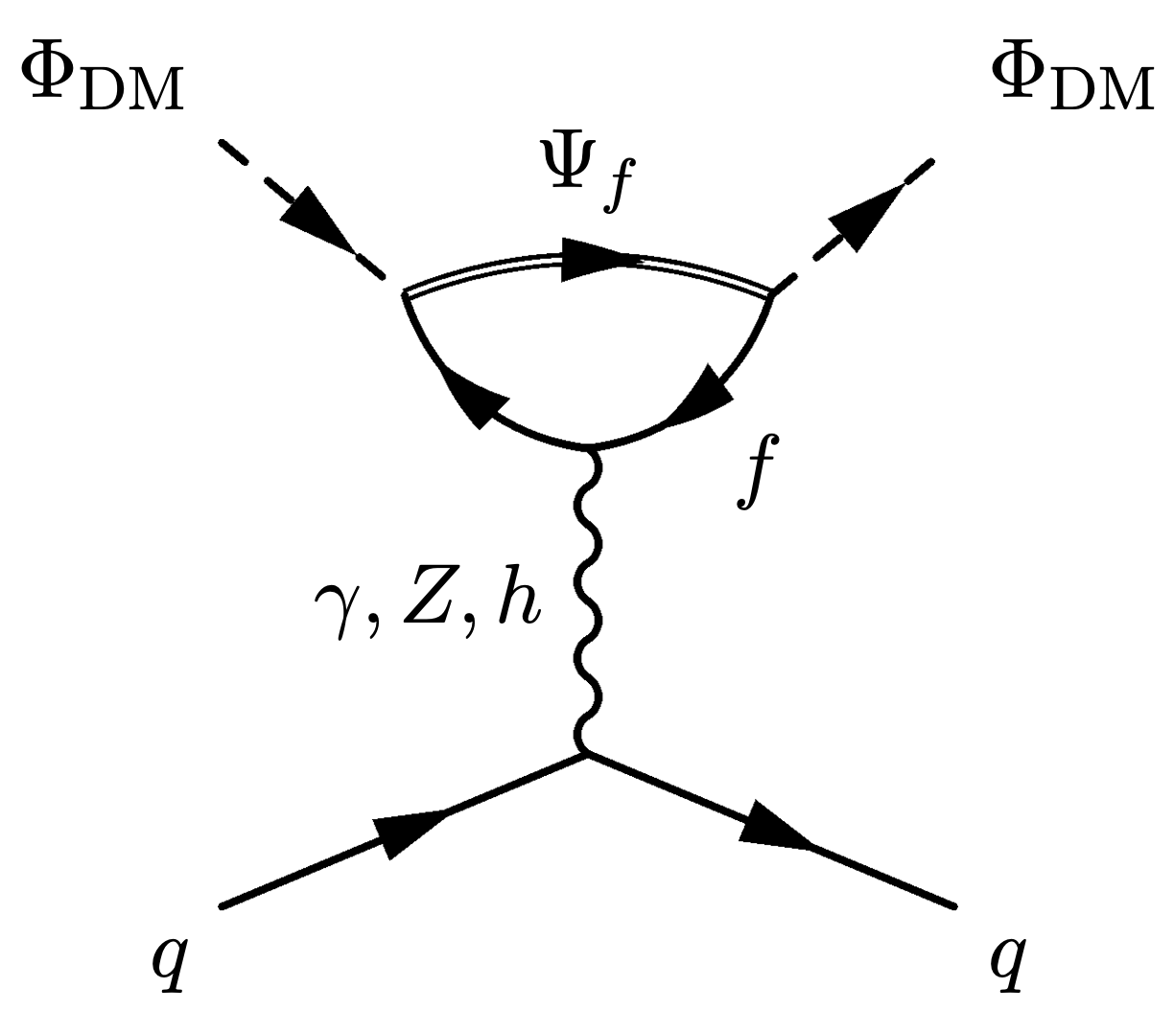}}
&
\subfloat[]{\includegraphics[width=0.3\textwidth]{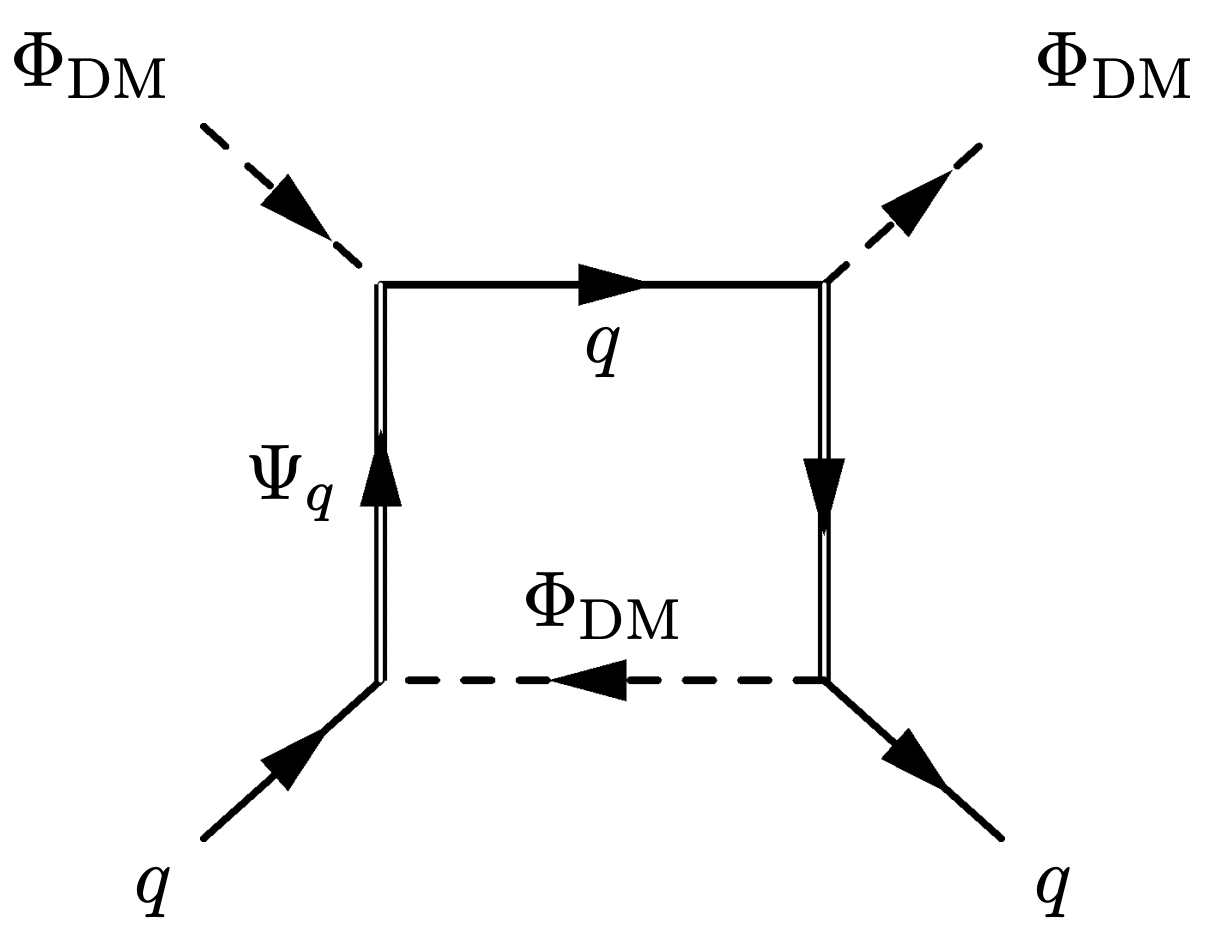}}
\\
\subfloat[]{\includegraphics[width=0.3\textwidth]{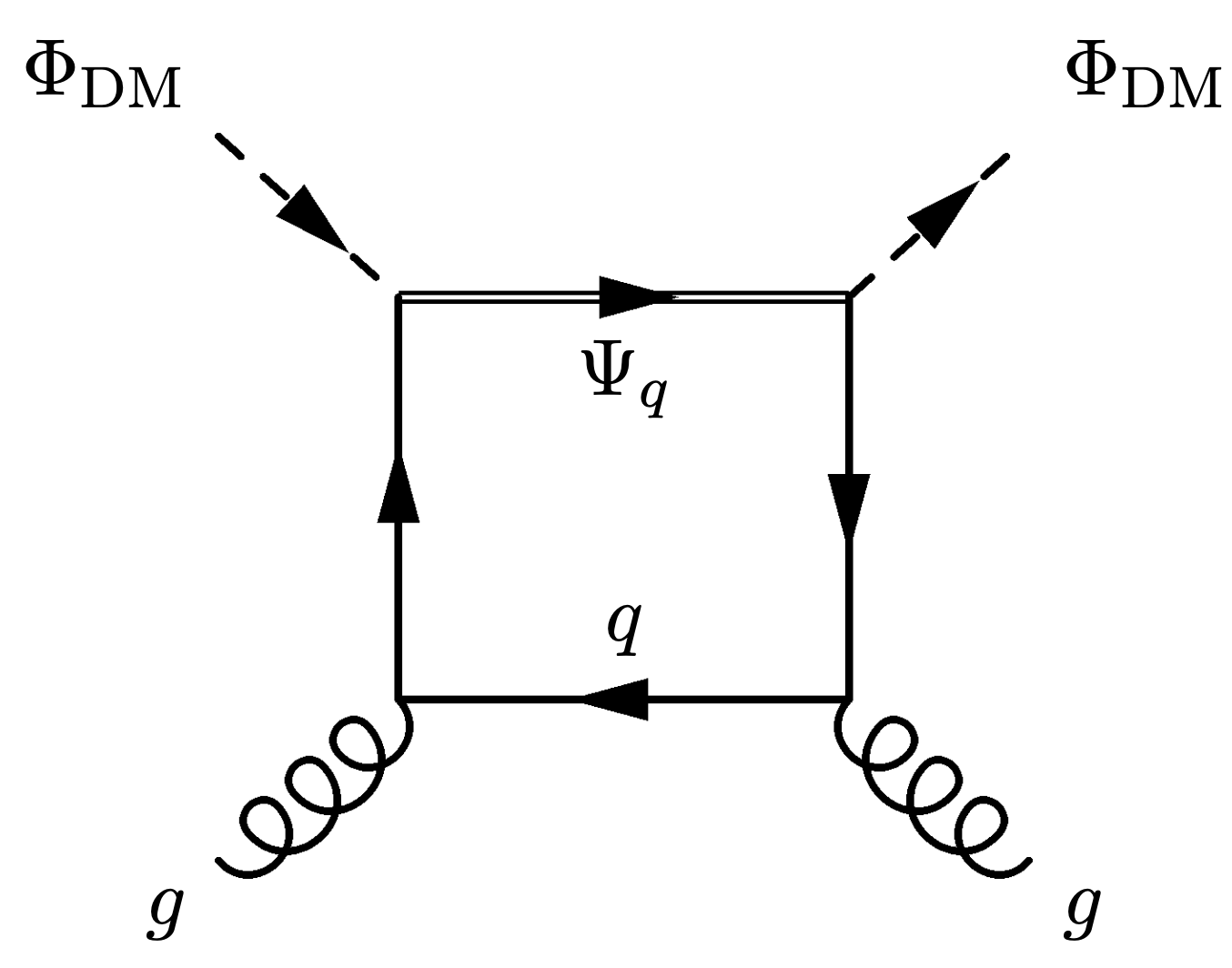}}
&
\subfloat[]{\includegraphics[width=0.3\textwidth]{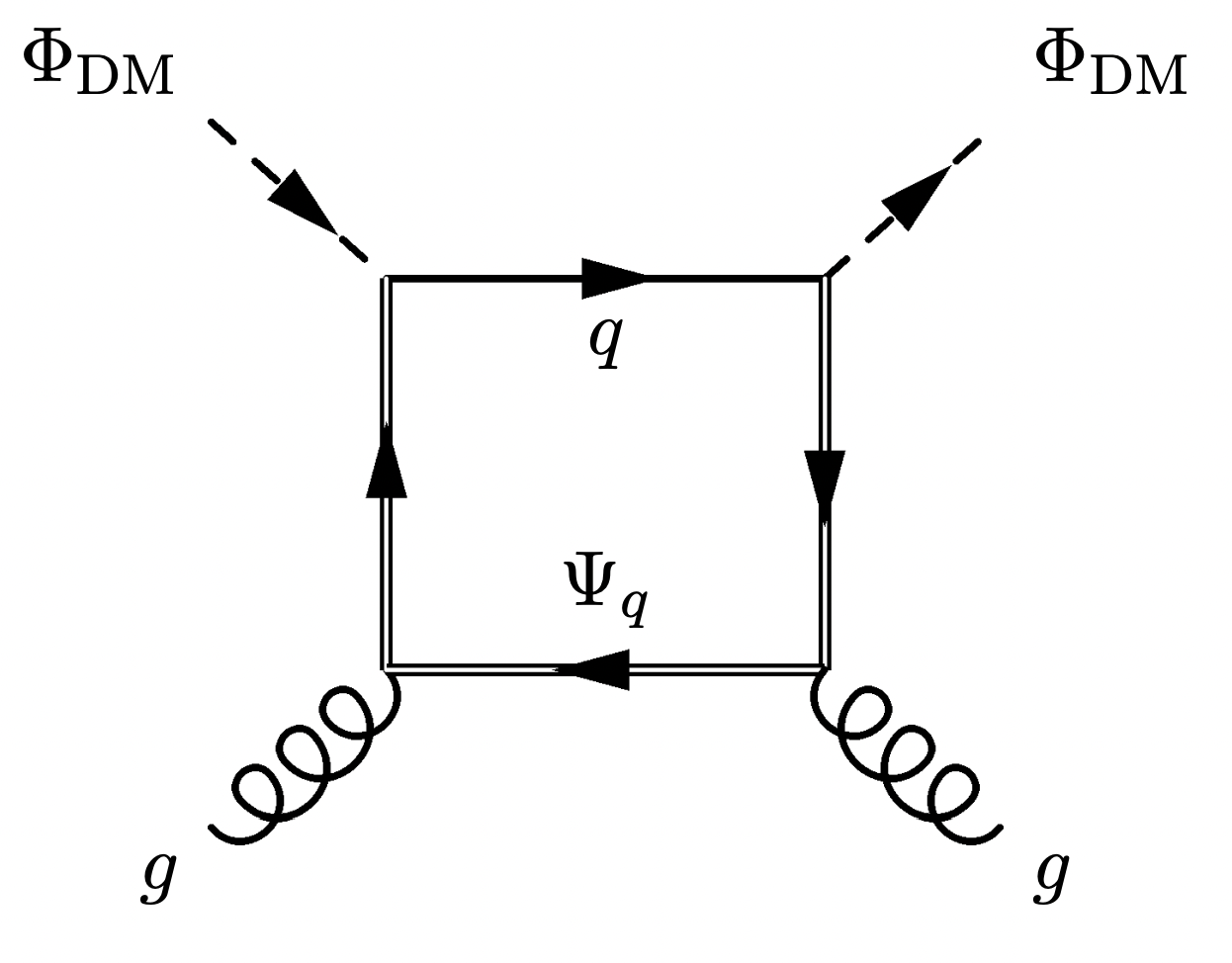}}
&
\subfloat[]{\includegraphics[width=0.3\textwidth]{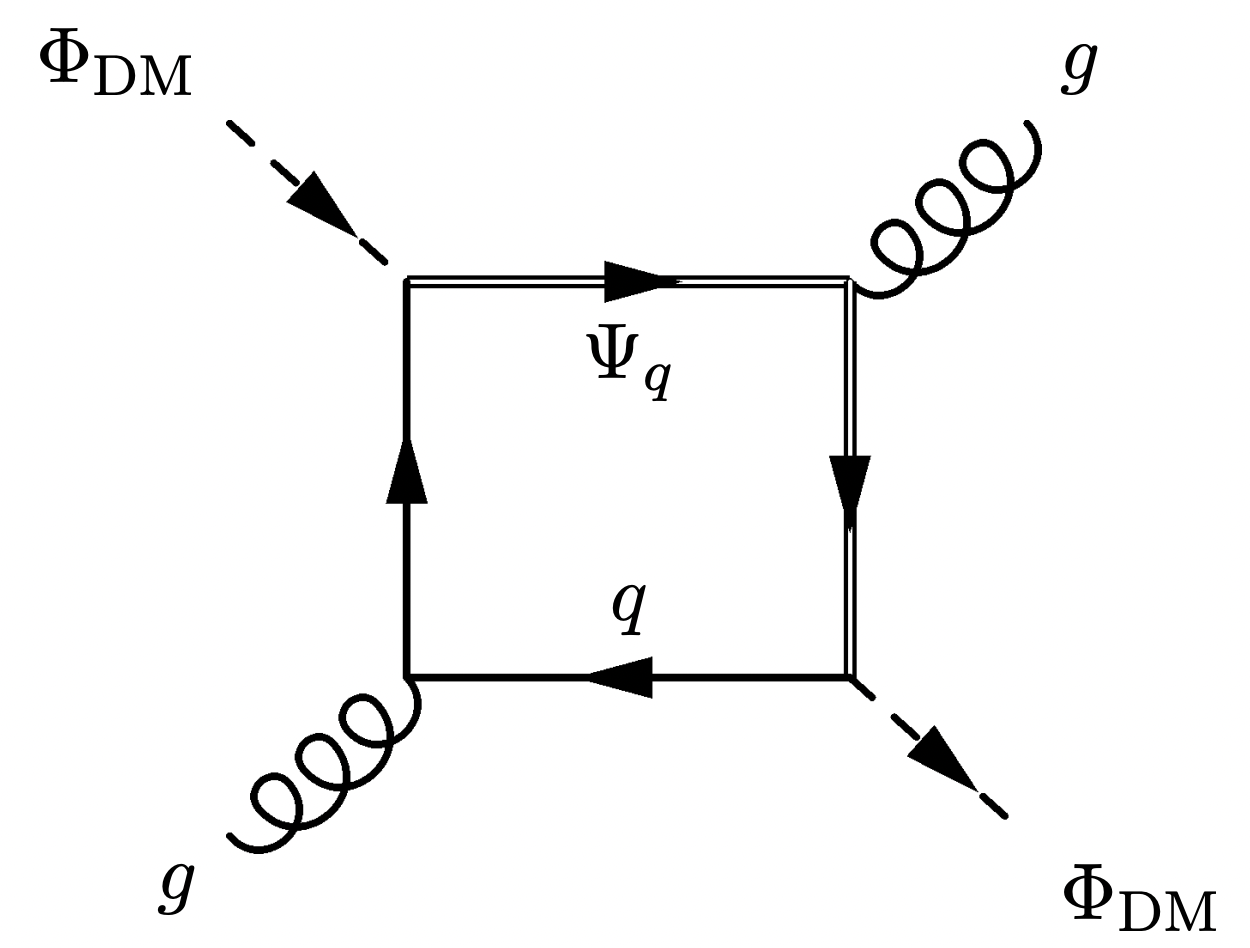}}
\\
\end{tabular}
\end{center}
\caption{{\it Feynman diagrams contributing to the Wilson coefficients in the effective Lagrangian for the DM DD in the case of a real/complex scalar DM $\Phi_{\rm DM}$. Diagram (b) has a partner (not shown) interchanging $\Psi_{f_i}$ with $f_i$. }}
\label{diags}
\end{figure*}

The operator proportional to the quark current receives for kind of contributions:
\begin{align}
\label{eq:Wilson_current}
c^{u,d} = c^{u,d}_{\text{tree}} +  c^{u,d}_{Z} + c^{u,d}_{\gamma}   + c^{u,d}_{\text{box}}\ .
\end{align}
The first, dubbed tree, is obtained just by integrating out the fermionic mediator: 
\begin{equation}
c^{q=u,d}_{\rm tree}
= -\dfrac{|\Gamma_{L,R}^q|^2}{ 4(M_{\Phi_{\rm DM}}^2-M_{\Psi_f}^2)}.
\end{equation}
Since the quark current should be evaluated only for the valence quark, the tree-level contribution to the Wilson coefficient exists only if the DM is coupled with the up and/or down quark. In the absence of such a coupling, loop contributions to the Wilson coefficient become important. In this context, $\gamma$ and $Z$ penguin diagrams generate the coefficients called,  $c^{u,d}_\gamma$ and $c^{u,d}_Z$, respectively:
\begin{align}
\label{eq:Sgamma}
& c^{q=u,d}_{\gamma} = \sum_{f}\frac{\alpha_{\rm em} |\Gamma_{L,R}^f|^2 n^f_c Q_f Q_q}{24 \pi M^2_{\Psi_f} } \mathcal{F}_\gamma
\Bigg[\frac{m_f^2}{M_{\Psi_f}^2},\frac{M_{\Phi_{DM}}^2}{M_{\Psi_f}^2}\Bigg]\ ,
\end{align}
with 
\begin{align}
\mathcal{F}_\gamma (x_f,x_\phi) &=
\frac{1}{x_\phi^2 }
\bigg\{
\frac{(1+x_{f}+2 x_{\phi})\log (x_{f})}{2}\nonumber\\
& -\frac{(x_f-1)(1+x_f-x_\phi)x_\phi}{\Delta}
\nonumber\\
& + \frac{(1-x_f)}{\Delta^{3/2}} \Big(x_{f}^{3}-x_{f}^{2}(1+x_{\phi})\nonumber\\
& +(1-x_{\phi})^{2}(1+x_{\phi})-x_{f}(1+10 x_{\phi}+x_{\phi}^{2})\Big)\nonumber\\
&\times \log\bigg[\frac{1+x_f-x_{\phi}+\sqrt{\Delta}}{2\sqrt{x_{f}}}\bigg] \bigg\}\ ,
\end{align}
where
\begin{equation}
\Delta= x_{f}^{2}-2 x_{f}(1+x_{\phi})+(1-x_{\phi})^{2}\ .
\label{Delta}
\end{equation}
Similarly,
\begin{align}
 c_Z^{q=u,d}=&\frac{G_F}{4\sqrt{2}}\sum_f \frac{T_3^f (T_3^q-2Q_q s_W^2)n^f_c| \Gamma_{L,R}^f|^2}{\pi^2}\,\frac{m_f^2}{M_{\Psi_f}^2} \nonumber \\
& \times \mathcal{F}_Z \bigg[\frac{m_f^2}{M_{\Psi_f}^2},\frac{M_{\Phi_{DM}}^2}{M_{\Psi_f}^2}\bigg]\,,
\end{align}
with $G_F$ being the Fermi constant while $T_3^f$ and $Q_f$ are, the weak isospin and electric charge, respectively, of the fermions in the internal and/or external lines in the loop diagram. Again $s_W=\sin \theta_W$.
\begin{align}
\mathcal{F}_Z(x_f,x_\phi)&=
\frac{1}{x_\phi}+\frac{1-x_f+x_\phi}{2 x_\phi^2}\log{x_f}\nonumber\\
& +\frac{1-2x_f + (x_f-x_\phi)^2}{x_\phi^2\sqrt{\Delta}}\nonumber\\
&\times\log{\bigg(\frac{1+x_f-x_\phi+\sqrt{\Delta}}{2\sqrt{x_f}}\bigg)}.
\end{align}
The coefficients $c_{\gamma, Z}^q$ are particularly relevant as they are present even if the quantum numbers of the $t$-channel mediator allow only couplings with the SM leptons.
The final loop contribution to $c^{q=u,d}$ comes from a box topology and it is written as:
\begin{align}
 c^{q=u,d}_{\text{box}} =&\sum_{f}\frac{1}{4}\dfrac{n_c^f |\Gamma_{L,R}^f|^2 |\Gamma_{L,R}^q|^2}{32 \pi^2 M_{\Phi_{\rm DM}}^2}\,  \mathcal{F}_\text{box}\Bigg[\frac{m_f^2}{M_{\Psi_f}^2},\frac{M_{\Phi_{\rm DM}}^2}{M_{\Psi_f}^2}\Bigg],
\end{align}
with $f$ or $f_i$, as per Eq. (60).
\begin{align}
& \mathcal{F}_\text{box}\left(x_f,x_\phi\right)
=
\frac{x_f-x_\phi}{x_\phi-1}+\frac{\beta_2 x_f^2(x_f-3 x_\phi)}{x_\phi(x_\phi-1)^2}\log \Bigg[\frac{\sqrt{x_f}(1+\beta_2)}{2\sqrt{x_\phi}}\Bigg]
\nonumber\\
& -\frac{x_f+x_\phi}{2x_\phi}\log{\left[x_f\right]}+\frac{(x_f^3-5x_f^2 x_\phi+4 x_f x_\phi^2+2x_\phi^3)}{2x_\phi(x_\phi-1)^2}\log\left[x_\phi\right]
\nonumber\\
& +\frac{1}{x_\phi \sqrt{\Delta}(x_\phi-1)^2}\Big[-x_f^4+x_\phi(x_\phi-1)^3+x_f(x_\phi-1)^2(2x_\phi-1)\nonumber\\
&+x_f^3(1+6x_\phi) x_f^2(1-x_\phi(5+8x_\phi))\Big]\log\Bigg[\frac{1+x_f-x_\phi+\sqrt{\Delta_\phi}}{2\sqrt{x_f}}\Bigg],
\end{align}
with
\begin{equation}
\beta_2=\sqrt{1-4 \frac{x_\phi}{x_f}}\ .
\end{equation}
The operator proportional to the quark scalar bilinear $\bar q q$ is originated by two classes of Feynmann diagrams. The first is the Higgs penguin:
\begin{align}
& d_H^{q}=\sum_f \frac{g^2 |\Gamma^f_{L,R}|^2 M_{\Psi_f}^2}{32\pi^2 M_h^2 M_W^2}\mathcal{F}_H\left(\frac{m_f^2}{M_{\Psi_f}^2},\frac{M_{\Phi_{\rm DM}}^2}{M_{\Psi_f}^2}\right),
\end{align}
with 
\begin{align}
& \mathcal{F}_H (x_f,x_\phi)=
2 x_f \nonumber\\
& +2x_f\frac{x_f^2+(x_\phi-1)x_\phi-x_f(1+2x_\phi)}{x_\phi \sqrt{\Delta}}\log{\left[\frac{1+x_f-x_\phi+\sqrt{\Delta}}{2 \sqrt{x_f}}\right]}
\nonumber\\
& + \frac{x_f(x_\phi-x_f)}{x_\phi}\log{x_f}+\frac{32 \pi^2 M_W^2\lambda_{1H\Phi}^{(f)}(\mu)}{g^2 M_{\Psi_f}^2 |\Gamma^f_{L,R}|^2}
+ 2 x_f \log{\frac{\mu^2}{m_f^2}}.
\end{align}
where  $\lambda_{1H\Phi}^{(f)}$ is defined by the relation:
\begin{align}
    & \lambda_{1H\Phi}(\mu)=\lambda_{1H\Phi}(M)-\sum_f \lambda_{1H\Phi}^{(f)}\,\,\,\,\mbox{with} \nonumber\\
    & \lambda_{1H\Phi}^{(f)}=\log\frac{\mu^2}{M^2} \frac{g_2^2 m_f^2 |\Gamma_{L,R}|^2}{16 \pi^2 M_W^2}
\end{align}
with $\mu$ being the scale at which the RG evolved coupling is computed while $M$ is the scale at which the initial condition of the RG evolution is set. We assume $M=M_{\Psi_f}$.
The feasible diagrams are depicted in the top row of Fig. \ref{diags}. As evident, the Wilson coefficient $d_H^{q}$, depends explicitly on the DM-Higgs coupling $\lambda_{1H\Phi}$, computed at the renormalization scale $\mu$. As shown in Ref.~\cite{Arcadi:2023imv}, the presence of such coupling is necessary to make the Wilson coefficient finite. Alternatively stated, the Wilson coefficient $d^q_H$ is interpreted as a radiative correction to the $\Phi^\dagger_{\rm DM}\Phi_{\rm DM} H^\dagger H$ coupling.
The second diagram topology relies on the box diagrams, this time without the presence of the DM particle in the internal lines, as shown in the bottom row of Fig. \ref{diags}. The corresponding Wilson coefficient is written as:
\begin{align}
& d_{\rm QCD}^g=\frac{1}{2}\frac{|\Gamma_{L,R}^q|^2 }{24  M_{\Psi_q}^2}\,
\mathcal{F}_{gg}^{(1)}\Bigg[\frac{m_q^2}{M_{\Psi_q}^2},\frac{M_{\Phi_{\rm DM}}^2}{M_{\Psi_q}^2}\Bigg],
\end{align}
with
\begin{align}
& \mathcal{F}_{gg}^{(1)}\left(x_f,x_\phi\right)=
\frac{12 \log \left[\frac{1+x_{f}-x_{\phi}+\sqrt{\left(x_\phi-1\right)^{2}-2\left(1+x_{\phi}\right) x_{f}+x_{f}^{2}}}{2 \sqrt{x_{f}}}\right]}{\left(x_{f}^{2}+\left(-1+x_{\phi}\right)^{2}-2 x_{f}\left(1+x_{\phi}\right)\right)^{5 / 2}}
\nonumber\\
& \times x_{f}x_\phi\left(1+x_{f}-x_{\phi}\right)\nonumber\\
& -x_\phi \frac{x_{f}^{2}-2 x_{f}\left(-5+x_{\phi}\right)+\left(-1+x_{\phi}\right)^{2}}{\left(x_{f}^{2}+\left(-1+x_{\phi}\right)^{2}-2 x_{f}\left(1+x_{\phi}\right)\right)^{2}},
\end{align}
Moving finally to the coefficients associated with the twist-2 operators we have:
\begin{align}
& g_{1,\rm QCD}^q =
|\Gamma_{L,R}^q|^2\frac{ M_{\Phi_{\rm DM}}^2}{{\left(M_{\Psi_q}^2-M_{\Phi_{\rm DM}}^2\right)}^2} \ ,
\nonumber\\
& g_{1,\rm QCD}^g =
\frac{1}{2}\frac{|\Gamma_{L,R}^q|^2 \alpha_s M_{\Phi_{DM}}^2}{6\pi M_{\Psi_q}^4}\,
\mathcal{F}_{gg}^{(2)}\Bigg[\frac{m_q^2}{M_{\Psi_q}^2},\frac{M_{\Phi_{DM}}^2}{M_{\Psi_q}^2}\Bigg] ,
\end{align}
with
\begin{align}
    & \mathcal{F}_{gg}^{(2)}\left(x_f,x_\phi\right)=\frac{3\left(x_{f}^{2}+\left(-1+x_{\phi}\right)^{2}-2 x_{f}\left(3+x_{\phi}\right)\right)}{\left(x_{f}^{2}+\left(-1+x_{\phi}\right)^{2}-2 x_{f}\left(1+x_{\phi}\right)\right)^{2}}\nonumber\\
&  -\frac{4 \log \left[\frac{1+x_{f}-x_{\phi}+\sqrt{\left(x_{f}-1\right)^{2}-2\left(1+x_{f}\right) x_{\phi}+x_{\phi}^{2}}}{2 \sqrt{x_{f}}}\right]}{\left(x_{f}^{2}+\left(x_{\phi}-1\right)^{2}-2 x_{f}\left(1+x_{\phi}\right)\right)^{5 / 2}}\nonumber\\
& \times \left(1+x_{f}-x_{\phi}\right)\left(x_{f}^{2}+\left(x_{\phi}-1\right)^{2}-x_{f}\left(5+2 x_{\phi}\right)\right) .
\end{align}

We remark again that the expressions above have been determined under the hypothesis that the DM is an SM gauge singlet. In case the DM is part of a $SU(2)_L$ multiplet, additional contributions might arise at the loop level from the couplings of the DM with the SM gauge bosons. To our best knowledge, no complete computation has been performed for this scenario to date. not be accounted for here. We already said this before. 
Another important remark is that in the case of a real scalar DM, there is no contribution to the cross-section from the Wilson coefficient $c^q$, as the corresponding operator automatically vanishes.

\begin{figure*}
\begin{center}
\begin{tabular}{ccc}
\subfloat[]{\includegraphics[width=0.3\textwidth]{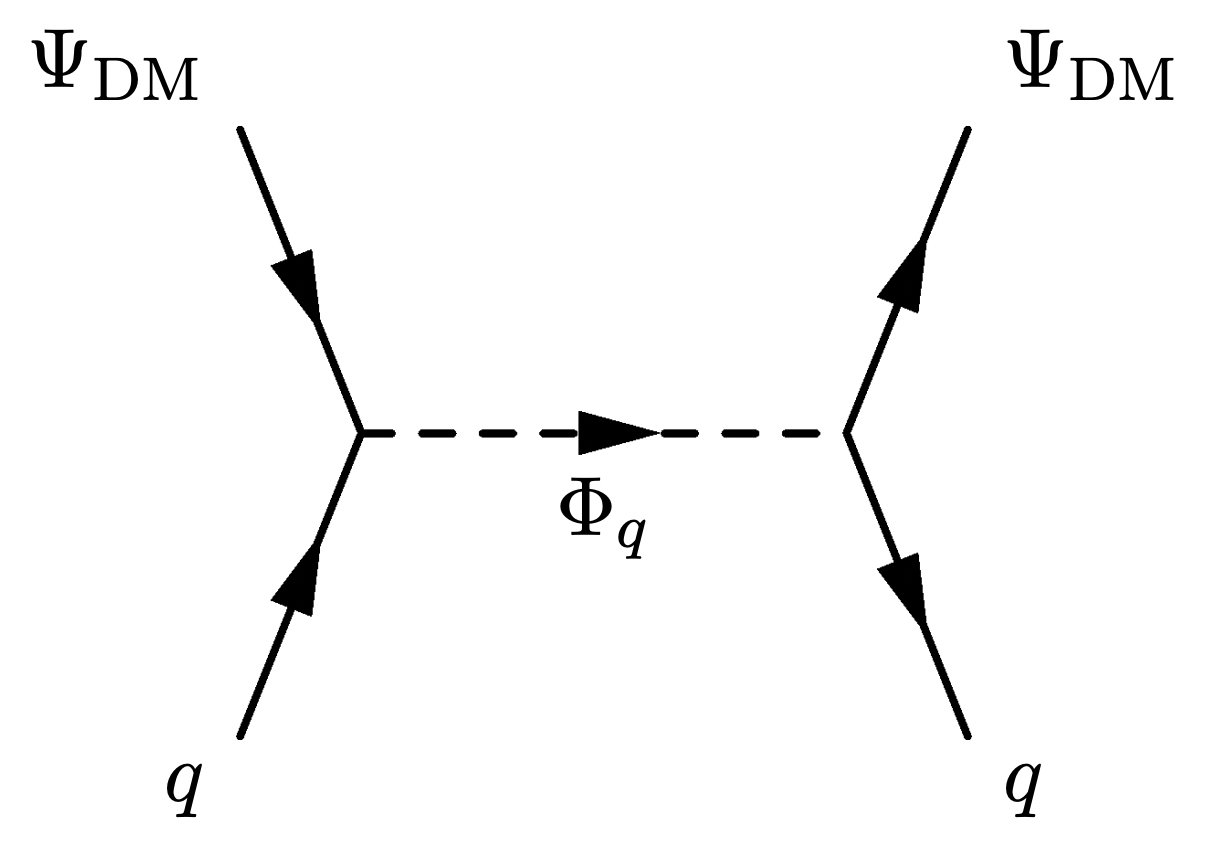}}
&
\subfloat[]{\includegraphics[width=0.3\textwidth]{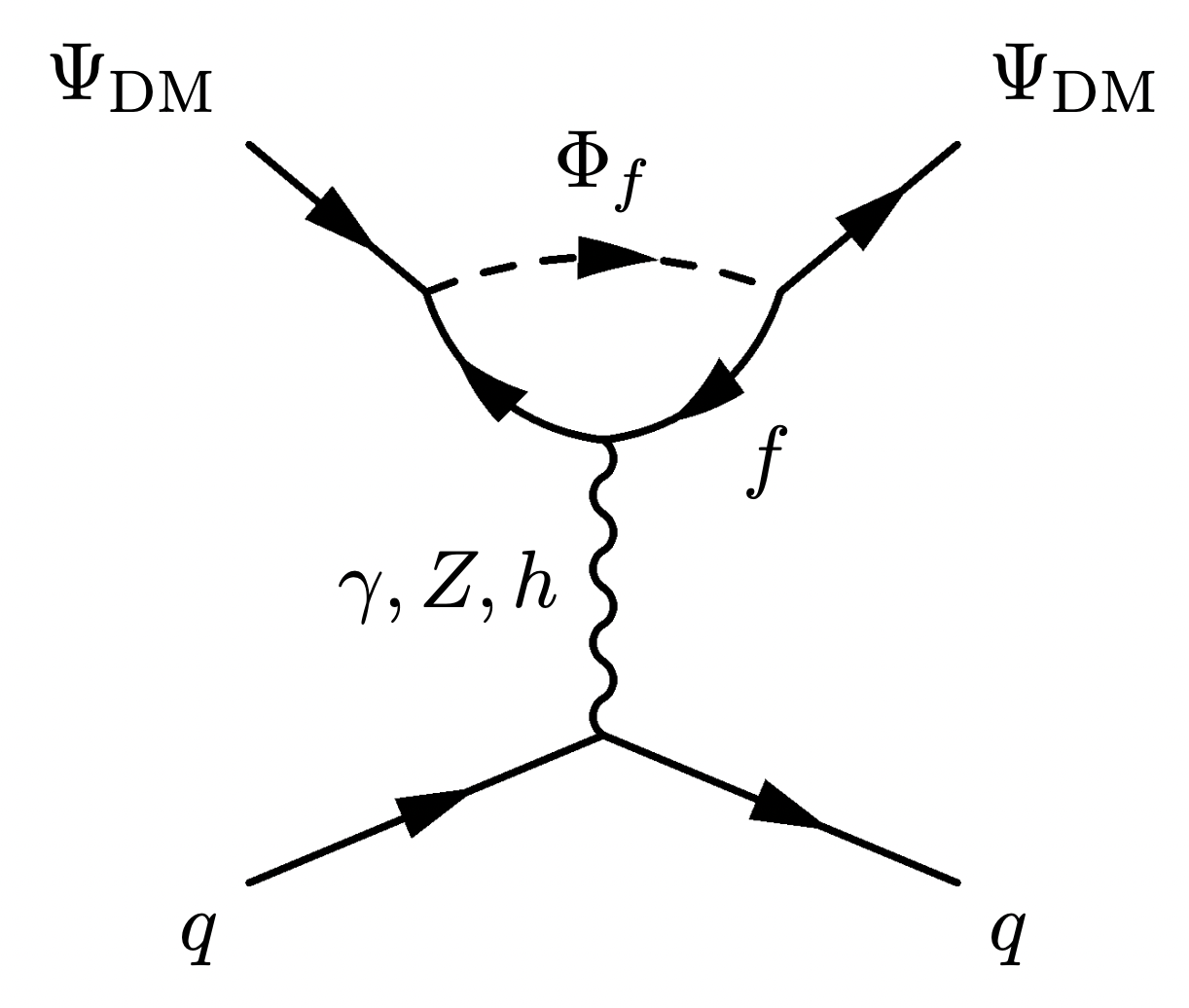}}
&
\subfloat[]{\includegraphics[width=0.3\textwidth]{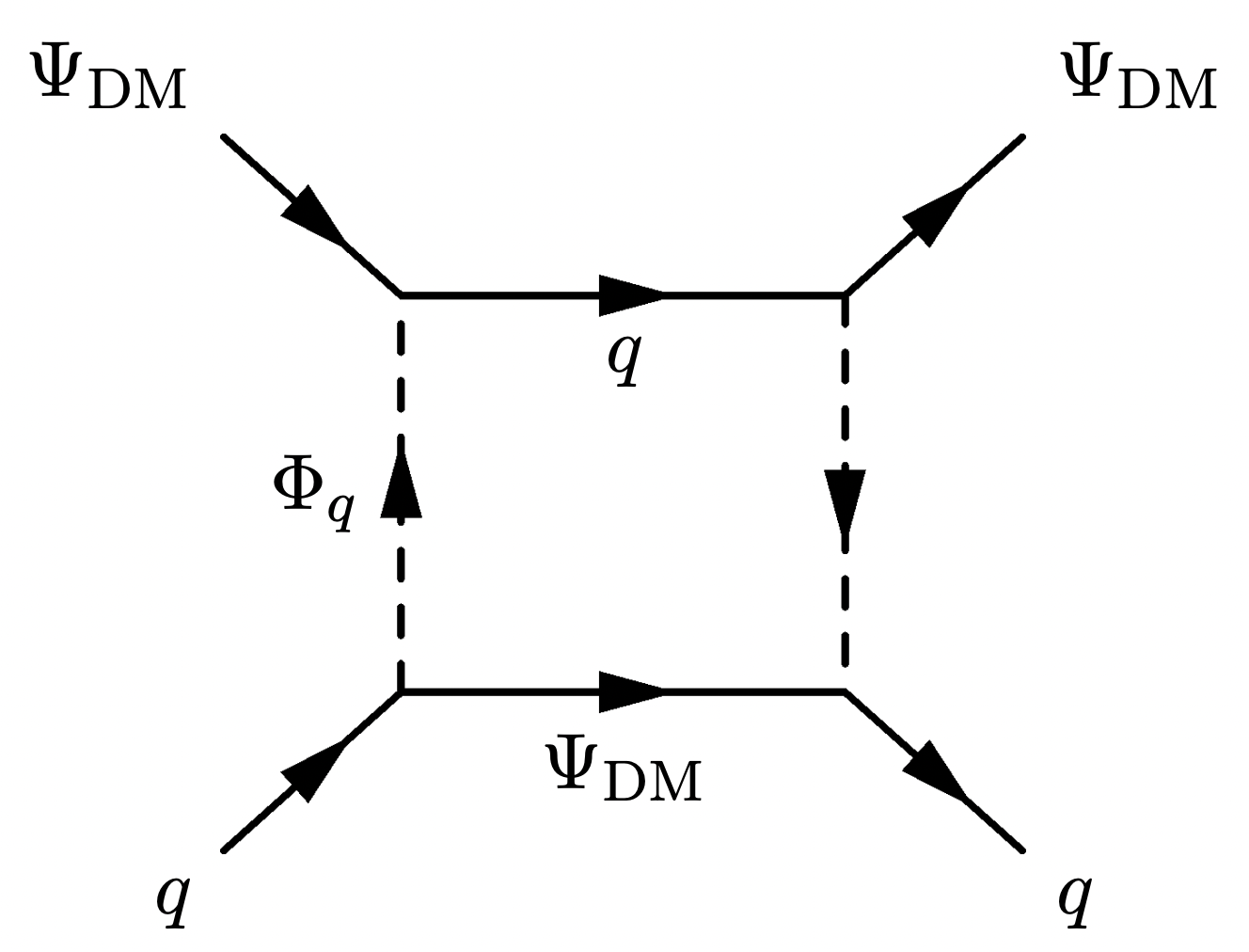}}
\\
\subfloat[]{\includegraphics[width=0.3\textwidth]{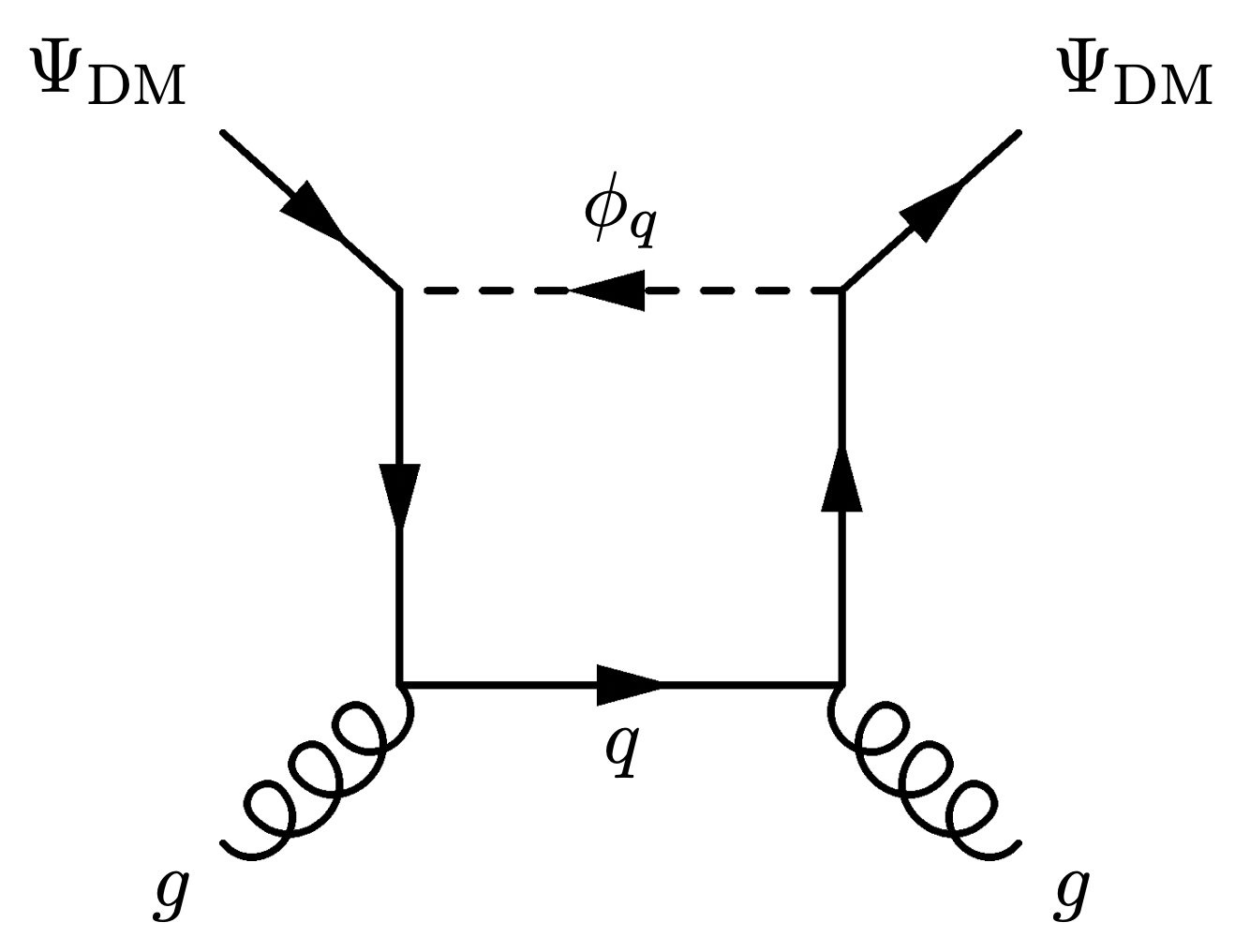}}
&
\subfloat[]{\includegraphics[width=0.3\textwidth]{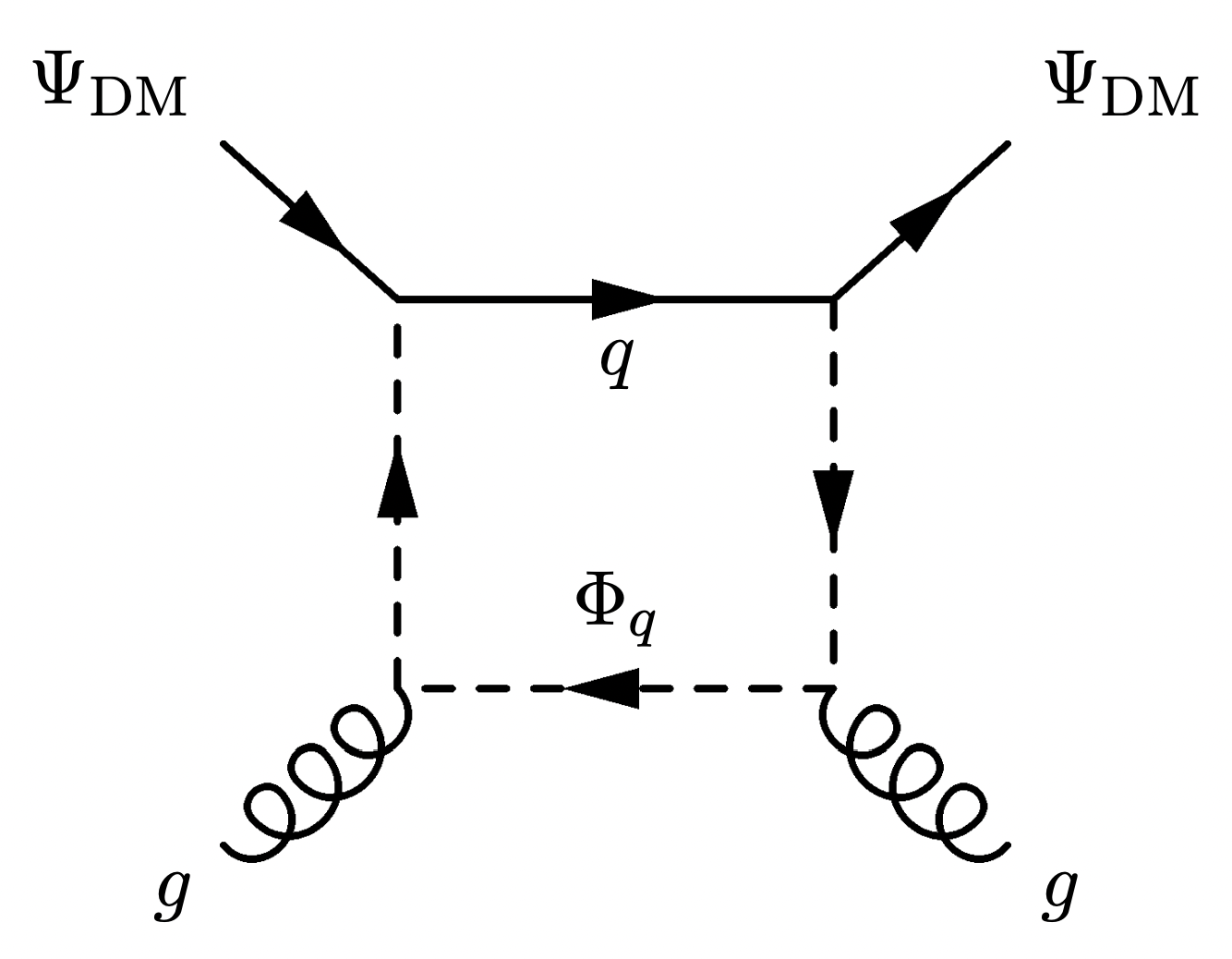}}
&
\subfloat[]{\includegraphics[width=0.3\textwidth]{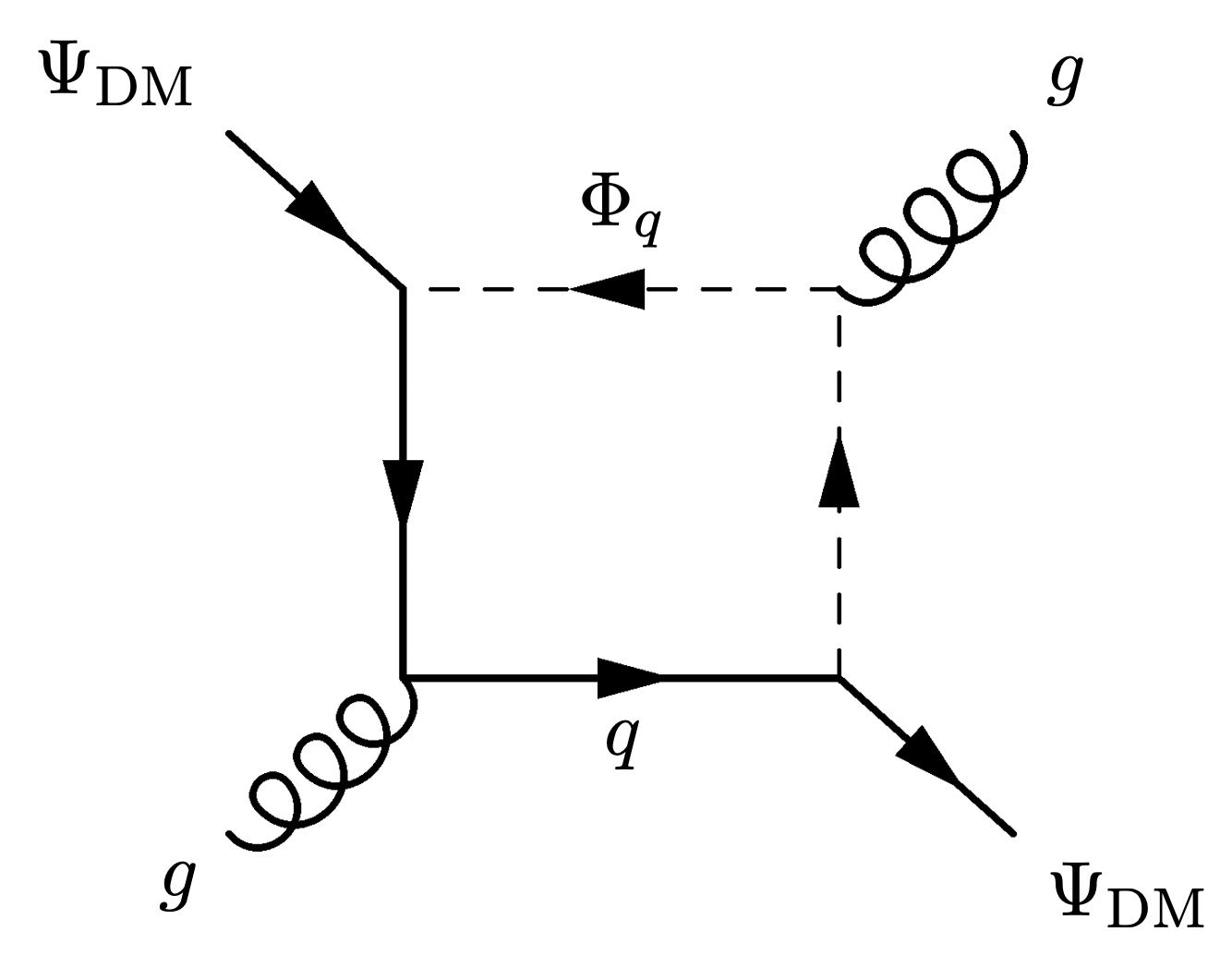}}
\\
&\subfloat[]{\includegraphics[width=0.3\textwidth]{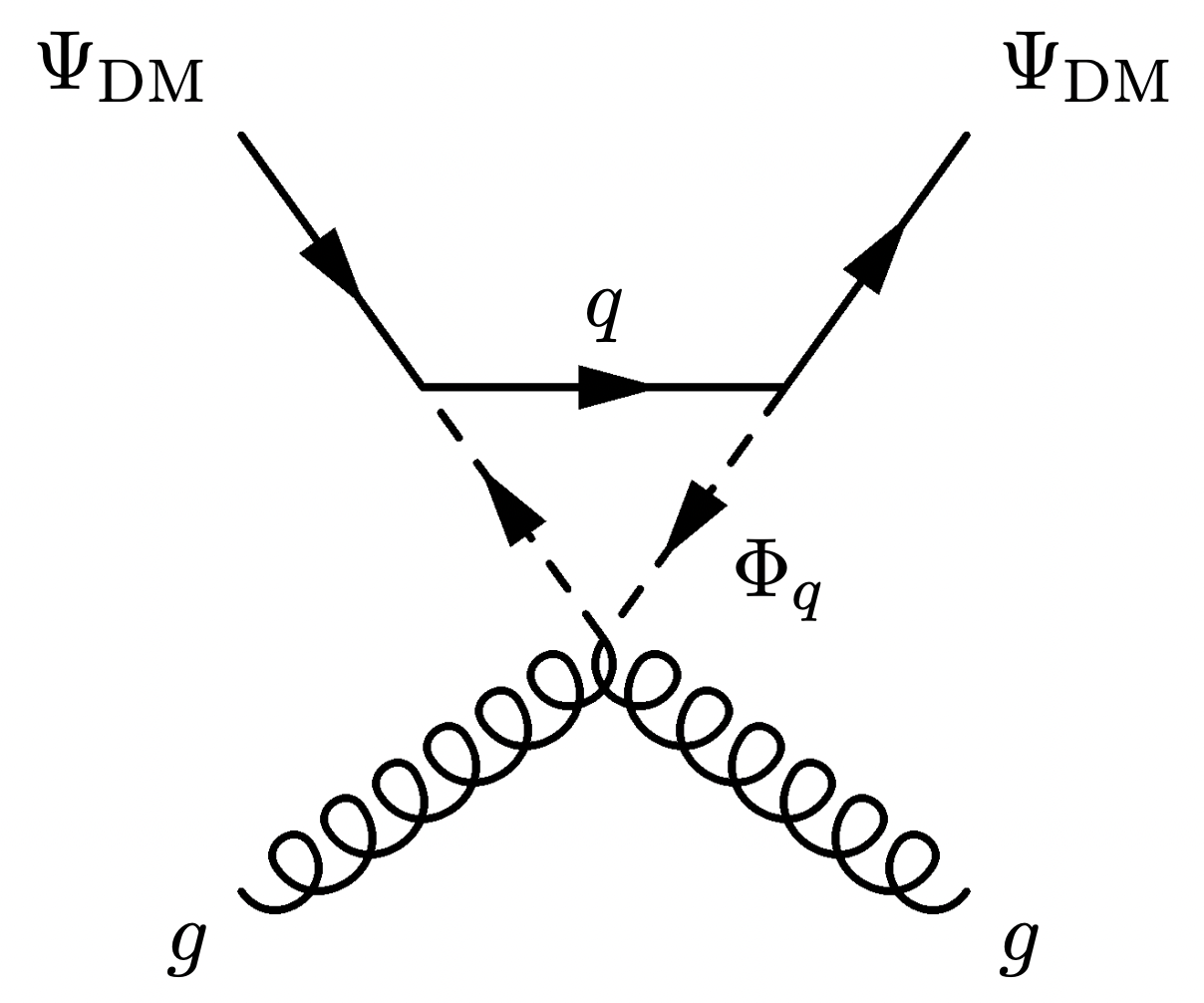}}
&
\\
\end{tabular}
\end{center}
\caption{\it Feynman diagrams contributing to the Wilson coefficients in the effective Lagrangian for the DM DD in the case of a Dirac fermion DM $\Psi_{\rm DM}$.}
\label{diags3}
\end{figure*}

Let us now consider the case of a Dirac fermion DM. We can follow the same reasoning as the scalar DM and express the general low energy EFT Lagrangian (the Feynmann diagrams leading to the Wilson coefficients are show in Fig.\ref{diags3}) as:
\begin{equation}
\begin{aligned}
& L_{\rm eff}^{\text{Dirac},q}=
\sum_{q=u,d} c^q\, \ovl\Psi_{\rm DM} \gamma_\mu \Psi_{\rm DM}\,\bar q \gamma^\mu q \nonumber\\ 
& +\sum_{q=u,d,s}\tilde{c}^q\, \ovl\Psi_{\rm DM} \gamma_\mu \gamma_5 \Psi_{\rm DM} \,\bar q \gamma^\mu \gamma_5 q
\nonumber\\
& + \sum_{q=u,d,s} d^q\, m_q\ovl \Psi_{\rm DM} \Psi_{\rm DM}\,\bar q q   +\sum_{q=c,b,t} d_q^g\, \ovl \Psi_{\rm DM}\,\Psi_{\rm DM} G^{a\mu \nu}G^a_{\mu \nu}
\nonumber\\
&
+ \sum_{q=u,d,s} \ovl \Psi_{\rm DM} \Bigg( g_{1}^{q} \frac{i \partial^\mu \gamma^\nu  }{M_{\Psi_{\rm DM}}}  +  g_{2}^{q} \frac{ \left( i  \partial^\mu \right)\left(i \partial^\nu \right) }{M_{\Psi_{\rm DM}}^2}\Bigg) \Psi_{\rm DM} \mathcal{O}^q_{\mu \nu}
\nonumber\\
&
+ \sum_{q=c,b,t} \ovl \Psi_{\rm DM} \Bigg( g_{1}^{g,q} \frac{ i \partial^\mu \gamma^\nu }{M_{\Psi_{\rm DM}}}+ g_{2}^{g,q} \frac{\left( i  \partial^\mu \right)\left(i \partial^\nu \right)}{M_{\Psi_{\rm DM}}^2}\Bigg) \Psi_{\rm DM} \mathcal{O}^g_{\mu \nu}\nonumber\\
& + \frac{\tilde{b}_\Psi}2 \,\ovl \Psi_{\rm DM} \sigma^{\mu \nu}\Psi_{\rm DM} F_{\mu \nu}+b_\Psi \ovl \Psi_{\rm DM} \gamma^\mu \Psi_{\rm DM} \partial^\nu F_{\mu \nu}.
\label{eq:Lfermion}
\end{aligned}
\end{equation}
Contrary to the case of a scalar DM, the effective coupling with the photon, emerging at the loop level, cannot be incorporated in a contact operator but appears explicitly in the low energy Lagrangian via two long-range terms dubbed, respectively, magnetic dipole moment and charge radius operators. 

As pointed out before, a rigorous assessment of the DD prospect hence requires the full computation of the DM scattering rate:
\begin{align}
\label{eq:DM_scattering_rate}
& \frac{d\sigma}{dE_R}=
\left(\frac{M_T}{2\pi v^2}|f^T|^2 + \alpha_{\rm em}\tilde{b}_\Psi^2 Z^2 \left(\frac{1}{E_R}-\frac{M_T}{2 \mu_{T }^2 v^2}\right)\right)|F_{\rm SI}(E_R)|^2\nonumber\\ 
& +\tilde{b}_\Psi^2 \frac{\mu_T^2 M_T}{\pi v^2}\frac{J_T+1}{3 J_T}|F_{\rm D}(E_R)|^2\ ,
\end{align}
where $F_{\rm SI}$ is the conventional $SI$ form factor while $F_D$ is the form factor associated to the dipole-dipole scattering \cite{Banks:2010eh}. $J_T$ is the spin of the target nucleus.
Finally, the parameter $f^T=Z f_p+(A-Z) f_n$ can be decomposed, in terms of the Wilson coefficients, in a similar fashion as the case of a scalar DM (see Eq. (\ref{eq:ffsDM})):
\begin{align}
\label{eq:fpfn}
& f_{N=p,n} =c^N-e b_\Psi-\frac{e \tilde{b}_\Psi}{2 M_\Psi}\nonumber\\
& +m_N\sum_{q=u,d,s} \left( f_q^M d_q+\frac{3}{4}\left(q(2)+\bar{q}(2)\right)\left(g_1^q+g_2^q\right)\right)
\nonumber\\
& +\frac{3}{4}m_N\sum_{q=c,b,t}G(2)\left(g_1^{g,q}+g_2^{g,q}\right)
-\frac{8}{9}f_{TG}f_G.
\end{align}
The coefficient $c^N$ is the sum of a tree-level contribution\footnote{The operator $\frac{1}{M_{\Phi_f}}\ovl \psi \gamma^\mu \left(1\pm\gamma_5\right) \psi \bar f \gamma_\mu (1\pm \gamma_5)f$ is obtained from $\ovl \psi P_{L,\, R} f \Phi_f \bar f P_{R, L} \psi \Phi_f^{*}$ (see Eq. (\ref{eq:tchannel_fermion_lagrangian})) by integrating out the scalar mediator and then by performing a Fierz transformation.}
\begin{equation}
c_{\rm tree}^q=\frac{|\Gamma_{L,R}^q|^2}{8 (M_{\Phi_q}^2-M_{\Psi_{\rm DM}}^2)}\ , and
\end{equation}
a loop-induced contribution from $Z$-penguin diagrams:
\begin{align}
c_Z^{q}=& \big[T^3_q-2Q_q \sin^2{(\theta_W)}\big]
\sum_f \frac{G_F}{\sqrt{2}}\frac{n^f_c}{16 \pi^2}\big(Q_f-\frac{1}{6}\big)  \nonumber\\
&\times \big[\big(\Gamma_{L}^f\big)^2-\big(\Gamma_{R}^f\big)^2 \big] F_Z\left(\frac{m_f^2}{M_{\Phi_f}^2},\frac{M_{\Psi_{\rm DM}}^2}{M_{\Phi_f}^2}\right),
\end{align}
with
\begin{align}
 F_Z(x_f,x_\psi) = &-\frac{2 x_f \left(x_f-x_\psi-1\right)}{\sqrt{\Delta}} \log\left[\frac{x_f+1-x_\psi+\sqrt{\Delta}}{2 \sqrt{x_f}}\right]
\nonumber\\
& + x_f \log (x_f)\ ,
\end{align}
and
\begin{align}
& \Delta=x_f^2-2 x_f \left(x_\psi+1\right)+(x_\psi-1)^2.
\end{align}
(see (c), (d) of Fig. \ref{diags3})  is given by
\begin{align}
& c_{\rm box}^q =
\sum_f \frac{|\Gamma_{L,R}^u|^2 |\Gamma_{L,R}^f|^2 }{16\pi^2 M_{\Phi_f}^2 } F_{\rm box}\left(\frac{m_f^2}{M_{\Phi_f}^2},\frac{M_{\Psi_{\rm DM}}^2}{M_{\Phi_f}^2}\right),
\end{align}
with 
\begin{align}
 F_{\rm box}(x_f,x_\psi) =& \frac{1}{4(x_f-1)^2(x_\psi-1)^2(x_f-x_\psi)}\nonumber\\ \times & \Big[x_f^2(x_\psi-1)^2 \log x_f - x_\psi^2 (x_f-1)^2 \log x_\psi \nonumber\\
&  + (x_f-1)(x_\psi-1)(x_f-x_\psi)\Big].
\end{align}
The operators proportional to the bilinear $\bar q q$ are again originated by a combination of QCD boxes and Higgs penguins (see Fig. \ref{diags3}):
\begin{align}
& d_{\rm QCD}^q=\frac{M_{\Psi_{\rm DM}}|\Gamma_{L,R}^q|^2}{16 {\big(M_{\Phi_q}^2-(m_q+M_{\Psi_{\rm DM}})^2\big)}^2},  
\end{align}
\begin{align}
 d_{\rm H}^q=
\frac{m_q}{32 \pi^2 M_h^2}\sum_f & n_c^f \Bigg[\frac{m_f^2}{M_{\Psi_{\rm DM}}v^2}|\Gamma_{L,R}^f|^2 F_{H,1}\bigg(\frac{m_f^2}{M_{\Phi_f}^2},\frac{M_{\Psi_{\rm DM}}^2}{M_{\Phi_f}^2}\bigg) \nonumber\\ +& \frac{\lambda_{H \Phi_f \Phi_f}}{2\sqrt{2} M_{\rm \Psi_{\rm DM}}}|\Gamma_{L,R}^f|^2 F_{H,2}\bigg(\frac{m_f^2}{M_{\Phi_f}^2},\frac{M_{\Psi_{\rm DM}}^2}{M_{\Phi_f}^2}\bigg) \Bigg],
\end{align}
with
\begin{align}
& F_{H,1}(x_f,x_\psi)=
\frac{x_f-1}{x_\psi}\log{x_f} - 2\nonumber\\
& - 2 \frac{(x_f-1)^2-x_\psi-x_f x_\psi}{x_\psi \sqrt{\Delta}}\log\left[\frac{1+x_f-x_\psi+\sqrt{\Delta}}{2 \sqrt{x_f}}\right],\nonumber\\
\end{align}
and
\begin{align}
   & F_{H,2}(x_f,x_\psi)= -F_{H,1}(x_f,x_\psi)\nonumber\\
& -\log{x_f}
+\frac{2 (x_\psi+x_f-1)}{\sqrt{\Delta}}\log\left[\frac{1+x_f-x_\psi+\sqrt{\Delta}}{2 \sqrt{x_f}}\right].
\end{align}
Moving to the coefficients associated with the scalar operator coupling the DM and the gluons this can be written as \cite{Gondolo:2013wwa}:
\begin{align}
& d_{q,\rm QCD}^{g}=
\frac{\alpha_s}{96\pi}\frac{M_{\Psi_{\rm DM}}}{M_{\Phi_q}^4} f_S^q,
\end{align}
with 
\begin{align}
& f_S^q=
\frac{\Delta_{\rm QCD} (x_\psi-1-x_q)-6 x_q \left(x_q-1-x_\psi\right)}{2 \Delta_{\rm QCD}^4}\nonumber\\
& +\frac{3 x_q (x_q^2-1+x_\psi)}{\Delta_{\rm QCD}^2}L_{\rm QCD},
\end{align}
where
\begin{align}
 & L_{\rm QCD}=\nonumber\\
 & \left \{
\begin{array}{cc}
\frac{2}{\sqrt{\Delta_{\rm QCD}}}\arctan\left[\frac{|\Delta_{\rm QCD}|}{x_q+1-x_\psi}+\theta(x_\psi-1-x_q)\right],  \\
\frac{1}{\sqrt{|\Delta_{\rm QCD}|}}\log \left[\frac{x_q+1-x_\psi+\sqrt{|\Delta_{\rm QCD}|}}{x_q+1-x_\psi-\sqrt{|\Delta_{\rm QCD}|}}\right]+2\pi i \theta(x_\psi-1-x_q)  ,   
\end{array}
\right.
\end{align}
for $\Delta_{\rm QCD} \geq 0$ and $\Delta_{\rm QCD}<0 $, respectively,
and
\begin{align}
    \Delta_{\rm QCD} = 2 x_\psi (x_q+1)-x_\psi^2-(1-x_q)^2.
\end{align}
Passing to the coefficients of the twist-2 operator:
\begin{align}
    g_1^{q,g}+g_2^{q,g}=\frac{1}{8} M_{\Psi_{\rm DM}} g_S^q
\end{align}
with
\begin{align}
 g_S^q=&
-\frac{\alpha_s\,\log{x_q}}{4\pi M_{\Psi_{\rm DM}}^4}
-\frac{\alpha_s}{3\pi M_{\Phi_q}^4}\Bigg[
\frac{3 x_q (x_q-1-x_\psi)}{\Delta_{\rm QCD}^2}\nonumber\\
& +\frac{2 x_q^2-x_q-1-4 x_q x_\psi-4 x_\psi+2 x_\psi^4}{2 \Delta_{\rm QCD}x_\psi}
+\frac{1}{x_\psi}
\nonumber\\
& -L_{\rm QCD} \left(\frac{3 (x_q-1+x_\psi)}{4 x_\psi^2}+\frac{3 x_q^2-3 x_q x_\psi-2 x_\psi+2 x_\psi^4}{2 \Delta_{\rm QCD} x_\psi^2}\right.\nonumber\\
& \left. +\frac{3 x_q (x_q^2-x_q-2 x_q x_\psi-x_\psi+x_\psi^2)}{\Delta^2 _{\rm QCD}}\right)\Bigg].
\end{align}
Finally, the coefficients of the dipole and charge radius operators are given by:
\begin{align}
& b_\Psi=
\frac{\alpha_{\rm em}}{8\pi M_{\Psi_{\rm DM}}^2}\sum_f n_c^f Q_f |\Gamma_{L,R}^f|^2 F_\gamma \left(\frac{M_{\Psi_{\rm DM}}^2}{M_{\Phi_f}^2},\frac{m_f^2}{M_{\Phi_f}^2}\right),\nonumber
\\
& \tilde{b}_\Psi=
\frac{\alpha_{\rm em}}{8\pi M_{\Psi_{\rm DM}}}\sum_f n_c^f Q_f |\Gamma_{L,R}^f|^2 \widetilde{F}_\gamma \left(\frac{M_{\Psi_{\rm DM}}^2}{M_{\Phi_f}^2},\frac{m_f^2}{M_{\Phi_f}^2}\right),
\end{align}
with
\begin{align}
 F_\gamma (x_f,x_\psi)=
\frac{1}{12}\Big(&-\frac{8 (1-x_f)+x_\psi}{x_\psi}\log x_f \nonumber\\
& \left. -\frac{4}{\Delta} \left[4 \Delta+x_\psi (1+3 x_f)-x_\psi^2\right]\right.
\nonumber\\
& \left.-\frac{1}{x_\psi \Delta}\left[8 \Delta^2+(9-5 x_\psi+7 x_f)x_\psi \Delta\right.\right.\nonumber\\
&  \left. -4 x_f x_\psi^2 (3-x_\psi+x_f)\right]L_{\rm EW} \Big),
\end{align}
and
\begin{align}
& \widetilde{F}_\gamma (x_f,x_\psi)=
1+\frac{1-x_f}{2 x_\psi}\log x_f+\frac{\Delta +x_\psi (1-x_\psi+x_f)}{2 x_f}L_{\rm EW}.
\end{align}
where $L_{\rm EW}(x_f,x_\psi)=L_{\rm QCD}(x_f,x_\psi)$.
In the case of a Majorana DM, the bilinear $\ovl \psi \gamma^\mu \psi$ as well as the dipole and charge radius operators automatically vanish. In such a case we can again just compare the DM scattering cross-section with the corresponding experimental limits.

\begin{figure*}
    \centering
    \subfloat{\includegraphics[width=0.43\linewidth]{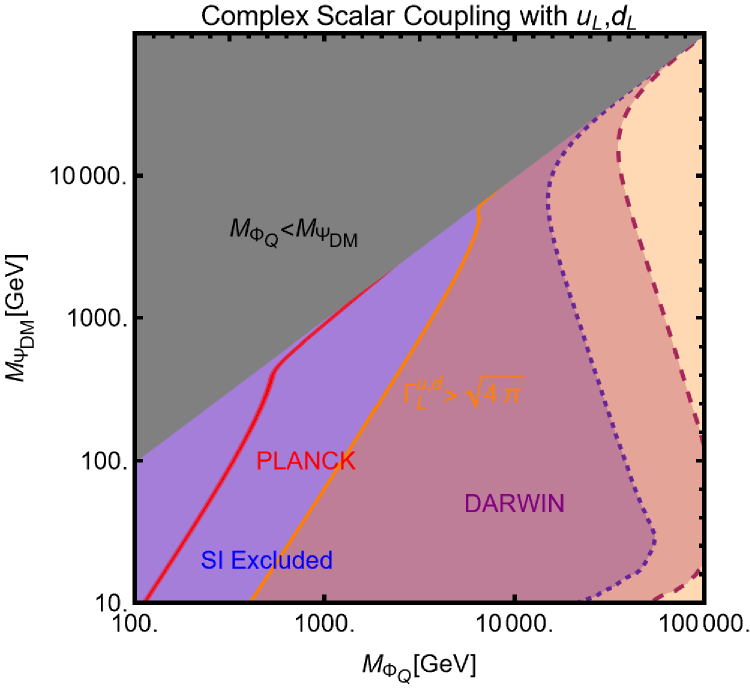}}
    \subfloat{\includegraphics[width=0.43\linewidth]{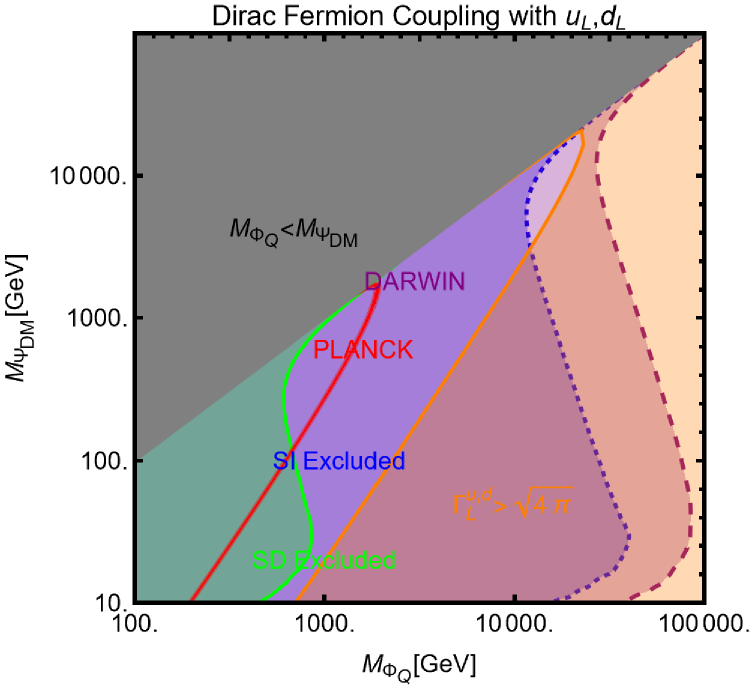}}
    \caption{Summary of constraints for models of $t$-channel portals in which SI interactions arise mostly at the tree level when a complex scalar (Dirac fermion) DM couples to the first generation of quarks via a coloured mediator, charged under the $SU(3)_C$, is depicted via the left (right) plot. For simplicity, only the case of coupling with the $SU(2)_L$ doublet has been accounted for. According to the customary colour coding, we show red coloured isocontours corresponding to the correct DM relic density while the blue, purple coloured regions correspond to the current, projected exclusion on the SI interactions. In the case of a Dirac fermion DM, the SD bounds (green coloured) have also been reported. The gray-coloured region corresponds to unstable DM and hence, is excluded. In the orange coloured region, the correct DM relic density cannot be achieved with perturbative couplings. }
    \label{fig:tchannelq}
\end{figure*}

\begin{figure*}
    \centering
    \subfloat{\includegraphics[width=0.25\linewidth]{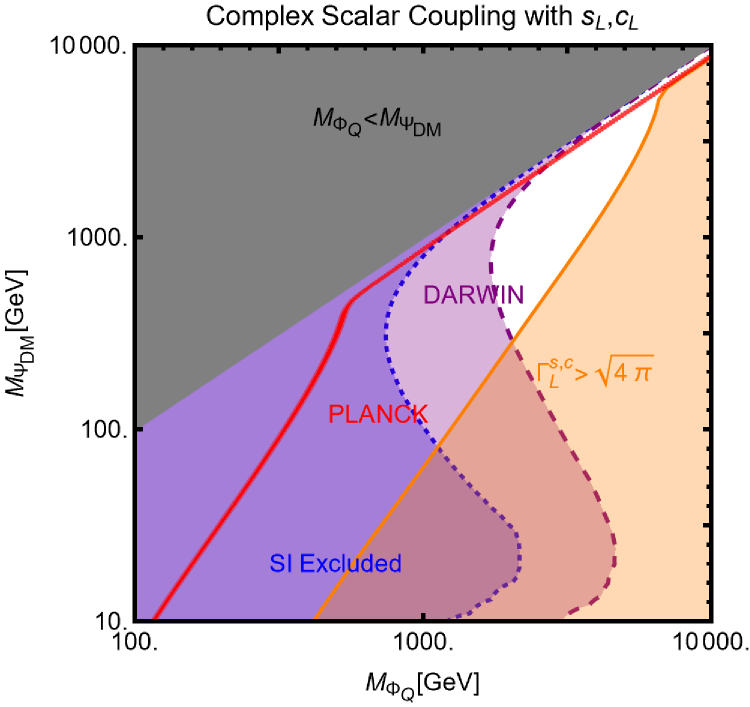}}
    \subfloat{\includegraphics[width=0.25\linewidth]{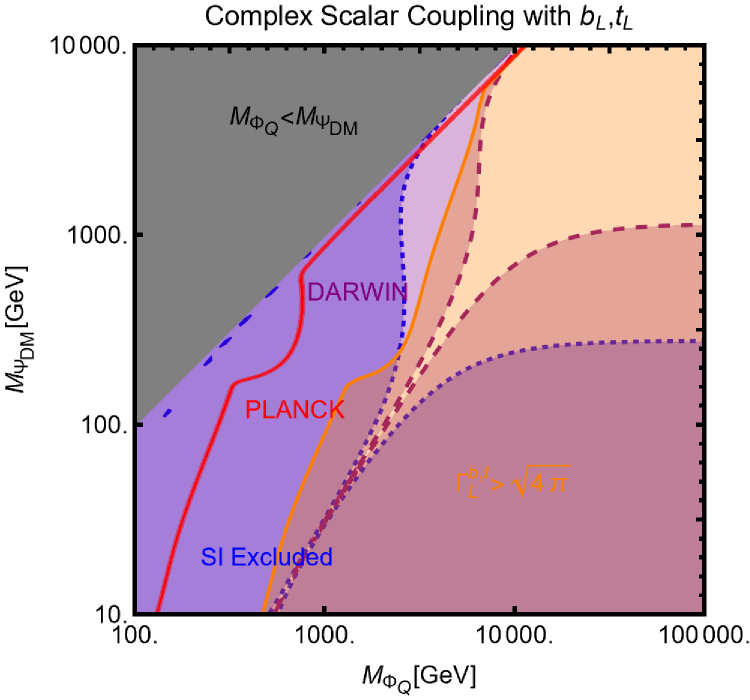}}
    \subfloat{\includegraphics[width=0.25\linewidth]{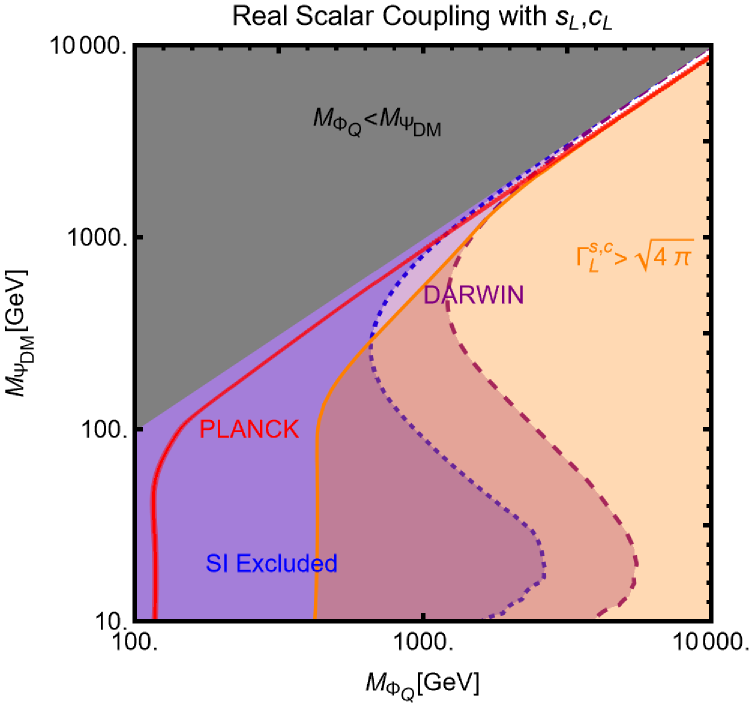}}
    \subfloat{\includegraphics[width=0.25\linewidth]{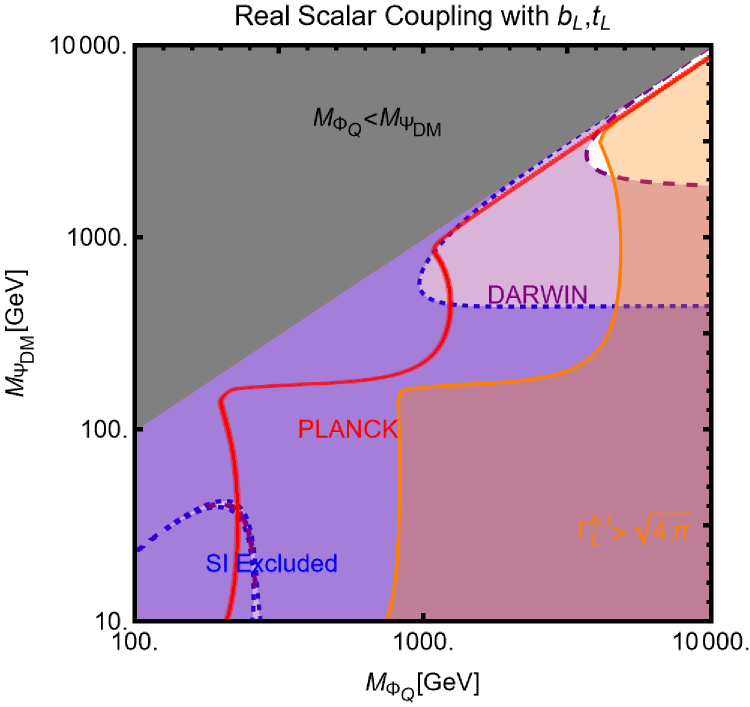}}\\
     \subfloat{\includegraphics[width=0.25\linewidth]{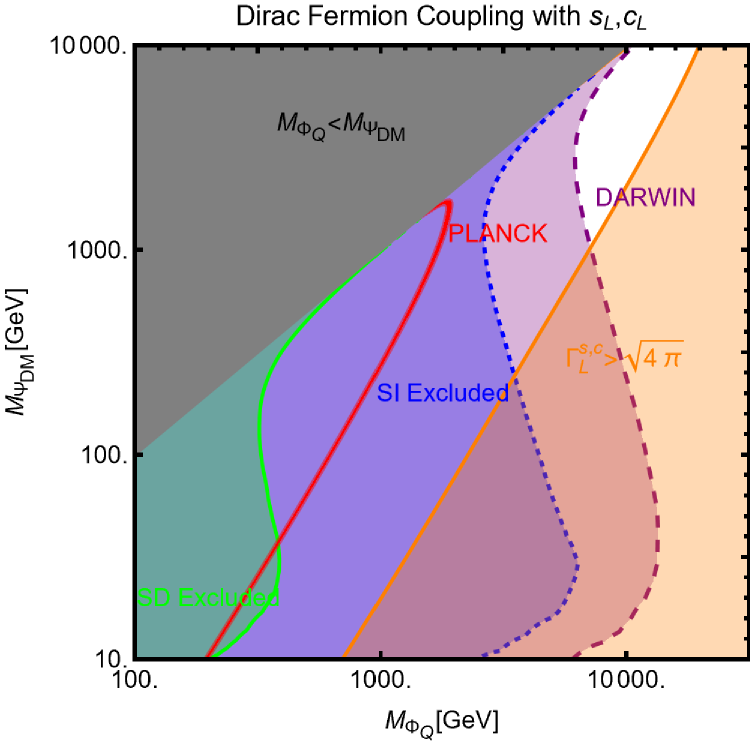}}
    \subfloat{\includegraphics[width=0.25\linewidth]{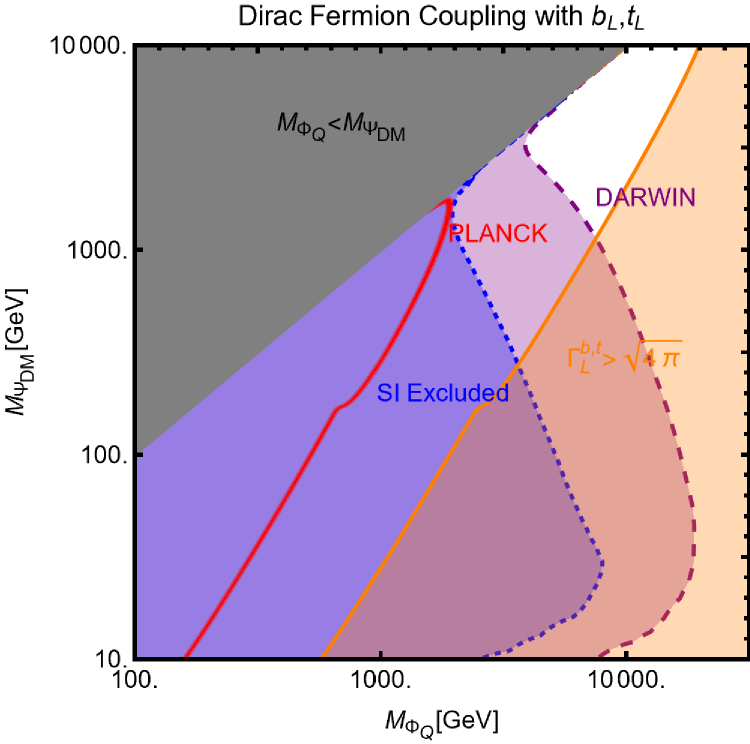}}
    \subfloat{\includegraphics[width=0.25\linewidth]{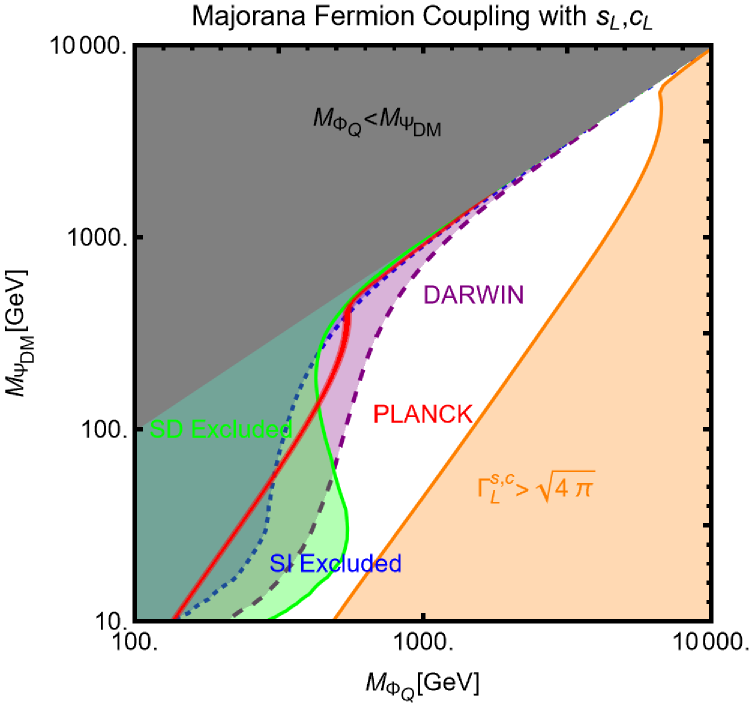}}
    \subfloat{\includegraphics[width=0.25\linewidth]{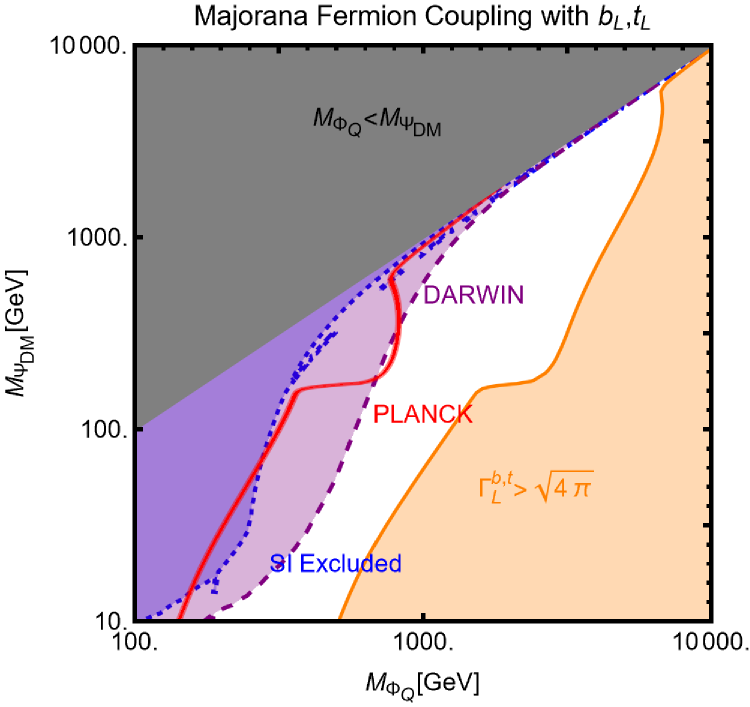}}
    \caption{Same as the Fig.~\ref{fig:tchannelq} but considering $t$-channel portals in which SI interactions arise at the one-loop level. The first (last) two plots of the top row depict
    the cases when a complex (real) scalar DM couples to the 
    second and third generations of $SU(2)_L$ quarks. The bottom row represents the same for a Dirac (Majorana) fermion DM, respectively. To correct wrong axis labels and text on the grey-coloured regions, see my comments in Fig. 8. The Last two plots appear smaller in size.}
    \label{fig:tchannelb}
\end{figure*}

In the simplest realization, a $t$-channel portal model has just three free parameters, the DM and mediator masses (i.e., $M_{\Phi_{\rm DM}},\, M_{\Psi_{f_i}}$ or $M_{\Psi_{\rm DM}},\, M_{\Phi_{f_i}}$) and a relevant single coupling $\Sigma^{f_i}_{L/R}$. Without the loss of generality, thus, the combination of the DM constraints can be shown in the bidimensional plane ($M_{\Phi_{\rm DM}},\, M_{\Psi_{f_i}}$) or ($M_{\Psi_{\rm DM}},\, M_{\Phi_{f_i}}$)) for a fixed assignation of the concerned coupling. Illustrating all the possible variants of this setup is beyond the scope of this work and we refer to Ref. \cite{Arcadi:2023imv} for a more comprehensive study. For this study, we just demonstrate
a few simple examples, as depicted in Figs. \ref{fig:tchannelq}
and ~\ref{fig:tchannelb}.

We start with a scenario where a DM (complex scalar or Dirac fermion) couples to the first generation of left-handed quarks, i.e., $u_L, \, d_L$, through a mediator, charged under the $SU(3)_C$.  This is the scenario where the strongest constraints are expected as the interactions relevant to the DD arise at the tree level. The results shown in Fig.~\ref{fig:tchannelq} confirm, indeed this expectation as the constraints on the SI interaction exclude the mass of the DM and the mediator $\sim O(10$ TeV), much beyond the parameter space compatible with the thermal relic density.

In the next level, for Fig.~\ref{fig:tchannelb}, we considered
a similar scenario, but now the DM couples to the second and third
generations of the $SU(2)_L$ doublet quarks.  Here the result of the combined DM constraints depends on the spin and Lorentz representation of the DM. Scenarios of either a complex scalar or Dirac fermionic DM are substantially ruled out also in this cases, although now the relevant DD couplings are loop-induced. This is mostly due to the contribution of photon penguin diagrams. For a complex scalar DM, coupled with the third generation quarks, we also see that DM masses below $100$-$200$ GeV are ruled out regardless of the mass of the $t$-channel mediator, as a consequence for the radiatively induced Higgs portal coupling (see also the next section). Such coupling is present for both the complex and real DM, hence, the latter scenario is also strongly disfavoured. For the real scalar DM, the case of coupling with the second-generation quarks also appears to be ruled out (second last plot of the top row of Fig.~\ref{fig:tchannelb}). Even if effective coupling of the photon is not present and the radiative Higgs coupling is suppressed by the small second-generation Yukawas, the $d$-wave suppression of the DM annihilation cross-section drastically reduces the allowed parameter space. To get the correct relic density one should rely on the high values of the couplings or coannihilation, falling again in the regions excluded by DD experiments.  In synthesis, the only scenario with the potential to evade DD bounds is the one with a Majorana DM (the last and the second last plots of the bottom row of Fig.~\ref{fig:tchannelb}). However, the next-generation experiments, like DARWIN, have a high capability of testing also the case of a Majorana DM.

\begin{figure*}
    \centering
    \subfloat{\includegraphics[width=0.43\linewidth]{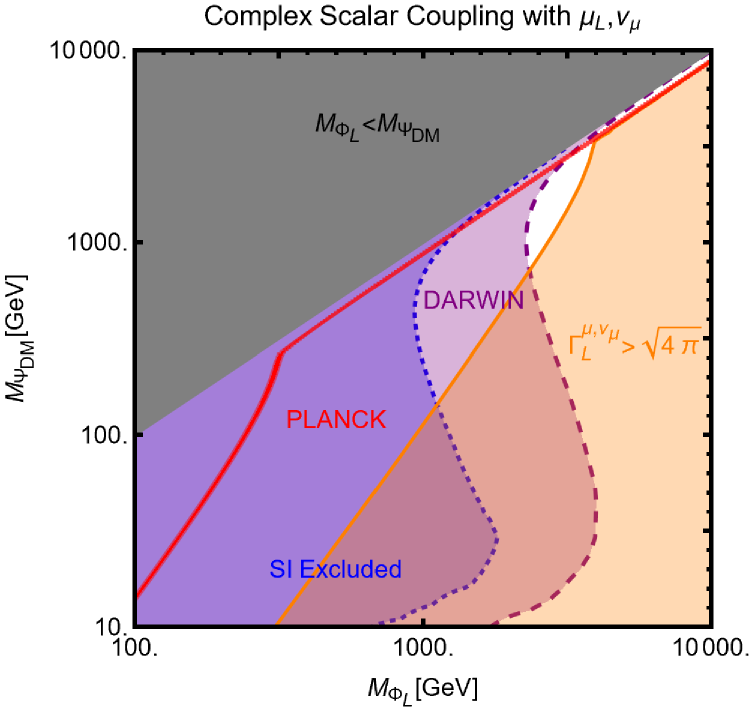}}
    \subfloat{\includegraphics[width=0.43\linewidth]{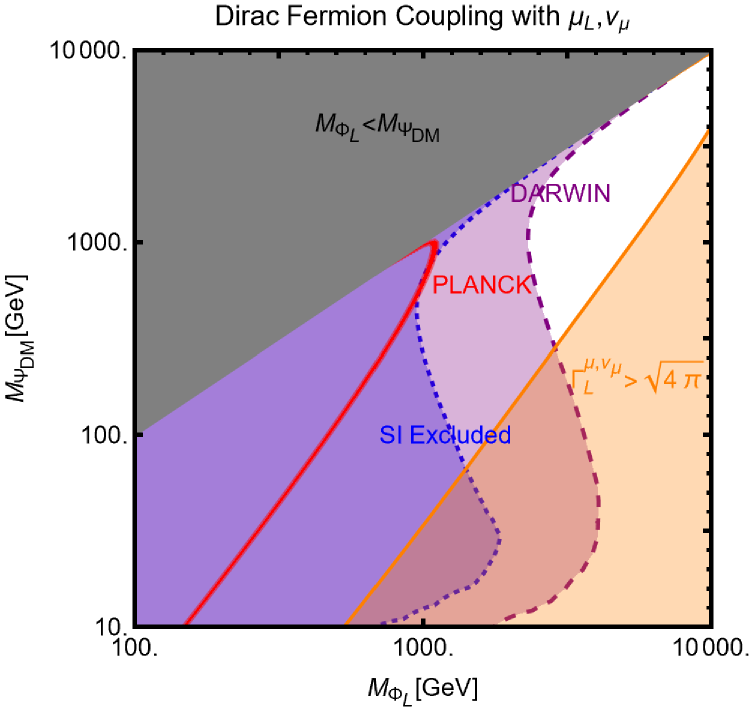}}
    \caption{Summary of the DM constraints for two sample $t$-channel models in which the  states are coupled only with the second generation of the $SU(2)_L$ leptons. The left (right) plot depicts the case of a complex (real) scalar DM. The colour coding is the same as Fig.~\ref{fig:tchannelq}.}
    \label{fig:tchannell}
\end{figure*}

As a final illustration for the case of a SM singlet DM, we show in Fig.~\ref{fig:tchannell}, two representative cases of DM coupling to the second $SU(2)_L$ family of lepton, i.e., $\mu_L, \nu_\mu$. Such couplings are present only for either a complex scalar or a Dirac fermion. The outcome shown in the figure strongly resembles the one of coupling with the second generation of quark flavours. This happens because the effective coupling with the photons plays the most relevant role. 

As concluding remark we mention that collider limit are as well relevant of t-channel portal, especially in the case of color charged mediators which might be efficiently produced a LHC. Recent studies have been presented in Refs.~\cite{Arina:2023msd} (see also \cite{Mohan:2019zrk}). We do not explicitly account these results here as we preferred, for the simplified models, to keep the focus on dedicated search DM experiments.

\subsection{Direct Detection of EW Interacting DM}

A very simple realization of a WIMP model is to consider the DM as the lightest neutral component of a $SU(2)$ multiplet. Here, no extra $s$- or $t$-channel mediator field is needed to connect the DM to the SM as the former has gauge couplings with the $Z$ and $W^\pm$ bosons via the heavier components of the same multiplet where it lies. 
Further, for some specific multiplets, there is no need to introduce ad hoc discrete symmetries to  assure a cosmologically stable DM. In such a case, on realizes a so called minimal DM model \cite{Cirelli:2005uq,Cirelli:2009uv,Escudero:2016gzx}.
Following ref. \cite{Hisano:2011cs}, we consider a simplified Lagrangian of the form:
\begin{align}
\label{eq:leff_EW}
    \mathcal{L}_{\rm eff}&=\left[\frac{g_2}{4}\sqrt{n^2-(2Y+1)^2}\ovl{\widetilde{\chi}^0}\gamma^\mu \widetilde{\psi}^{-}W^+_\mu\right.\nonumber\\
    &\left. +\frac{g_2}{4}\sqrt{n^2-(2Y-1)^2}\ovl{\widetilde{\chi}^0}\gamma^\mu \widetilde{\psi}^{+}W^-_\mu+\mbox{H.c.}\right.\nonumber\\
    &\left. +\frac{ig_2 (-Y)}{\cos \theta_W}\ovl{\widetilde{\chi}^0}\gamma^\mu \widetilde{\eta}^0 Z_\mu \right],
\end{align}
with $Y$ as the hypercharge and $g_2$ as the $SU(2)_L$ gauge coupling. This effective Lagrangian is obtained under the hypothesis that the DM candidate is a Majorana fermion $\widetilde{\chi}^0$. The coupling with the $W^{\pm}$ boson is ensured by the presence of an electrically charged Dirac fermion $\widetilde{\psi}^{\pm}$, while the coupling with the $Z$-boson, for $Y \neq 0$, is ensured by an additional electrically neutral Majorana fermion, heavier than the DM. 
To our  knowledge, no complete computations are present in the literature for the case of a scalar DM in this scenario. One should notice, in addition, that in the case of a scalar DM one expects, in general, the presence of a coupling with the SM Higgs boson, see e.g., Ref.~\cite{Arakawa:2021vih} as well as the discussion in the previous section. 
DM relic density is determined by annihilation processes into gauge bosons final state. Once the quantum numbers of the DM under the EW group, i.e. the pair $(n,Y)$, are fixed, the only free parameter is the DM mass, so that the correct relic density is achieved for a specific value of such parameter. The computation of the $\Omega_{\rm DM}$ is, however, more complicated, with respect to the previously discussed simplified models, due to the presence of Sommerfeld enhancement and bound state formation \cite{Mitridate:2017izz}. Updated results on the DM relic density for different values of $n$ have been provided in \cite{Bottaro:2021snn}\footnote{Ref. \cite{Bottaro:2021snn} considered also the relic density for real scalar DM. Ref- \cite{Bottaro:2022one} considered scenarios of complex scalar and dirac fermionic DM. However, to overcome constraints from DD, DM elastic scattering has been forbidden by introducing a sizable mass splitting between the DM and its charged and neutral partners.}. 

The effective Lagrangian (see Eq.~(\ref{eq:leff_EW}) leads, at the scale relevant for the DD, to the following interaction Lagrangian between the DM, the quarks and the gluons:
\begin{align}
\label{eq:lag_EW_loop}
    \mathcal{L}&=d_q \overline{\tilde{\chi_0}}\gamma^\mu \gamma_5 \tilde{\chi}_0 \bar q \gamma_\mu \gamma_5 q+f_q m_q \overline{\tilde{\chi}}_0 \tilde{\chi}_0 \bar q q \nonumber\\
    & \frac{g_q^{(1)}}{m_{\rm DM}}\overline{\tilde{\chi}}_0 i \partial^\mu \gamma^\nu \tilde{\chi}_0 \mathcal{O}_{\mu \nu}^q+\frac{g_2^{(q)}}{m_{\rm DM}^2}\overline{\tilde{\chi}}_0 (i \partial^\mu) (i \partial^\nu)\tilde{\chi}_0 \mathcal{O}_{\mu \nu}^q \nonumber\\
    & +f_G \overline{\tilde{\chi}}_0 \tilde{\chi}_0 G_{\mu \nu}^a G^{\mu \nu\,\,a},
\end{align}
to which SI and SD scattering cross-sections correspond:
\bea
    \sigma_{\rm DM N=p,n}^{\rm SI}&&=\frac{4 \mu_{\rm DM N}^2}{\pi}|f_N|^2,\nonumber\\
    \sigma_{\rm DM N=p,n}^{\rm SD}&&=\frac{12 \mu_{\rm DM N}^2}{\pi}|a_N|^2,
\eea
The couplings $f_N,\,a_N$ of the DM with the nucleons are given, in an analogous way as the $t$-channel portals, by a combination of form factors and the Wilson coefficients (the diagram topologies leading to such Wilson coefficients are shown in fig. \ref{diags3EW}) in Eq.~(\ref{eq:lag_EW_loop}):
\begin{align}
    \frac{f_N}{m_N}&=\sum_{q=u,d,s}f_q f_q^N+\sum_{u,d,s,c,b}\frac{3}{4}(q(2)+\bar{q}(2))(g_q^{(1)}+g_q^{(2)})\nonumber\\
    & -\frac{8\pi}{9\alpha_s}f_{TG}f_G,\nonumber\\
    a_N&=\sum_{q=u,d,s}d_q \Delta_q^N,
\end{align}
with
\begin{align}
     f_q&=\frac{\alpha_2}{4 M_h^2}\left[\frac{n^2-(4Y^2+1)}{8 M_W}g_H\left(\frac{M_W^2}{m_{\rm DM}^2}\right)\right.\nonumber\\
    & \left. +\frac{Y^2}{4 M_Z \cos^4 \theta_W}g_H\left(\frac{M_Z^2}{m_{\rm DM}^2}\right)\right]\nonumber\\
    & +\frac{\left((g_{Zq}^V)^2-(g_{Zq}^A)^2\right)Y^2}{\cos^4 \theta_W}\frac{\alpha_2^2}{M_Z^3}g_S\left(\frac{M_Z^2}{m_{\rm DM}^2}\right),\nonumber\\
    g_q^{(1)}&=\frac{n^2-(4Y^2+1)}{8}\frac{\alpha_2^2}{M_W^2}g_{T1}\left(\frac{M_W^2}{m_{\rm DM}^2}\right)\nonumber\\
    & + \frac{2 \left((g_{Zq}^V)^2+(g_{Zq}^A)^2\right)Y^2}{\cos^4 \theta_W}\frac{\alpha_2^2}{M_Z^3}g_{T1}\left(\frac{M_Z^2}{m_{\rm DM}^2}\right),\nonumber\\
    g_q^{(2)}&=\frac{n^2-(4Y^2+1)}{8}\frac{\alpha_2^2}{m_W^2}g_{T2}\left(\frac{m_W^2}{m_{\rm DM}^2}\right)\nonumber\\
    & + \frac{2 \left((g_{Zq}^V)^2+(g_{Zq}^A)^2\right)Y^2}{\cos^4 \theta_W}\frac{\alpha_2^2}{M_Z^3}g_{T2}\left(\frac{M_Z^2}{m_{\rm DM}^2}\right).
\end{align}
Here, $g_{Zq}^{V,\, A}$ are the vectorial and axial-vectorial couplings of the Z-boson with the SM quarks: 
\begin{equation}
    g_{Zq}^V=\frac{1}{2}T_{3q}-Q_q \sin^2 \theta_W,\,\,\,\,\,g_{Zq}^A=-\frac{1}{2}T_{3q},
\end{equation}
The Wilson coefficient of the DM-gluon effective couplings is decomposed into three contributions $f_G=f_G^{(a)}+f_G^{(b)}+f_G^{(c)}$:

\begin{figure*}
\begin{center}
\begin{tabular}{ccc}
\subfloat[]{\includegraphics[width=0.3\textwidth]{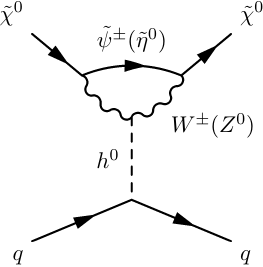}}
&
\subfloat[]{\includegraphics[width=0.3\textwidth]{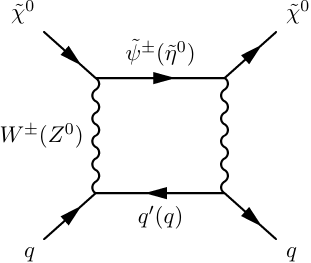}}
&
\subfloat[]{\includegraphics[width=0.3\textwidth]{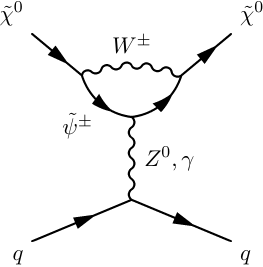}}
\\
\subfloat[]{\includegraphics[width=0.3\textwidth]{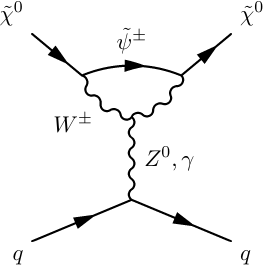}}
&
\subfloat[]{\includegraphics[width=0.3\textwidth]{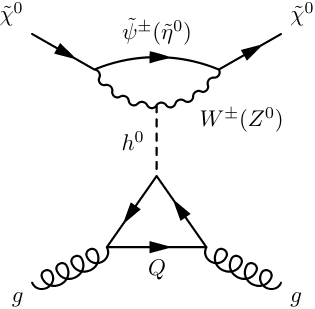}}
&
\subfloat[]{\includegraphics[width=0.3\textwidth]{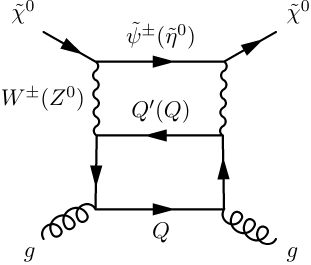}}
\\
\end{tabular}
\end{center}
\caption{\it Feynman diagrams contributing to the Wilson coefficients in the effective Lagrangian \ref{eq:lag_EW_loop}.}
\label{diags3EW}
\end{figure*}

\begin{align}
    f_G^{(a)}&=-\frac{\alpha_s \alpha_2}{48 \pi m_h^2}\sum_{Q=c,b,t}c_Q \left[\frac{n^2-(4Y^2+1)}{8 M_W}g_H \left(\frac{M_W^2}{m_{\rm DM}^2}\right)\right.\nonumber\\
    & \left. +\frac{Y^2}{4 M_Z \cos^4 \theta_W}g_H\left(\frac{;_Z^2}{m_{\rm DM}^2}\right)\right],\nonumber\\
    f_G^{(b)}&+f_G^{(c)}=\frac{\alpha_s \alpha_2^2}{4\pi}\left[\frac{n^2-(4Y^2+1)}{8 M_W^3}g_W\left(\frac{M_W^2}{m_{\rm DM}^2},\frac{m_t^2}{m_{\rm DM}^2}\right)\right.\nonumber\\
    & \left. +\frac{Y^2}{4 M_Z^3 \cos^4 \theta_W}g_Z\left(\frac{M_Z^2}{m_{\rm DM}^2},\frac{m_t^2}{m_{\rm DM}^2}\right)\right],
\end{align}
where $c_Q=1+\frac{11 \alpha_s (m_Q)}{4\pi}$. Following Ref.~\cite{Hisano:2011cs} we have adopted the following input values:
\begin{equation}
    \alpha_s(m_Z)=0.118,\,\,\,\,c_c=1.32,\,\,\,\,c_b=1.19,\,\,\,\,c_t=1 .
\end{equation}
Contrary to the previous coefficients, we do not report explicitly the form of the functions $g_{W, Z}$ as they are very lengthy. Besides, for the case of $g_Z$, some contributions can be expressed only in terms of integrals which are to be evaluated numerically. The interested reader can refer to the appendix of Ref. \cite{Hisano:2011cs}.
For what concern the SD cross-section, the effective coefficient is given by:
\begin{align}
     d_q&=\frac{n^2-(4Y^2+1)}{8 m_W}\frac{\alpha_2^2}{m_W^2}g_{AV}\left(\frac{m_W^2}{m_{\rm DM}^2}\right) \nonumber\\
    & + \frac{2 \left((g_{Zq}^V)^2+(g_{Zq}^A)^2\right)Y^2}{\cos^4 \theta_W}\frac{\alpha_2^2}{m_Z^2}g_{AV}\left(\frac{m_Z^2}{m_{\rm DM}^2}\right),\nonumber\\
    g_{AV}(x)&=\frac{1}{24 b_x}\sqrt{x}(8-x-x^2){\tan}^{-1}\left(\frac{2 b_x}{\sqrt{x}}\right)\nonumber\\
    & -\frac{1}{24}x(2-(3+x)\log(x)).
\end{align}
Where $b_x=\sqrt{1-x/4}$.

\begin{figure*}
    \centering
    \includegraphics[width=0.45\linewidth]{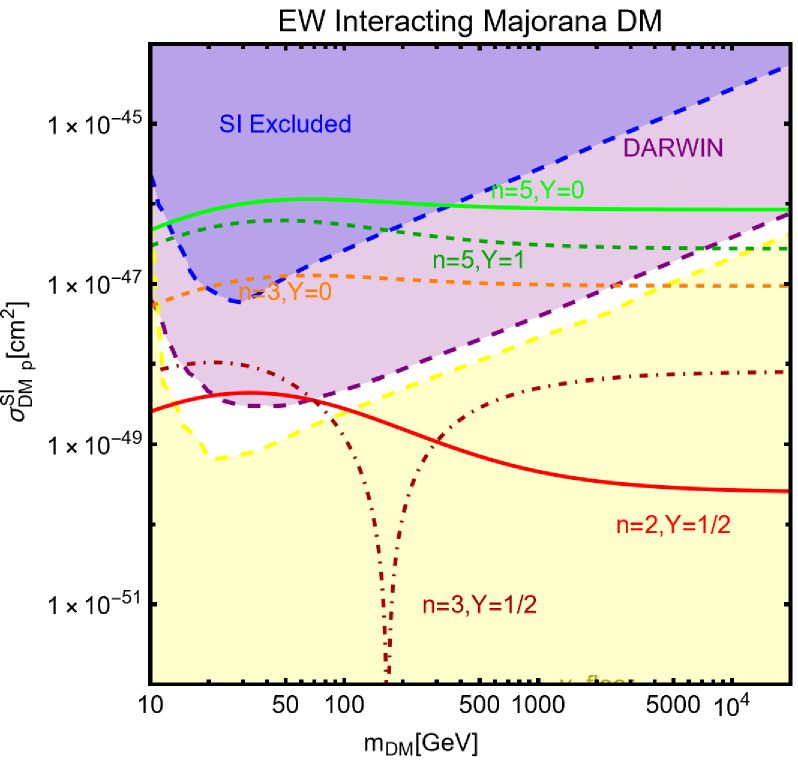}
    \caption{The SI Cross-section, as a function of the DM mass, for a Majorana DM candidate having gauge interactions with the EW SM gauge bosons. The various curves correspond to different assignations of the parameter pair $(n, Y)$ (see text for details). The blue coloured region is excluded by the current experimental limits given by LZ. The purple coloured region is the one which will be excluded in case of negative detection by DARWIN. The yellow coloured regions correspond, finally, to the $\nu$ floor.}
    \label{fig:pEWmajo}
\end{figure*}
Fig. \ref{fig:pEWmajo} shows the DD prospects of the scenario under consideration. Indeed, the DM SI cross-section, as a function of the DM mass, is shown for the cases of different $SU(2)_L$-multiples, identified by the parameter $n$, and different assignations of the hypercharge $Y$. The curves corresponding to the different values of $(n, Y)$ are compared with the current experimental exclusions, as given by LZ (blue coloured), as well as the projected limits (purple coloured) by DARWIN and the $\nu$-floor (yellow coloured). The light green coloured solid (dark green coloured dashed) line corresponds to $Y=0~(1)$ for an $SU(2)_L$ quintuplet while the orange coloured dashed (dark red coloured dot-dashed) line corresponds to $Y=0~(1/2)$ for an $SU(2)_L$ triplet. Finally, the case of a $SU(2)_L$ doublet with $Y=1/2$ is depicted by red coloured solid line.


\section{Higgs portal (s)}

In this section, we will present a series of models based on the idea of coupling a DM candidate to the SM Higgs doublet $H$. We will start from one of the simplest realizations, conventionally dubbed Higgs Portal, in which the  SM particle content is augmented only by the DM candidate. More realistic and complex realizations will be discussed subsequently. 

\subsection{The EFT Realization}
Even if it is classified as a further example of the $s$-channel portal, we dedicate special attention to this so-called Higgs portal.
This class of models represents the most minimal option to couple an SM gauge singlet DM candidate with the SM Higgs doublet $H$. The Higgs portal can be formulated for a real scalar $(\chi)$, vector $(V_\mu)$, and fermionic $(\psi)$ DM according to the following Lagrangians \cite{McDonald:1993ex,Burgess:2000yq,Kim:2006af,Andreas:2010dz,Kanemura:2010sh,Djouadi:2011aa,Djouadi:2012zc,Lebedev:2011iq,Mambrini:2011ik,LopezHonorez:2012kv,Goodman:2010ku,Fox:2011pm,Buckley:2014fba,Abdallah:2015ter,Baglio:2015wcg,Alanne:2017oqj,Biekotter:2022ckj}:
\begin{eqnarray} 
\label{Lag:DM}  
\!&&\Delta {\cal L}_\chi = -\frac12 m_\chi^2 \chi^2 -\frac14 \lambda_\chi \chi^4 - \frac14 \lambda_{H\chi \chi}  H^\dagger H \chi^2 \;, \nonumber \\ 
\!&&\Delta {\cal L}_V = \frac12 m_V^2 V_\mu V^\mu\! +\! \frac14 \lambda_{V}  (V_\mu V^\mu)^2\! +\! \frac14 \lambda_{HVV}  H^\dagger H
V_\mu V^\mu , \nonumber \\  
\!  &&\Delta {\cal L}_\psi = - \frac12 m_\psi \ovl \psi \psi - \frac14 {\lambda_{H\psi\psi}\over \Lambda} H^\dagger H \ovl \psi \psi \;,
\end{eqnarray} 
where $\Lambda$ is the scale of some NP. Also, for a complex scalar DM, $\chi^2$ will be replaced by $\chi^* \chi \equiv |\chi|^2$. Note that we have assumed CP-conservation as well as the presence of a $Z_2$ or $U(1)$ symmetry, i.e., whether the DM belongs to a real or complex representation of the Lorentz group.
This symmetry assures that the DM states interact only in pairs and hence ensures its stability.
Decomposing the SM Higgs doublet as $H ={\left(0\,\,\,\frac{v_h+h}{\sqrt{2}}\right)}^T$ in the unitary gauge, we can obtain, from Eq.~\eqref{Lag:DM}, interaction vertex between the DM pairs and the physical Higgs state $h$. 
By giving a closer look to Eq.~\eqref{Lag:DM}, we notice that  Higgs portal fermionic DM  model relies on $D>4$ operator, which indeed explicitly depends on an unknown NP scale $\Lambda$. Thus it can be regarded only as an EFT valid up to energies $\sim O(\Lambda)$. The dependence of the DM related observables on the scale $\Lambda$ can nevertheless be hidden by the redefinition $\lambda_{H \psi \psi}\rightarrow \lambda_{H \psi \psi}\frac{v_h}{\Lambda}$.
Despite the $H^\dagger H V^\mu V_\mu$ operator has mass dimension four, the Higgs portal vector DM case should be regarded as an EFT as the aforementioned operator is non-renormalizable and leads to perturbative unitarity violation issues as pointed out in Refs. \cite{Lebedev:2011iq,Baek:2012se,Arcadi:2020jqf,Baek:2021hnl}.
In light of this reasoning, we will call the models described by Eq.~\eqref{Lag:DM}, including also the case of a scalar DM for simplicity, as the EFT Higgs portal. In the next subsections, we will discuss the phenomenology of possible realistic completions of the minimal Higgs portal models.

\begin{figure*}[!t]	
\begin{center}
\subfloat{\includegraphics[width=0.33\textwidth]{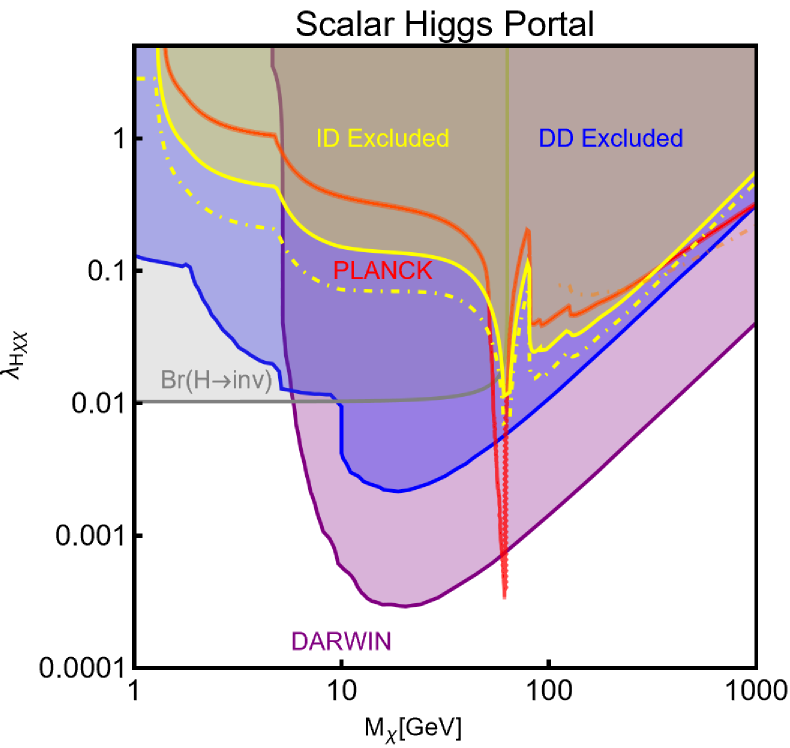}}
\subfloat{\includegraphics[width=0.33\textwidth]{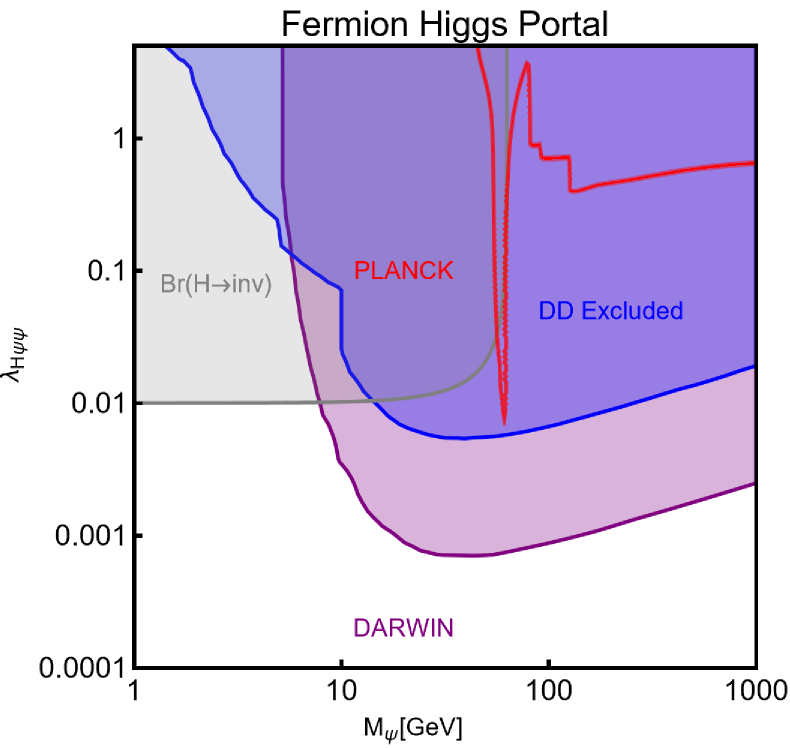}}
\subfloat{\includegraphics[width=0.33\textwidth]{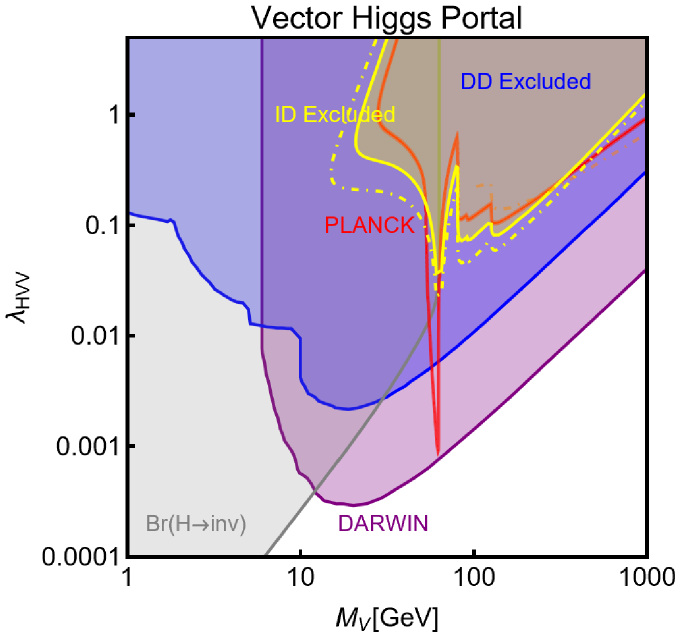}}
\end{center}
\caption{Illustration of the DM constraints for the SM Higgs portal in the relevant bidimensional planes for a
scalar (left panel), fermionic (middle panel) 
and vectorial (right panel) DM. In each plot, the red coloured line represents the model points featuring the correct DM relic 
density. The blue coloured region is excluded by the current SI DD limits while the yellow coloured region correspond to exclusion form ID. The purple  region would be excluded in the absence of signals at the DARWIN experiment. The dot-dashed yellow and orange lines represent the projected sensitivities by FERMI and CTA respectively. Finally, the grey coloured region represents the complementary exclusion, for light DM masses, by searches of the invisible Higgs decay at the LHC.
}
\label{fig:EFTHportal}
\end{figure*}

The EFT Higgs portal has only two free parameters, namely the DM mass and its coupling with the Higgs. The relevant phenomenology can then be encoded in simple bidimensional plots as the ones shown in Fig.~\ref{fig:EFTHportal}.

%

Each panel of the figure shows, in analogy with the models discussed in the previous section, in red colour the isocontours of the correct relic density, while the blue (purple) coloured regions represent the present (near) future bounds from the DD of DM SI interactions. Contrary to the previous models, the Higgs portal features an additional constraint as the Higgs boson might decay into a DM pair if kinematically allowed. The possibility of an invisible Higgs decay, i.e., non-zero branching ratio $(Br)$ for Higgs to a DM pair is widely explored by the LHC, see e.g., Ref.~\cite{ATLAS:2023tkt}.  In panels of Fig.~\ref{fig:EFTHportal}, the grey coloured region represents cases for which $Br(h\rightarrow \mbox{DM}\,\mbox{DM})>0.11$. Thus, one gets a new exclusion zone
besides the ones excluded by the DD. Visible production modes of the Higgs boson can provide more stringent constraints on the invisible branching ratio~\cite{Biekotter:2022ckj}. However, estimating precisely these bounds requires dedicated model-dependent analyses that are beyond the scope of this work.  Since these constraints are not the most important for the relevant parameter space in most cases, we will use the bounds $Br(h\rightarrow \mbox{DM}\,\mbox{DM}) \equiv Br(h\rightarrow \mbox{invisible}) >0.11$ in the following.

Analytical expressions for the DM annihilation and scattering cross-sections can be straightforwardly adapted from the ones of the generic spin-$0$ portals, hence we do not report them explicitly here. The shape of the relic density contours can be explained as follows. For $m_{\rm DM} \leq M_h/2$ the DM annihilates mostly into SM fermions with the cross-section progressively increasing as heavier final states quark thresholds get open. At $m_{\rm DM} \sim M_h/2$ the DM annihilation cross-section fermion pairs (mostly $\bar b b$) encounters an $s$-channel resonance so that the correct relic density is achieved for suppressed couplings. After the resonance, there is another sharp enhancement of the cross-section due to the opening of the annihilation channels into gauge bosons. In the high DM mass regime, finally, the relic density is mostly accounted for annihilations into $hh$ final state. The case of the fermionic DM  is comparatively more constrained as in this case, the DM annihilation cross-section is $p$-wave, i.e., velocity, suppressed. Fig.~\ref{fig:EFTHportal} shows that indeed the fermion Higgs portal is already completely ruled out by current constraints on the DD.  The scalar and vector scenarios currently survive only around the $s$-channel resonance regions and will be as well completely ruled out in the absence of positive signals at the DARWIN experiment.

Even if not competitive with DD, we have shown for completeness, the ID constraints/future prospects which apply in the cases of scalar and vector DM.

\subsection{Towards complete realizations: Singlet extension}

The Higgs portal models, as depicted in Eq. (\ref{Lag:DM}), are built by combining the Lorentz and gauge invariant bilinear $H^\dagger H$ with a DM bilinear. As pointed out before, this kind of construction leads to effective theories whose theoretical factuality might remain questionable, especially in the case of vectorial DM candidates. For this reason, it is appropriate to work out more theoretically consistent setups. There are typically two strategies to achieve a renormalizable coupling between the Higgs doublet and pairs of DM candidates: i) mixing of the Higgs doublet with a scalar SM gauge singlet which in turn couples to a pair of SM gauge singlet DM states;  ii) the DM appears charged, at least partially, under the $SU(2)_L$.  Let us start with the first scenario. On general grounds, one can consider a theory with a scalar sector composed of the SM Higgs doublet and a singlet $S$ and consider the following potential:
%
\begin{align}
\label{eq:general_potential}
     V(H,S)=&\frac{\lambda_H}{4}(H^\dagger H)^2+\frac{\lambda_{HS}}{4}H^\dagger H S^2+\frac{\lambda_S}{4}S^4\nonumber\\
    & +\frac{1}{2}\mu_H^2 H^\dagger H+\frac{1}{2}\mu_S^2 S^2 \, . 
\end{align}
The singlet scalar $S$ can be interpreted as the real component of a complex field breaking a new gauge symmetry (unless differently stated we will implicitly assume the simplest case, i.e., a $U(1)$ symmetry) via a vacuum expectation value (VEV) $v_S$.

Once EW symmetry is also broken, the scalar potential leads to a mixed mass term for the doublet and the singlet. It is then needed to diagonalize a mass matrix of the form:
\begin{equation}
    {\mathcal{M}}^2=\left(
    \begin{array}{cc}
    2 \lambda_H v_h^2 & \lambda_{HS}v_h v_S \\     \lambda_{HS}v_h v_S & 2 \lambda_S v_S^2
    \end{array}
    \right) \, . 
\end{equation}
This task is achieved via an orthogonal matrix: 
\begin{equation}
    O=\left(
    \begin{array}{cc}
    \cos\theta & \sin\theta  \\
    -\sin\theta & \cos\theta 
    \end{array}
    \right)
~~~{\rm with}~~
\tan 2 \theta = \frac{\lambda_{HS}v_h v_S}{\lambda_S v_S^2-\lambda_H v_h^2} \, , 
\end{equation}
so that:
\begin{equation}
O^T {\mathcal{M}}^2 O=\mbox{diag}\left(M_{H_1}^2,M_{H_2}^2\right),
\end{equation}
with:
\begin{equation}
    M_{H_1,H_2}^2=\lambda_H v_h^2+\lambda_S v_S^2 \mp \frac{\lambda_S v_S^2-\lambda_H v_h^2}{\cos 2 \theta} \, .
\end{equation}
Unless differently stated, the state $H_1$ will be identified with the 125 GeV SM-like Higgs boson.
By inverting the previous relation we can express the quartic couplings in terms of the physical masses:
\begin{align}
    & \lambda_H=\frac{M_{H_1}^2}{2v_h^2}+\sin^2 \theta \frac{M_{H_2}^2-M_{H_1}^2}{2v_h^2} \, , \\
    & \lambda_S=\frac{2\lambda_{HS}^2}{\sin^2 2 \theta}\frac{v_h^2}{M_{H_2}^2-M_{H_1}^2}\left(\frac{M_{H_2}^2}{M_{H_2}^2-M_{H_1}^2}-\sin^2 \theta\right) \, . 
\end{align}
Notice that, assuming all the parameters of the scalar potential to be real, to ensure that $v_{h, S}>0$, we need to require:
\begin{equation}
    \lambda_H > \frac{\lambda_{HS}^2}{4 \lambda_S},\,\,\,\,\lambda_S>0 \, ,
\end{equation}
On similar grounds we can write:
\begin{equation}
    v_S^2=\frac{M_{H_1}^2 \sin^2 \theta +M_{H_2}^2 \cos^2 \theta}{2 \lambda_S}
\end{equation}

In the mass basis, the interaction Lagrangian between $H_{1,2}$ and the SM fermions and gauge bosons is given by:
\begin{align}
    \mathcal{L}_{\rm scalar,SM}&=\frac{H_1 \cos\theta + H_2 \sin\theta}{v_h}\left(2 M_W^2 W^{+}_{\mu}W^{-\,\mu}\right.\nonumber\\
    & \left. +M_Z^2 Z_\mu Z^\mu -m_f \bar f f\right) \, , 
\label{eq:Lag-mix-SM}
\end{align}
while the trilinear scalar couplings relevant to the DM phenomenology are:
\begin{align}\label{eq:Lag-scalar-trilinear}
    \mathcal{L}_{\rm scalar,trilinear}&=-\frac{\kappa_{111}}{2}v_h H_1^3-\frac{\kappa_{112}}{2}H_1^2 H_2 v_h \sin\theta\nonumber\\
    & -\frac{\kappa_{221}}{2}H_1 H_2^2 v_h \cos\theta-\frac{\kappa_{222}}{2}H_2^3 v_h \, , 
\end{align}
with the various $\kappa$ factors given by:
\begin{align}
\label{eq:trilinear}
    & \kappa_{111}=\frac{M_{H_1}^2}{v_h^2 \cos\theta}\left(\cos^4 \theta+\sin^2\theta \frac{\lambda_{HS}v_h^2}{M_{H_1}^2-M_{H_2}^2} \right) , \nonumber\\ 
    & \kappa_{112}=\frac{2 M_{H_1}^2+M_{H_2}^2}{v_h^2}\left(\cos^2 \theta +\frac{\lambda_{HS}v_h^2}{M_{H_2}^2-M_{H_1}^2}\right),  \nonumber\\
    & \kappa_{221}=\frac{2 M_{H_2}^2+M_{H_1}^2}{v_h^2}\left(\sin^2 \theta +\frac{\lambda_{HS}v_h^2}{M_{H_1}^2-M_{H_2}^2}\right) , \nonumber\\
    & \kappa_{222}=\frac{M_{H_2}^2}{v_h^2 \sin\theta}\left(\sin^4 \theta+\cos^2\theta \frac{\lambda_{HS}v_h^2}{M_{H_2}^2-M_{H_1}^2} \right) \, . 
\end{align}
The aforesaid framework just allow us to introduce the SM singlet fermionic and vector DM candidates in a consistent way. They will initially interact with the singlet field $S$. Subsequently a double portal with the SM will be established by the mass mixing between the $S$ and the Higgs. Notice however that such mixing would imply a deviation of the couplings of the 125 GeV Higgs, from the SM prediction, against the experimental evidence \cite{Falkowski:2015iwa}.

\paragraph{Fermion DM:}
A fermionic DM can be introduced in the aforementioned framework as a fermion of the new dark sector, dynamically getting its mass from the VEV of the singlet:
\begin{equation}
    \mathcal{L}_{\chi}=-y_\chi \ovl \chi \chi S \, , \,\,\,\,\,y_\chi={M_\chi}/{v_S}.
\end{equation}
In the physical basis, the relevant Lagrangian for the DM phenomenology is written as:
\begin{align}
     \mathcal{L}=&-y_\chi \bar \chi (-H_1 \sin\theta +H_2 \cos \theta) \chi\nonumber\\
    & +\mathcal{L}_{\rm scalar,SM}+\mathcal{L}_{\rm scalar,trilinear},
\end{align}
where $\mathcal{L}_{\rm scalar,SM},\,\mathcal{L}_{\rm scalar,trilinear}$ are given by Eqs. (\ref{eq:Lag-mix-SM}) and (\ref{eq:Lag-scalar-trilinear}).
Note that, if the scalar $S$ is interpreted as the "dark Higgs" of an additional gauge symmetry, then one should consider as well the associated gauge boson. We will implicitly assume that the latter is not relevant for the DM phenomenology. We will come back to this point later again. We now have all the elements to discuss the DM phenomenology for the concerned setup. For what DM relic density is concerned, the DM annihilation cross-section can be described via very similar expressions as the case of the Higgs portal; consequently, we will not show it explicitly. The most interesting feature compared to the case of the SM Higgs portal is the presence of a new annihilation channel $\ovl \chi \chi \rightarrow H_2 H_2$, if kinematically allowed,, as it can relax the strong correlation between relic density and the DD. Moving to the DD, the presence of a second scalar mediator leads to the following modification of the DM scattering cross-section over protons:  
\begin{align}\label{eq:h1h2medfermiondm}
    \sigma_{\chi p}^{\rm SI}&=\frac{\mu_{\chi p}^2}{\pi}\frac{y_{\chi}^2 \sin^2 \theta \cos^2 \theta m_p^2}{v_h^2}{\left(\frac{1}{M_{H_1}^2}-\frac{1}{M_{H_2}^2} \right)}^2|f_p|^2\nonumber\\ 
    & f_p =\sum_{q=u,d,s}f_q^p+\frac{6}{27}f_{TG}\approx 0.3
\end{align}

\begin{figure*}
    \centering
    \subfloat{\includegraphics[width=0.33\linewidth]{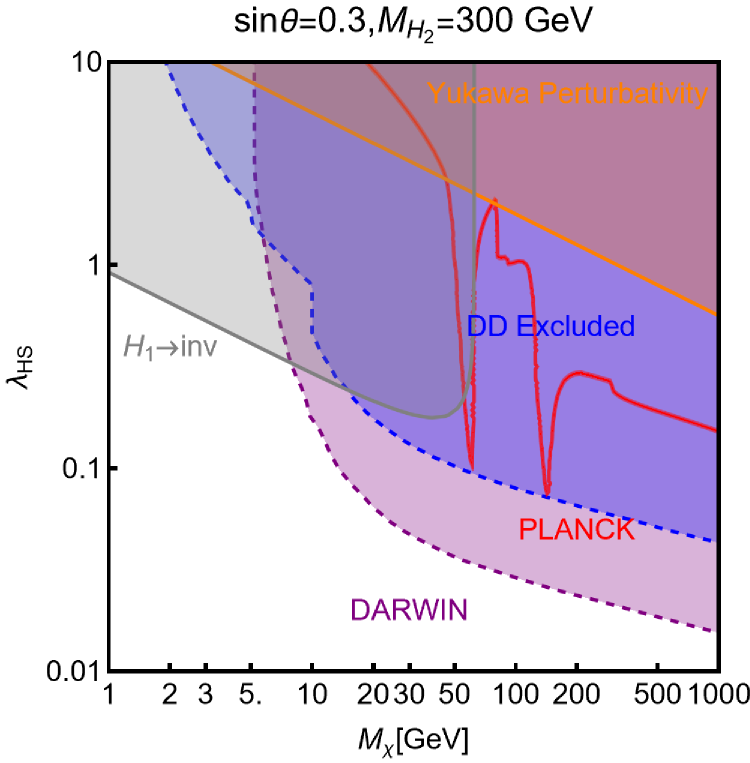}}
    \subfloat{\includegraphics[width=0.33\linewidth]{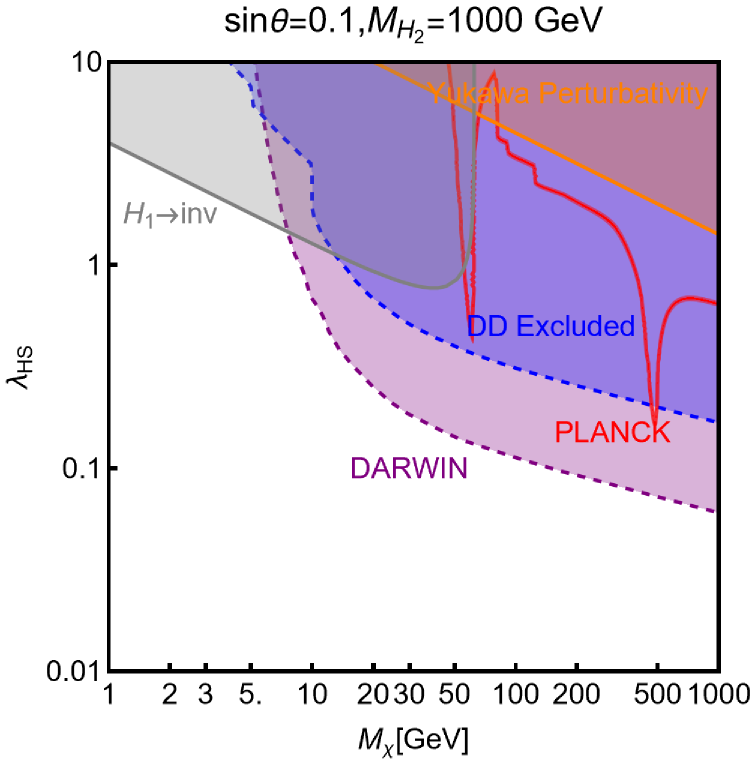}}
    \subfloat{\includegraphics[width=0.34\linewidth]{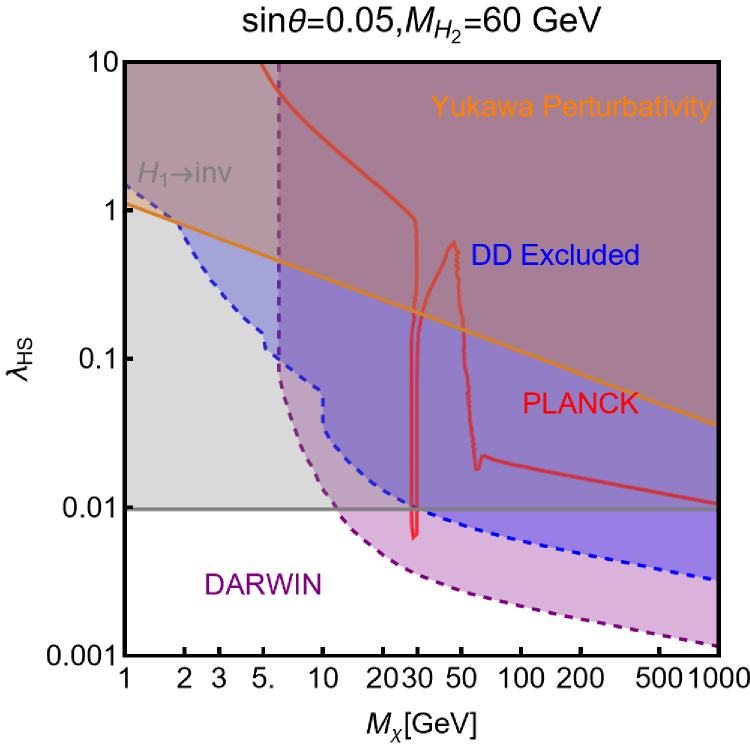}}
    \caption{Combination of the DM constraints for a model with a fermionic DM interacting via a singlet extension of the SM Higgs sector. The three panels differ for the assignations of the $(M_{H_2},\,\sin\theta)$ ($H_1$ is identified with the 125 GeV SM-like Higgs boson), as mentioned on the top. In each plot, the orange coloured region corresponds to the non-perturbative values of the DM Yukawa coupling. The remaining colour coding is the same as of Fig. \ref{fig:EFTHportal}.}
    \label{fig:fermionU1}
\end{figure*}

To provide a first illustration of the constraints of the scenario under scrutiny we have selected three benchmark assignations of the $(M_{H_2},\sin\theta)$ pair. The first, $(300\,\mbox{GeV},0.3)$ corresponds to a mass of the second Higgs not too far from the SM-like state while the mixing angle $\theta$ has the maximal allowed value by bounds on the Higgs signal strengths. The second assignation corresponds to a heavy ($1$ TeV) second Higgs and $\sin\theta=0.1$. For this choice, the considered model should feature the EFT Higgs portal as the consistent limit \cite{Arcadi:2020jqf, Arcadi:2021mag}. The last choice 
$(60\,\mbox{GeV},0.05)$ corresponds to a light mediator configuration. Having set the $(M_{H_2},\sin\theta)$ pair, the combination of the DM constraints is shown in Fig.~\ref{fig:fermionU1}, in terms of the remaining two free parameters, i.e., $M_\chi$ and $\lambda_{HS}$. The results can be interpreted along a similar philosophy as the simplified models illustrated in the previous section. Each panel of the figure shows red coloured isocontours highlighting the narrow regions of the parameter space for which the correct DM relic density is obtained. Such regions survive provided the isocontours lie outside the coloured areas corresponding to both theoretical and experimental exclusions. Looking at more detail at the different panels we see that for the benchmarks having $M_{H_2}>M_{H_1}$,  the shape of the relic density isocontours strongly resembles the one of the EFT Higgs as long portals as $M_\chi < M_{H_2}/2$. When $M_\chi \sim M_{H_2}/2$, a second s-channel resonance occurs, shown an additional "cusp" in the DM isocontour. In the case of the light additional Higgs (last panel of Fig.~\ref{fig:fermionU1}), the corresponding pole is very narrow, due to the small decay width of $H_2$. Further, we notice that for high DM masses, the correct relic density is matched for very suppressed $\lambda_{HS}$. This is due to the efficient annihilation process $\chi \chi \rightarrow H_2 H_2$. Nevertheless, the high sensitivity of XENONnT/LZ experiments seem to entirely rule out the three benchmarks. To provide a complete picture, we have also included in Fig.~\ref{fig:fermionU1} the region excluded by the Higgs invisible decay bound (grey coloured regions) and the one corresponding to a non-perturbative value of the DM Yukawa coupling (orange coloured). Concerning the bound on the invisible decay of the Higgs, we notice that for the third benchmark the corresponding curve is flat with the DM mass $M_\chi$. This is because the decay of the Higgs receives an additional contribution from the $H_1 \rightarrow H_2 H_2$ process throughout irrespective of $M_\chi$. 
 
\begin{figure*}
    \centering
    \subfloat{\includegraphics[width=0.45\linewidth]{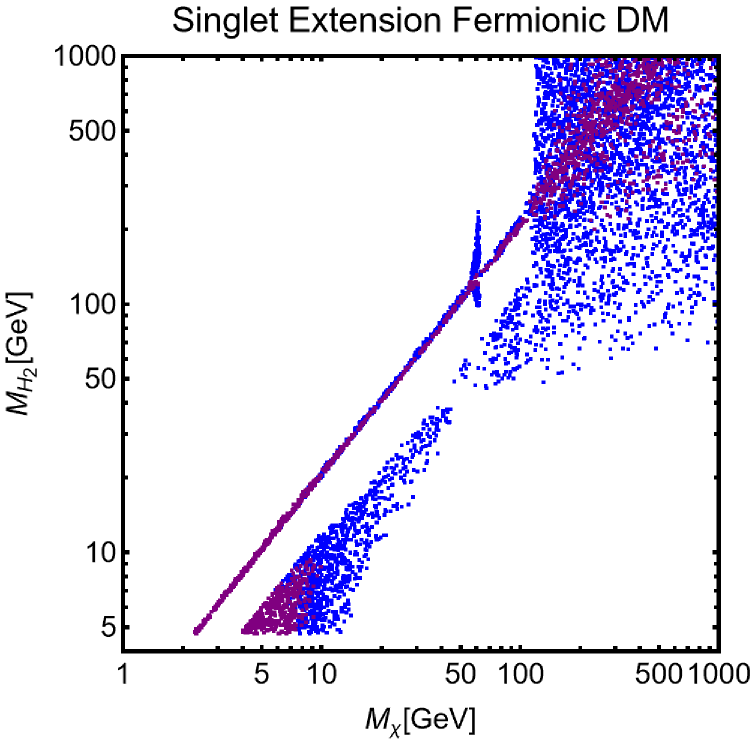}}
    \subfloat{\includegraphics[width=0.45\linewidth]{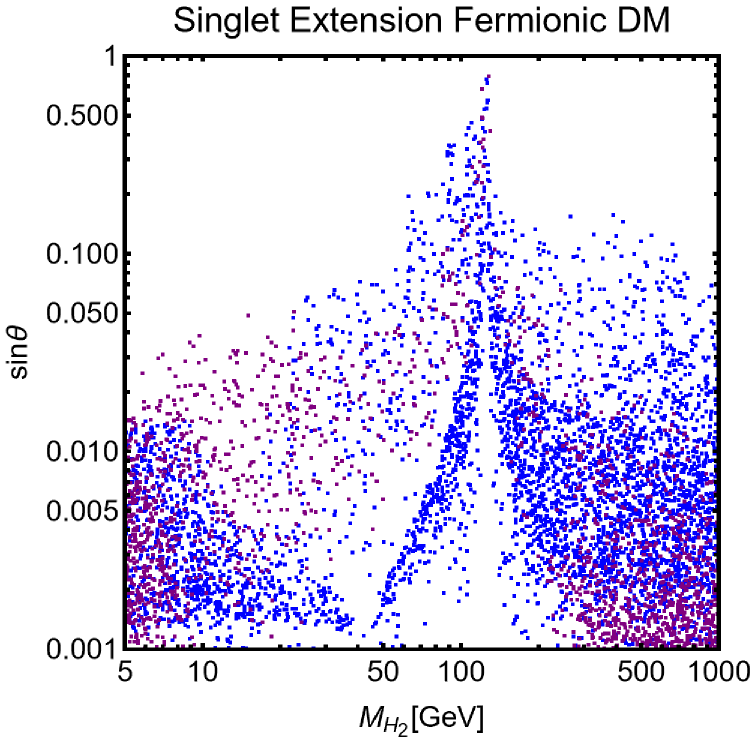}}
    \caption{Outcome of a parameter scan (see main text for details) for the model with a fermionic DM coupled to a Higgs sector made by the SM doublet $H$ and a real SM gauge singlet $S$ through the mass mixing. This framework is dubbed Singlet Extension Fermionic DM as mentioned on the top of each plot.
    The two panels show the $(M_\chi, M_{H_2})$ (left) and $(M_{H_2},\sin\theta)$ (right) planes. In each panel, the blue coloured points feature the correct DM relic density and pass all the present experimental constraints. The purple coloured points correspond to the parameter assignation compatible with an eventual future bound by the DARWIN experiment.}
    \label{fig:scanFU1}
\end{figure*}

To assess, in a more sensible way, whether there are viable region of parameter space, we have complemented the study of the three benchmarks with a parameter scan conducted over the following ranges:
%
\begin{align}
   & M_\chi \in [1,1000]\,\mbox{GeV},\,\,\,\,M_{H_2}\in [1,3000]\,\mbox{GeV},\nonumber\\
   & \lambda_{HS}\in \left[10^{-3},1\right],\,\,\,\,\, \sin\theta \in \left[10^{-3},1\right].
\end{align}
%
The outcome of such a parameter scan has been shown in Fig.~\ref{fig:scanFU1}. The figure displays in the $(M_\chi,\, M_{H_2})$ (left plot of Fig.~\ref{fig:scanFU1}) and $(M_{H_2},\,\sin\theta)$ (right plot of Fig.~\ref{fig:scanFU1}) planes the model points featuring the correct relic density and scattering cross-section on nucleons below the current limits (blue coloured points) as well as the parameter assignations (purple coloured points) which would comply with a negative result from the DARWIN experiment. By the inspection of the figure, we notice in particular a very strong bound on the mixing angle, namely $\sin\theta <0.1$ except the region $M_{H_1} \simeq M_{H_2}$. As evident from the analytical expressions, in such a region we have a suppression of the DM scattering cross-section in the NR limit due to the destructive interference between the contribution of the two mediators $H_1,\, H_2$ (see Eq. (\ref{eq:h1h2medfermiondm})). Remarkably, the DARWIN experiment will have the capability of probing almost the entire parameter space of the model, but the pole $M_\chi \sim M_{H_2}/2$, for parameter assignations corresponding to very narrow widths, and a small region with $M_\chi \lesssim 5\,\mbox{GeV}$, due to the energy threshold limitations. 

\paragraph{Vectorial DM from a dark $U(1)$:}
In the absence of any extra fermions, the gauge boson associated with the extra gauge symmetry can itself be the DM candidate. Starting from a Lagrangian of the form:
\begin{equation}
    \mathcal{L}_{\rm hidden}=-\frac{1}{4}V_{\mu \nu}V^{\mu \nu}+{\left(D^\mu S \right)}^{\dagger}\left(D_\mu S\right)-V(S,H) \, , 
\end{equation}
where $D_\mu=\partial_\mu +i \tilde{g} V_\mu$, $\tilde{g}$ being the gauge coupling of the new $U(1)$ symmetry. The vector DM dynamically gets a mass $M_V=\frac{1}{2}\tilde{g}v_S$ after the spontaneous breaking of the associated $U(1)$ symmetry via the VEV of $S$. This setup automatically features a $Z_2$ symmetry under which $V_\mu \rightarrow -V_\mu$ \cite{Gross:2015cwa}, hence there is no need to introduce an ad-hoc symmetry to make the DM stable. The mass mixing between $S$ and $H$ allows us to write a portal Lagrangian of the form:
\begin{align}
    \mathcal{L}&=\frac{\tilde{g}M_V}{2}\left(-H_1 s_\theta + H_2 c_\theta\right)V_\mu V^\mu \nonumber\\
    & +\frac{\tilde{g}^2}{8}\left(H_1^2 s^2_\theta -2 H_1 H_2 s_\theta c_\theta+H_2^2 c^2_\theta\right)V_\mu V^\mu \nonumber\\
    & + \mathcal{L}_{\rm scalar, SM}+\mathcal{L}_{\rm scalar, trilinear},
\end{align}
where $s_\theta,\, c_\theta=\sin\theta,\,\cos\theta$, and $\mathcal{L}_{\rm scalar,SM},\,\mathcal{L}_{\rm scalar,trilinear}$ are given by Eqs. (\ref{eq:Lag-mix-SM}) and
(\ref{eq:Lag-scalar-trilinear}).

\begin{figure*}[!t]
    \begin{center}    \subfloat{\includegraphics[width=0.33\textwidth]{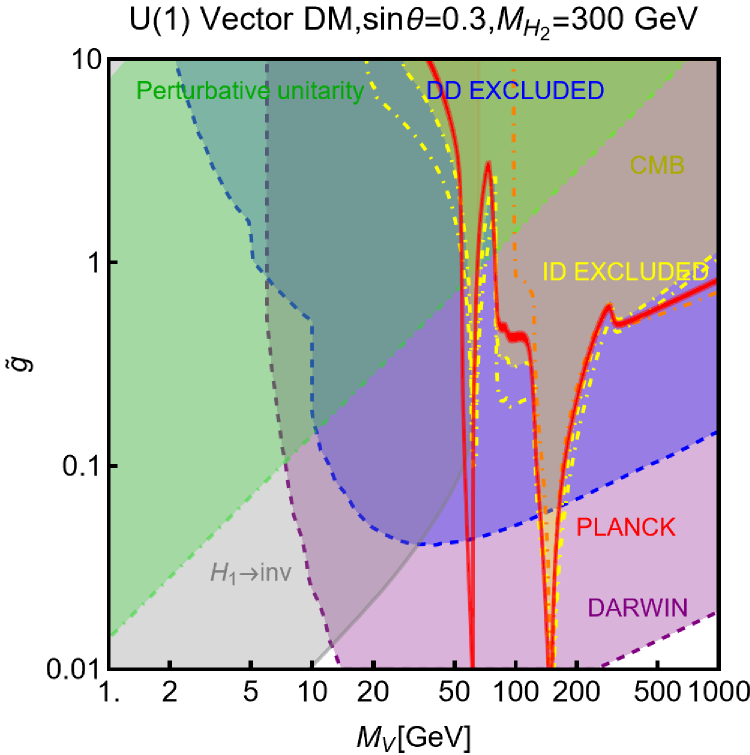}}
\subfloat{\includegraphics[width=0.33\textwidth]{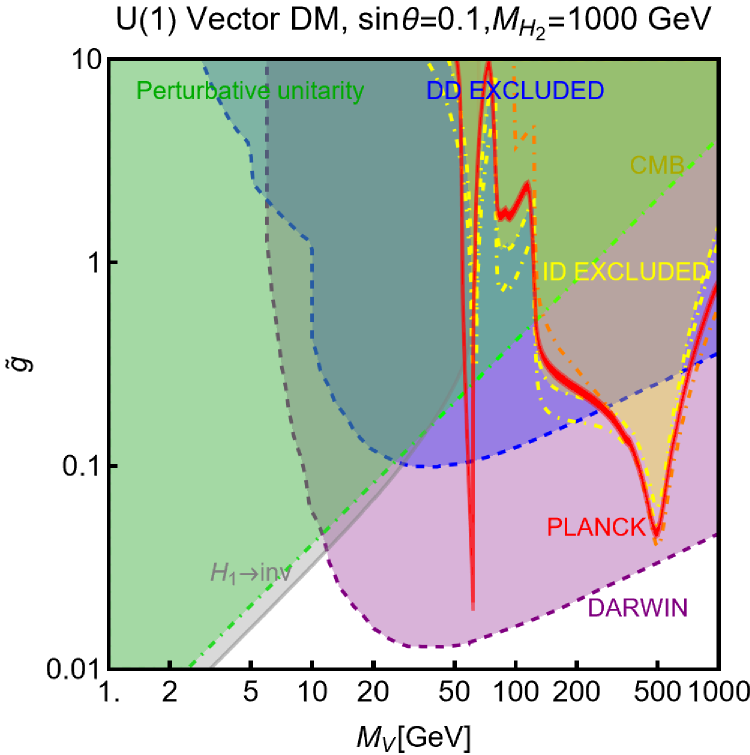}}    \subfloat{\includegraphics[width=0.33\textwidth]{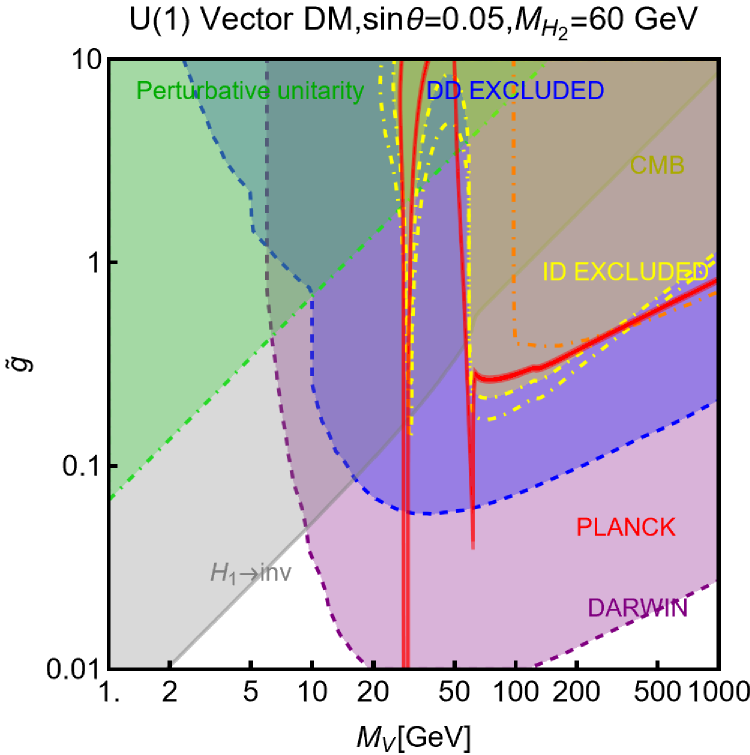}}
\end{center}
    \caption{Summary of the DM constraints for the dark $U(1)$ vector DM model. The colour codes and the $(M_{H_2},\,\sin\theta)$ values are the same as the fermionic DM model described in Fig.~\ref{fig:fermionU1}. In addition to these regions, exclusion regions from the ID constraints are shown in yellow colour while the green coloured regions represent the parameter space not compatible with the perturbative unitarity of the concerned scalar sector couplings.}
    \label{fig:pU1}
\end{figure*}
We have repeated the same analysis performed for the case of a fermionic DM, considering the same three assignations of the $(M_{H_2},\sin\theta)$ pair.
This time the results have been shown in the $(M_V,\tilde{g})$ plane, as depicted in Fig.~\ref{fig:pU1}. The relic density, DD and $H_1$ invisible decay constraints have been shown with the same colour convention as the case of a fermionic DM shown in Fig.~\ref{fig:fermionU1}. Contrary to the fermionic DM scenario, we see a new green coloured region in Fig.~\ref{fig:pU1}, corresponding to the parameter space not compatible with the perturbative unitarity constraints $\lambda_{H,\, S,\, {HS}} \leq \mathcal{O}(4\pi/3)$ \cite{Chen:2014ask}. For completeness, the regions excluded by the ID constraints have been marked as well with yellow colour, evidencing that the latter are competitive with their DD counterparts in the resonant regions, since they follow the shape of the relic contours

\begin{figure*}
    \centering
    \subfloat{\includegraphics[width=0.45\linewidth]{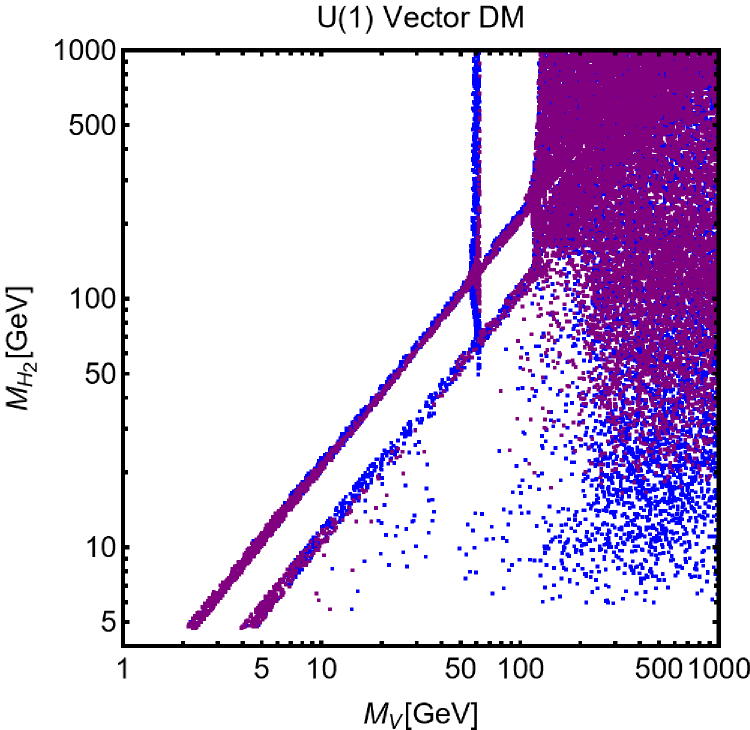}}
    \subfloat{\includegraphics[width=0.45\linewidth]{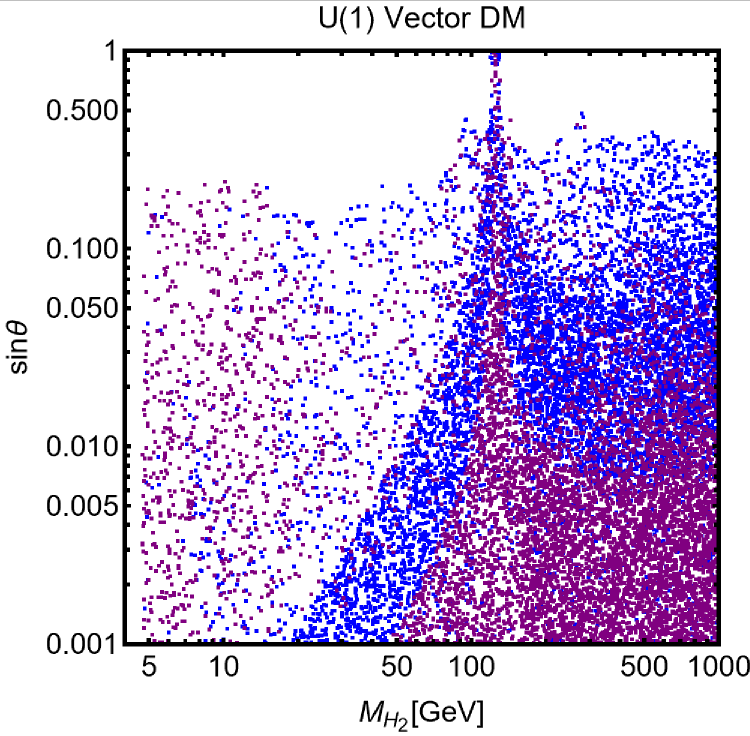}}
    \caption{Parameter scan for the dark $U(1)$ vector DM model. The colour convention is the same as Fig.~\ref{fig:scanFU1}.}
    \label{fig:scanU1}
\end{figure*}

Along the same footing of the fermion DM scenario, we complemented Fig.~\ref{fig:pU1} with a parameter scan over the ranges:
\begin{align}
\label{eq:U1scan}
   & M_V \in [1,1000]\,\mbox{GeV},\,\,\,\,M_{H_2}\in [1,3000]\,\mbox{GeV},\nonumber\\
   & \tilde{g}\in \left[10^{-3},1\right],\,\,\,\,\, \sin\theta \in \left[10^{-3},1\right],
\end{align}
and reported the result in Fig.~\ref{fig:scanU1} following the same colour conventions of the fermionic DM scenario shown in Fig.~\ref{fig:scanFU1}.

\paragraph{Vector DM from larger gauge groups:}
A double Higgs portal framework can also be obtained, at least as a limiting scenario, by embedding the vector DM candidate in larger gauge groups. 
Along this reasoning, the first possibility would be represented by a dark $SU(2)$. The starting Lagrangian is given by:
\begin{equation}
    \mathcal{L}_{\rm SU(2)}=-\frac{1}{4}F_{\mu \nu}^a F^{a \mu \nu}+\left(D_\mu \phi\right)^\dagger D^\mu \phi-V(\phi,H),
\end{equation}
with $\phi$ begin a dark $SU(2)$ Higgs doublet decomposed as:
\begin{equation}
    \phi=\frac{1}{\sqrt{2}}\left(
    \begin{array}{c}
         0  \\
         v_S+S 
    \end{array}
    \right).
\end{equation}
Again, after the spontaneous breaking of the dark $SU(2)$ gauge symmetry, a residual discrete symmetry appeared naturally. This time we will have a $Z_2 \times Z_2'$ and the following transformation properties for the DM fields:
\begin{align}
    & Z_2: V_\mu^1 \rightarrow -V_\mu^1,\,\,\,\, V_\mu^2 \rightarrow -V_\mu^2,\,\,\,\,V_\mu^3 \rightarrow V_\mu^3 \nonumber\\
    & Z_2': V_\mu^1 \rightarrow -V_\mu^1,\,\,\,\,V_\mu^2 \rightarrow V_\mu^2\,\,\,\, V_\mu^3 \rightarrow -V_\mu^3.
 \end{align}
In the physical basis, the relevant Lagrangian for DM phenomenology is written as follows:
\begin{align}
    \mathcal{L}&=\frac{\tilde{g} M_{V}}{2}\left(-s_\theta H_1 +c_\theta H_2\right)\sum_{a=1}^3V_{\mu}^a V^{\mu\,a}\nonumber\\
    & +\tilde{g}\epsilon_{abc}\partial_\mu V_\nu^a V^{b\,\mu}V^{c\,\nu}\nonumber\\
    &-\frac{\tilde{g}^2}{4}\left[ {\left(V_\mu^a V^{a\,\mu}\right)}^2-\left(V_\mu^a V_\nu^a V^{b\,\mu} V^{b\,\nu}\right)\right]\, . 
\end{align} 
As discussed e.g., in Refs. \cite{Gross:2015cwa, Arcadi:2021mag} (model of $SU(2)$ DM have been presented as well in \cite{Hambye:2007vf,Hambye:2009fg}), the phenomenology almost resembles the one of the $U(1)$ case just upon the rescaling $\tilde{g}\rightarrow \sqrt{3}\tilde{g}$. For this reason, we will not show results of this scenario explicitly.
A different phenomenology is instead achieved by further enlarging the dark gauge group to $SU(3)$. The model is built starting from the following Lagrangian:
\begin{align*}
    & \mathcal{L}_{\rm Higgs}=-\frac{\lambda_H}{2}|H^\dagger H|^2-m_H^2 H^\dagger H \, , \nonumber\\
    & \mathcal{L}_{\rm portal}=-\lambda_{H11} H^\dagger H \phi_1^\dagger \phi_1-\lambda_{H22} H^\dagger H \phi_2^\dagger \phi_2 \nonumber\\
    & +\left(H^\dagger H \phi_1^{\dagger}\phi_2+\mbox{H.c}\right) \, ,  
    \end{align*}
    \begin{align}
    & \mathcal{L}_{\rm hidden}=-\frac{1}{2}\mbox{Tr}\left\{V_{\mu \nu}V^{\mu \nu}\right\}+|D_\mu \phi_1|^2+|D_\mu \phi_2|^2-V_{\rm hidden} \, , \nonumber\\
    & V_{\rm hidden}=m_{11}^2 \phi_1^\dagger \phi_1 +m_{22}^2 \phi_2^\dagger \phi_2-m_{12}^2\left(\phi_1^{\dagger}\phi_2+\mbox{H.c.}\right) \, ,  \nonumber\\
    & +\frac{\lambda_1}{2}|\phi_1^\dagger \phi_1|^2+\frac{\lambda_2}{2}|\phi_2^\dagger \phi_2|^2+\lambda_3 \left(\phi_1^\dagger \phi_1\right) \left(\phi_2^\dagger \phi_2\right)+\lambda_4 |\phi_1^{\dagger}\phi_2|^2 \, ,  \nonumber\\
    & +\left[\frac{\lambda_5}{2}\left(\phi_1^{\dagger}\phi_2\right)^2+\lambda_6 \left(\phi_1^\dagger \phi_1\right) \left(\phi_1^{\dagger}\phi_2\right)\right.\nonumber\\
    &\left. +\lambda_7 \left(\phi_2^\dagger \phi_2\right)\left(\phi_1^{\dagger}\phi_2\right)+\mbox{H.c.}\right] \, , 
\end{align}
where we use the usual notation $V_{\mu \nu}=\partial_\mu V_\nu-\partial_\nu V_\mu +i \tilde{g} \left[V_\mu,V_\nu\right]$, $D_\mu \phi_i =\left(\partial_\mu +i \tilde{g}V_\mu \right)\phi_i$.

Besides the SM Higgs doublet $H$, the Higgs sector of the concerned model is made by two $SU(3)$ triplets with misaligned VEVs which, in the unitary gauge, can be decomposed as:
\begin{equation}
    \phi_1= \frac{1}{\sqrt{2}}\left(
    \begin{array}{c}
         0  \\
         0  \\
         v_1+h_1
    \end{array}
    \right),\,\,\,\,\phi_2= \frac{1}{\sqrt{2}}\left(
    \begin{array}{c}
         0  \\
         v_2+h_2  \\
         v_3+h_3+i \left(v_4+ h_4\right)
    \end{array}
    \right) \, . 
\end{equation}
As discussed in Ref. \cite{Gross:2015cwa}, this represents the minimal choice for the Higgs sector ensuring a complete breaking of the $SU(3)$ symmetry. The complete breaking of the symmetry is a crucial requirement as it prevents the presence of massless degrees of freedom, very dangerous from the cosmological perspective. The breaking of the dark gauge symmetry leaves, also in this case, a relic $Z_2 \times Z_2'$ symmetry. The gauge bosons satisfy the following transformation relations:
\begin{equation}
    V_\mu^a \rightarrow \eta(a) V_\mu^a,
\end{equation}
with:
\begin{align}
     Z_2: & \eta(a)=1,\,\,\,\mbox{for}\,\,\,a=1,2,4,5 \nonumber\\
    & \eta(a)=-1\,\,\,\,\mbox{for}\,\,\,a=3,6,7,8\nonumber\\
    Z_2': & \eta(a)=-1\,\,\,\mbox{for}\,\,\,a=1,3,4,6,8\nonumber\\
    & \eta(a)=1\,\,\,\mbox{for}\,\,\,a=2,5,7.
\end{align}

After the $SU(3)$ and EW breaking and assuming CP conservation (i.e. $v_4=0$), the Higgs mass Lagrangian is written as:
\begin{equation}
    \mathcal{L}=-\frac{1}{2}\Phi^T \mathcal{M}_{\rm CP-even}^2 \Phi-\frac{1}{4}\left(\lambda_4-\lambda_5\right) (v_1^2+v_2^2) \psi^2 \, , 
\end{equation}
where $\psi \equiv h_4$ is a CP-odd state while $\Phi={\left(H,h_1,h_2,h_3\right)}^T$ are, instead, CP-even states. The mass matrix of the latter is given by
\begin{align}
    & \mathcal{M}_{\rm CP-even}^2=\nonumber\\
    & \left(\begin{array}{cccc}
    \lambda_H v_h^2  &  \lambda_{H11}v_h v_1  &  \lambda_{H22}v_h v_2     &  0  \\
     \lambda_{H11}v_h v_1  &  \lambda_1 v_1^2   &  \lambda_3 v_1 v_3  & 0 \\
     \lambda_{H22}v_h v_2 &  \lambda_3 v_1 v_3  & \lambda_2 v_2^2   & 0 \\
     0 & 0 & 0 & \frac{1}{2}\left(\lambda_4+\lambda_5\right)(v_1^2+v_2^2)
    \end{array}
    \right) \, .
\end{align}
As already pointed out that we want to reduce the $SU(3)$ to a portal model as the ones previously discussed. This task can be achieved by assuming $\lambda_{H11}=\lambda_3 \ll 1$  (see e.g., Refs.~\cite{Gross:2015cwa,Arcadi:2016kmk,Arcadi:2016qoz} for a more extensive discussion). In such a setup, one has two CP-even mass eigenstates $H_3 \simeq h_3$ and $H_4 \simeq h_1$ having negligible mixing with the SM doublet Higgs and, consequently, negligible interactions with the visible sectors. The only CP-even mass eigenstates relevant for the DM phenomenology will be again dubbed $H_{1,2}$ and will be a mixture of the SM and the  dark components, weighted by an angle $\theta$:
\begin{align}\label{eq:H1H2newSU3}
    H_1 \simeq \cos\theta h- \sin\theta h_2 \,\,\,\,\, ,
     H_2 \simeq \sin \theta h + \cos\theta h_2,
\end{align}
with
\begin{equation}
    \tan 2\theta \simeq \frac{2 \lambda_{H22}v_h v_2}{\lambda_2 v_2^2-\lambda_H v_h^2} \, . 
\end{equation}
As customary, we adopt the convention that $H_1$ is identified as the 125 GeV SM-like Higgs.
To complete the characterisation of the model we need to discuss the vector sector. $SU(3)$ being completely broken, we have eight massive gauge bosons. Six of them, which we call $V^{1,2},V^{4,5}, V^{6,7}$ group in three mass degenerate pairs with masses:
\begin{align}\label{eq:vmass1}
    & M_{V^1}^2=M_{V^2}^2=\frac{1}{4}\tilde{g}^2v_2^2,\,\,\,\, M_{V^4}^2=M_{V^5}^2=\frac{1}{4}\tilde{g}^2v_1^2,\,\,\,\,\nonumber\\
    & M_{V^6}^2=M_{V^7}^2=\frac{1}{4}\tilde{g}^2(v_1^2+v_2^2),\,\,\,\,
\end{align}
while the remaining two combine into two mass eigenstates $V^3$ and $V^8$ with masses: 
\begin{equation}
\label{eq:primed_vectors}
    M_{V^{3}}^2= \frac{\tilde{g}^2 v_2^2}{4}\left(1-\frac{\tan\alpha}{\sqrt{3}}\right),\,\,\,\,\, M_{V^{8} }^2= \frac{\tilde{g}^2 v_1^2}{4}\frac{1}{\left(1-{\tan\alpha}/{\sqrt{3}}\right)} \, , 
\end{equation}
with:
\begin{equation}\label{eq:vmass2}
    \alpha= \frac{1}{2} \arctan\left(\frac{\sqrt{3}v_2^2}{2 v_1^2+v_2^2}\right) \, . 
\end{equation}
The breaking of the dark $SU(3)$ leaves, as the remnant, a discrete $Z_2 \times Z_2^\prime$ symmetry, under which the new particle sector has the following charges as summarized in Tab.~\ref{tab:dmSU3}
\begin{table}[!h]
\begin{center}
\renewcommand{\arraystretch}{1.6}
\begin{tabular}{ccc}\hline
gauge eigenstates & mass eigenstates &
 $Z_2 \times Z_2^{\prime}$
\\\hline\hline
$h,h_1,h_2,h_3,V_\mu^7$ &
$H_1, H_2, H_3, H_4,V_\mu^7$ & $(+,+)$\\
$V_\mu^1,V_\mu^4$ &
$V_\mu^1,V_\mu^4$ & $(-,-)$\\
$V_\mu^2,V_\mu^5$ &
$V_\mu^2,V_\mu^5$ & $(-,+)$\\
$h_4,V_\mu^3,V_\mu^6,V_\mu^8$ &
$\psi,V_\mu^{\prime3},V_\mu^6,V_\mu^{\prime8}$ & $(+,-)$\\\hline
\end{tabular}
\caption{$Z_2 \times Z_2^{\prime}$ assignments of the various fields of the $SU(3)$ dark model.}
\label{tab:dmSU3}
\end{center}
\end{table}
By taking $v_1 \gg v_2$, we can decouple at a high energy scale five of the eight vectors. In such a setup, the Lagrangian relevant for the DM phenomenology resembles again the one of the Higgs portal and is written as:

\begin{align}
\label{eq:SU3_lagrangian}
     \mathcal{L}& =\frac{\tilde{g} M_{V}}{2}\left(-\sin\theta H_1 +\cos\theta H_2\right)\left(\sum_{a=1,2}V_{\mu}^a V^{\mu\,a}\right. \nonumber\\
     &\left. +{\left(\cos\alpha-\frac{\sin\alpha}{\sqrt{3}}\right)}^2V_\mu^3 V^{\mu\,3}\right)\nonumber\\
    & +\tilde{g}\cos\alpha \sum_{a,b,c}\epsilon_{abc}\partial_\mu V_\nu V_\nu^a V^{b\,\mu}V^{c\,\nu}\nonumber\\
    & -\frac{\tilde{g}^2}{2}\cos^2 \alpha \sum_{a=1,2}\left(V_\mu^a V^{a\,\mu}V_\nu^3 V^{3\,\nu}-{\left(V_\mu^a V^{a\,\mu}\right)}^2\right)\nonumber\\
    & -\frac{1}{2}m_\psi^2 \psi^2 +\left[\frac{\tilde{g}}{2 M_V}\left(-\sin\theta H_1 +\cos\theta H_2\right)\right.\nonumber\\
    & \left. -\frac{1}{4}\left(\lambda_{\psi \psi 11}H_1^2+2 \lambda_{\psi \psi 12}H_1 H_2+\lambda_{\psi \psi 22}H_2^2\right)\right]\psi^2\nonumber\\
    & -\frac{\kappa_{111}}{2}v_h H_1^3-\frac{\kappa_{112}}{2}H_1^2 H_2 v_h \sin\theta\nonumber\\
    & -\frac{\kappa_{221}}{2}H_1 H_2^2 v_h \cos\theta-\frac{\kappa_{222}}{2}H_2^3 v_h \nonumber\\
     & +\mathcal{L}_{\rm scalar,SM}\, , 
\end{align}
where $M_V$ represents a generic mass term for $V^1,\, V^2$ and $V^3$, taking the $v_1 \gg v_2$ limit, as can be seen from Eqs. (\ref{eq:vmass1})-(\ref{eq:vmass2}), $\kappa$'s are given by Eq.~(\ref{eq:trilinear}) and $\mathcal{L}_{\rm scalar,SM}$
is given in Eq. (\ref{eq:Lag-mix-SM}) with $H_1,\,H_2$ now defined by Eq. (\ref{eq:H1H2newSU3}). Notice that, to simplify the notation, we have renamed $V'^3 \rightarrow V^3$ (The $V'^3$ and $V^3$ states actually coincide in the limit $v_1 \gg v_2$).
The coupling of the latter with the Higgs bosons $H_{1,2}$ are given by:
\begin{align}
    & \lambda_{\psi \psi 11}=\frac{\tilde{g}}{2 M_V v_h}\sin\theta\left(\cos^3 \theta \left(M_{H_2}^2-M_{H_1}^2\right)\right.\nonumber\\
    & \left. +\frac{\tilde{g}}{2 M_V v_h}\sin\theta\left(\sin^2 \theta M_{H_1}^2+\cos^2 \theta M_{H_2}^2\right)\right) ,  \nonumber \\
    & \lambda_{\psi \psi 12}=\frac{\tilde{g}}{2 M_V v_h}\sin\theta \cos\theta\left(\sin\theta \cos\theta \left(M_{H_2}^2\!-\!M_{H_1}^2\right)\! \right.\nonumber\\
    & \left. -\!\frac{\tilde{g}}{2 M_V v_h}\sin\theta\left(\sin^2 \theta M_{H_1}^2\!+\!\cos^2 \theta M_{H_2}^2\right)\right) , \nonumber\\
    & \lambda_{\psi \psi 22}=\frac{\tilde{g}}{2 M_V v_h}\cos\theta \left(\sin^3 \theta \left(M_{H_2}^2-M_{H_1}^2\right)\right.\nonumber\\
    & \left. +\frac{\tilde{g}}{2 M_V v_h}\cos\theta\left(\sin^2 \theta M_{H_1}^2+\cos^2 \theta M_{H_2}^2\right)\right) \, . 
\end{align}
If CP is preserved in the scalar sector, the lightest CP-odd state is stable, together with the degenerate pair $V^{1,2}$. The case in which $V^3$ is the lightest state, as will be discussed below, the DM phenomenology will strongly resemble the case of a single vector DM candidate, in the limit $v_2 \ll v_1$,\, $\alpha \ll 1$ and consequently, the mass and coupling of $V^3$ substantially coincide with the ones of $V^{1,2}$. The case of a light $\Psi$, instead, will represent a scenario of two-component DM. Together with these two cases, making a small exception to our convention of the strict CP conservation, we will consider as well the case of a tiny CP violation in the scalar sector of the model. While this will not affect the description of the mass spectrum just provided but will render both $V^3$ and $\Psi$ cosmologically unstable, restoring a strict one-component DM setup and opening some interesting perspectives for the DM phenomenology.

To illustrate the phenomenological results, we start discussing the scenario in which the DM is composed of $V^{1,2,3}$ including together the cases of cosmologically stable and unstable $V^3$.

\begin{figure*}
    \begin{center}    
\subfloat{\includegraphics[width=0.43\textwidth]{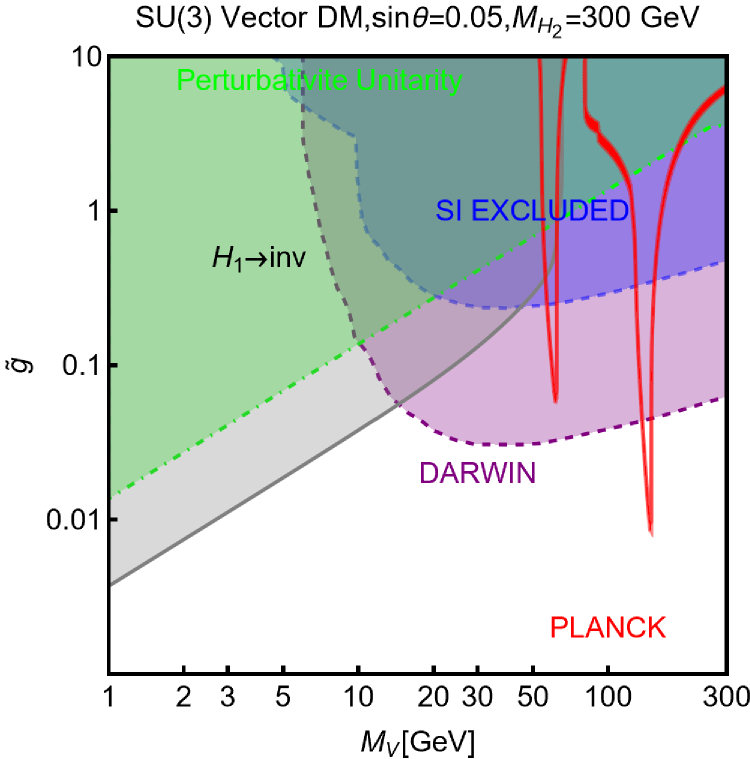}}    \subfloat{\includegraphics[width=0.43\textwidth]{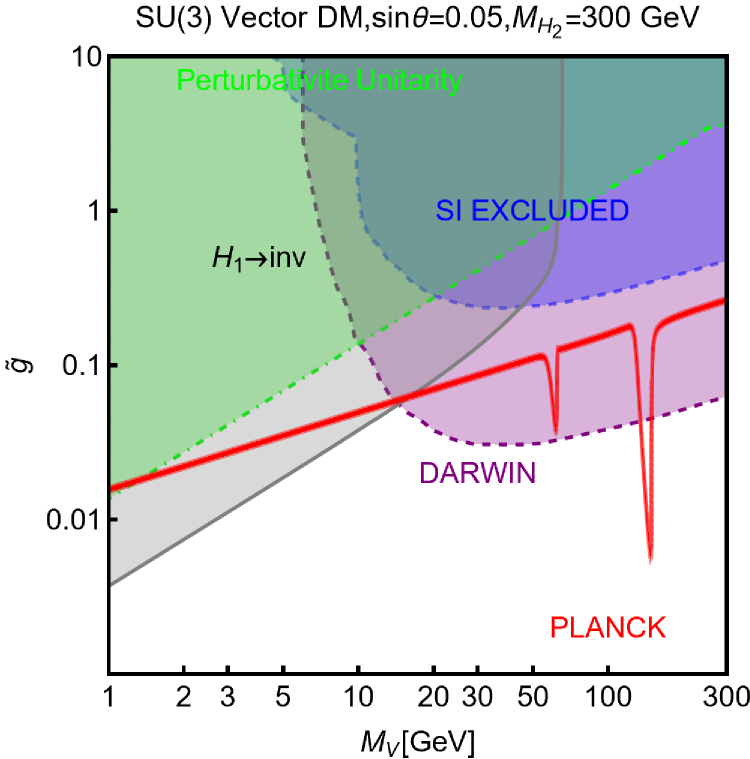}}\\
\subfloat{\includegraphics[width=0.43\textwidth]{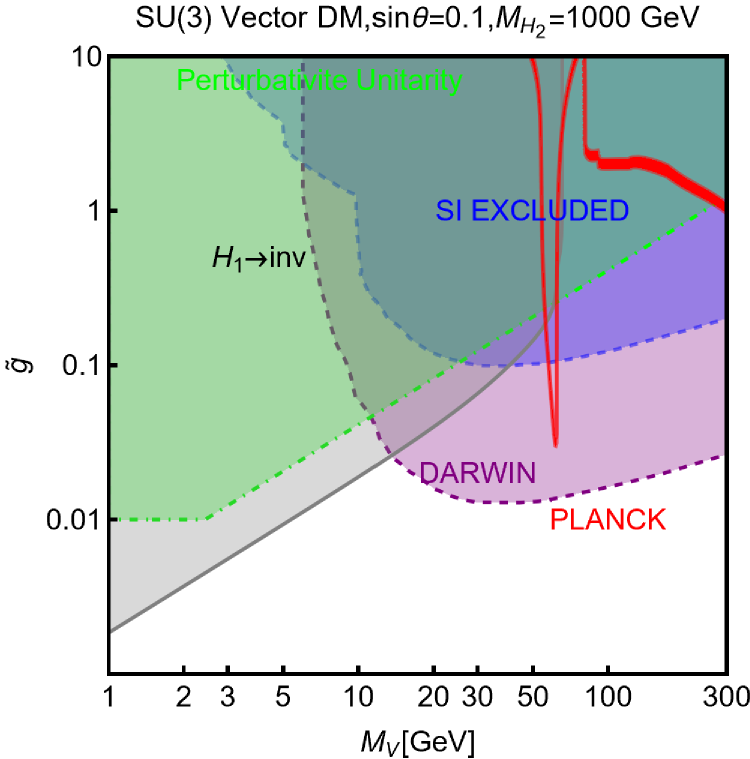}}    \subfloat{\includegraphics[width=0.43\textwidth]{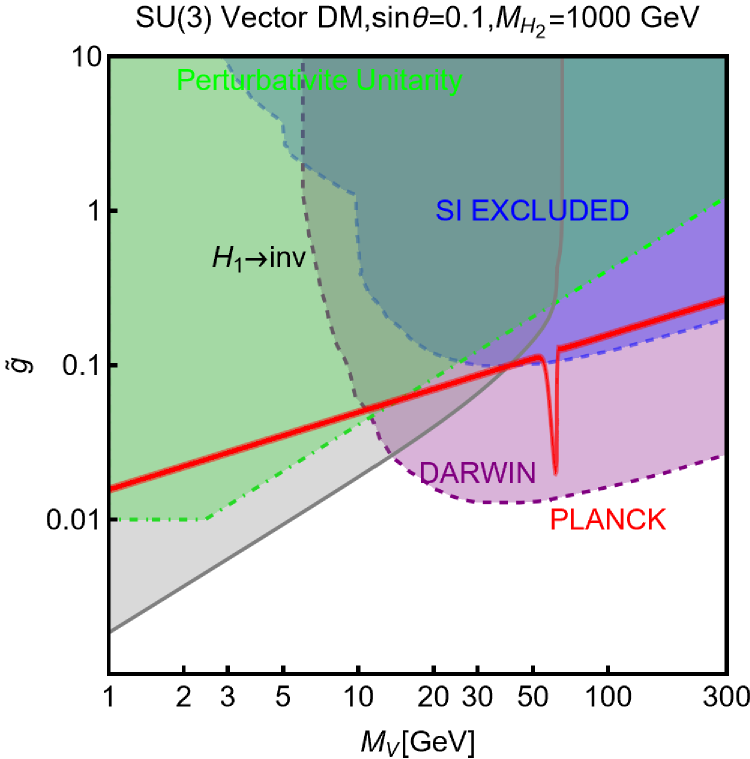}}\\
\end{center}
    \caption{Summary of the DM constraints for $SU(3)$ vector DM model. The colour convention is the same as Fig.~\ref{fig:pU1}. The top (bottom) row corresponds to $M_{H_2}=300$ GeV, $\sin\theta=0.05$ ($M_{H_2}=1000$ GeV, $\sin\theta=0.1$). The left column refers to the case where the DM is composed of the three cosmologically stable almost mass degenerate vectors $V^{1,2,3}$. The right column considers the case in which only $V^3$ is not cosmologically stable. }
    \label{fig:pSU3}
\end{figure*}
As customary for this kind of models, we start showing the combined constraints in the $(M_V,\tilde{g})$ plane (notice that $V\equiv V^1\equiv V^2 \equiv V^3$) for fixed $(M_{H_2},\sin\theta)$ values as shown in Fig.~\ref{fig:pSU3}. This time we have considered only two assignations of $(M_{H_2},\sin\theta)$ and masses of the DM up to a value of $300$ GeV, corresponding to our assignation $M_\Psi=300$ GeV. The left column of the both rows of Fig.~\ref{fig:pSU3} considers the case in which all three vectorial DM are stable. The combination of the DM constraints is analogous to the other vector DM models showing that the setup under consideration is currently viable mostly in the vicinity of the "pole", $M_V \sim M_{H_2}/2$. Very different is, instead, the scenario emerging from the right column of the both rows of the same figure, where the case of an unstable $V^3$ is considered. Most of the relic density  (red coloured) indeed, lie outside the excluded region (DARWIN will turn out to be an effective probe though). We are in front of one of the possible solutions for the tension, in WIMP models, between relic density and the DD constraints, i.e., the case in which the DM can annihilate into the light-dark sector states. This additional annihilation channel can reduce the DM relic density, making it easier to match the experimentally favoured value, without affecting the DM scattering rate on nucleons. Note that in the case of a small CP violation, considered here, $V^3$ is not a DM component but can be long-lived, compared to the collider scales. For this reason, the limit on the invisible decay of the Higgs, which receives a contribution from all three vectors, is the same for all four plots of Fig.~\ref{fig:pSU3}.

\begin{figure*}
    \centering
    \subfloat{\includegraphics[width=0.43\linewidth]{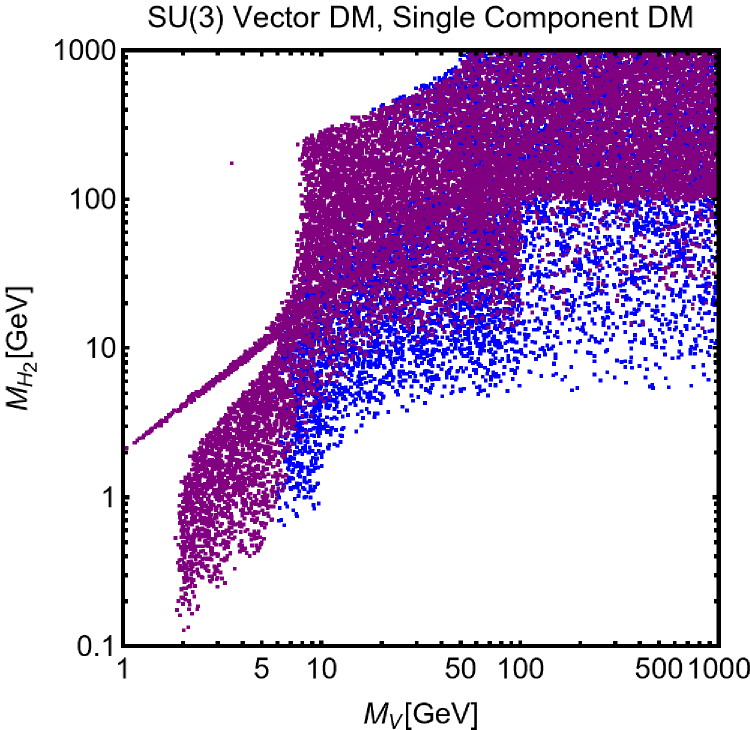}}
    \subfloat{\includegraphics[width=0.43\linewidth]{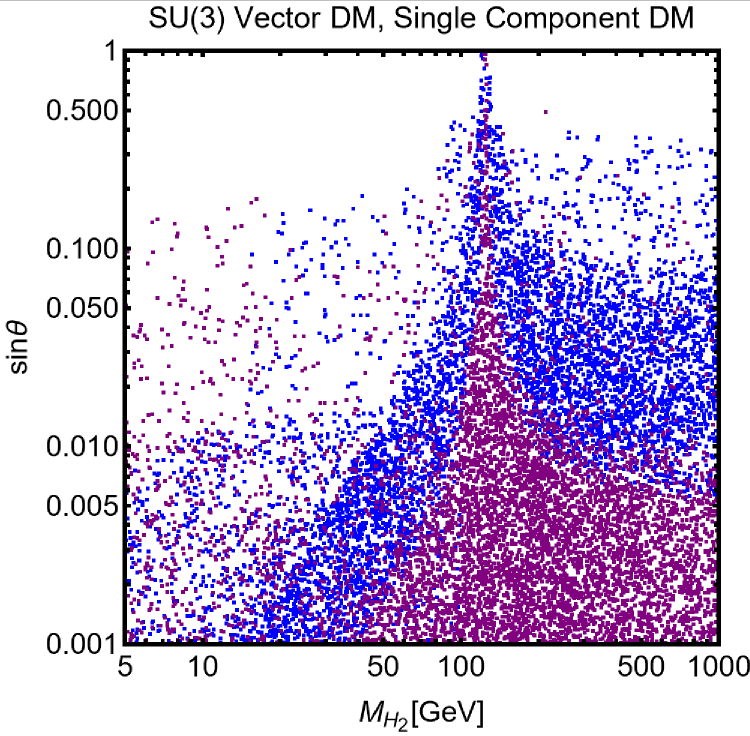}}\\
    \subfloat{\includegraphics[width=0.43\linewidth]{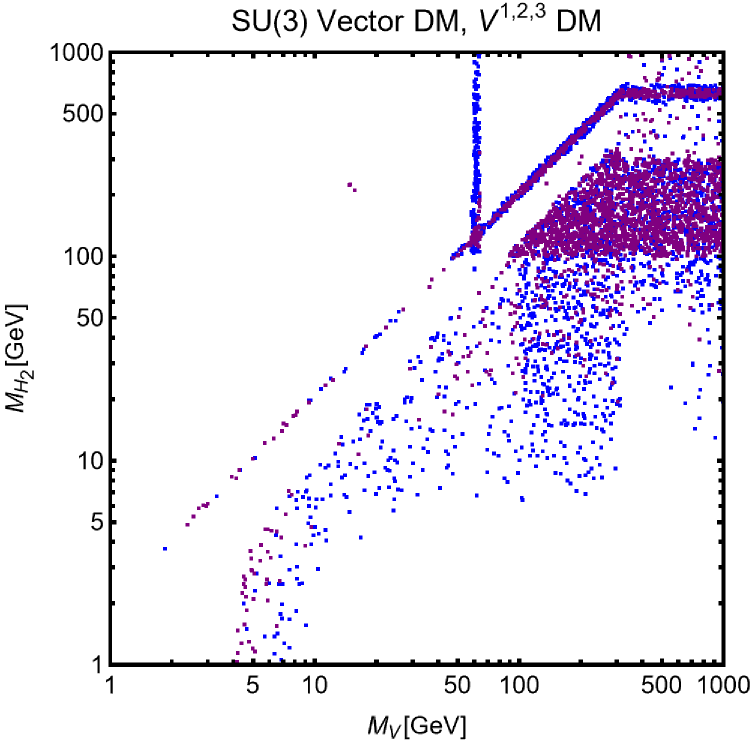}}
    \subfloat{\includegraphics[width=0.43\linewidth]{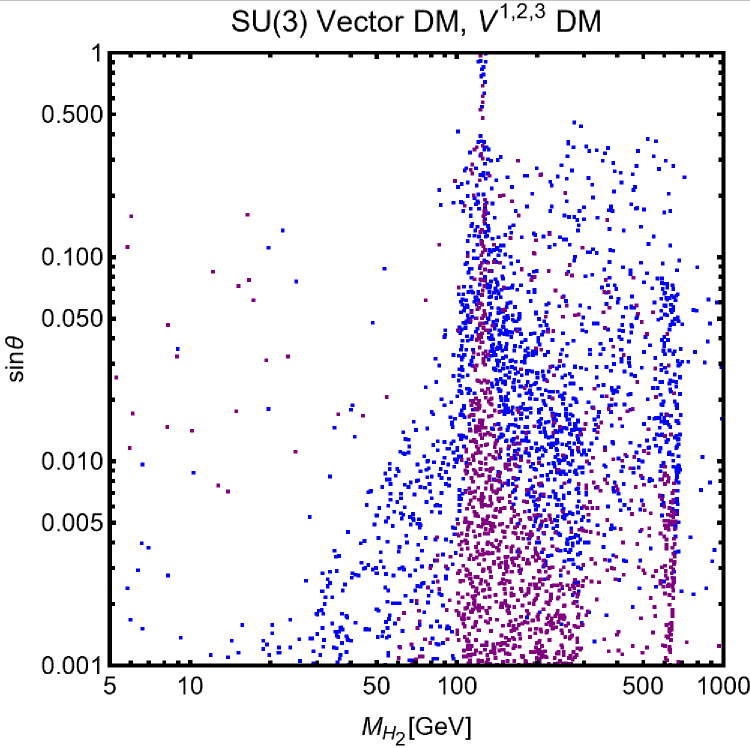}}
    \caption{Parameter scan for the dark $SU(3)$ vector model considering the cases of both single component (top row) and two-component (bottom row) DM. For both scenario the results are shown in the $(M_V,M_{H_2})$ (left) and $(M_{H_2},\sin\theta)$ planes. As usual, blue coloured points correspond to parameter assignations compatible with the current constraints, as given by XENON1T/XENONnT/LZ, but ruled out by negative results at the DARWIN experiment. Purple coloured points correspond to a viable parameter space even  no DM signals gets detected by DARWIN. }
    \label{fig:scanSU3}
\end{figure*}
%
The analysis has then been continued through a more general parameter scan. The ranges of the parameters are the same as eq. \ref{eq:U1scan}. (We remind again $M_V\equiv M_{V^1}=M_{V^2}=M_{V^3}$.) The corresponding outcome is shown in Fig.~\ref{fig:scanSU3} where the usual colour codes are used to highlight the parameter assignation compatible with the current and the near future DD constraints. Again both the cases of stable $V^3$ (top row) and unstable $V^3$ (bottom row) have been accounted for.

Let us now  finally move to the scenario of a scalar/vector two component DM. In such a case the DM relic density is described by a system of coupled Boltzmann's equations:
\begin{align}
    & \frac{dY_V}{dx}=-\frac{\langle \sigma v \rangle (VV \rightarrow X X)s}{Hx}\left(Y_{V}^2-Y_{V,eq}^2\right)\nonumber\\
    & -\frac{\langle \sigma v \rangle (VV \rightarrow \Psi \Psi) s}{Hx}\left(Y_V^2-\frac{Y_{V,eq}^2}{Y_{\Psi,eq}^2}Y_\Psi^2\right)\nonumber\\
    & - \frac{\langle \sigma v \rangle (VV \rightarrow V^3 H_{1,2})s}{Hx}\left(Y_V^2-\frac{Y_{\Psi,eq}}{Y_\Psi}Y_V^2\right)
    \end{align}
    \begin{align}
    & \frac{dY_\Psi}{dx}=-\frac{\langle \sigma v \rangle (\Psi \Psi \rightarrow X X)s}{Hx}\left(Y_{\Psi}^2-Y_{\Psi,eq}^2\right)\nonumber\\
    & +\frac{\langle \sigma v \rangle (VV \rightarrow \Psi \Psi) s}{Hx}\left(Y_V^2-\frac{Y_{V,eq}^2}{Y_{\Psi,eq}^2}Y_\Psi^2\right)\nonumber\\
    & -\frac{\langle \sigma v \rangle (V V^3 \rightarrow V H_{1,2})s}{Hx}Y_V Y_{V^3,eq}\left(\frac{Y_{\Psi}}{Y_{\Psi,eq}}-1\right)\nonumber\\
    & +\frac{\langle \sigma v \rangle (VV \rightarrow V^3 H_{1,2})s}{Hx}\left(Y_V^2-\frac{Y_{\Psi,eq}}{Y_\Psi}Y_V^2\right)
\end{align}
where $X$ represent either a SM state or a $H_{1,2}$ boson while $x=M_V/T$. Besides ordinary pair annihilations, the abundances of the two DM components are determined by the conversion process $VV \rightarrow \Psi \Psi$ as well be the semi-annihilations $VV \rightarrow V^3 H_{1,2}$ and semi-coannihilations $VV^3 \rightarrow V H_{1,2}$. The latter two processes contribute to the abundance of the particle $\Psi$ as it is produced in the decay of $V^3$. The numerical study performed in \cite{Arcadi:2016kmk} showed that semi-annihilation and semi-coannhilations contribute to the Boltzmann equations to a negligible extent. The total DM relic density can, in good approximation, be written as:
\begin{align}
    & \Omega_{\rm DM}h^2=\Omega_{V}h^2+\Omega_{\Psi}h^2\nonumber\\
    & \approx 8.8 \times 10^{-11}{\mbox{GeV}}^{-2}\left[{\left(\overline{g}_{\rm eff,V}^{1/2}\int_{T_0}^{T_{\rm f.o,V}}\langle \sigma v \rangle_V \frac{dT}{M_V} \right)}^{-1}\right.\nonumber\\
    & \left. +\left(\overline{g}_{\rm eff,\Psi}^{1/2}\int_{T_0}^{T_{\rm f.o,\Psi}}\langle \sigma v \rangle_\Psi \frac{dT}{M_\Psi} \right)^{-1}\right]
\end{align}
where:
\begin{align}
    & \langle \sigma v \rangle_V \equiv \langle \sigma v \rangle (VV \rightarrow XX)+\langle \sigma v \rangle (VV \rightarrow \Psi \Psi) \nonumber\\
    & \langle \sigma v \rangle_\Psi \equiv \langle \sigma v \rangle (\Psi \Psi \rightarrow XX).
\end{align}
$T_{f.o.,V,\Psi}$ are the Standard freeze-out temperatures of the two candidates, considered individually, computing according to the pair annihilation rates defined above. The DM fraction retained by the two DM components depends on relative size of their annihilation rates. For example we can define:
\begin{equation}
\label{eq:DMfraction}
    f_V=\frac{\Omega_V}{\Omega_{DM}}\approx \frac{1}{1+\frac{\langle\sigma v \rangle_V}{\langle \sigma v \rangle_\Psi}}
\end{equation}

The main interest in this two-component scenario relies, however, on DD. Indeed it provides an interesting example of how it is possible to "naturally" evade the DD constraints in a WIMP model\footnote{Other examples of mechanism to overcome DD bounds in WIMP models can be found in \cite{Gross:2017dan,Karamitros:2019ewv,Cai:2021evx}.}.
On general grounds, one would expect that the DM scattering over nucleons is described, for both the scalar and vector DM candidate, by Feynman's diagrams with $t$-channel exchange of the $H_{1,2}$ states. One would then expect, for the two DM candidates, scattering cross-section given by analogous expressions as the Higgs portal scenarios described in the previous sections. While this is the case for the vector DM candidate, if one noticed the analytic form of the coupling of a pair of $\Psi$ with $H_{1,2}$:
\begin{align}
    & \left[\lambda_2 v_2 \left(-\sin \theta H_1+\cos \theta H_2\right)+\lambda_{H22}\left(\cos \theta H_1+\sin\theta H_2\right)\right] \Psi^2\nonumber\\
    & =\frac{\tilde{g}^2}{4 M_V}\sin\theta \cos\theta \left(-M_{H_1}^2 \sin\theta H_1+ M_{H_2}^2 \cos\theta H_2\right)\Psi^2.
\end{align}
One would conclude that here a destructive interference between diagrams involving $H_{1}$ and $H_2$ occurs, in the NR limit, relevant for the DD. Thus, one gets an exact cancellation of the two contributions.
To understand the origin of this "blind spot", one can rewrite the effective four-field operator responsible for the scattering of $\Psi$ in the interaction basis. It will be given by a product of the form $A^\dagger (m^2)^{-1} B$ with $A$ and $B$ representing, respectively, the couplings of $h,h_1,h_2$ with the DM and the SM fermion pairs:
\begin{align}
    & A=\left(
    \begin{array}{c}
        g_{h\Psi \Psi} \\
        g_{h_1 \Psi \Psi}\\
        g_{h_2 \Psi \Psi}
    \end{array}
    \right)
    \propto \left(
    \begin{array}{c}
           v_h \lambda_{H22}\\
           v_1 \left(\lambda_3+\lambda_4-\lambda_5\right)\\
           v_2 \lambda_2
    \end{array}
    \right),\nonumber\\
    & B=\left(
    \begin{array}{c}
         g_{hff}  \\
         g_{h_1 ff}\\
         g_{h_2 ff}
    \end{array}
    \right)\propto \left(
    \begin{array}{c}
         k  \\
         0 \\
         0
    \end{array}
    \right),
\end{align}
while $m^2$ is the subset of the CP-even mass matrix corresponding to $(h,h_1,h_2)$:
\begin{equation}
    m^2=\left(
    \begin{array}{ccc}
       \lambda_H v_h & \lambda_{H11}v_h v_1 & \lambda_{H22}v_h v_2  \\
        \lambda_{H11} v_h v_1 & \lambda_1 v_1^2 & \lambda_3 v_1 v_2 \\
        \lambda_{H22}v_h v_2 & \lambda_3 v_1 v_2 & \lambda_2 v_2^2
    \end{array}
    \right).
\end{equation}
Combining the previous expression one can write the effective coupling between $\Psi$ and a nucleon $N$ as:
\begin{equation}
    g_{\psi \psi NN}\propto A^\dagger (m^2)^{-1} B \propto \lambda_{H11}\lambda_2-\lambda_{H22}\lambda_3,
\end{equation}
which is put automatically to zero by the assumption $\lambda_{H11}=\lambda_3 \simeq 0$, i.e., negligible mixing between $h_1$ and the other two scalars.
This result shows that the scalar DM component is COY \cite{Boehm:2014hva}. If it retains most of the DM relic density, we achieve a WIMP setup with relaxed constraints from the DD. As discussed in Ref. \cite{Arcadi:2016kmk}, this is indeed the case over larger regions of the parameter space. The main reason is the conversion process $V^{1,2}V^{1,2}\rightarrow \Psi \Psi$. The corresponding rate is very efficient as it involves vertices not suppressed by the mixing angle $\theta$, which is forced to be small by Higgs signal strength constraints. We then have $\langle \sigma v \rangle_V \gg \langle \sigma v \rangle_\Psi$. From eq.\ref{eq:DMfraction} we can conclude that the scalar DM component hence tends to retain most of the DM fraction. To have $\langle \sigma v \rangle_V \sim \langle \sigma v \rangle_\Psi$ one would need and enhancement of the annihilation cross-section of $\Psi$ into SM state, as can occur in correspondence of $s$-channel poles.

\begin{figure*}
    \centering
    \subfloat{\includegraphics[width=0.45\linewidth]{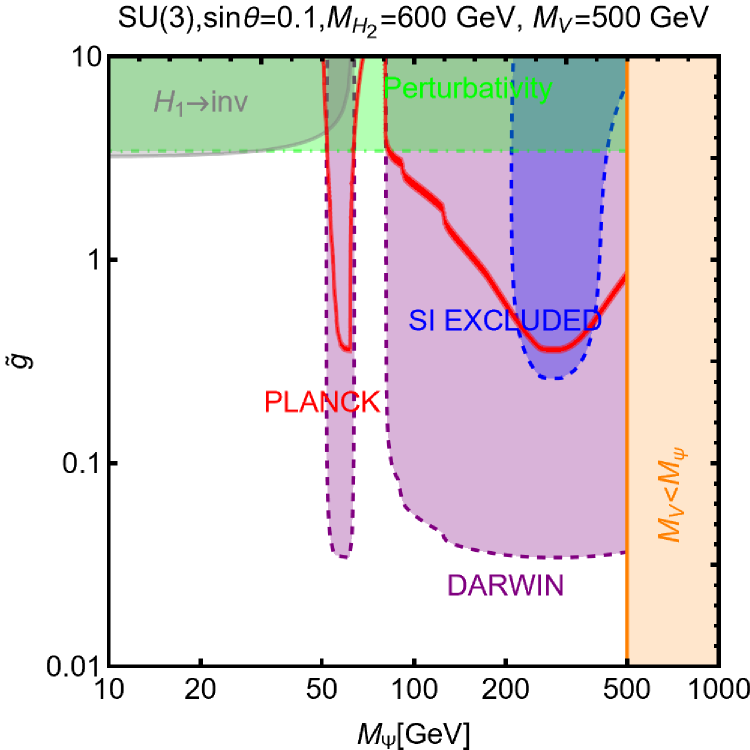}}
    \subfloat{\includegraphics[width=0.45\linewidth]{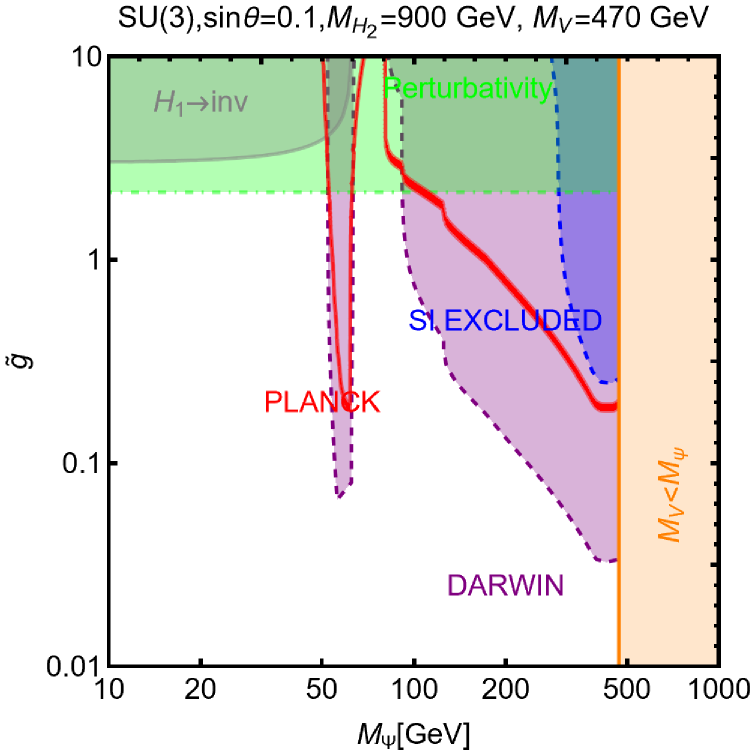}}
    \caption{The summary of the DM constraints on the dark $SU(3)$ model in the configuration with a scalar/vector two component DM in the $(M_\psi,\tilde{g})$ bidimensional plane. The orange-coloured region is excluded from the fact that $M_V<M_\psi$ in this part. The remaining colour codes are the same as of Fig. \ref{fig:pSU3}.  The other relevant parameter assignations are shown on top of the two panels.}
    \label{fig:pmPsiSU3}
\end{figure*}

This expectation is confirmed by Fig.~\ref{fig:pmPsiSU3}. It shows the DM constraints in the $(M_\psi,\tilde{g})$ plane. We see from Fig.~\ref{fig:pmPsiSU3} that most of the contours of the correct relic density evade the current DM constraints, but the $M_\psi \sim M_{H_2}/2$ pole. Remarkably, the expected sensitivity reach of DARWIN will be capable of probing, and possibly ruling out the SI interactions of the subdominant $V$-component of the DM relic density.

\begin{figure*}
    \centering
    \subfloat{\includegraphics[width=0.35\linewidth]{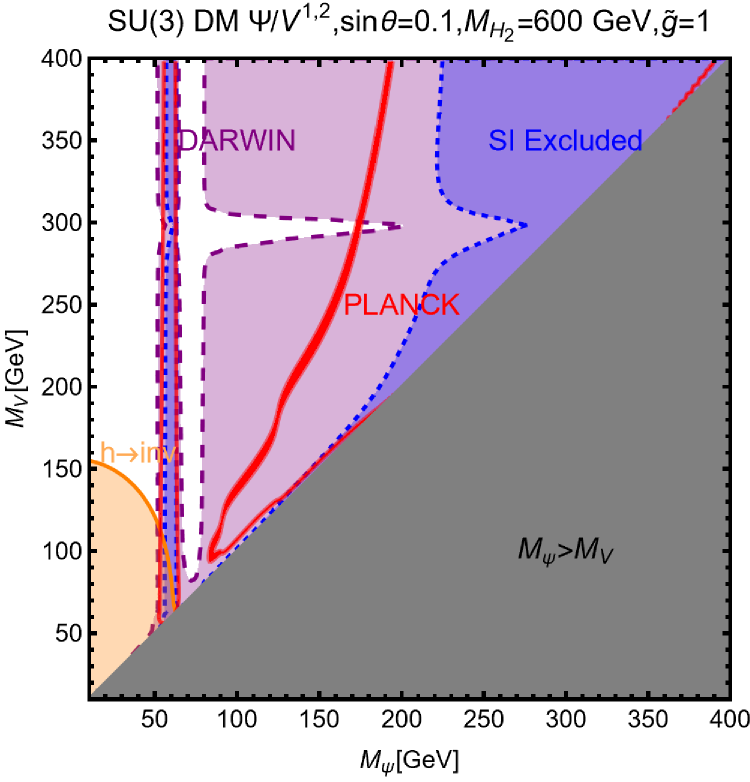}}
    \subfloat{\includegraphics[width=0.35\linewidth]{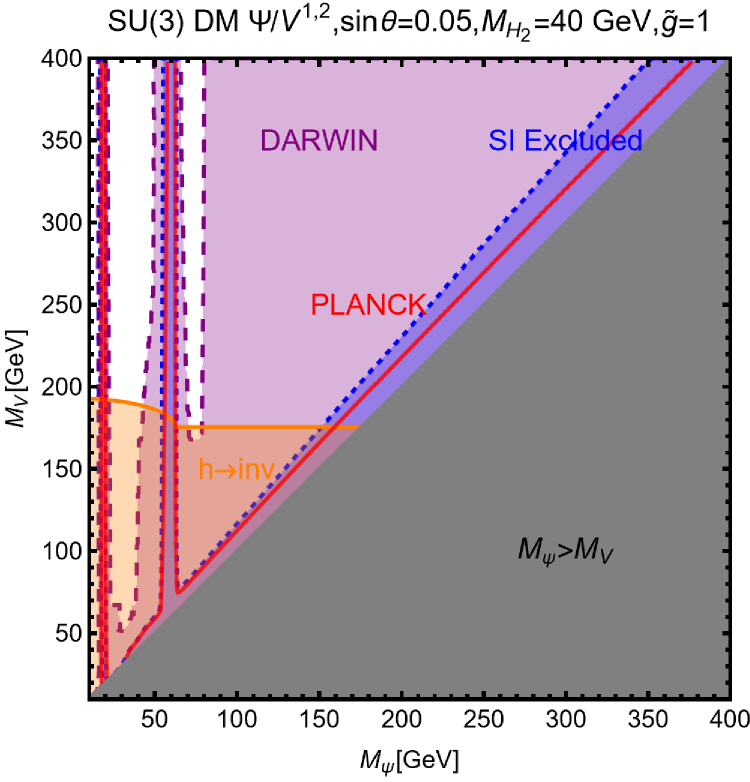}}
    \subfloat{\includegraphics[width=0.35\linewidth]{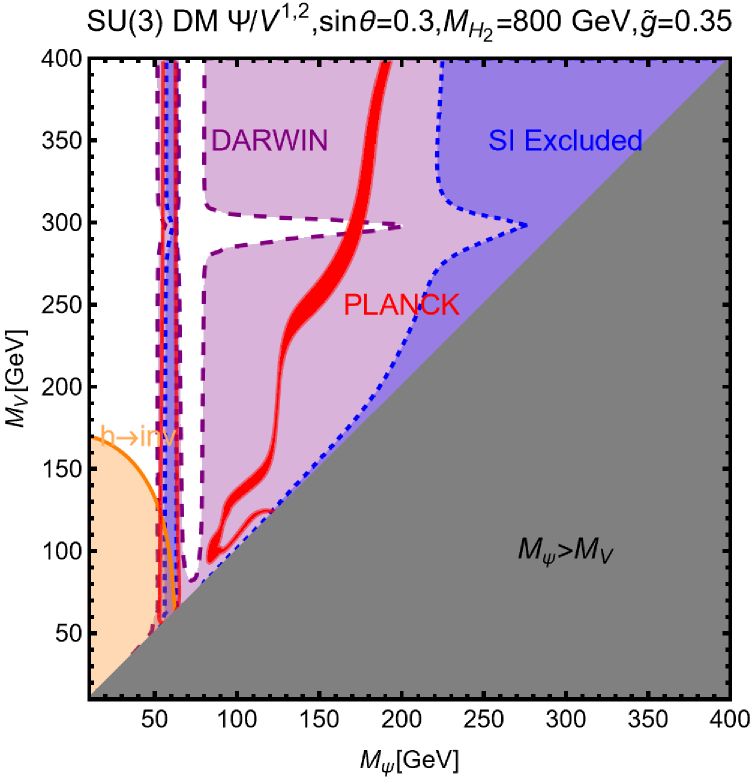}}
    \caption{Summary of the DM constraints in the $(M_\psi, M_V)$ plane for the dark $SU(3)$ model in the regime of a two component scalar/vector DM for three benchmark parameter assignations, summarised on the top of the panels.
    Here the orange-coloured region is excluded from the bound on the invisible Higgs decay while the grey-coloured area is excluded from the fact that $M_\Psi > M_V$ for this region. The remaining colour codes are the same as of Fig.     \ref{fig:pmPsiSU3}}
    \label{fig:SU32CDMVchia}
\end{figure*}
Fig. \ref{fig:SU32CDMVchia} shows a further illustration of the results in the $(M_\Psi,M_V)$ bidimensional plane.

\paragraph{Generalization to $SU(N)$:}
The scenario described before can be generalised to an arbitrary dark $SU(N)$. The scalar potential is built by considering $N-1$ fields in the fundamental representation which can be decomposed as \cite{Gross:2015cwa}:
\begin{align}
    & \phi_1=\left(\begin{array}{c}
         0  \\
         0  \\
         \cdots \\
         0\\
         \rho_1          
    \end{array}
    \right),\,\,\,\,
    \phi_2=\left(\begin{array}{c}
         0  \\
         0  \\
         \cdots \\
         \rho_2^{(1)}\\
         \rho_2^{(2)}e^{i\xi_2}
    \end{array}
    \right),\,\,\,\,
    \cdots, \nonumber\\
    & \phi_{N-1}=\left(
    \begin{array}{c}
         0  \\
         \rho_{N-1}^{(1)}\\
         \cdots\\
         \rho_{N-1}^{(N-2)}e^{i\xi_{N-1}^{(N-3)}}\nonumber\\
         \rho_{N-1}^{(N-1)}e^{i\xi_{N-1}^{(N-2)}}
    \end{array}
    \right),
\end{align}
with $\rho_i^{(j)}$ and $\xi_{i}^{(j)}$ stemming, respectively, for the radial fields and phases. With this choice, the scalar and gauge sector will maintain a $Z_2 \times Z_2^{'}$ symmetry under which a set of dark $SU(2)$ gauge fields remain stable. We can again identify a set of vector $V_{\mu}^{1,2,3}$ with $m_{V^1}=m_{V^2}\neq m_{V^3}$ as the DM candidates and reduce the phenomenology to the one of the dark $SU(3)$ model discussed in the previous subsection.

\subsection{Inert Doublet Model}

As already pointed out, one possibility to concretely realise the coupling between the Higgs bosons and the DM pairs is to assume the latter to be (at least partially) charged under $SU(2)_L$. One popular realisation of this idea, for a scalar DM, is the so-called inert doublet model (IDM) \cite{Deshpande:1977rw,LopezHonorez:2006gr,Barbieri:2006bg,Ma:2006km,Arhrib:2013ela}. The particle spectrum of the SM is then enlarged with an additional $SU(2)_L$ doublet. By introducing a $Z_2$ symmetry appropriately, the coupling between the new doublet and the SM fermions is made forbidden. Further, the new doublet does not participate in the EWSB as it does not acquire a VEV. The scalar potential of the theory reads:

\bea
&&V=\mu_1^2 H_1^\dagger H_1+\mu_2^2 H_2^\dagger H_2+\lambda_1 |H_1^\dagger H_1|^2+\lambda_2 |H_2^\dagger H_2|^2\nonumber\\
&&+\lambda_3 \left(H_1^\dagger H_1\right) \left(H_2^\dagger H_2\right)+\lambda_4 |H_1^{\dagger}H_2|^2\nonumber\\
&& +\frac{\lambda_5}{2}\left[(H_1^{\dagger}H_2)^2+\mbox{H.c.}
\right],~~~
\eea
with $H_1$ and $H_2$ being, respectively, the SM and the new scalar doublet. Besides guaranteeing that only one doublet gets a VEV upon minimization after EWSB, the parameters of scalar potential must satisfy constraints from the perturbative unitary and boundness from below which can be summarized as follows:
\bea
\label{eq:IDMbounds}
 |\lambda_i|&&\leq 4 \pi ~~\forall~i, \nonumber\\
 \lambda_{1,2}&&>0, \nonumber\\
 \lambda_3, \lambda_3+\lambda_4-|\lambda_5|&&>-2 \sqrt{\lambda_1 \lambda_2},
\eea
After the EWSB, the scalar spectrum is made of four states, the two CP-even Higgses, $h$ and $H^0$, with the former being identified with the $125$ GeV SM-like Higgs boson, one CP-odd Higgs $A$ and, finally an electrically charged state $H^{\pm}$. Their masses can be expressed in terms of the potential parameters as:

\bea
\label{eq:IDM_masses}
 M_{H^{\pm}}^2&&=\mu_2^2+\frac{\lambda_3 v_h^2}{2},\nonumber\\
 M_{H^0}^2&&=\mu_2^2+\frac{1}{2}(\lambda_3+\lambda_4+\lambda_5)v_h^2,\nonumber\\
 M_{A^0}^2&&=\mu_2^2+\frac{1}{2}(\lambda_3+\lambda_4-\lambda_5)v_h^2.
\eea
It is also useful to define the following combination $\lambda_L=\frac{1}{2}(\lambda_3+\lambda_4+\lambda_5)$ as the coupling of the DM with the Higgs is proportional to the latter.

DM relic density is determined, in a similar fashion as the scalar Higgs portal, by annihilation processes into $\bar f f$, $W^+ W^-$, $ZZ$ and $hh$ final states. We do not provide analytical approximations as coannihilation effects between the DM and the other BSM states are often relevant. Furthermore, for $m_{\rm DM}\lesssim 100 \,\mbox{GeV}$ one has to take into account the $W^{\pm}W^{\mp\,*}$ final state \cite{LopezHonorez:2006gr}, with $W^{*}$ being an off-shell boson. Also in this case an analytical expression for the annihilation cross-section would be particularly complicated and not very useful for the reader. An up-to-date determination of the DM relic density in the IDM has been provided in Refs.\cite{Banerjee:2021oxc,Banerjee:2021anv,Banerjee:2021xdp,Banerjee:2021hal}.  
For what the DD is concerned, it is well described by the analytical expression:
\begin{equation}
\sigma_{H^0p}^{\rm SI}=\frac{\mu_{H^0 p}^2}{4\pi}\frac{m_p^2}{M_{H^0}^2 M_h^4}\lambda_L^2
\left[f_p \frac{Z}{A}+f_n \left(1-\frac{Z}{A}\right)\right]^2,
\end{equation}
where $m_p$ is the mass of the proton, $\mu_{H^0 p}$ is the reduced mass of the DM-proton system, the coefficients $f_p$ and $f_n$ are the ones defined in Eq.~(\ref{eq:formfac1}), and $A\,(Z)$ 
is the atomic mass (number) of the target nucleus.

\begin{figure}
\begin{center}
\subfloat{\includegraphics[width=0.95\linewidth]{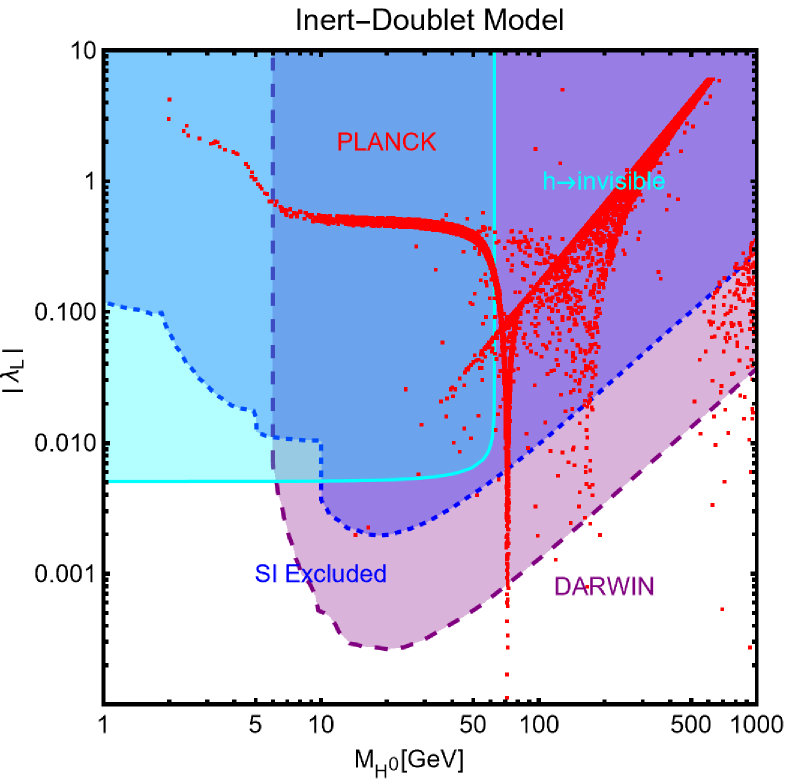}}
\end{center}
\caption{Model points with the correct DM relic density (red coloured) obtained from a scan over the 
parameters of the IDM (see main text for details) in the $(m_{H^0},|\lambda_L|)$ bidimensional plane. 
The blue coloured region is excluded by the current constraints on the DM SI interactions while the purple coloured region is the projected sensitivity reach of the DARWIN experiment. The cyan coloured region corresponds to the parameter space giving $Br(h\rightarrow \mbox{invisible})>0.11$.}
\label{fig:pIDMtot}
\end{figure}

The IDM has been already reviewed in, e.g., Refs. \cite{Arcadi:2017kky, Arcadi:2021mag}, hence we will just provide, in Fig.~\ref{fig:pIDMtot}, an overview of the updated constraints of the model.
To obtain them we have performed a parameter scan over the following ranges:
\begin{align}
    & |\mu_2| \in \left[1\,\mbox{GeV}, 1000\,\mbox{TeV}\right], \,\,\,\,\lambda_{i=1,2}\in \left[0, 4 \pi\right],\nonumber\\
    & |\lambda_{i=3,4,5}|\in \left[0,4\pi\right],
\end{align}
retaining only the parameter assignations complying with Eq.~\eqref{eq:IDMbounds} and featuring the correct DM relic density. Such assignations are shown, via scatter plot in the $(M_{H^0},|\lambda_L|)$ bidimensional plane, in fig. \ref{fig:pIDMtot}. To be viable, the model points should lie outside the blue region, corresponding to the exclusion from present DD bounds, as well as the cyan region, associated to the bound from invisible Higgs decays. As usual, the region which will be probed by the DARWIN experiment, has been highlighted in red.
We see that experimental bounds are overcome in basically three scenarios: at $M_{H^0}\sim M_h/2$, i.e. in presence of s-channel enhancement of DM annihilation into SM fermion final states; for $M_h/2 \lesssim M_{H^0} \lesssim 100\,\mbox{GeV}$, where the relic density is mostly accounted for by annihilations into $W^{\pm} W^{\mp\,*}$ final states; for $M_{H^0} \gtrsim 500\,\mbox{GeV}$. As described in detail in Ref.~\cite{LopezHonorez:2006gr}. While for such masses, the annihilation of the DM into gauge bosons is still very efficient, an enhancement of the relic density, driving the latter towards the thermally favoured value, is provided by the decays into DM of the pseudoscalar $A$ and of the charged Higgs, which decouple from the primordial plasma shortly after the freeze-out~\cite{Queiroz:2015utg}.



\subsection{Singlet-Doublet Model}\label{ss:SDM}

A realistic completion of the Higgs portal with a fermionic DM, without the introduction of an extra scalar mediator, is represented by the so-called Singlet-Doublet model \cite{Cheung:2013dua, Calibbi:2015nha}. This model considers the coupling of the SM Higgs doublet with a non-trivial fermionic sector composed of a pure SM gauge singlet $S$ and two Weyl fermions $D_{L, R}$, doublet under $SU(2)_{L,\, R}$ and with hypercharge $Y=\pm 1/2$. The concerned piece of Lagrangian can be written as:
\bea
\mathcal{L}&&=-\frac{1}{2}m_S S^{2}-m_D D_L D_R \nonumber\\
&&-y_1 D_L H S-y_2 D_R \widetilde{H} S+\mbox{H.c.},
\eea
where we have assumed the definition:
\begin{equation}
D_L=\left(
\begin{array}{c}
N_L \\
E_L
\end{array}
\right),\,\,\,\,\,
D_R=\left(
\begin{array}{c}
-E_R \\
N_R
\end{array}
\right),
\end{equation}
The BSM fermions, i.e., $N_{L, R}$,\, $D_{L, R}$, just introduced, are assumed to be odd under an ad hoc $Z_2$ symmetry. In such a way the lightest of them is ensured to be cosmologically stable and hence, if electrically neutral, work as the DM candidate.

Analogously to what was done in Ref.~\cite{Calibbi:2015nha}, it is possible to trade the parameters $y_1,\,y_2$ with the pair $y,\theta$, given by:
\begin{equation}
y_1=y\cos\theta,\,\,\,\,\,y_2=y \sin\theta.
\end{equation} 
After the EWSB we can define the following $3 \times 3$ matrix for the electrically neutral states: 

\begin{equation}
M=\left(
\begin{array}{ccc}
m_S & \frac{y_1 v_h}{\sqrt{2}} & \frac{y_2 v_h}{\sqrt{2}} \\
\frac{y_1 v_h}{\sqrt{2}} & 0 & m_D \\
\frac{y_2 v_h}{\sqrt{2}} & m_D & 0
\end{array}
\right),
\end{equation}
which is diagonalized through a unitary transformation, using the matrix $U$, leading to three (Majorana) mass eigenstates:
\begin{equation}
N_i=S U_{i1}+D_L U_{i2}+D_R U_{i3},\,\,i=1,2,3,
\end{equation}
that are defined according the convention $m_{N_1}< m_{N_2} < m_{N_3}$. The charged components of the $SU(2)$ multiplets form a Dirac fermion $E^{\pm}$ with a mass $m_{E^{\pm}}\approx m_D$. Thanks to the presence of an ad hoc $Z_2$ symmetry that makes the lightest of the new fermionic states stable. If this is the electrically neutral state $N_1$, the model has a DM candidate.

In the mass basis, the interaction Lagrangian of the new fermions can be written as:
\begin{align}
\label{eq:physical_SD}
& \mathcal{L}=g_{h N_i N_j}h\ovl N_i N_j +\mbox{H.c.}\nonumber\\
&+\ovl N_i \gamma^\mu \left(g_{Z N_i N_j}^V -g_{Z N_i N_j}^A \gamma_5\right) N_j Z_\mu\nonumber\\
& +\ovl {E^{-}}\gamma^\mu \left(g_{W^{\mp} E^{\pm} N_i}^V-g_{W^{\mp} E^{\pm} N_i}^A \gamma_5 \right) W^{-}_\mu N_i \nonumber\\
& -e \ovl {E^-} \gamma^\mu E^- A_\mu -\frac{g_2}{2 \cos^2 \theta_W}(1-2 \sin^2\theta_W) \ovl {E^-} \gamma^\mu E^- Z_\mu.  \; 
\end{align}
Contrary to the EFT portal, here the fermionic DM candidate couples with both the Higgs and the gauge bosons. The couplings depend on the elements of the mixing matrix $U$ and can be written as:
\begin{equation}
g_{h N_i N_j}=\frac{1}{\sqrt{2}}\left(y_1 U_{i2}^{*}U_{j1}^{*}+y_2 U_{j3}^{*}U_{i1}^{*}\right),
\end{equation}
for the Higgs $h$, while in the case of the gauge bosons we have:
\bea
 g^V_{Z N_i N_j}&&=c_{Z N_i N_j}-c^{*}_{Z N_i N_j}
,\,\,\,\,g^A_{Z N_i N_j}=c_{Z N_i N_j}+c^{*}_{Z N_i N_j},\nonumber\\
c_{Z N_i N_j}&&=\frac{g}{4 \cos\theta_W}\left(U_{i3}U_{j3}^{*}-U_{i2}U_{j2}^{*}\right),
\eea
and:
\bea
g^{V}_{W^\mp E^{\pm} N_i}&&=\frac{g}{2\sqrt{2}}\left(U_{i3}-U_{i2}^{*}\right), \nonumber\\
g^{A}_{W^\mp E^{\pm} N_i} &&=\frac{g}{2\sqrt{2}}\left(U_{i3}+U_{i2}^{*}\right),
\eea
with the labels $V$ and $A$ stemming for, respectively, vectorial and axial couplings. 

Before going ahead we remark again that the model conventionally dubbed Singlet-Doublet model has a Majorana DM. A variant with a Dirac DM has been, however, proposed in Ref. \cite{Yaguna:2015mva} and reviewed in Ref. \cite{Arcadi:2017kky}.
These studies conclude that it is already ruled out by
the DD because of the strong coupling between the DM and the $Z$-boson. For this reason,model with a Dirac fermion DM will not be explicitly discussed here.  

To get a better insight about the DM phenomenology in this setup, it is useful to inspect  couplings of the DM with the $Z$ and the Higgs boson in more detail. Their analytical expressions are~\cite{Cohen:2011ec,Cheung:2013dua}:
\begin{align}
     & g_{h N_1 N_1}=-\frac{(\sin 2 \theta  m_D +m_{N_1})y^2 v_h}{m_D^2+\frac{v_h}{2}y^2+2 m_S m_{N_1}-3 m_{N_1}^2}\ , \nonumber \\
     & g_{Z N_1 N_1}=\nonumber\\
     & -\frac{m_Z v_h y^2 \cos 2 \theta\left(m_{N_1}^2-m_D^2\right)}{2 (m_{N_1}^2-m_D^2)^2+v_h^2 (2 y^2 \sin 2 \theta m_{N_1} m_D+y^2 (m_{N_1}^2+m_D^2))}
\, . 
\end{align}
In the limit $m_D > y_{1,2}v_h \gg m_S$, the two heaviest neutral fermions $N_{2,3}$ and the charged state $E^{\pm}$ can be decoupled from the relevant DM phenomenology. In such a limit, the DM couplings read:
\begin{equation}
   g_{h N_1 N_1} = -\frac{ y^2 \sin 2 \theta v_h}{m_D}\, ,   \ \ \ 
   g_{Z N_1 N_1}^A \simeq  \frac{M_Z v_h y^2 \cos 2\theta }{2 m_D^2} \, .
\end{equation}
As can be seen, the coupling of the DM with the Higgs resembles one of the EFT models with $\Lambda=m_D/(y \sin 2\theta)$ (see Eq.~(\ref{Lag:DM} )). 

The other relevant insight is that it is possible to find a combination of the parameters, typically dubbed ``blind spots'', for which the couplings $g_{Z N_1 N_1}$ and $g_{h N_1 N_1}$ can be set to zero. More specifically, the blind spot in the DM Higgs coupling appears when the condition:
\begin{equation}
\label{eq:Hblind_spots}
m_D\,\sin2\theta+m_{N_1}=0,
\end{equation}
is satisfied. It requires, for the chosen sign convention, that $\sin 2 \theta <0$. 
In the case of the coupling between the DM and the $Z$ we have instead:
\begin{equation}
\label{eq:Zblind_spots}
\tan\theta=\pm 1,\,\,\mbox{and/or}\,\,m_D=m_{N_1},
\end{equation}
which corresponds to $|U_{12}|^2=|U_{13}|^2$. 

Moving to DM phenomenology, the DM relic density is mostly accounted for the DM annihilation processes into the SM fermion pairs, pairs of gauge bosons and $Zh$ final states. The corresponding cross-sections are described by the expressions below:
\begin{align}
 & \langle \sigma v \rangle_{ff} =\frac{1}{2\pi}\sum_f n_c^f \sqrt{1-\frac{m_f^2}{m_{N_1}^2}}\left[\frac{m_f^2}{M_Z^4}|g_{ZN_1 N_1}^A|^2 |g_{Zff}^A|^2\right. \nonumber\\
& \left. +\frac{2 v^2}{3 \pi}|g_{ZN_1 N_1}^A|^2 \left(|g_{Zff}^V|^2+|g_{Zff}^A|^2\right) {\left(1-\frac{m_f^2}{M_{N_1}^2}\right)}^{-1} \frac{m_{N_1}^2}{(4 m_{N_1}^2-M_Z^2)^2}\right.\nonumber\\
&\left.+\frac{v^2}{2\pi}|g_{hN _1 N_1}|^2 \frac{m_f^2}{v_h^2}\left(1-\frac{m_f^2}{m_{N_1}^2}\right)\frac{m_{N_1}^2}{(4 m_{N_1}^2-M_H^2)^2} \right],
\end{align}

\begin{align}
 & \langle \sigma v \rangle_{WW} =\frac{1}{4 \pi}\sqrt{1-\frac{M_W^2}{m_{N_1}^2}}\frac{1}{M_W^4 (M_W^2-m_{N_1}^2-m_{E^{\pm}}^2)^2}\nonumber\\
 & \big[ (|g_{W^\mp E^\pm N_1}^V|^2+|g_{W^\mp E^\pm N_1}^A|^2)^2 
(2 M_W^4 (m_{N_1}^2-M_W^2))\nonumber\\
& +2 |g_{W^\mp E^\pm  N_1}^V|^2 |g_{W^\mp E^\pm N_1}^A|^2 m_{E^{\pm}}^2 (4 m_{N_1}^4+3 M_W^4-4 m_{N_1}^2 M_W^2))\big], 
\end{align}

\begin{align}
& \langle \sigma v \rangle_{ZZ} =\frac{1}{4 \pi}\sqrt{1-\frac{M_Z^2}{m_{N_1}^2}}\sum_{i=1,3}\frac{1}{(M_Z^2-m_{N_1}^2-m_{N_i}^2)^2}\nonumber\\
& (|g_{ZN_1 N_i}^V|^2+|g_{ZN_1 N_i}^A|^2) \nonumber \\
& (|g_{ZN_1 N_j}^V|^2+|g_{ZN_1 N_j}^A|^2) (m_{N_1}^2-M_Z^2),
\end{align}

\begin{align}
 & \langle \sigma v \rangle_{Zh} =\frac{1}{\pi}\sqrt{1-\frac{(M_h+M_Z)^2}{4 m_{N_1}^2}}\sqrt{1-\frac{(M_h-M_Z)^2}{4 m_{N_1}^2}}\nonumber\\
 & \frac{1}{256 m_{N_1}^2 M_Z^6}\lambda_{hZZ}^2 |g_{ZN_1 N_1}^A|^2\nonumber\\
& \times \big( M_h^4+(M_Z^2-4 m_{N_1}^2)^2-2 M_h^2 (M_Z^2-4 m_{N_1}^2)\big).
\end{align}
$\lambda_{hZZ}$ is the SM coupling between the Higgs and two Z-bosons.
The annihilation cross-section into the SM fermions is $p$-wave dominated, due to the helicity suppression of the $s$-wave contribution. Hence, it is not very efficient far from the $m_{N_1} \sim m_{h}/2, m_Z/2$ "poles". The annihilation cross-sections into gauge bosons are instead $s$-wave dominated. The cross-section, however, depends on the masses of the extra electrically neutral and charged fermions. As expected, they become suppressed as the hierarchy between the mass of the DM and the ones of the other fermions becomes more pronounced, i.e., when the EFT limit is recovered.
On the side of the DD, the DM can scatter, in the NR limit, with nuclei both via SI interactions (mediated by the Higgs) and via SD interactions (mediated by the $Z$). The corresponding cross-sections can be written as:
\begin{equation}
\sigma_{N_1 p}^{\rm SI}=\frac{\mu_{N_1 p}^2}{\pi M_H^4}|g_{h N_1 N_1}|^2\frac{m_p^2}{v_h^2}
{\left[f_p \frac{Z}{A}+f_n \left(1-\frac{Z}{A}\right)\right]}^2,
\end{equation}
and 
\begin{equation}
\sigma_{N_1 p}^{\rm SD}=\frac{\mu_{N_1 p}^2}{\pi m_Z^4}|g_{Z N_1 N_1}^A|^2 
{\left[A_u^{Z} \Delta_u^p+ A_d^Z \left(\Delta_d^p+\Delta_s^p\right)\right]}^2.
\end{equation}
As evident, the presence of blind spots in $g_{h N_1 N_1}$ or $g_{Z N_1 N_1}^A$ is very relevant as they can substantially impact the DD prospects of the model. 
If the DM candidate is light enough, it can also alter the invisible decays of the Higgs and $Z$ bosons by adding the following contributions:
\begin{align}
    & \Gamma (h \rightarrow N_1 N_1)=\frac{M_h}{16\pi}|g_{hN_1 N_1}|^2 {\left(1-\frac{m_{N_1}^2}{M_h^2}\right)}^{3/2},\nonumber\\
    & \Gamma(Z \rightarrow N_1 N_1)=\frac{M_Z}{6\pi}|g_{ZN_1 N_1}^A|^2 {\left(1-\frac{m_{N_1}^2}{M_Z^2}\right)}^{3/2}.
\end{align}
As already pointed out, the former should comply with the LHC bound $Br(h \rightarrow \mbox{invisible}) \lesssim 0.11$. At the same time, the latter should satisfy the bound from precision measurement $\Gamma(Z \rightarrow N_1 N_1)\leq 2.3\,\mbox{MeV}$ \cite{Workman:2022ynf}.
While not strictly related to the DM, one has also to consider that new fermions coupled with the Higgs and the gauges bosons are constrained by the Electroweak Precision Tests (EWPT) as we have a deviation, from the SM prediction, of the custodial symmetry parameter $\rho$ given by $\Delta \rho \propto (y_1^2-y_2^2)^2=y^4 (1-\tan^2 \theta)^2$ \cite{Calibbi:2015nha,Barbieri:2006bg,DEramo:2007anh}.

\begin{figure*}
\begin{center}
\subfloat{\includegraphics[width=0.4\linewidth]{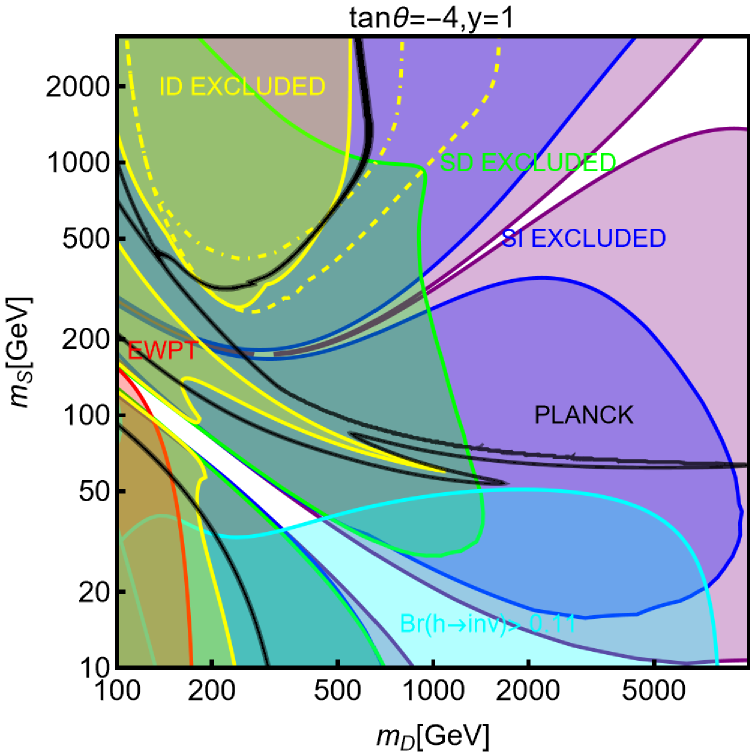}}
\subfloat{\includegraphics[width=0.4\linewidth]{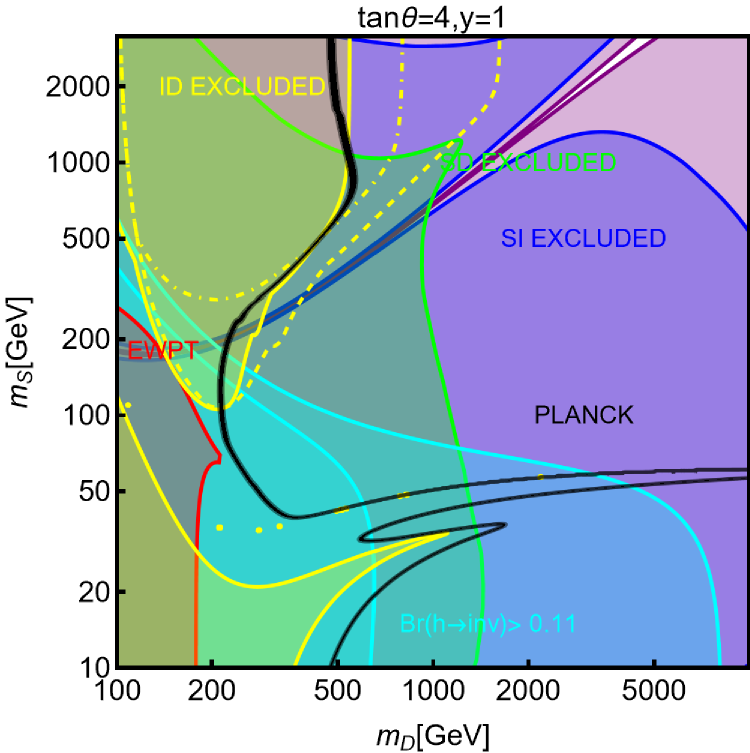}}\\
\subfloat{\includegraphics[width=0.4\linewidth]{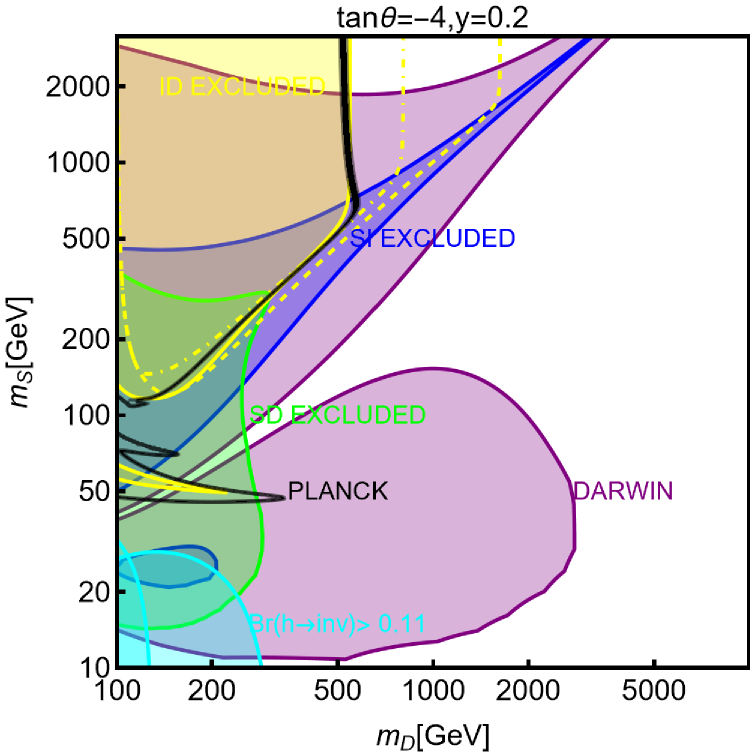}}
\subfloat{\includegraphics[width=0.4\linewidth]{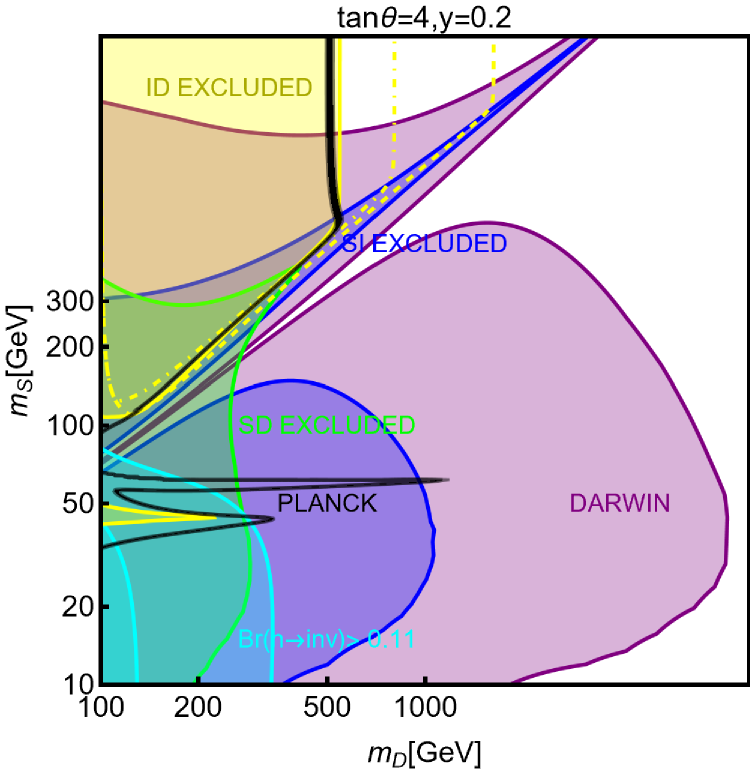}}
\end{center}
\caption{Summary of the DM constraints for the Singlet-Doublet model with a Majorana fermion DM in the bidimensional plane 
$(m_D,m_S)$ by taking $y=1(0.2)$ in the top (bottom) row and $\tan\theta=\pm 4$. The black coloured lines 
are the isocontours corresponding to the correct DM relic density. The blue coloured (green coloured) region is excluded by the LZ limit on the SI (SD) interactions while the purple coloured region corresponds to the expected sensitivity of the DARWIN experiment. The cyan coloured regions correspond to the exclusions from the invisible decays of the SM Higgs and Z bosons. Finally, the red-coloured region (top row only) depicts the exclusion area from the EWPT. }
\label{fig:pSDmajo}
\end{figure*}

The SD model has been already widely reviewed in Refs. \cite{Arcadi:2017kky, Arcadi:2019lka}, hence we just show in Fig.~\ref{fig:pSDmajo} an illustration of the updated constraints.
As usual, the figure compares the requirement of the correct relic density with several experimental exclusions in a bidimensional plane $(m_S,\,m_D)$. This time, the parameters which are varied, are the singlet and doublet mass parameters $m_S$ and $m_D$ while the $(y,\tan\theta)$ have been limited to a few benchmark assignations, i.e., $(y,\tan\theta)=(1,\pm 4),(0.2,\pm 4)$. Both signs for $\sin\theta$ have been considered, to better highlight the impact of the blind spots in the DD. 
The correct relic density is, as usual, associated with very narrow black-coloured isocontours. The strongest experimental constraints are represented by the SI interactions which, in the case of $\tan\theta >0$ entirely ruled out the parameter space corresponding to the correct relic density, up to masses of the order of the TeV. A substantial relaxation of the DD constraints from the SI is achieved while changing the sign of the $\tan\theta$ parameter; complementary bounds are nevertheless provided by the SD interactions (green coloured regions in the plot). It is, indeed important to bear in mind that blind spots in the coupling of the DM with the Higgs do not correspond to blind spots in the interactions with the Z-boson and vice versa. Although subdominant, the figure shows, for completeness (1) bounds from the invisible decays of the $Z$ and Higgs bosons (cyan coloured regions) and (2), exclusion region (red coloured) from the EWPT for $y=1$. Note that the ID limits exclude the remaining viable region of the multi-TeV regime of $m_S$ for $y=0.2$ regardless of the sign of $\sin \theta$

\section{Extension of Higgs Sector}
\label{sec:2HDMetal}

Many popular BSM setups rely on the extension of the Higgs sector with an additional doublet (for a general review see e.g., Ref. \cite{Branco:2011iw}). Considering the extended Higgs sector as a portal for the DM interactions is an interesting possibility. In the following, we will then illustrate a series of models considering this option. Unless differently stated, we will consider models having in common an interaction potential for the two $SU(2)$ doublets, called $\Phi_{1,2}$ of the form:

\begin{align}
    & V_{2HDM}=m_{11}^2 \Phi_1^\dagger \Phi_1+m_{22}^2 \Phi_2^\dagger \Phi_2-m_{12}^2 \left(\Phi_1^\dagger \Phi_2+\mbox{H.c.}\right)\nonumber\\
    & +\frac{\lambda_1}{2}{\left(\Phi_1^\dagger \Phi_1\right)}^2+\frac{\lambda_2}{2}{\left(\Phi_2^\dagger \Phi_2\right)}^2\nonumber\\ 
    & +\lambda_3 \left(\Phi_1^\dagger \Phi_1\right) \left(\Phi_2^\dagger \Phi_2\right)+\lambda_4 \left(\Phi_1^\dagger \Phi_2\right)\left(\Phi_2^\dagger \Phi_1\right)\nonumber\\
    & +\frac{\lambda_5}{2}\left[{\left(\Phi_1^\dagger \Phi_2\right)}^2+\mbox{H.c.}\right].
    \label{eq:2HDM_potential}
\end{align}
This class of models are conventionally dubbed 2HDM. The two doublets can be decomposed as:
\begin{equation}
    \Phi_i=\left(
    \begin{array}{c}
         \phi_i^+  \\
         (v_i+\rho_i+i \eta_i)/\sqrt{2}
    \end{array}
    \right),
\end{equation}
where $v_{i=1,2}$ are the VEVs acquired after the EWSB, which satisfy $\sqrt{v_1^2+v_2^2}=v_h= 246\,\text{GeV}$ and $v_2/v_1=\tan\beta$. Assuming that CP is conserved in the Higgs sector, 
one can define a set of physical states, customarily dubbed $(h, H, A, H^{\pm}$), related to the components of the doublets by:
\begin{align}\label{eq:2HDMdecomose}
    & \left(\begin{array}{c}
        \phi_1^+   \\
         \phi_2^+
    \end{array}
    \right)=\mathcal{R}_\beta \left(
    \begin{array}{c}
         G^+  \\
         H^+ 
    \end{array}
    \right),\,\,\,\,\,
    \left(\begin{array}{c}
        \eta_1   \\
         \eta_2
    \end{array}
    \right)=\mathcal{R}_\beta \left(
    \begin{array}{c}
         G^0  \\
         A 
    \end{array}
    \right),\nonumber\\
    & \left(\begin{array}{c}
        \rho_1   \\
         \rho_2
    \end{array}
    \right)=\mathcal{R}_\alpha \left(
    \begin{array}{c}
         H  \\
         h 
    \end{array}
    \right),
\end{align}
with $h, H$ are the two electrically neutral CP-even bosons, $A$ is the CP-odd state, $H^\pm$ is a charged Higgs, and,  
$\mathcal{R_{\alpha,\,\beta}}$ relate the physical states with the $\Phi_{1,2}$. Unless differently stated we will identify $h$ as the $125$ GeV SM-like Higgs discovered at the LHC. One might nevertheless also consider the opposite scenario, see e.g., Ref.~\cite{Cacciapaglia:2016tlr} where $H$, the heavier one, appears to be the SM-like Higgs.  The other states $G^+,\, G^0$ present in Eq. (\ref{eq:2HDMdecomose}) are the charged, neutral Goldstone bosons, respectively, which become the longitudinal $dof$ for the gauge bosons.
The masses of the physical Higgs bosons can be related to the scalar potential parameters via the following relations:
\begin{align}
\label{eq:2HDM_lambda_M}
    & \lambda_1=\frac{1}{v_h^2}\left[-M^2 \tan^2 \beta+\frac{\sin^2 \alpha}{\cos^2 \beta}M_h^2+\frac{\cos^2 \alpha}{\cos^2 \beta}M_H^2\right], \nonumber\\
    & \lambda_2=\frac{1}{v_h^2}\left[-\frac{M^2}{\tan^2 \beta}+\frac{\cos^2 \alpha}{\sin^2 \beta}M_h^2+\frac{\sin^2 \alpha}{\cos^2 \beta}M_H^2\right],\nonumber\\
    & \lambda_3 =\frac{1}{v_h^2}\left[-M^2+2 M_{H^\pm}^2+\frac{\sin 2\alpha}{\sin 2 \beta}\left(M_H^2-M_h^2\right)\right],\nonumber\\
    & \lambda_4=\frac{1}{v_h^2}\left[M^2+M_A^2-2 M_{H^\pm}^2\right],\nonumber\\
    & \lambda_5=\frac{1}{v_h^2}\left[M^2-M_A^2\right],
\end{align}
where we have defined $M^2=m_{12}^2/(\sin\beta \cos\beta)$. Thanks to these relations, it is possible to adopt the physical masses of the Higgs bosons, together with the $M$ parameter and $\tan\beta$ as the free parameters. As will be discussed in the following subsection, the $\lambda_i$ parameters are subject to constraints from the theoretical consistency of the scalar potential. The latter constraints can then be rephrased as constraints on the masses of the BSM Higgs boson. In particular their relative mass splitting will be constrained while very weak limits are imposed on the overall mass scale. 

The other relevant information to characterize a 2HDM  scenario is Yukawa Lagrangian:
\begin{align}
-{\cal L}_{\rm Yuk}& =\sum\limits_{f=u,d,l} \frac{m_f}{v_h} \left[g_{hff} \bar{f}f h +g_{Hff} \bar{f}f H-i g_{Aff} \bar{f} \gamma_5 f A \right] \notag \\
&- \frac{\sqrt{2}}{v_h} \left[ \bar{u} \left(m_u g_{Auu} P_L + m_d g_{Add} P_R \right)d H^+ \right.\nonumber\\
& \left. +  m_l g_{All} \bar \nu  P_R \ell H^+  +  \mathrm{H.c.} \right],
\end{align}

As evident the Yukawa couplings have been defined as combinations of the SM Yukawa coupling $m_f/v_h$ and of the reduced couplings $g_{\phi ff},\, \phi=h,\, H,\, A$.  Four specific sets of assignations for the couplings, conventionally called Type-I, Type-II, Type-X and Type-Y, are adopted in the literature \cite{Branco:2011iw} which prevent the emergence of tree-level flavour-changing neutral currents (FCNC). 
Even if flavor violation processes can be forbidden at the tree level, this is not the case at one loop. An illustration of the main constraints is provided, for example, by Ref. \cite{Enomoto:2015wbn}. Among them we stress $b \rightarrow s \gamma$ transitions, tested via the B-meson decay process $B\rightarrow X_s \gamma$ \cite{Belle:2016ufb,HFLAV:2016hnz}. The experimental constrains can be translated into bounds on $M_{H^{\pm}}$ and $\tan\beta$ as \cite{Misiak:2017bgg,Misiak:2020vlo}:
\begin{eqnarray}
\text{ Type \ II \ or \ Y}\ &:& ~~~ M_{H^\pm} \gsim 800 \ \text{GeV \ for \ any \ }  \tan\beta  \, , \nonumber \\
\text{Type \ I \ or \ X}\ &:& ~~~ M_{H^\pm} \gsim 500 \  \text{GeV \ for \ } \tan\beta \lsim 1  \, . 
\end{eqnarray}
The bounds on the mass of the charged Higgs translates as well into constraints on the masses of the other BSM scalars due to the relations imposed by theoretical constraints.

\begin{table*}
\renewcommand{\arraystretch}{1.55}
\begin{center}
\begin{tabular}{|c|c|c|c|c|}
\hline
~~~~~~ &  ~~Type I~~ & ~~Type II~ & Type-X/Lepton-specific & Type-Y/Flipped \\
\hline\hline 
$g_{huu}$ & $ \frac{\cos \alpha} { \sin \beta} \rightarrow 1$ & $\frac{ \cos \alpha} {\sin \beta} \rightarrow 1$ & $\frac{ \cos \alpha} {\sin\beta} \rightarrow 1$ & $ \frac{ \cos \alpha}{ \sin\beta}\rightarrow 1$ \\ \hline
$g_{hdd}$ & $\frac{\cos \alpha} {\sin \beta} \rightarrow 1$ & $-\frac{ \sin \alpha} {\cos \beta} \rightarrow 1$ & $\frac{\cos \alpha}{ \sin \beta} \rightarrow 1$ & $-\frac{ \sin \alpha}{ \cos \beta} \rightarrow 1$ \\ \hline
$g_{hll} $ & $\frac{\cos \alpha} {\sin \beta} \rightarrow 1$ & $-\frac{\sin \alpha} {\cos \beta} \rightarrow 1$ & $- \frac{ \sin \alpha} {\cos \beta} \rightarrow 1$ & $\frac{ \cos \alpha} {\sin \beta} \rightarrow 1$  \\ \hline\hline
$g_{Huu}$ & $\frac{\sin \alpha} {\sin \beta} \rightarrow -\frac{1}{\tan\beta}$ & $\frac{ \sin \alpha} {\sin \beta} \rightarrow -\frac{1}{\tan\beta}$ & $ \frac{\sin \alpha}{\sin \beta} \rightarrow -\frac{1}{\tan\beta}$ & $\frac{ \sin \alpha}{ \sin \beta} \rightarrow -\frac{1}{\tan\beta}$ \\ \hline
$g_{Hdd}$ & $ \frac{ \sin \alpha}{\sin \beta} \rightarrow -\frac{1}{\tan\beta}$ & $\frac{\cos \alpha}{\cos \beta} \rightarrow {\tan\beta}$ & $\frac{\sin \alpha} {\sin \beta} \rightarrow -\frac{1}{\tan\beta}$ & $\frac{ \cos \alpha} {\cos \beta} \rightarrow {\tan\beta}$ \\ \hline
$g_{Hll}$ & $\frac{ \sin \alpha} {\sin \beta} \rightarrow -\frac{1}{\tan\beta}$ & $\frac{\cos \alpha} {\cos \beta} \rightarrow {\tan\beta}$ & $\frac{ \cos \alpha} {\cos \beta} \rightarrow {\tan\beta}$ & $\frac{\sin \alpha} {\sin \beta} \rightarrow -\frac{1}{\tan\beta}$ \\
\hline\hline
$g_{Auu}$ & $\frac{1}{\tan\beta}$ & $\frac{1}{\tan\beta}$ & $\frac{1}{\tan\beta}$ & $\frac{1}{\tan\beta}$
\\
\hline
$g_{Add}$ & $-\frac{1}{\tan\beta}$ & ${\tan\beta}$ & $-\frac{1}{\tan\beta}$ & ${\tan\beta}$
\\
\hline
$g_{All}$ & $-\frac{1}{\tan\beta}$ & ${\tan\beta}$ & ${\tan\beta}$ & $-\frac{1}{\tan\beta}$
\\
\hline
\end{tabular}
\caption{Couplings of the 2HDM Higgs bosons to the SM fermions as a function of the angles $\alpha$ and $\beta$. For the CP--even states, the values are normalized to those of the SM-like Higgs boson and are also given in the alignment limit $\beta \!-\! \alpha \rightarrow \frac{\pi}{2}$.}
\label{table:2hdm_type}
\end{center}
\vspace*{-6mm}
\end{table*}

The assignation of the various $g_{\phi ff}$ couplings for the four 2HDM variants mentioned above are summarised in Tab.~\ref{table:2hdm_type}. Another relevant constraint, on the coupling $g_{hff}$ comes from the measurements of the $125$ GeV SM-like Higgs signal strengths (see e.g., Ref. \cite{ATLAS:2021vrm} for the most updated constraints) strongly favouring the SM prediction, i.e., $g_{hff}=1$ in our parametrization. It is possible to automatically comply with the latter constraints via the so-called alignment limit, i.e., $\beta-\alpha=\pi/2$, which will represent also the default choice in our study.
Extension of the Higgs sector should comply as well with the theoretical constraints (i.e., perturbative unitarity) on the parameters of the scalar potential, the EWPT as well as collider searches. Such constraints will vary in the different models considered in this work. Hence, they will be discussed in the following subsections case-by-case.

\subsection{Singlet-Doublet + 2HDM}
This model is discussed in detail, for example, in Refs.~\cite{Berlin:2015wwa,Arcadi:2018pfo} and represents a straightforward extension, within a 2HDM setup, of the Singlet-Doublet model previously discussed in this work. We hence consider again an SM gauge singlet fermion field $S$ and two $SU(2)$ doublet Weyl fermions $D_{L, R}$ which, this time coupled with the two doublets $\Phi_{1,2}$, introduced in the subsection \ref{ss:SDM}, as:
\begin{equation}
\mathcal{L}=-\frac{1}{2}m_S S^{2}-m_S D_L D_R -y_1 D_L \Phi_a S-y_2 D_R \widetilde{\Phi}_b S+\mbox{H.c.},
\end{equation}
with $a,b=1,2$.
A $Z_2$ symmetry needs to be enforced to ensure the stability of the DM candidate. While the new fermionic states can, in principle, couple arbitrarily with each doublet, it is useful to introduce specific coupling schemes, analogous to the Type-I,-II,-X,-Y configurations. In such a way the fields $D_{L,R}$ will couple selectively with the Higgs states $\Phi_{1,2}$. In this work, we will adopt the four schemes introduced in Ref.~\cite{Berlin:2015wwa}, dubbed "$dd$", "$du$", "$ud$", "$uu$". The passage from the interaction basis $(S, D_L, D_R)$ to the physical basis $N_i, E^\pm$ is performed in the same way as the original Singlet Doublet model with the only difference being the fact that the mixing matrix $U$ now will depend on the $\tan\beta$ parameter as well. In the physical basis, the relevant Lagrangian for the DM phenomenology can be written as:  
\begin{align}
 \mathcal{L} &=\overline{E^-} \gamma^\mu \left(g^V_{W^{\mp}E^{\pm}N_i}-g^A_{W^{\mp}E^{\pm}N_i}\gamma_5\right)N_i W_\mu^{-}+\mbox{H.c.}\nonumber\\
& +\frac{1}{2}\sum_{i,j=1}^3 \overline{N_i}\gamma^\mu \left(g_{Z N_i N_j}^V-g_{Z N_i N_j}^A \gamma_5\right) N_j Z_\mu \nonumber\\
& +\frac{1}{2}\sum_{i,j=1}^{3}\overline{N_i}\left(y_{h N_i N_j}h+y_{H N_i N_j}H+y_{A N_i N_j}\gamma_5 A\right)N_j \nonumber\\
& +\overline{E^-} \left(g^S_{H^{\pm}E^\mp N_i}-g^P_{H^{\pm}E^\mp N_i}\gamma_5\right)N_i H^{-}+\mbox{H.c.}\nonumber\\
& -e A_\mu \overline{E^{-}}\gamma^\mu E^{-}-\frac{g}{2 c_W}(1-2 s^2_W) Z_\mu \overline{E^{-}}\gamma^\mu E^{-}+\mbox{H.c.} , 
\end{align}
\noindent
where the couplings in the case of $\phi=h,H,A$ and $H^\pm$ are given by
\begin{align}
\label{eq:SD2HDM_couplings}
& y_{ \phi N_i N_j}=\frac{\delta_\phi}{2\sqrt{2}}\left[U_{i1}\left(y_1 R_a^\phi U_{i2}+y_2 R_b^\phi U_{i3}\right)+(i \leftrightarrow j)\right] ,  \nonumber\\
& g^{S/P}_{H^{\pm} E^\mp N_i}=\frac{1}{2}U_{i1}\left(y_1 R_1^{H^{\pm}} \pm y_2 R_2^ {H^{\pm}}\right) ,
\end{align}
%
%
with $\delta_h=\delta_H=-1$,  $\delta_A=-i$,  and we have considered the following decomposition of the $\Phi_{1}$ and $\Phi_{2}$ doublets in terms of the physical Higgs states $h,H,A,H^{\pm}$:
\begin{equation}
\Phi_{1,2}=\frac{1}{\sqrt{2}}
\left(
\begin{array}{c}
\sqrt{2} R^{H^\pm}_{1,2} H^\pm \\
v_{1,2}+R_{1,2}^h h+ R_{1,2}^H H+i R_{1,2}^A A
\end{array}
\right),
\end{equation}
with the parameters $R^\phi_{1,2},\, R^{H^\pm}_{1,\,2}$ being entries of the rotation matrices $\mathcal{R}_{\alpha,\beta}$ for the Higgs states introduced earlier in Eq.~(\ref{eq:2HDMdecomose}.)

To facilitate the understanding of the phenomenological results, it is useful to provide analytical expressions of the couplings of the DM pairs with the electrically neutral Higgs bosons in the four couplings scheme mentioned before:\\

{\it uu}:
\begin{align}
    & y_{hN_1 N_1}=y^2 v_h \sin^2 \beta \frac{m_{N_1}+m_D \sin 2 \theta}{2 m_D^2+4 m_S m_{N_1}-6 m_{N_1}^2+y^2 v_h^2 \sin^2 \beta}, \nonumber\\
    & y_{HN_1 N_1}=-\frac{1}{2}y^2 v_h \sin 2 \beta \frac{m_{N_1}+m_D \sin 2 \theta}{2 m_D^2+4 m_S m_{N_1}-6 m_{N_1}^2+y^2 v_h^2 \sin^2 \beta}, \nonumber\\
    & y_{A N_1 N_1}=-\frac{1}{2}y^2 v_h \sin 2 \beta \frac{m_{N_1}\cos 2 \theta}{2 m_D^2+4 m_S m_{N_1}-6 m_{N_1}^2+y^2 v_h^2 \sin^2 \beta}.
\end{align}
{\it ud:}
\begin{align}
    & y_{hN_1 N_1}=\nonumber\\
    & y^2 v_h \frac{m_{N_1} \left(\sin^2 \beta \cos^2 \theta+\cos^2 \beta \sin^2 \theta\right)+\frac{1}{2}m_D \sin 2 \beta \sin 2 \theta}{2 m_D^2+4 m_S m_{N_1}-6 m_{N_1}^2+\frac{1}{2}y^2 v_h^2 \left(1-\cos 2 \beta \cos 2 \theta\right)},\nonumber\\
    & y_{H N_1 N_1}=\nonumber\\
    & -\frac{1}{2}y^2 v_h \frac{m_{N_1}\sin 2 \beta \cos 2 \theta+m_D \cos 2 \beta \sin 2 \theta}{2 m_D^2+4 m_S m_{N_1}-6 m_{N_1}^2+\frac{1}{2}y^2 v_h^2 \left(1-\cos 2 \beta \cos 2 \theta\right)},\nonumber\\
    & y_{A N_1 N_1}=\nonumber\\
    & -\frac{1}{2}y^2 v_h \frac{m_{N_1} \sin 2 \beta+m_D \sin 2\theta}{2 m_D^2+4 m_S m_{N_1}-6 m_{N_1}^2+\frac{1}{2}y^2 v_h^2 \left(1-\cos 2 \beta \cos 2 \theta\right)}.
\end{align}
{\it du:}
\begin{align}
    & y_{hN_1 N_1}=\nonumber\\
    & \frac{1}{2}y^2 v_h \frac{m_{N_1}\left(1+\cos 2 \beta \cos 2 \theta\right)+m_D \sin 2 \beta \sin 2\theta}{2 m_D^2 + 4 m_S m_{N_1}-6 m_{N_1}^2+\frac{1}{2}y^2 v_h^2 \left(1+\cos 2 \beta \cos 2 \theta\right)},\nonumber\\
    & y_{H N_1 N_1}=\nonumber\\
    & \frac{1}{2}y^2 v_h \frac{m_{N_1}\sin 2 \beta \cos 2 \theta-m_D \cos 2 \beta \sin 2 \theta}{2 m_D^2+ 4 m_S m_{N_1} -6 m_{N_1}^2 +\frac{1}{2}y^2 v_h^2 \left(1+\cos 2 \beta \cos 2 \theta\right)},\nonumber\\
    & y_{A N_1 N_1}=\nonumber\\
    & \frac{1}{2}y^2 v_h \frac{m_{N_1} \sin 2 \beta+m_D \sin 2 \theta}{2 m_D^2 4 m_S m_{N_1}-6 m_{N_1}^2+\frac{1}{2}y^2 v_h^2 \left(1+\cos 2 \beta \cos 2 \theta\right)}.
\end{align}
{\it dd:}
\begin{align}
    & y_{h N_1 N_1}=y^2 v_h \cos^2 \beta \frac{m_{N_1}+m_D \sin 2 \theta}{2 m_D^2+4 m_S m_{N_1}-6 m_{N_1}^2+y^2 v_h^2 \cos^2 \beta},\nonumber\\
    & y_{H N_1 N_1}=\frac{1}{2}y^2 v_h \sin 2 \beta \frac{m_{N_1}+m_D \sin \theta}{2 m_D^2 +4 m_S m_{N_1}-6 m_{N_1}^2+y^2 v_h^2 \cos^2 \beta},\nonumber\\
    & y_{A N_1 N_1}=\frac{1}{2}y^2 v_h \sin 2 \beta \frac{m_{N_1}\cos 2 \theta}{2 m_D^2 +4 m_S m_{N_1}-6 m_{N_1}^2+y^2 v_h^2 \cos^2 \beta}.
\end{align}

Now let us consider the constraints on the undertaken model. For what concerns the Higgs sector, the standard constraints on the 2HDM apply. First of all, 
one needs to check that the parameters of the scalar potential ensure that the concerned potential remains bounded from below and no violation of perturbative unitarity occurs. Such requirements are satisfied via the following conditions \cite{Becirevic:2015fmu}:
\begin{align}
    & \lambda_{1,2} > 0,\,\,\, \lambda_3 > -\sqrt{\lambda_1\lambda_2},\,\,\,\, \lambda_3 + \lambda_4 - \left|\lambda_5\right| > -\sqrt{\lambda_1\lambda_2}, \nonumber\\
    & {\rm and,~} \left| a_{\pm} \right|, \left| b_{\pm} \right|, \left| c_{\pm} \right|, 
\left| d_\pm \right| , \left| e_\pm \right| , 
\left| f_{\pm} \right|  < 8\pi,\,\, {\rm with} \nonumber\\
& a_{\pm} = \frac{3}{2}(\lambda_1 + \lambda_2) \pm \sqrt{\frac{9}{4}(\lambda_1-\lambda_2)^2 + (2\lambda_3 + \lambda_4)^2}, \nonumber \\
& b_{\pm} = \frac{1}{2}(\lambda_1 + \lambda_2) \pm \sqrt{(\lambda_1-\lambda_2)^2 + 4\lambda_4^2}, \nonumber \\
& c_{\pm} = \frac{1}{2}(\lambda_1 + \lambda_2) \pm \sqrt{(\lambda_1-\lambda_2)^2 + 4\lambda_5^2}, \nonumber \\
& d_{\pm} = \lambda_3 + 2\lambda_4 \mp 3\lambda_5, \ e_\pm = \lambda_3 \mp \lambda_5, \  f_\pm = \lambda_3 \pm  \lambda_4.
\end{align}
In addition to these conditions, one has to ensure that $v_{1,2}$ corresponds to a global minimum of the potential and that the EW vacuum is stable. These requirements can be used to fix the mass parameters appearing in the scalar potential:
\begin{align}
    & m_{12}^2 \left(m_{11}^2-m_{22}^2 \sqrt{\lambda_1/\lambda_2}\right)\left(\tan\beta-\sqrt[4]{\lambda_1/\lambda_2}\right)>0,\nonumber\\
    & m_{11}^2+\frac{\lambda_1 v_h^2 \cos^2\beta}{2}+\frac{\lambda_3 v_h^2 \sin^2\beta}{2}\nonumber\\
    & =\tan\beta \left[m_{12}^2-(\lambda_4+\lambda_5)\frac{v_h^2 \sin 2\beta}{4}\right] , \nonumber\\
    & m_{22}^2+\frac{\lambda_2 v_h^2 \sin^2\beta}{2}+\frac{\lambda_3 v_h^2 \cos^2\beta}{2}\nonumber\\
    & =\frac{1}{\tan\beta} \left[m_{12}^2-(\lambda_4+\lambda_5)\frac{v_h^2 \sin2\beta}{4}\right] . 
\end{align}
The relations written above, combined with Eq.~\eqref{eq:2HDM_lambda_M}, provide constraints on the masses of the BSM Higgs states.
The other general constraint to account for appears from the EWPT. The presence of an extended Higgs sector affects the custodial symmetry parameter $\rho$ making it different from the SM prediction, $\rho=1$, by an amount:
\begin{align}
& \Delta \rho = \frac{\alpha_{\rm em}}{16 \pi^2 M_W^2 (1 -M_W^2/M_Z^2)} \big[ 
A(M^2_{H\pm},m^2_H)\nonumber \\
& + A(M^2_{H\pm},M^2_A) -  A(M^2_A,M^2_H)  \big] \, , 
\end{align}
with
\begin{equation}\label{eq:Afnrho}
    A(x,y)=x+y-f(x,\,y),~~{\rm with}~~f(x,\,y)=\frac{2xy}{x-y}\log \frac{x}{y}\,.
\end{equation}
It is useful to notice that $f(x,y)\rightarrow 0$ as $ x\rightarrow y$ and $A(x,0)=x$. From this we see that $\Delta \rho=0$ for $M_H=M_{H^\pm}$ and/or $M_A=M_{H^{\pm}}$. Note that a more refined analysis can be based on the fit of the Peskin-Takeuchi parameters \cite{Peskin:1991sw}, done in a similar fashion as Refs.~\cite{Arcadi:2022lpp,Arcadi:2023qgf,Arcadi:2023imv}.

Finally, the collider constraints should also be accounted for. The extra BSM boson can indeed be resonantly produced at the LHC. Assuming degenerate masses, to automatically comply with the EWPT, the most relevant signatures are electrically neutral resonances decaying into $\tau^+ \tau^-$ and electrically charged resonances decaying into $\bar {t} b$ or $\tau \nu_\tau$ (see e.g., Refs. \cite{ATLAS:2020zms,ATLAS:2020jqj,CMS:2022rbd,CMS:2019pzc} for updated results). The collider constraints are mostly effective in the case of the Type-II 2HDM while being strongly relaxed in the other scenario as a consequence of $1/\tan\beta$ suppression in the production vertex and/or decay branching fraction of the resonances.

The parameter space of a general 2HDM can be explored systematically using publicly available tools such as 2HDMC \cite{Eriksson:2009ws} and HiggsTools \cite{Bahl:2022igd}.

Moving to the DM-related constraints, for what the relic density is concerned, the most relevant impact of the extended Higgs sector is the coupling of the DM with a pseudoscalar state. 

First of all, it lifts the velocity dependence of the DM annihilation cross-section:
\begin{align}
& \langle \sigma v \rangle_{ff}=\frac{1}{2\pi}\sum_f n_c^f \sqrt{1-\frac{m_f^2}{m_{N_1}^2}} \big[ \frac{\big|g_{Aff} \big|^2 \big| y_{AN_1 N_1} \big|^2 m_f^2 m_{N_1}^2}{v_h^2 (4 m_{N_1}^2-M_A^2)^2} \nonumber\\
& +\frac{m_f^2}{M_Z^4}|g_{ZN_1 N_1}^A|^2 |g_{Zff}^A|^2 \nonumber\\ 
&-2 \frac{m_f^2 m_{N_1}}{v_h \,M_Z^2 (4 m_{N_1}^2 -M_A^2)}
\mbox{Re}\left( g_{Aff}y_{AN_1 N_1}^{*}g_{ZN_1 N_1}^A g_{Zff}^A\right)\nonumber\\
& + \frac{v^2}{2\pi}|y_{hN _1 N_1}|^2 \frac{m_f^2}{v_h^2}\left(1-\frac{m_f^2}{m_{N_1}^2}\right)m_{N_1}^2\nonumber\\
& \left \vert \frac{y_{hN_1 N_1}}{(4 m_{N_1}^2-M_h^2)}+  \frac{y_{HN_1 N_1}}{(4 m_{N_1}^2-M_H^2)}\right \vert^2\bigg].
\end{align}

This will imply, in turn, the enhancement of the ID signal as well. Note that in the analytical expression below we have also explicitly accounted for the $p$-wave contribution associated with the CP-even boson as it can become dominant in the presence of $s$-channel resonances.
Further, it provides additional annihilation channels into $Ah$, $AA$ and $AZ$ final states with cross-sections which can be approximated as:
\begin{align} 
& \langle \sigma v \rangle_{ZA}  =\frac{v^2}{16 \pi M_Z^2}\sqrt{1-\frac{(M_A-M_Z)^2}{4 m_{N_1}^2}}\sqrt{1-\frac{(M_A+M_Z)^2}{4 m_{N_1}^2}}\nonumber\\
& \bigg( 16 m_{N_1}^4-8 m_{N_1}^2 (m_Z^2+M_A^2)+(M_Z^2-M_A^2)^2 \bigg)\nonumber\\
& \times {\left[\frac{\lambda_{hAZ}y_{hN_1 N_1}}{(4
m_{N_1}^2-M_h^2)}+\frac{\lambda_{HAZ}y_{HN_1 N_1}}{(4
m_{N_1}^2-M_H^2)}\right]}^2,
\end{align} 
\begin{align} 
& \langle \sigma v
\rangle_{hA}=\frac{1}{16\pi}\sqrt{1-\frac{(M_h+M_A)^2}{4
m_{N_1}^2}}\sqrt{1-\frac{(M_h-M_A)^2}{4 m_{N_1}^2}}\nonumber\\
& \left[\frac{\lambda_{hAA}^2 y_{AN_1 N_1}^2}{(4
m_{N_1}^2-M_A^2)^2}+\frac{1}{4}\frac{\lambda_{hAZ}^2 g_{ZN_1 N_1}^2
}{(4 m_{N_1}^2-M_Z^2)^2} (M_A^2-M_h^2)^2)\right.\nonumber\\
&\left. \sum^3_{i,j=1} \frac{y_{AN_1 N_i}y_{AN_1 N_j}^{*}y_{hN_1
N_i}y_{hN_1 N_j}^{*}}{m_{N_1}^2 (M_A^2+M_h^2-2
m_{N_1}^2-m_{N_i}^2)^2 (M_A^2+M_h^2-2 m_{N_1}^2-m_{N_j}^2)^2
}\right.\nonumber\\ 
&\left. \times\left(M_A^4+M_h^4-8
m_{N_1}m_{N_j}M_h^2+16 m_{N_i}m_{N_j}m_{N_1}^2\right.\right.\nonumber\\
&\left.\left. -2 M_A^2 (M_h^2-4
m_{N_1}m_{N_j}) \right)\right.\nonumber\\ &\left. \times
\mbox{Re}\left[\lambda_{hAA}^{*}y_{AN_1 N_1}^{*}y_{hN_1
N_1}^{*}\lambda_{hAZ}g_{ZN_1 N_1}^A\right]\frac{(M_A^2-M_h^2)}{M_Z^2
m_{N_1}}\right.\nonumber\\
&\left.+\frac{2}{m_{N_1}^2}\mbox{Re}\left[\lambda_{hAA}^{*}y_{AN _1
N_1}^{*}y_{hN_1 N_1}^{*}y_{hN _1 N_i}y_{AN _1
N_i}\right]\right.\nonumber\\
& \left. \frac{(M_A^2 m_{N_1}-M_h^2 m_{N_1}+4
m_{N_i}m_{N_1}^2)}{(M_A^2+M_h^2-2 m_{N_1}^2-2 m_{N_i}^2)(4
m_{N_i}^2-M_A^2)}\right.\nonumber\\ &\left.
+\frac{1}{2}\sum^3_{i=1}\mbox{Re}\left[\lambda_{hAZ}^{*}g_{ZN_1
N_1}^{*}y_{hN_1 N_i}y_{AN_1 N_i}\right]\right.\nonumber\\
& \left. \frac{(M_A^2-M_h^2)^2+4
m_{N_1}m_{N_i} (M_A^2-M_h^2)}{m_{N_1}^2 M_Z^2(M_A^2+M_h^2-2
m_{N_1}^2-2 m_{N_i}^2)}\right],
\end{align}
and,
\begin{align} 
& \langle \sigma v\rangle_{AA}  =\frac{v^2}{128\pi}\sqrt{1-\frac{M_A^2}{m_{N_1}^2}}
\left[{\left( \frac{\lambda_{hAA}y_{hN_1
N_1}}{(4 m_{N_1}^2-M_h^2)}+\frac{\lambda_{HAA}y_{HN_1 N_1}}{(
m_{N_1}^2-M_H^2)}\right)}^2 \right.\nonumber\\
&\left. +\frac{8}{3}|y_{AN_1
N_1}|^2  m_{N_1} \right.\nonumber\\ &
\left. \times    \bigg(2\frac{m_{N_1}^2 (m_{N_1}^2-M_A^2)^2}{(2
m_{N_1}^2-M_A^2)^4} -\frac{(m_{N_1}^2-M_A^2)}{(2
m_{N_1}^2-M_A^2)^2} \bigg) \right.\nonumber\\
&\left. \times \left(\frac{y_{hN_1 N_1}\lambda_{hAA}}{(4
m_{N_1}^2-M_h^2)}+\frac{y_{HN_1 N_1}\lambda_{HAA}}{(4
m_{N_1}^2-M_H^2)}\right)\right].
\end{align} 
The parameters $\lambda_{HAA, hAA, hAZ, HAZ}$ represent tri-linear couplings between two electrically neutral Higgs boson and between two Higgs and the Z-boson. Their analytical expressions are given, for example, in \cite{Arcadi:2019lka}.
The DM DD cross-section is influenced as well by the extended Higgs sector as both the CP-even bosons can contribute to the effective interactions of the DM with nucleons:
\begin{equation}
\sigma_{\chi p}^{\rm SI}=\frac{\mu_\chi^2}{\pi}\frac{m_p^2}{v_h^2}\bigg| \sum_{q}f_q \left(\frac{y_{hN_1 N_1}g_{hqq}}{M_h^2}+\frac{y_{HN_1 N_1}g_{Hqq}}{M_H^2}\right) \bigg|^2 \, .
\end{equation}

The SD interactions mediated by the Z-boson are still present as well. The cross-section has the same expression as the minimal SD model, hence we will not rewrite it explicitly.

\begin{figure*}
    \centering
    \subfloat{\includegraphics[width=0.25\linewidth]{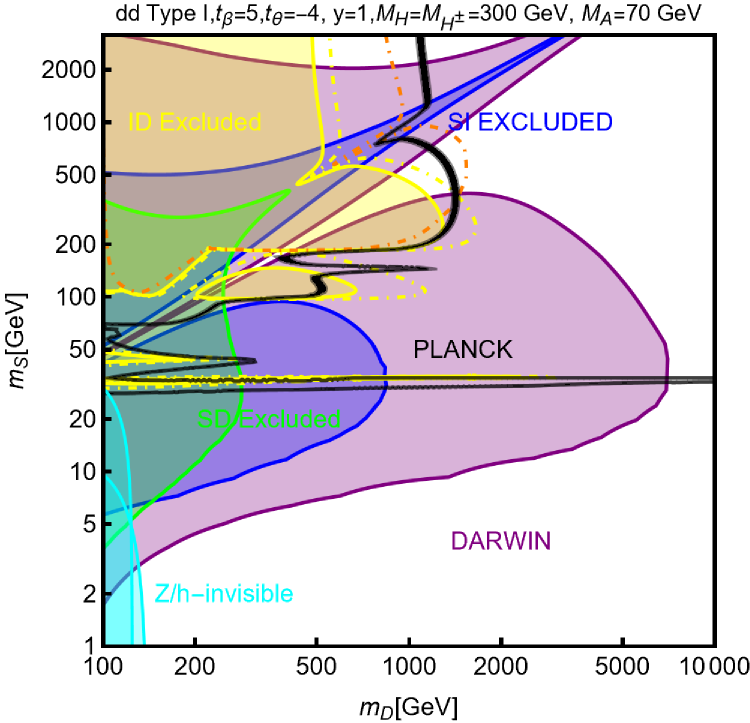}}
    \subfloat{\includegraphics[width=0.25\linewidth]{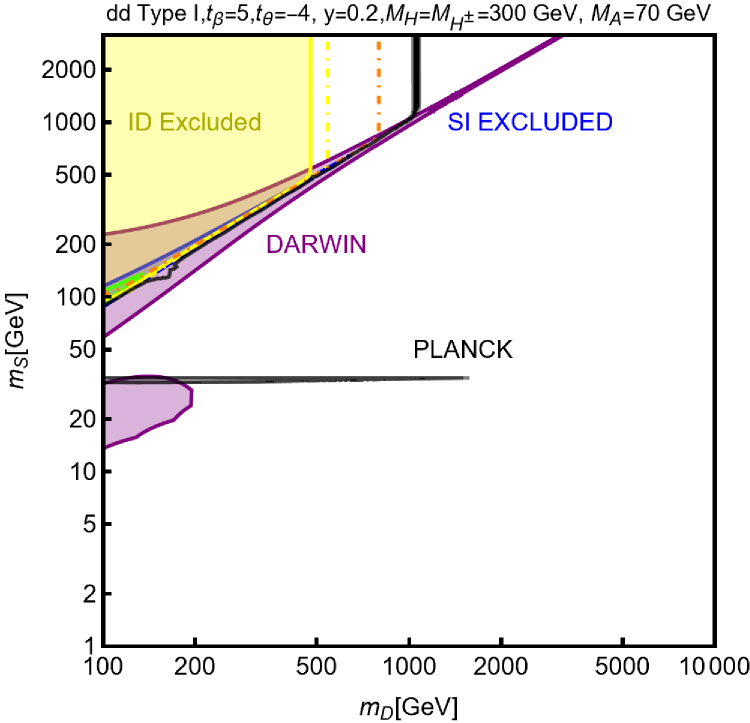}}
    \subfloat{\includegraphics[width=0.25\linewidth]{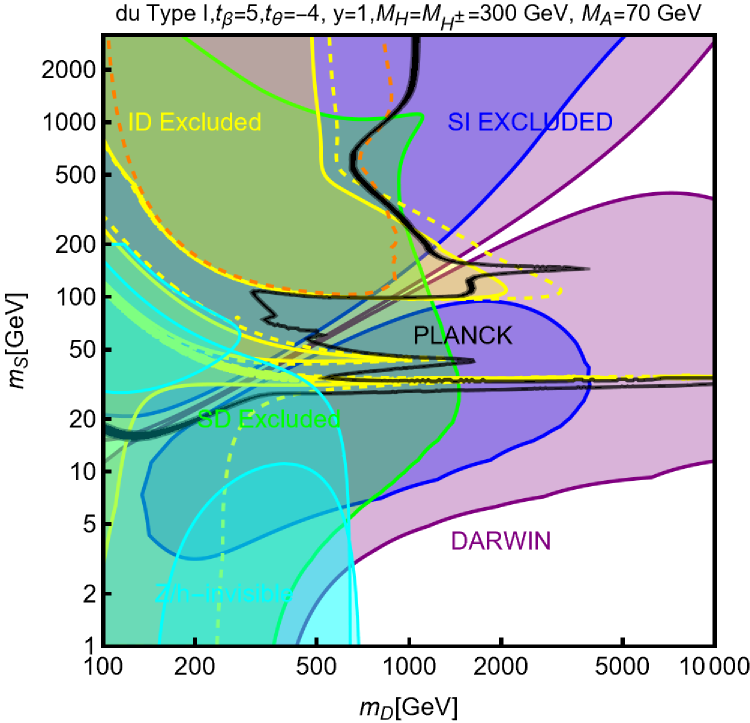}}
    \subfloat{\includegraphics[width=0.25\linewidth]{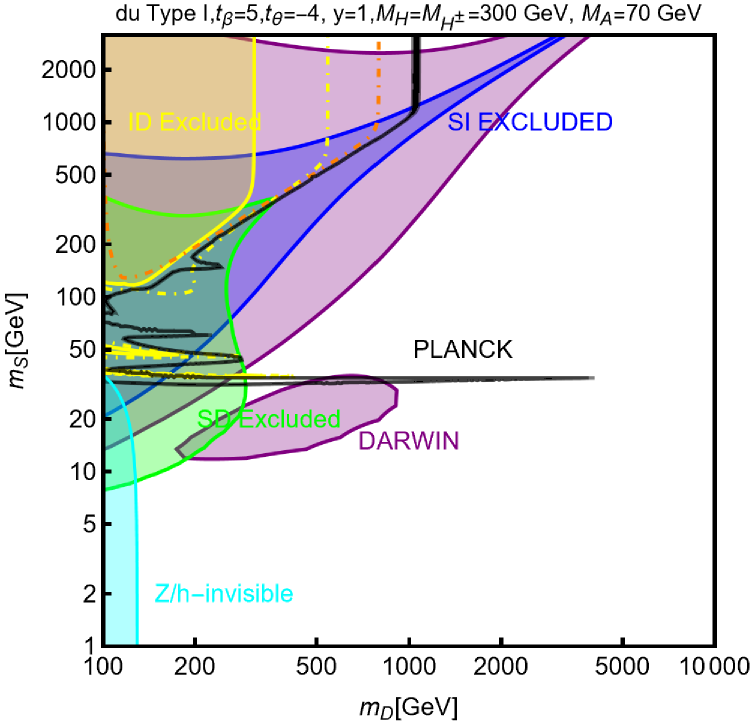}}\\
    \subfloat{\includegraphics[width=0.25\linewidth]{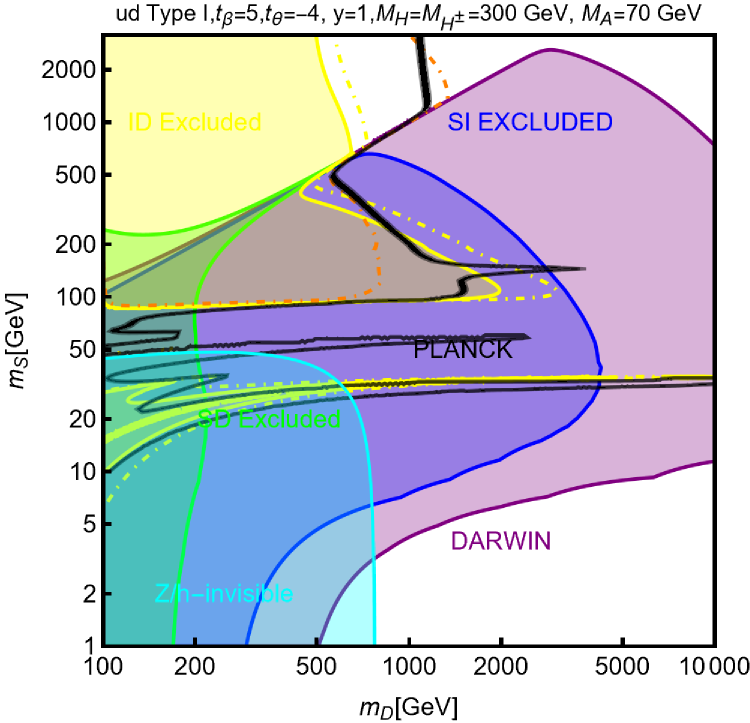}}
    \subfloat{\includegraphics[width=0.25\linewidth]{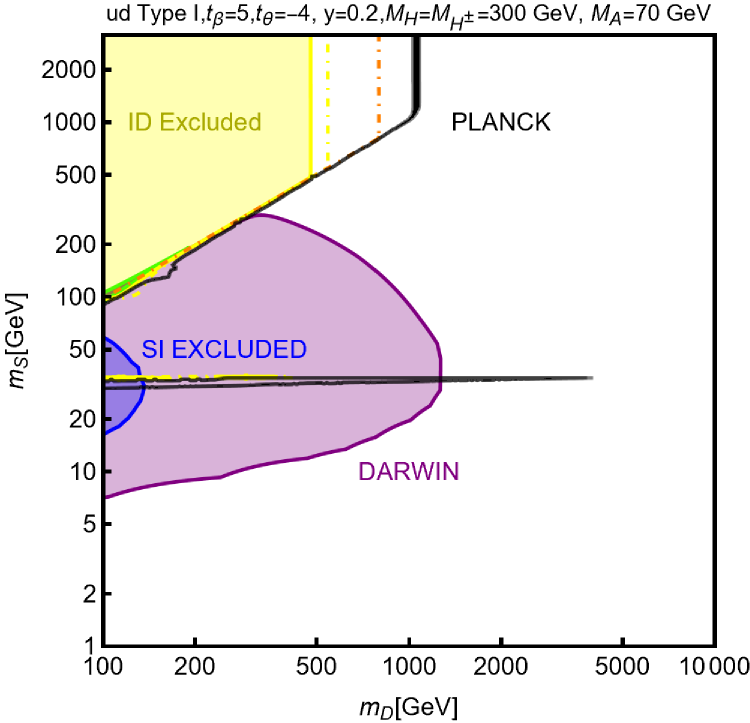}}
    \subfloat{\includegraphics[width=0.25\linewidth]{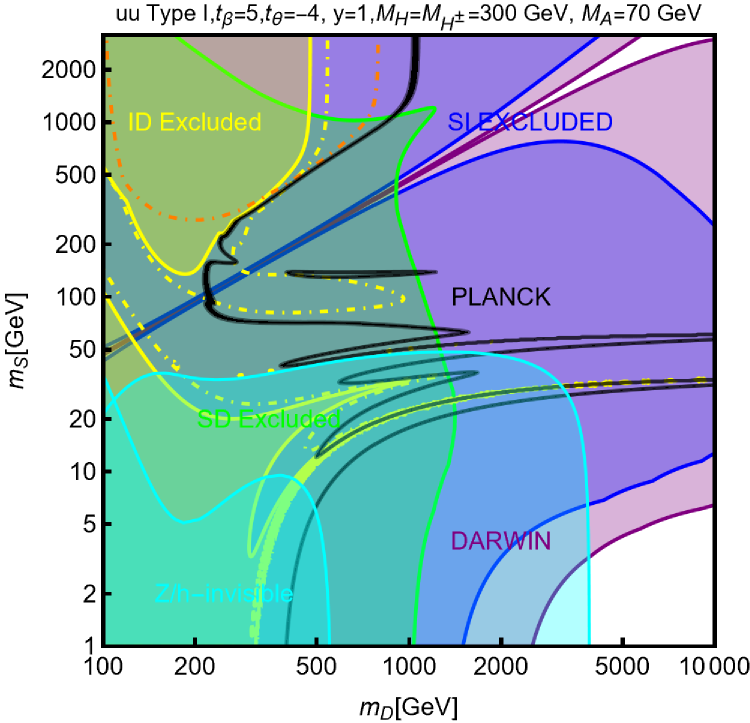}}
    \subfloat{\includegraphics[width=0.25\linewidth]{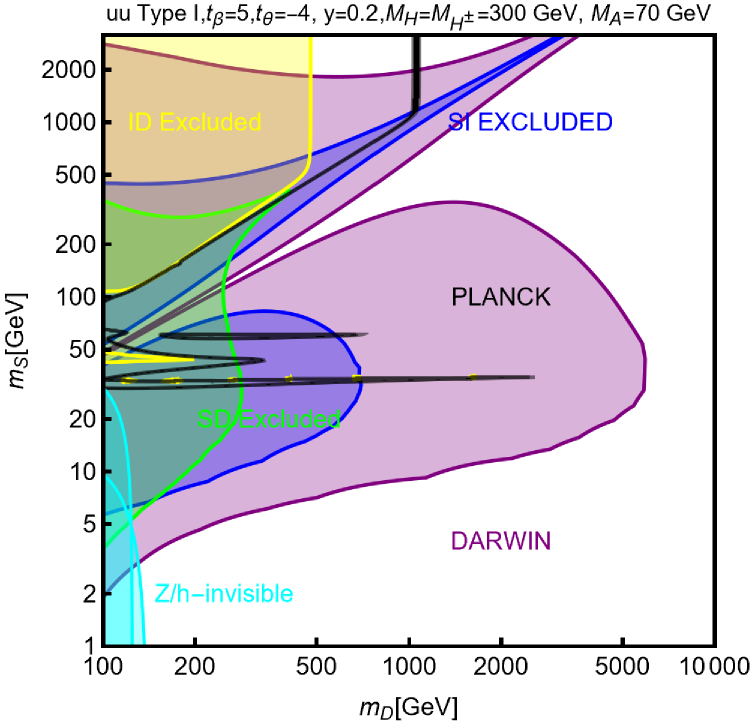}}
    \caption{Summary of the DM constraints for the Singlet-Doublet+2HDM model in the $(m_S,m_D)$ plane for benchmark assignations of the set $(\tan\beta\equiv t_\beta, \tan\theta\equiv t_\theta,\,y)$ and of the masses of the BSM Higgs states $M_H,\,M_{H^\pm},\,M_A$, reported on the top of each panel. These plots are for a Type-I 2HDM where four different types ($dd,\,du, ud, uu$) of couplings exist between the BSM fermionic DM and the Higgs states. The remaining colour codes are the same as of Fig. \ref{fig:pSDmajo}.}
    \label{fig:SD2HDMtypI}
\end{figure*}

\begin{figure*}
    \centering
    \subfloat{\includegraphics[width=0.25\linewidth]{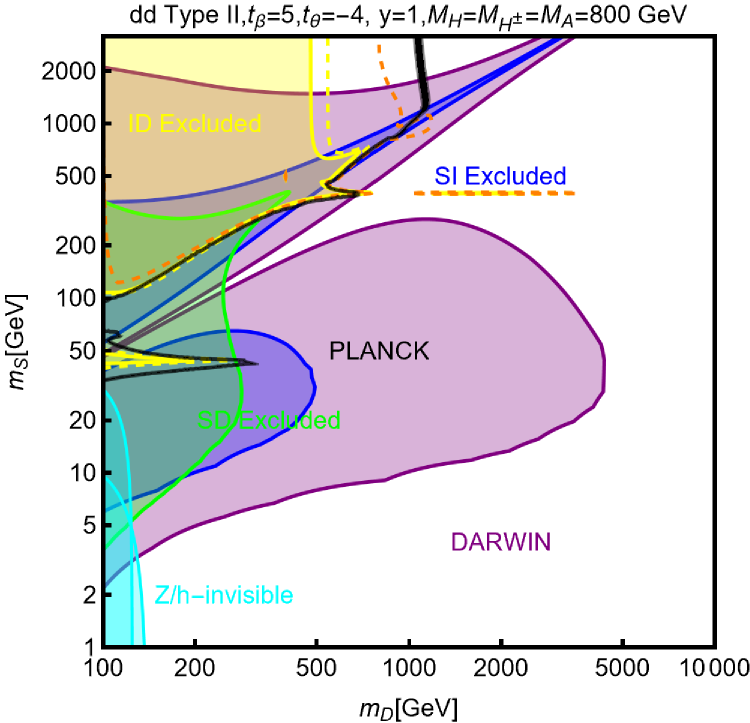}}
     \subfloat{\includegraphics[width=0.25\linewidth]{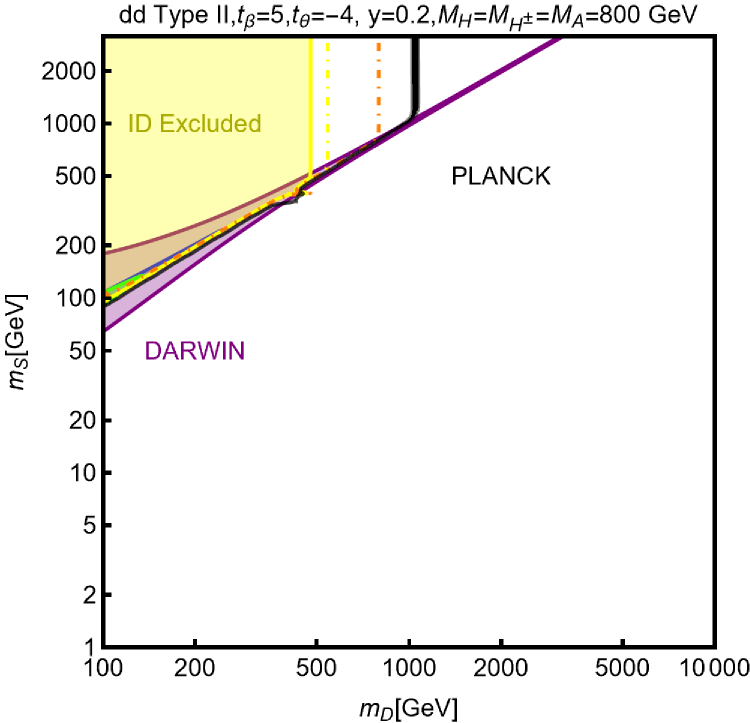}}
    \subfloat{\includegraphics[width=0.25\linewidth]{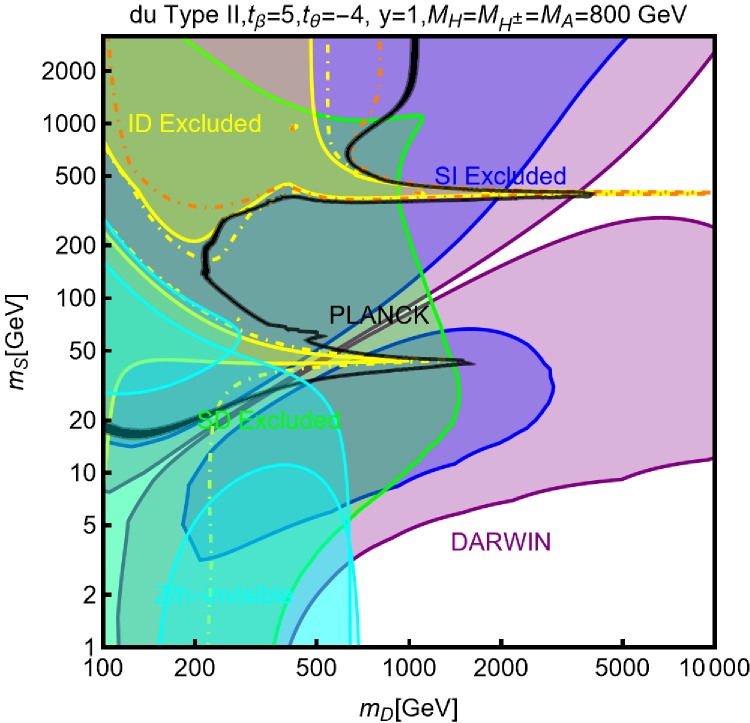}}
     \subfloat{\includegraphics[width=0.25\linewidth]{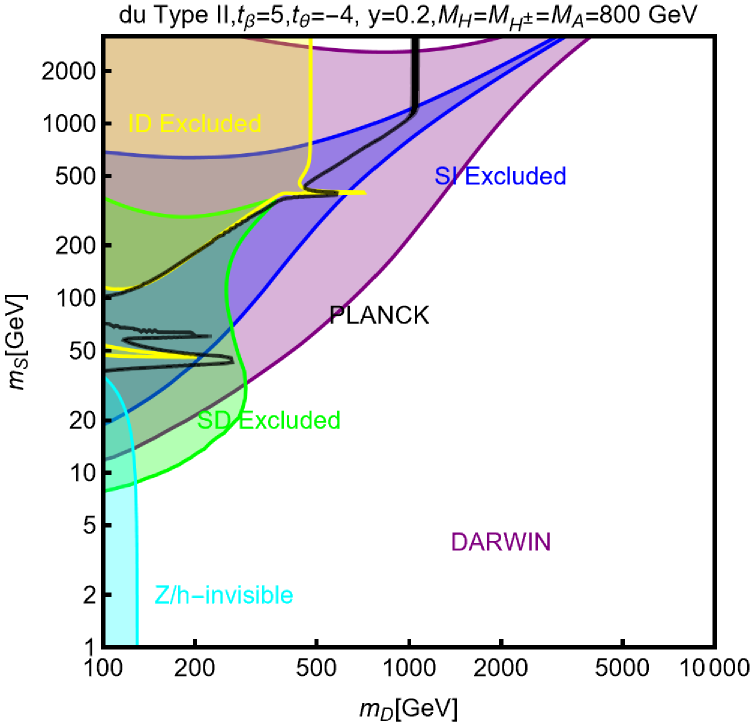}}\\
    \subfloat{\includegraphics[width=0.25\linewidth]{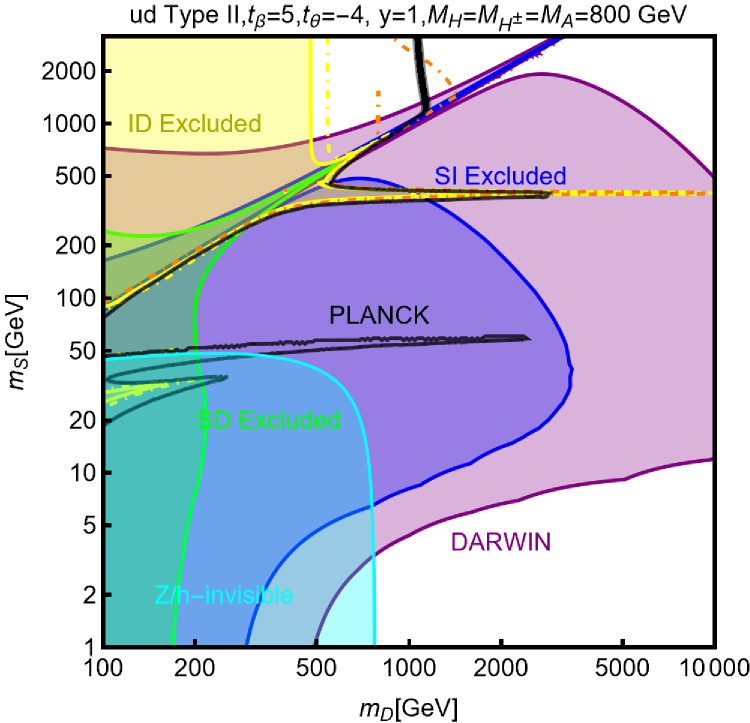}}
     \subfloat{\includegraphics[width=0.25\linewidth]{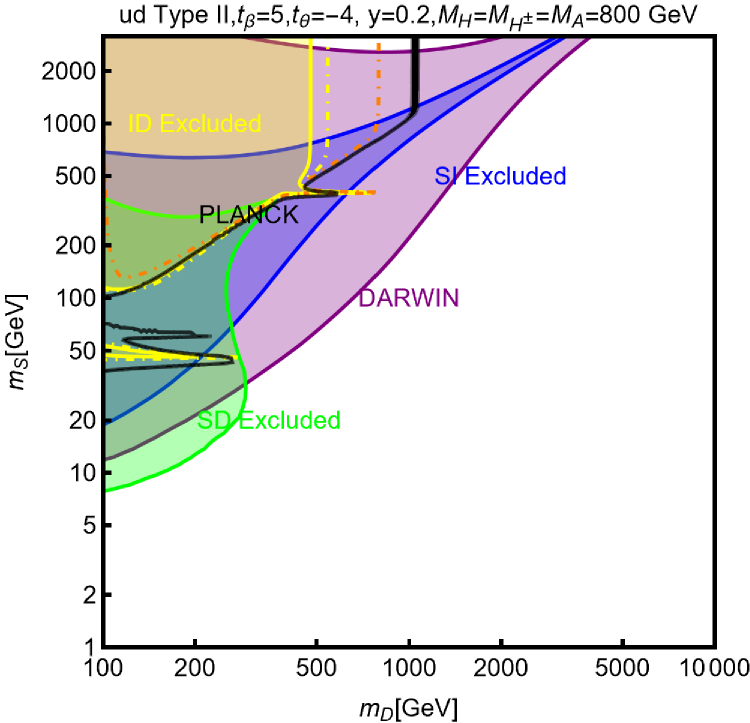}}
    \subfloat{\includegraphics[width=0.25\linewidth]{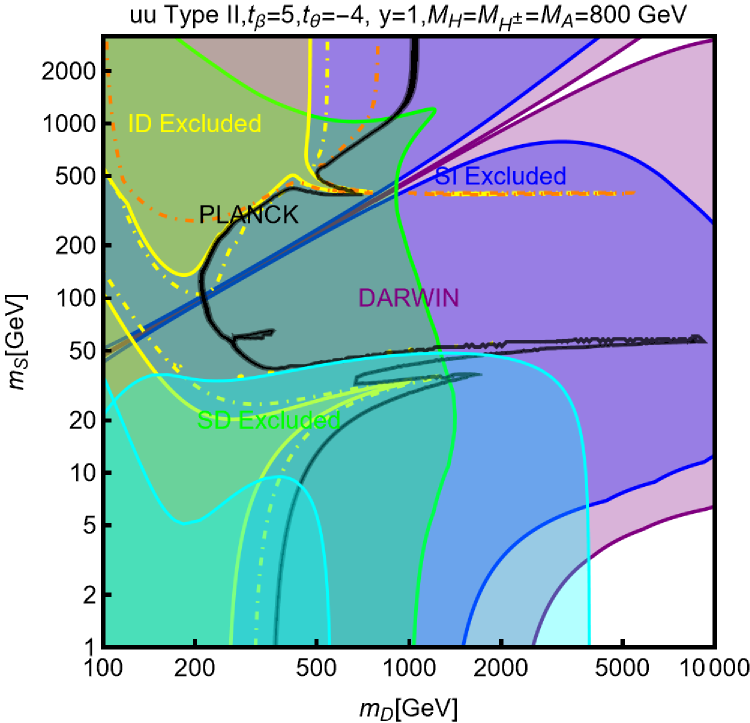}}
     \subfloat{\includegraphics[width=0.25\linewidth]{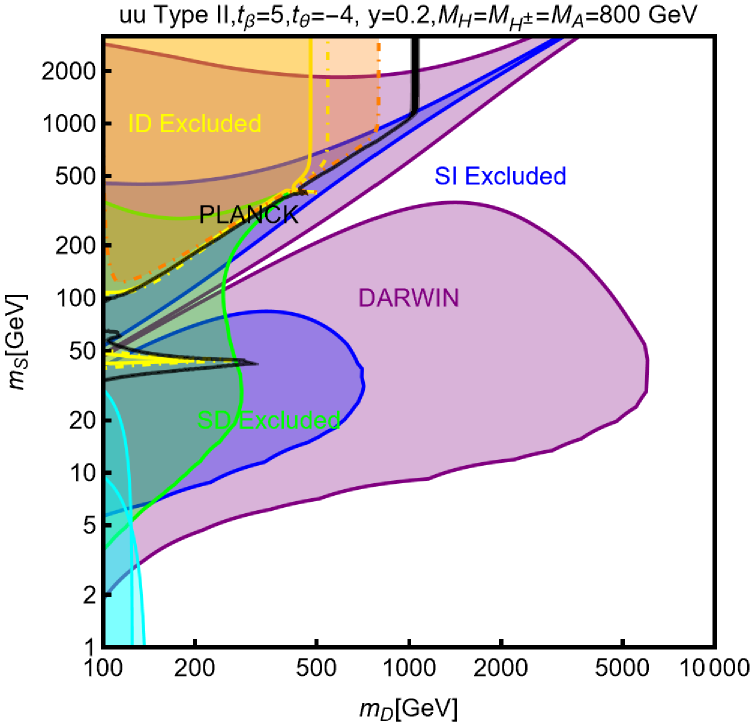}}
    \caption{This is analogous to Fig.~\ref{fig:SD2HDMtypI}, but for a Type-II 2HDM scenario fixing $M_H=M_{H^\pm}=M_A=800$ GeV.}
    \label{fig:SD2HDMtypII}
\end{figure*}

\begin{figure*}
    \centering
    \subfloat{\includegraphics[width=0.25\linewidth]{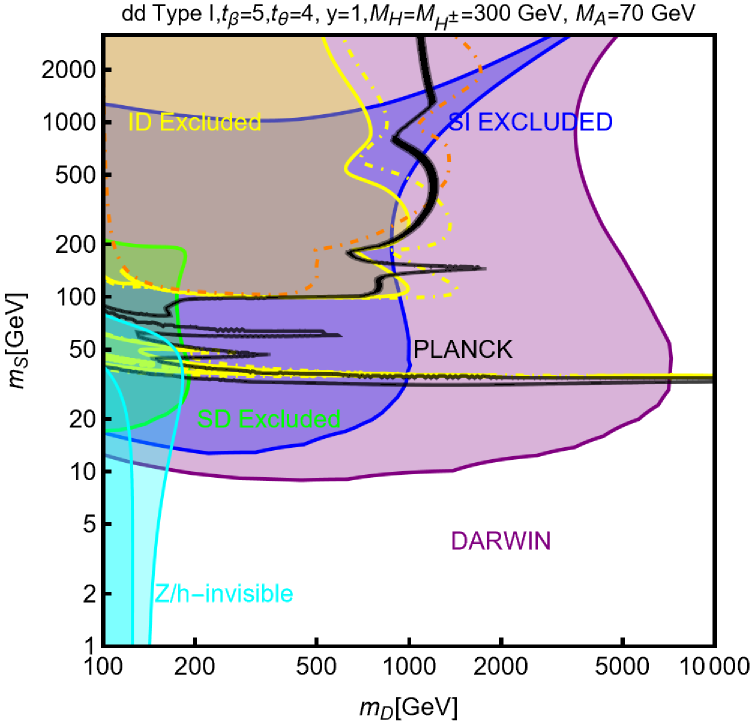}}
    \subfloat{\includegraphics[width=0.25\linewidth]{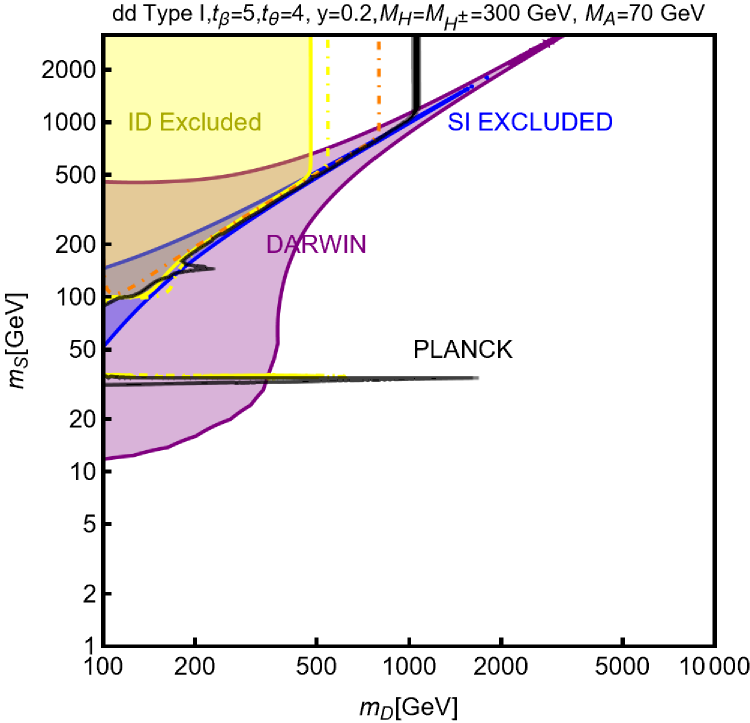}}
    \subfloat{\includegraphics[width=0.25\linewidth]{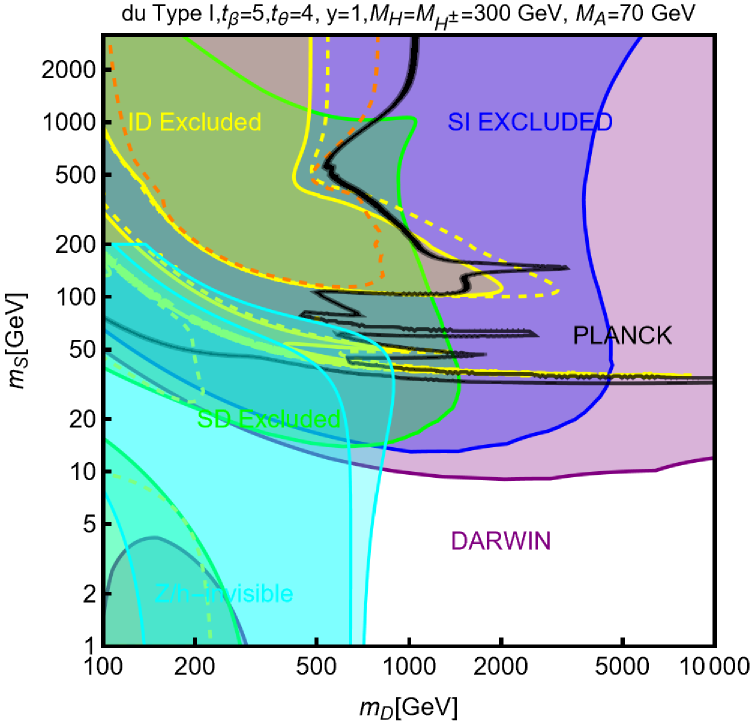}}
    \subfloat{\includegraphics[width=0.25\linewidth]{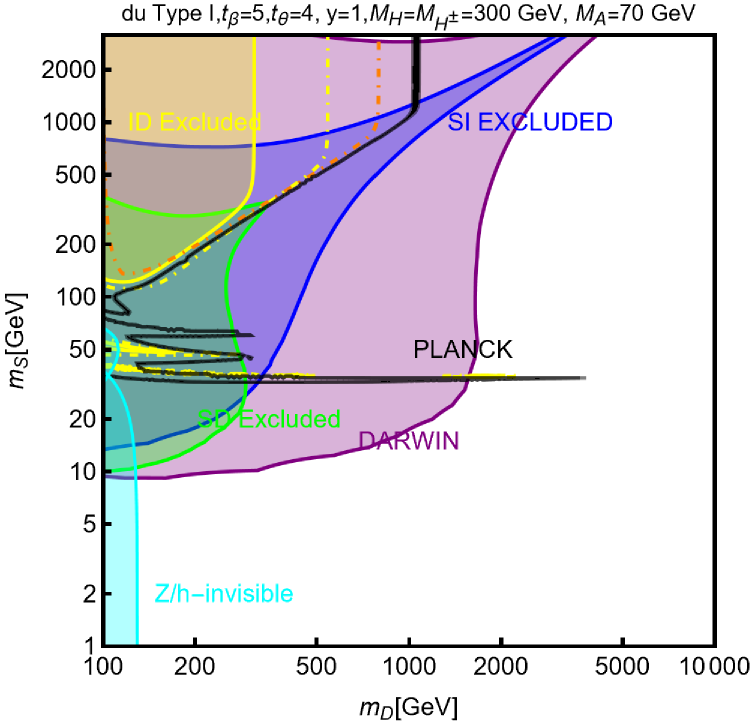}}\\
    \subfloat{\includegraphics[width=0.25\linewidth]{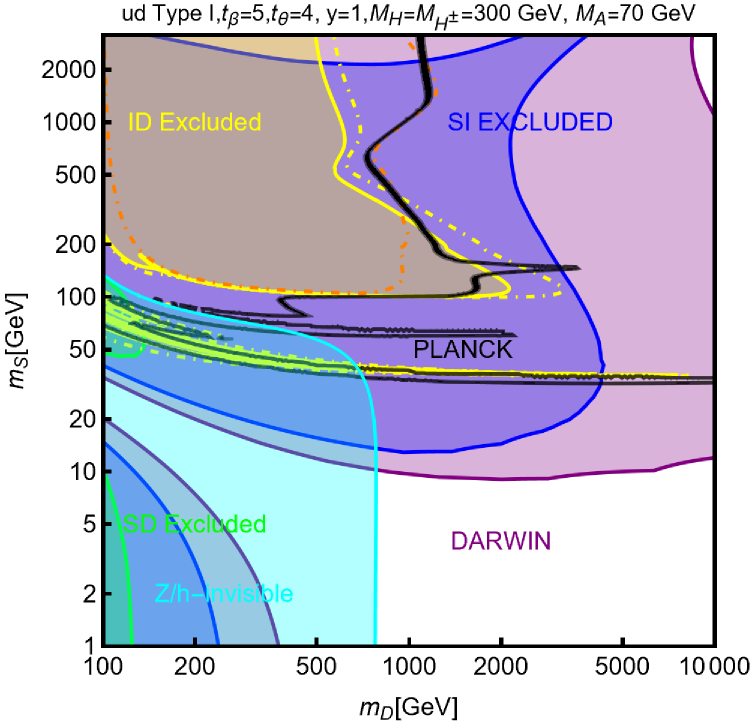}}
    \subfloat{\includegraphics[width=0.25\linewidth]{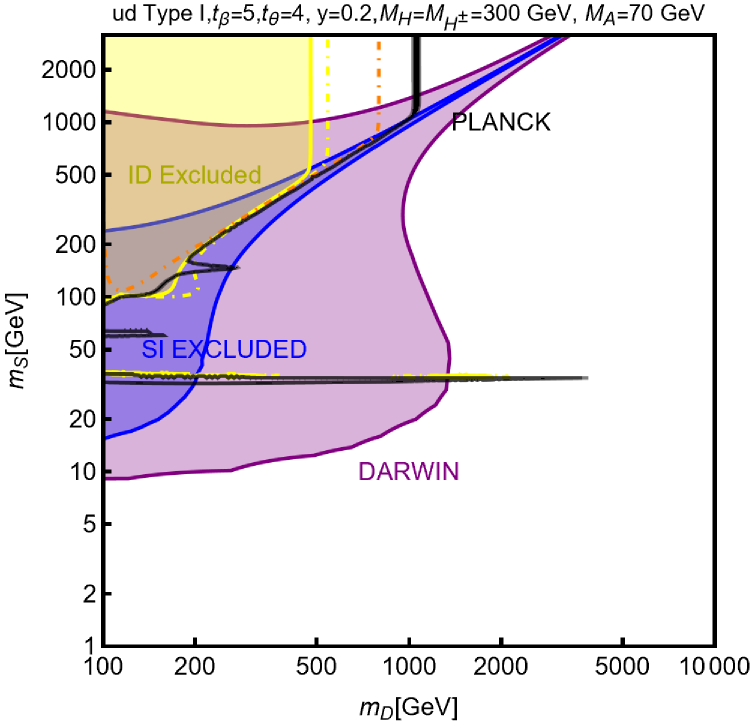}}
    \subfloat{\includegraphics[width=0.25\linewidth]{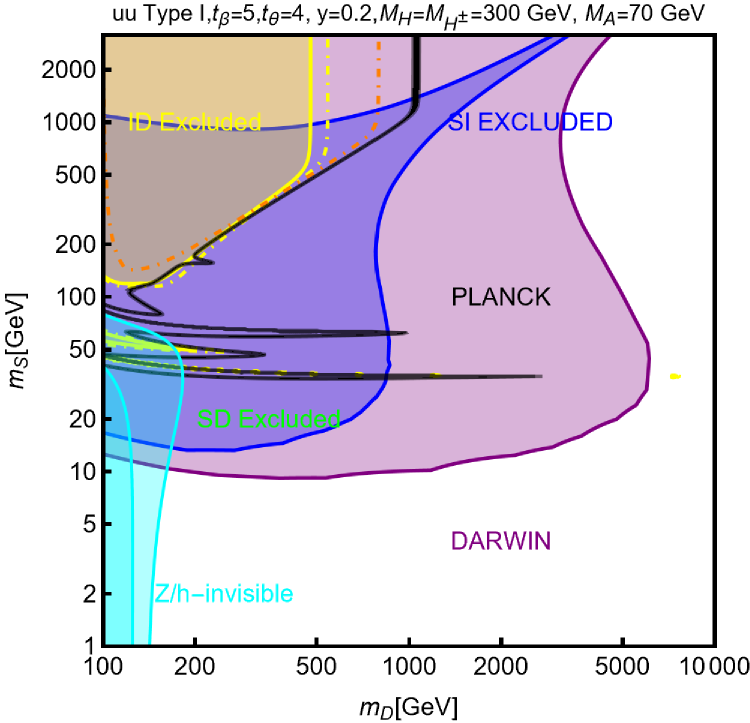}}
    \subfloat{\includegraphics[width=0.25\linewidth]{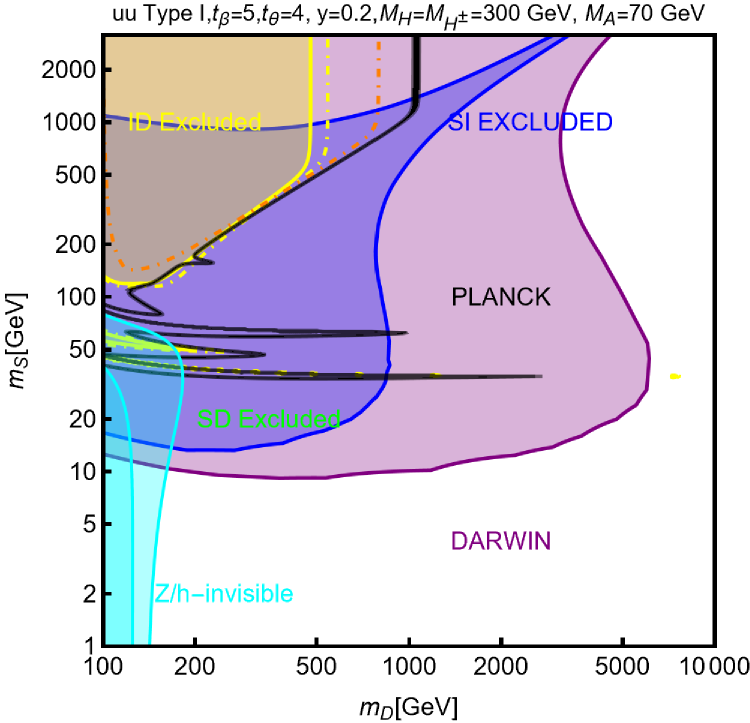}}
    \caption{Same benchmarks as Fig.~\ref{fig:SD2HDMtypI} but for positive $\tan\theta$.}
    \label{fig:SD2HDMtypIp}
\end{figure*}

\begin{figure*}
    \centering
    \subfloat{\includegraphics[width=0.25\linewidth]{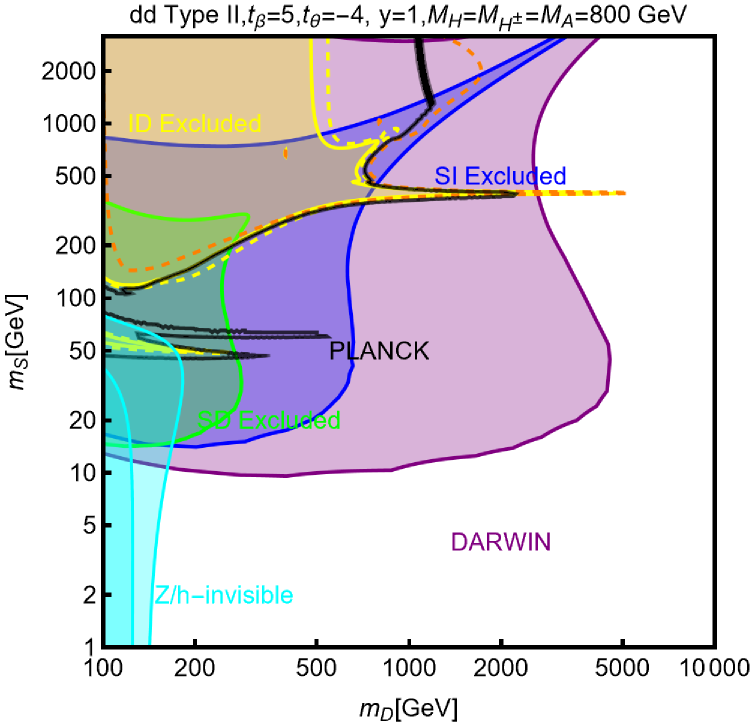}}
     \subfloat{\includegraphics[width=0.25\linewidth]{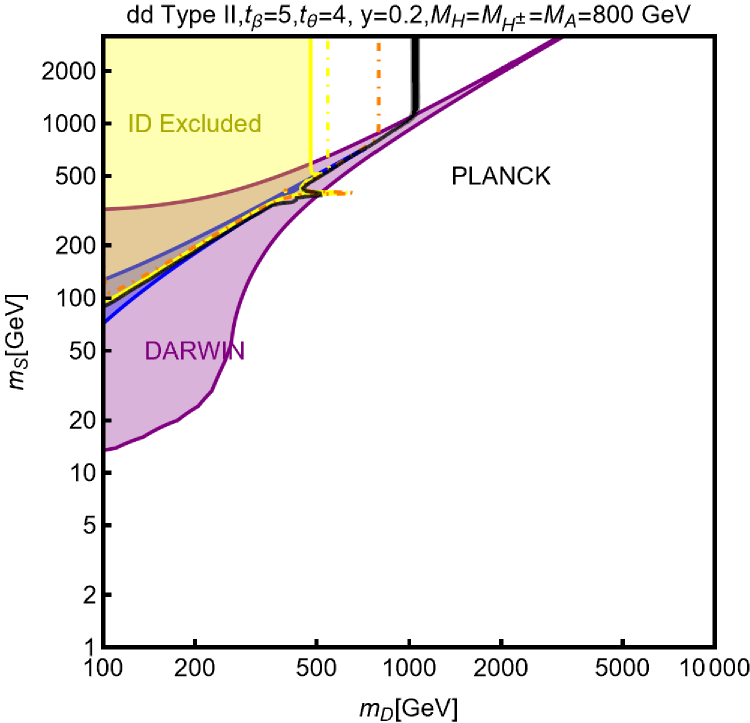}}
    \subfloat{\includegraphics[width=0.25\linewidth]{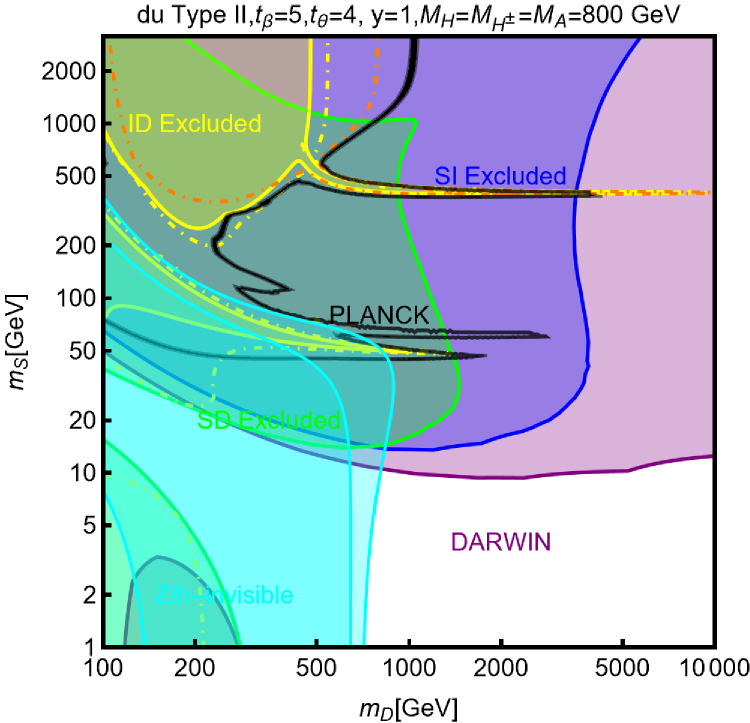}}
     \subfloat{\includegraphics[width=0.25\linewidth]{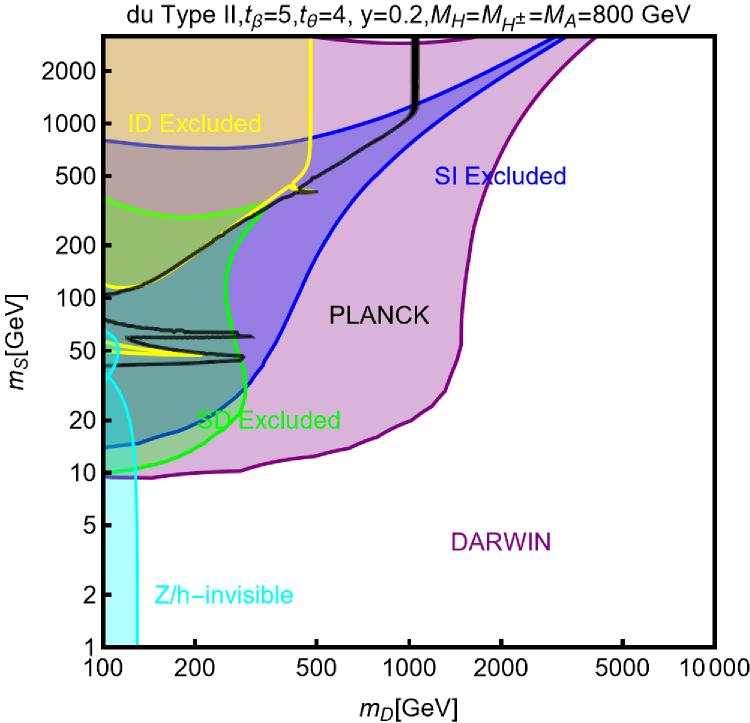}}\\
    \subfloat{\includegraphics[width=0.25\linewidth]{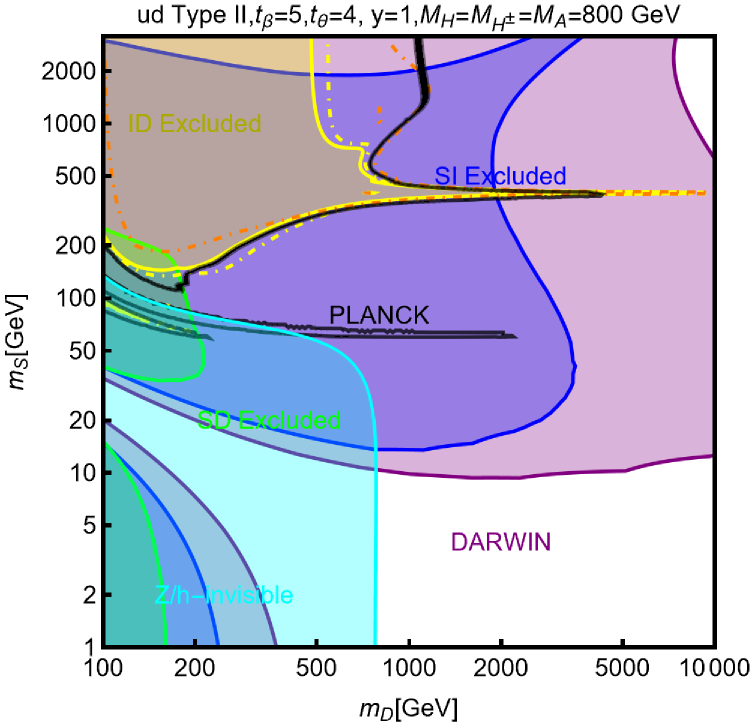}}
     \subfloat{\includegraphics[width=0.25\linewidth]{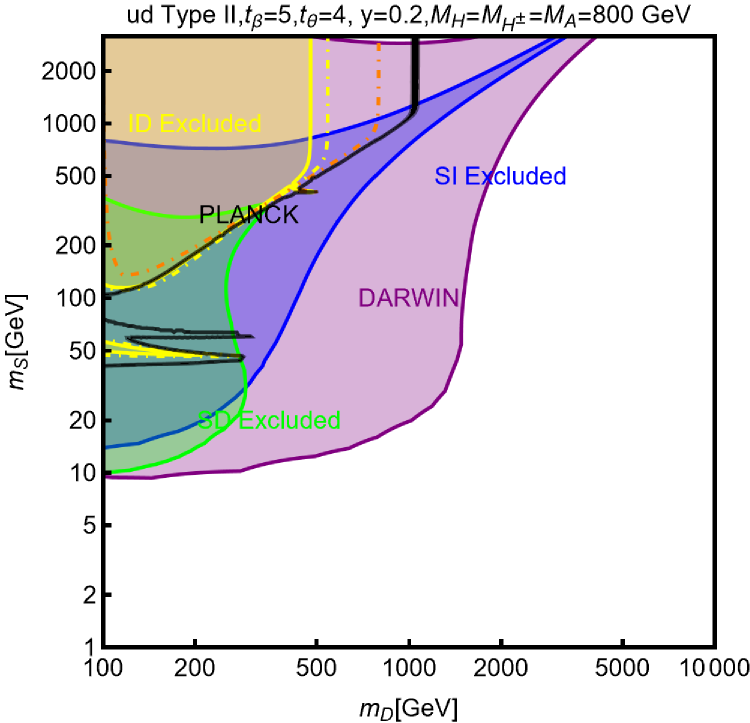}}
    \subfloat{\includegraphics[width=0.25\linewidth]{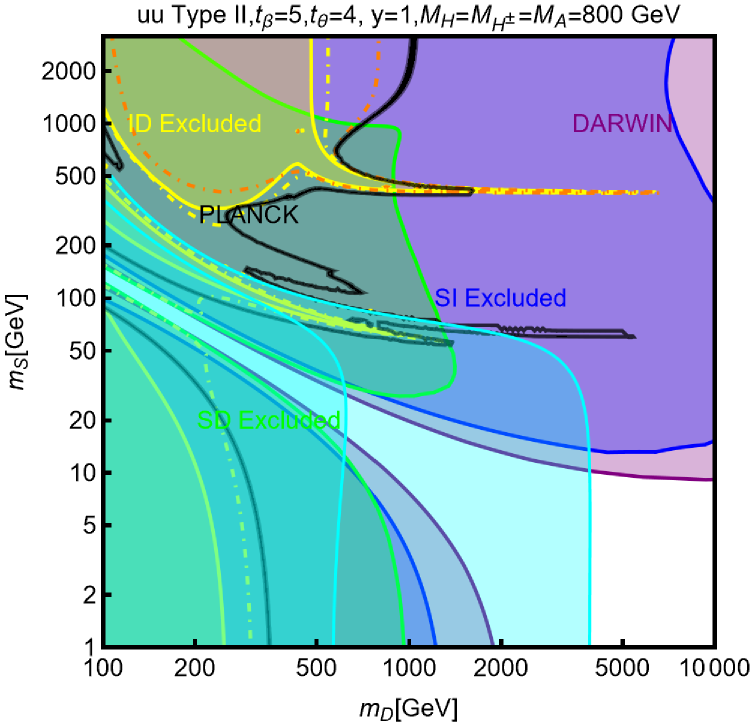}}
     \subfloat{\includegraphics[width=0.25\linewidth]{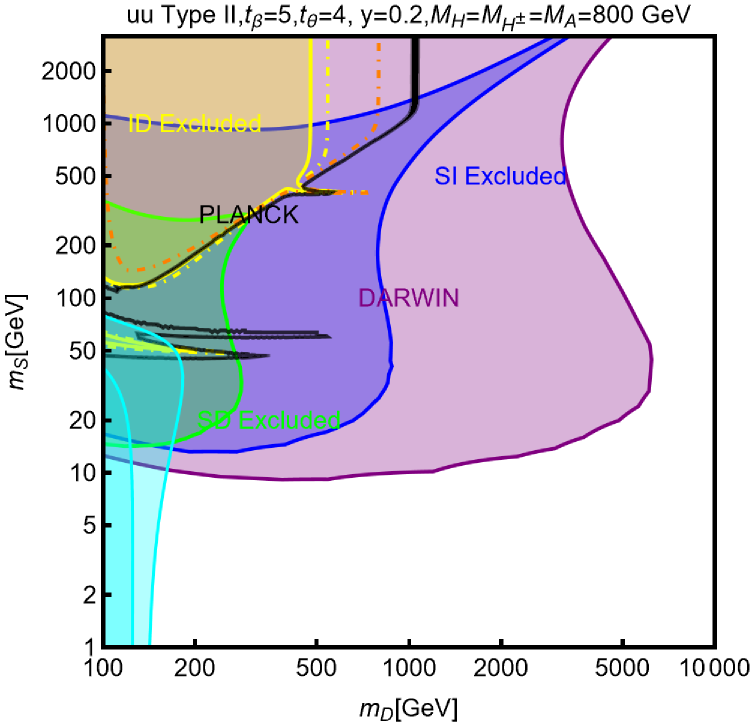}}
    \caption{Same as Fig.~\ref{fig:SD2HDMtypII}, but for positive $\tan\theta$.}
    \label{fig:SD2HDMtypIIp}
\end{figure*}

We have combined the constraints listed above in an analogous way as the minimal SD model in Fig. \ref{fig:SD2HDMtypI} and \ref{fig:SD2HDMtypIp} (\ref{fig:SD2HDMtypII} and \ref{fig:SD2HDMtypIIp}) adopting Type-I (II) scheme for
the $\phi f f$ couplings with $\phi=h,\,H,\,H$ and $f$ being any
SM fermions. For these figures, the first (next) two plots of the top row correspond to $dd$-type ($du$-type) $y_{\phi N_1 N_1}$ couplings whereas for the bottom row, the first (next) two plots correspond to $ud$-type ($uu$-type) $y_{\phi N_1 N_1}$ couplings.
Besides, we consider fixed assignations of $(\tan\beta\equiv t_\beta,\,\tan\theta\equiv t_\theta,\,y,\,M_H = M_{H^{\pm}},\,M_A)$ for these plots which are written on the top of each plot.  In the case of the Type-I 2HDM scenario, we have considered the possibility of light masses for the BSM Higgs states, namely $M_H=M_{H^{\pm}}=300\,\mbox{GeV}$ (the
degeneracy evades the dangerous contributions to the EWPT) and even $M_A=70\,\mbox{GeV}$. In the case of the Type-II 2HDM we have instead considered the assignation $M_H=M_{H^{\pm}}=M_A=800\,\mbox{GeV}$ to comply with bounds from the LHC and B-physics.  As evident from these figures, even in the presence of the extended Higgs sector,  this scenario remains very constrained. The DD constraints can be evaded mostly in correspondence of the poles, i.e., when the DM mass is approximately $1/2$ the mass of an electrically neutral boson.

\subsection{Singlet fermion+2HDM+s}

One can consider extending the Higgs sector of the theory further by considering the presence of an additional CP-even scalar $S$ singlet under $SU(2)$. The scalar potential of the theory can then be written as \cite{Bell:2017rgi}: 
\begin{align}\label{eq:2HDMSpotential}
    & V_{\rm 2HDMs}=V_{\rm 2HDM}+V_{s},~~~\mbox{where} \nonumber\\   
& V_S=\frac{1}{2}m_s^2 S^2+\frac{1}{3}\mu_S S^3+\frac{1}{4}\lambda_S S^4 \nonumber\\
& +\mu_{11S}(\Phi_1^\dagger \Phi_1)S + \mu_{22S}(\Phi_2^\dagger \Phi_2)S + (\mu_{12S} \Phi_2^\dagger \Phi_1 S + \mathrm{H.c.}) \nonumber\\
& +  \frac{\lambda_{11S}}{2}(\Phi_1^\dagger \Phi_1)S^2 +  \frac{\lambda_{22S}}{2}(\Phi_2^\dagger \Phi_2)S^2 + \frac{1}{2}(\lambda_{12S} \Phi_2^\dagger \Phi_1 S^2 + \mathrm{H.c.}),
\end{align}
with $V_{\rm 2HDM}$ is the two Higgs doublet potential already given in Eq.~\eqref{eq:2HDM_potential}. To be consistent with the perturbative unitarity and boundness from below, the couplings of the potential should satisfy the following conditions \cite{Klimenko:1984qx,Kanemura:2015ska,Bell:2016ekl}:
\begin{align}
     & |\lambda_3|+|\lambda_4| <1, \nonumber\\
    & |\lambda_3|+|\lambda_5| <1 ,\nonumber\\
    & \lambda_1+\lambda_2+\sqrt{\lambda_1^2-2 \lambda_1 \lambda_2 +\lambda_2^2+4 \lambda_5^2}<2, \nonumber\\
    & \lambda_1+\lambda_2+\sqrt{\lambda_1^2-2 \lambda_1 \lambda_2 +\lambda_2^2+4 \lambda_4^2}<2, \nonumber\\
    & \lambda_{11S}+\lambda_{22S}+\sqrt{\lambda_{11S}^2-2 \lambda_{11S}\lambda_{22S} +\lambda_{22S}^2+4 \lambda_{12S}^2}<2,\nonumber\\
    & \lambda_{1,2,S}>0, \nonumber\\
    & \sqrt{\lambda_1 \lambda_2}>-\lambda_3,\nonumber\\
    & \sqrt{2 \lambda_{1}\lambda_S}>-\lambda_{11S,}\nonumber\\
    & \sqrt{2 \lambda_{2}\lambda_S}>-\lambda_{22S},\nonumber\\
    & \sqrt{\lambda_1 \lambda_2}>|\lambda_5|-\lambda_3-\lambda_4.
\end{align}
For analysing this setup is convenient to the so-called Higgs-basis defined by the following relations:
\begin{align}
    & \Phi_h=\cos \beta \Phi_1+\sin \beta \Phi_2=\left(\begin{array}{c}
         G^+  \\
         \frac{v+h+iG^0}{\sqrt{2}} 
    \end{array}
    \right),\nonumber\\
    & \Phi_H=-\sin\beta \Phi_1+\cos \beta \Phi_2=\left(\begin{array}{c}
         H^+  \\
         \frac{H+iA}{\sqrt{2}} 
    \end{array}
    \right),
\end{align}
so that the scalar potential of Eq. (\ref{eq:2HDMSpotential}) can be reexpressed as \cite{Bell:2017rgi,Arcadi:2020gge}:
\begin{align}
    & V(\Phi_h,\Phi_H,S)=\hat{M}_{hh}^2 \Phi_h^\dagger \Phi_h+\hat{M}_{HH}^2 \Phi_H^\dagger \Phi_H\nonumber\\
    & +\left(\hat{M}_{hH}^2 \Phi_H^\dagger \Phi_h+\mbox{H.c.}\right)\nonumber\\
    & +\frac{\hat{\lambda}_h}{2}{\left(\Phi_h^\dagger \Phi_h\right)}^2+\frac{\hat{\lambda}_H}{2}{\left(\Phi_H^\dagger \Phi_H\right)}^2+\hat{\lambda}_3 \left(\Phi_h^\dagger \Phi_h\right)\left(\Phi_H^\dagger \Phi_H\right)\nonumber\\
    & +\hat{\lambda}_4 \left(\Phi_H^\dagger \Phi_h\right)\left(\Phi_h^\dagger \Phi_H\right)+\frac{\hat{\lambda}_5}{2}\left(\left(\Phi_H^\dagger \Phi_h\right)^2+\mbox{H.c.}\right)\nonumber\\
    & +\frac{1}{2}\hat{M}_{SS}^2 S^2+\frac{1}{4}\hat{\lambda}_S S^4\nonumber\\
    & +\frac{\hat{\lambda}_{HHS}}{2}\left(\Phi_H^\dagger \Phi_H\right)S^2+\frac{\hat{\lambda}_{hhs}}{2}\Phi_h^\dagger \Phi_h S^2+\frac{1}{2}\left(\hat{\lambda}_{hHS}\Phi_H^\dagger \Phi_h S^2+\mbox{H.c.}\right).
\end{align}

Similar to other scenarios considered in this work, the singlet field can be interpreted as the real component of a complex field; spontaneously breaking an extra $U(1)$ gauge symmetry via a VEV $v_S$. After the EWSB breaking, $V_{\rm 2HDMs}$ would lead to three CP-even states with arbitrary mass mixing. It is nevertheless convenient to consider also in this scenario an alignment limit which would lead to a pure SM-like CP-even state $h$ and two additional particles $S_{1,2}$, being a mixture between the $SU(2)$ singlet and doublet components.
The aforementioned alignment limit is achieved by imposing:
\begin{align}
& \hat{\lambda}_h=\hat{\lambda}_H=\hat{\lambda}_3+\hat{\lambda}_4+\hat{\lambda}_5,\nonumber\\
& \hat{\lambda}_{hHS}=0.
\end{align}
In this limit the physical BSM scalars $S_{1,2}$ are defined as:
\begin{align}
    & H=\cos \theta S_1 -\sin \theta S_2,\nonumber\\
    & S=v_S+\sin \theta S_1+\cos \theta S_2.
\end{align}
The angle $\theta$ weights the $SU(2)$ singlet and doublet components of the physical states and can be written as:
\begin{equation}
    \tan 2 \theta=\frac{2 \lambda_{hHs} v_h v_S}{M_{S_1}^2-M_{S_2}^2}.
\end{equation}
The rest of the physical scalar spectrum is constituted of the charged Higgs boson $H^\pm$ and the pseudoscalar $A$ as in the ordinary 2HDM models. Moving to Yukawa Lagrangian we have:
\begin{align}
\mathcal{L}_{\rm Yuk}=&\sum_f \frac{m_f}{v_h}\left[ g_{hff} h \bar f f+g_{S_1 ff}
S_1\bar f f+ g_{S_2 ff} S_2 \bar f f\right.\nonumber\\
& \left. -i g_{Aff} A \bar f \gamma_5 f  \right] \, , 
\end{align}
with:
\begin{equation}
    g_{S_1 ff}=\cos\theta g_{Hff},\,\,\,\,g_{S_2 ff}=-\sin\theta g_{Hff},
\end{equation}
while $g_{h,H,A,ff}$ are the reduced parameters already introduced before in Table (the alignment limit implies $g_{hff}=1$).

Also, the EWPT observables are affected by the extended Higgs sector. The custodial symmetry violation parameter $\Delta \rho$ reads:
\begin{align}
    &\Delta \rho=\frac{\alpha(M_Z^2)}{16\pi^2 M_W^2 (1-M_W^2/M_Z^2)}\left \{ A(M_{H^\pm}^2,M_A^2) \right.\nonumber\\
    & \left. +\cos^2 \theta A(M_{H^\pm}^2,M_{S_2}^2)\right.\nonumber\\
    &\left. +\sin^2 \theta A(M_{H^\pm}^2,M_{S_1}^2)-\cos^2 \theta A(M_{A}^2,M_{S_2}^2) \right.\nonumber\\
    & \left. -\sin^2 \theta A(M_{A}^2,M_{S_1}^2)\right \},
\end{align}
where $A(x,y)$ is given by Eq. (\ref{eq:Afnrho}).
$\Delta \rho$ can be straightforwardly set to zero by taking $M_A=M_{H^\pm}$.

From the DM perspective, one of the main reasons to include a $SU(2)$ singlet in the Higgs sector is to maintain a minimal exotic fermion sector, just composed by the SM gauge singlet DM candidate $\chi$: 
\begin{equation}
\mathcal{L}_{\rm DM}=y_\chi S \ovl \chi \chi=y_\chi \left(\sin\theta S_1+\cos\theta S_2\right) \ovl \chi \chi \, . 
\end{equation}
In analogy with \cite{Bell:2017rgi}, we interpret the singlet $S$ as the real component of the Higgs boson of a spontaneously broken $U(1)$-symmetry and then set: $y_\chi=m_\chi/v_S$, i.e. we interpret the DM as a chiral fermion charged under the new $U(1)$ with mass dynamically generated by the spontaneous breaking.

Let us now compare the requirement of the correct relic density with limits from DD. For what concerns the latter, SI interactions should be accounted for, as described by the following cross-section on nucleons:
\begin{align}
    & \sigma_{\chi p}^{\rm SI}=\frac{\mu_{\chi p}^2}{\pi}\frac{m_N^2}{v_h^2} y_\chi^2 \sin^2 \theta \cos^2 \theta {\left(\frac{1}{M_{S_1}^2}-\frac{1}{M_{S_2}^2}\right)}^2\nonumber\\
    & \times \left \vert  g_{Huu}f_u^p+\sum_{q=d,s}g_{Hqq}f_q^p+\frac{2}{9}f_{TG}\frac{2 g_{Huu}+g_{Hdd}}{3}\right \vert^2.
\end{align}

\begin{figure*}
    \centering
    \subfloat{\includegraphics[width=0.35\linewidth]{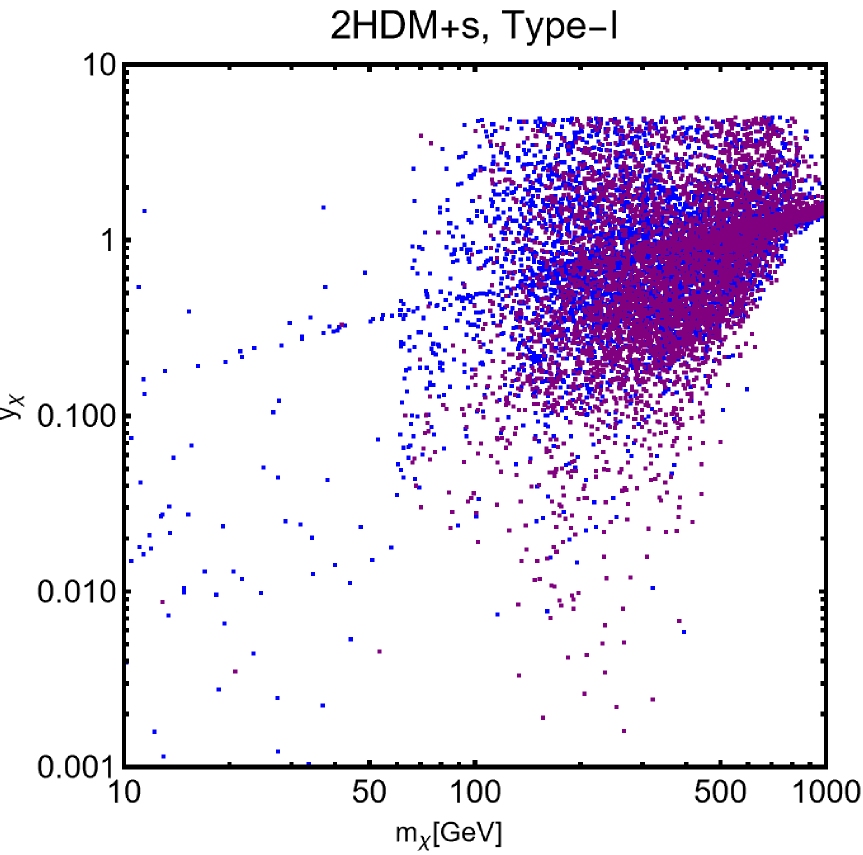}}
\subfloat{\includegraphics[width=0.35\linewidth]{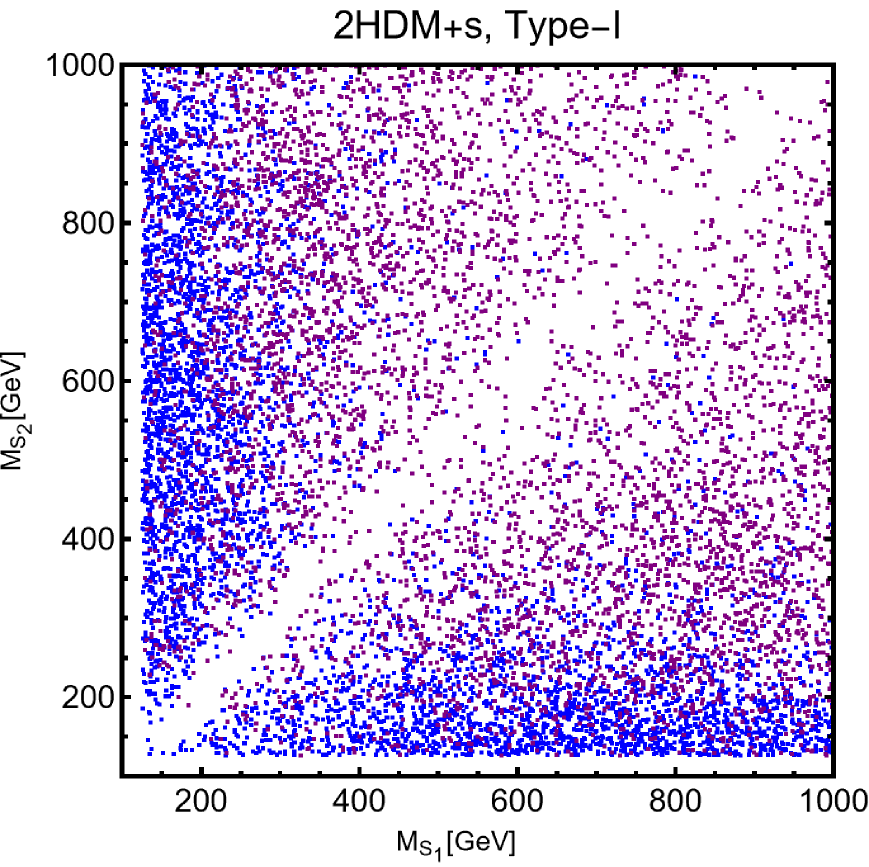}}
\subfloat{\includegraphics[width=0.35\linewidth]{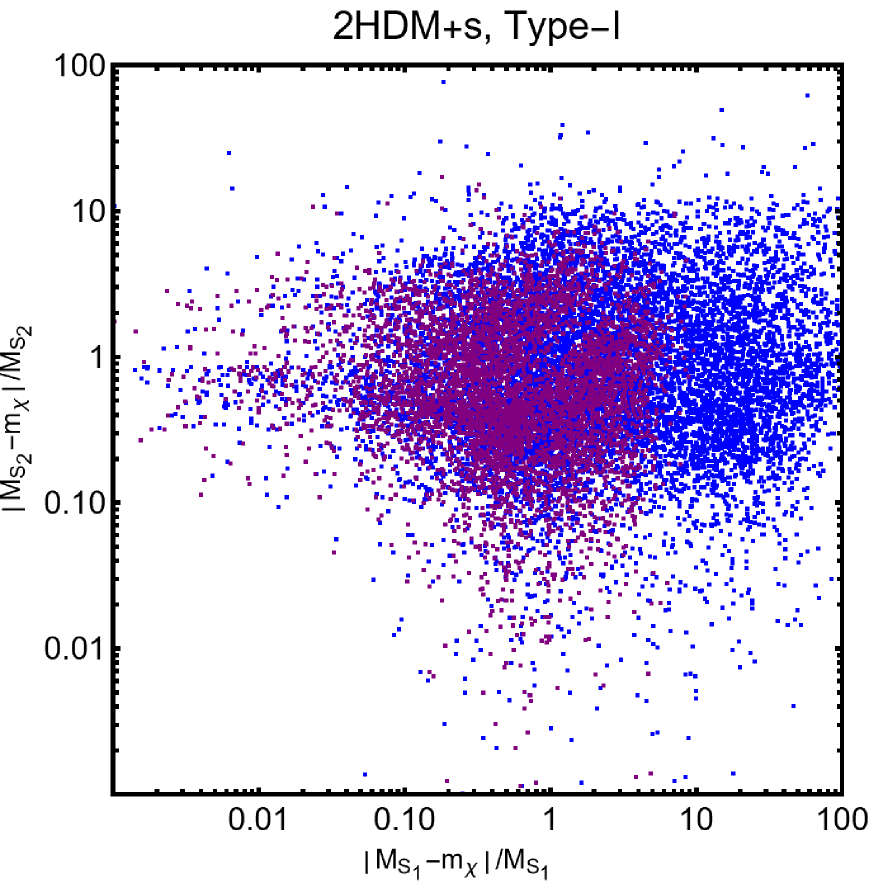}}\\    \subfloat{\includegraphics[width=0.35\linewidth]{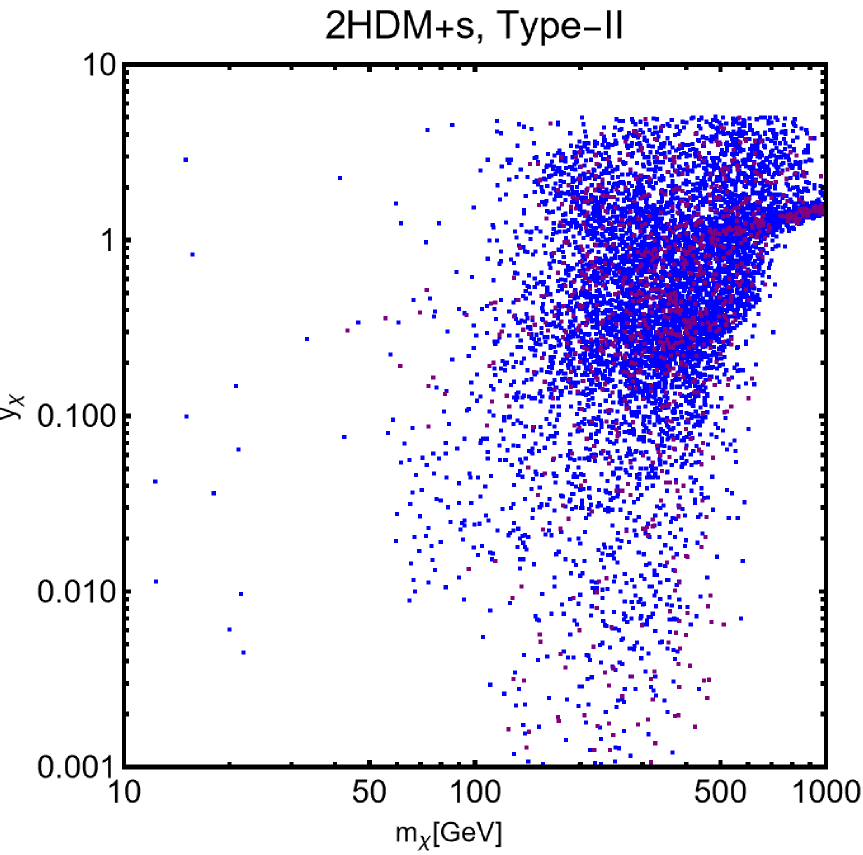}}
\subfloat{\includegraphics[width=0.35\linewidth]{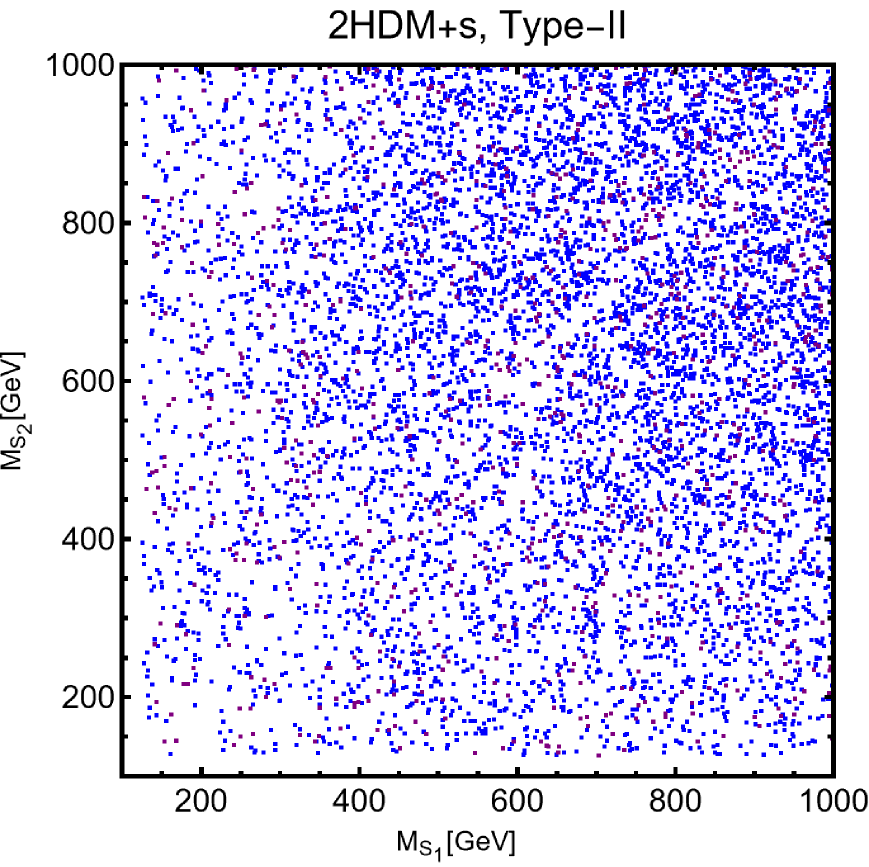}}
\subfloat{\includegraphics[width=0.35\linewidth]{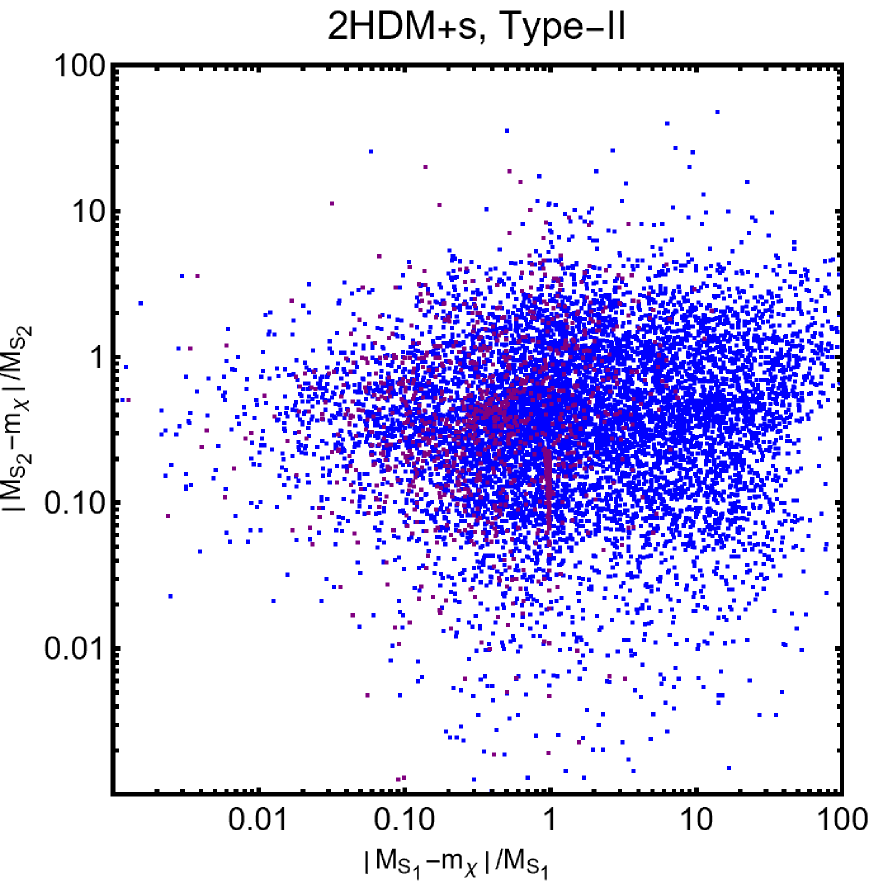}}
\caption{Parameter scan of the 2HDM+Singlet model (dubbed 2HDM+s) coupled to a $SU(2)$ singlet fermionic  DM (see main text for details). The top row refers to the Type-I 2HDM configuration of Yukawa couplings while the bottom row depicts the same for Type-II. The colour convention is the same as Fig. \ref{fig:scanSU3}.}
\label{fig:2HDMs_scan}
\end{figure*}

Contrary to other models discussed here, it is more complicated to identify a single pair of parameters for a bidimensional plot. For this reason, we will directly rely on a parameter scan to present the results relative to the model under consideration as depicted in Fig.~\ref{fig:2HDMs_scan}. The ranges of the scan are very similar to the ones considered in Ref. \cite{Bell:2017rgi}: 
\begin{align}
    & m_\chi \in [1,1000]\mbox{GeV},\,\,\,\,M_{S_{1,2}}\in [10,1500]\,\mbox{GeV},\nonumber\\
    & M_{A}=M_{H^{\pm}} \in [M_h, 1500]\,\mbox{GeV},\nonumber\\
    & \tan\beta \in [1,60] \nonumber\\
    & \sin \theta \in \left[-\frac{\pi}{4},\frac{\pi}{4}\right], \,\,\,\, |\lambda_{hHS}|<2\,\,\,\,|\lambda_{HHS}|<4.
\end{align}

The scan have been repeated twice, considering the Type-I and Type-II configurations. 
Adopting the conventional color code the panels of Fig.\ref{fig:2HDMs_scan}, show, in blue, the parameter assignation corresponding to the correct relic density and DM scattering cross-section below the current constraints while, the points with DD cross-section below the projected sensitivity by the DARWIN experiment are shown in purple. For what other constraints are concerned, ID have a negligible impact as most of the relevant annihilation channels for the DM have p-wave suppressed cross-section.  EWPT have been automatically accounted for by setting $M_{H^{\pm}}=M_A$. In the case of the Type-II configuration, we have set $M_{H^{\pm}}>800\,\mbox{GeV}$, in agreement with the lower bound set by $b\rightarrow s$ transitions. For what LHC is concerned, the 2HDM+s deserves dedicated study beyond the purposes of this review (see e.g. \cite{Bell:2017rgi,Arcadi:2020gge}). For our study we have just accounted for possible exotic decay processes 125 Higgs, as for example $h\rightarrow S_i S_j$ and $h \rightarrow S_{i,j} \bar \chi \chi$ applying a conservative bound $Br(h\rightarrow BSM)<0.2$. 
The outcome of the parameter scan is shown in the $(m_\chi,y_\chi)$, $M_{S_1},M_{S_2}$ and $(|M_{S_1}-2 m_\chi |/M_{S_1},|M_{S_2}-2 m_\chi |/M_{S_2})$ bidimensional planes. By comparing the results for the Type-I and Type-II scenario we see that DARWIN can play a crucial role in probing the Type-II model.

\subsection{Singlet fermion+2HDM+a}
In this subsection, we consider the case in which the two Higgs doublet sector is extended by a CP-odd $SU(2)$ singlet $a^0$. In such a case the scalar potential reads \cite{Bauer:2017ota} (see also \cite{Ipek:2014gua}):
\begin{align}
\label{eq:V2HDMa}
    & V_{\rm 2HDMa}=V_{\rm 2HDM}+V_{a^0},~~\mbox{\rm where} \nonumber\\
  & V_{a_0}=\frac{1}{2} m_{a^0}^2 (a^0)^2+ \frac{\lambda_a}{4} (a^0)^4+\nonumber\\
  & \left(i \kappa a^0 \Phi^{\dagger}_1\Phi_2+\mbox{H.c.}\right)
+ \left(\lambda_{1P}(a^0)^2 \Phi_1^{\dagger}\Phi_1 \!+\! \lambda_{2P}(a^0)^2 \Phi_2^{\dagger}\Phi_2\right),
\end{align}
and $V_{\rm 2HDM}$ is already given in Eq.~\eqref{eq:2HDM_potential}.

Contrary to the case of the CP-even extension, we will not account here for the possibility of assigning a VEV to $a^0$ as it would result in a spontaneous breaking of the CP symmetry.\footnote{This option would have interesting consequences leading, in particular, to potential Gravitational Wave signatures, see e.g., Ref.~\cite{Huber:2022ndk}.} After the EWSB the physical mass spectrum emerging from Eq.~(\ref{eq:V2HDMa}) is made of two CP-even Higgses, $h$ (the SM-like Higgs boson) and $H$, a charged Higgs $H^{\pm}$, and two pseudoscalar states $a, A$, being a mixture of $SU(2)$ singlet and doublet components:
\begin{equation}
\left(
\begin{array}{c} A^0 \\ a^0 \end{array} \right)= 
\mathcal{R}_\theta \left(
\begin{array}{c} A \\ a \end{array} 
\right)
\end{equation}
where $\mathcal{R_\theta}$ is the mixing matrix between the flavour and the physical states with:
\begin{equation}
\tan2\theta=\frac{2 \kappa v_h}{M_{A}^2-M_{a}^2}\;.
\end{equation}
Note that throughout this work we will always take $M_a < M_A$. 
The Yukawa Lagrangian can be also straightforwardly determined in terms of the couplings of the 2HDM as:
\begin{align}
&\mathcal{L}_{\rm Yuk}=\sum_f \frac{m_f}{v_h}\left[ g_{hff} h \bar f f+g_{Hff}
H\bar f f\right.\nonumber\\
& \left. - i g_{Aff} \cos\theta A \bar f \gamma_5 f+i g_{Aff} \sin\theta a \bar f \gamma_5 f  \right]. \, 
\end{align}
Following the same order, in the list of constraints, of the other models considered in this section we provide the constraints on the scalar potential from the perturbative unitarity, which are satisfied by the following relations \cite{Abe:2019wjw,Arcadi:2022lpp}:
\begin{align}
\label{eq:unitarity}
    & |x_i| < 8\pi \, , \ |\lambda_{1,2P}|< 4\pi,\,\,\,\,|\lambda_3\pm \lambda_4|< 4 \pi
    \, , \nonumber\\ & 
    \left \vert \frac{1}{2}\left(\lambda_1+\lambda_2 \pm \sqrt{(\lambda_1-\lambda_2)^2+4\lambda_k^2}\right)\right \vert < 8 \pi \, , \  k=4,5 \, , \nonumber\\
    & |\lambda_3+ 2 \lambda_4 \pm 3 \lambda_5|< 8 \pi,\,\,\,\,|\lambda_3 \pm \lambda_5| < 8 \pi \, , 
\end{align}
where the $x_i$'s are the solutions of the equation
\begin{align}
& x^3-3 (\lambda_a+\lambda_1+\lambda_2)x^2 \nonumber\\
& + (9 \lambda_1 \lambda_a+9 \lambda_2 \lambda_a -4 \lambda_{1P}^2-4 \lambda_{2P}^2-4 \lambda_3^2-4 \lambda_3 \lambda_4-\lambda_4^2\nonumber\\
& +9 \lambda_1 \lambda_2)x \nonumber\\
& +12 \lambda_{2P}^2 \lambda_1+12 \lambda_{1P}^2\lambda_2 
 -16 \lambda_{1P}\lambda_{2P}\lambda_3-8 \lambda_{1P}\lambda_{2P}\lambda_4\nonumber\\
 & +(-27 \lambda_1 \lambda_2+12 \lambda_3^2+12 \lambda_3 \lambda_4+3 \lambda_4^2)\lambda_a=0.
\end{align}
Notice that the couplings $\lambda_{4,5}$ do not coincide with the expressions given for the ordinary 2HDM
(see Eq.~(\ref{eq:2HDM_lambda_M}) but are modified by the presence of the additional pseudoscalar as:
\begin{align} 
    & \lambda_4 v_h^2 = M^2+M_A^2 \cos^2 \theta+M_a^2 \sin^2 \theta-2 M_{H^{\pm}}^2 \, , \nonumber\\
    & \lambda_5 v_h^2 =M^2-M_A^2 \cos^2\theta-M_a^2 \sin^2 \theta \, .
    \label{eq:quartic-phys-A}
    \end{align}
    
Additionally, we have constraints on the parameters from the boundness from below of the scalar potential:
\begin{align}
& \lambda_{1} > 0, \quad \lambda_{2} > 0, \quad \lambda_{a} > 0,   \nonumber\\
  & \bar{\lambda}_{12} \equiv \sqrt{\lambda_{1} \lambda_{2}}+\lambda_{3}+\min (0, \lambda_{4}- |\lambda_{5}|) > 0,   \nonumber\\
  & \bar{\lambda}_{1P} \equiv \sqrt{\frac{\lambda_{1} \lambda_{a}}{2}} + \lambda_{1P} > 0,  \nonumber\\  
  & \bar{\lambda}_{2P} \equiv \sqrt{\frac{\lambda_{2} \lambda_{a}}{2}} + \lambda_{2P} > 0,   \nonumber\\
  & \sqrt{\frac{\lambda_{1} \lambda_{2} \lambda_{a}}{2}}   + \lambda_{1P} \sqrt{\lambda_{2}} + \lambda_{2P} \sqrt{\lambda_{1}}\nonumber\\
  & + [\lambda_{3} + \min (0, \lambda_{4} - |\lambda_{5}|)] \sqrt{\frac{\lambda_{a}}{2}}
  + \sqrt{2} \sqrt{\bar{\lambda}_{12} \bar{\lambda}_{1P} \bar{\lambda}_{2P}} > 0.
\end{align}

Besides, as already stated, the assumption of the CP conservation demands that $a^0$ should not acquire any VEV.
This implies:
\begin{equation}
    m_{a^0}^2+\left(\lambda_{1P}v_1^2+\lambda_{2P}v_2^2\right)>0\,.
\end{equation}
Moving to the EWPT, the $\Delta \rho$ parameter is written, for this model, as:
\begin{align}
& \Delta \rho = \frac{\alpha_{\rm em}}{16 \pi^2 M_W^2 (1 -M_W^2/M_Z^2)} \big[ 
A(M^2_{H\pm},M^2_H)\nonumber \\
& + \cos^2\theta A(M^2_{H\pm},M^2_A) + \sin^2\theta A(M^2_{H\pm},M^2_a) \nonumber \\
& - \cos^2\theta A(M^2_A,M^2_H) - \sin^2\theta A(M^2_a,M^2_H) \big], \, 
\end{align}
where $A(x,y)$ is given by Eq. (\ref{eq:Afnrho}).
Thus to have $\Delta \rho=0$, we need to impose mass degeneracy for the three BSM Higgses, i.e., $M_H=M_A=M_{H^{\pm}}$. 

Moving to the DM, the fermionic SM gauge singlet DM $\chi$ will couple with the pseudoscalar Higgses as:
\begin{equation}
\mathcal{L}_{\rm DM}=i y_\chi a^0 \ovl \chi \gamma_5 \chi=i y_\chi \left(\cos\theta a+\sin\theta A\right) \ovl \chi  \gamma_5 \chi \, . 
\end{equation}
The DM relic density is mostly accounted for by the combination of three annihilation channels: (1). {\it $\bar f f$} with an $s$-wave dominated cross-section:
\begin{align}
& \langle \sigma v \rangle_{ff}=\frac{1}{2\pi}\sum_f n_f^c \sqrt{1-\frac{m_f^2}{m_\chi^2}} y_\chi^2 \sin^2 \theta \cos^2 \theta m_\chi^2 \nonumber\\
& \times \left \vert \frac{1}{(4 m_\chi^2-M_a^2)}-\frac{1}{(4 m_\chi^2-M_A^2)}\right \vert^2,
\end{align}
(2). {$ha$} as well with an $s$-wave dominated cross-section:
\begin{align}
& \langle \sigma v
\rangle_{ha}=\frac{1}{16\pi}\sqrt{1-\frac{(M_h+M_a)^2}{4
m_\chi^2}}\sqrt{1-\frac{(M_h-M_a)^2}{4 m_\chi^2}} y_\chi^2 \nonumber\\
& \left \vert \frac{\lambda_{haa}\cos \theta}{(4
m_\chi^2-M_a^2)}+\frac{\lambda_{hAa}\sin \theta}{(4
m_\chi^2-M_A^2)}\right \vert^2,
\end{align} 
where $\lambda_{haa},\lambda_{hHa}$ are trilinear coupling between the corresponding Higgs bosons (analytical expressions can be found, for example, in Refs. \cite{Arcadi:2019lka,Abe:2018emu,Arcadi:2022lpp}), and, finally (3) $aa$ with a $p$-wave suppressed cross-section
\begin{align}
    & \langle \sigma v \rangle_{aa}=\frac{v^2}{12\pi}{\left(1-\frac{M_a^2}{m_{\chi}^2}\right)}^{5/2}
|y_\chi|^4  \frac{m_\chi^6}{(M_a^2-2 m_\chi^2)^4}
\end{align}

\begin{figure*}
    \centering
    \subfloat{\includegraphics[width=0.43\linewidth]{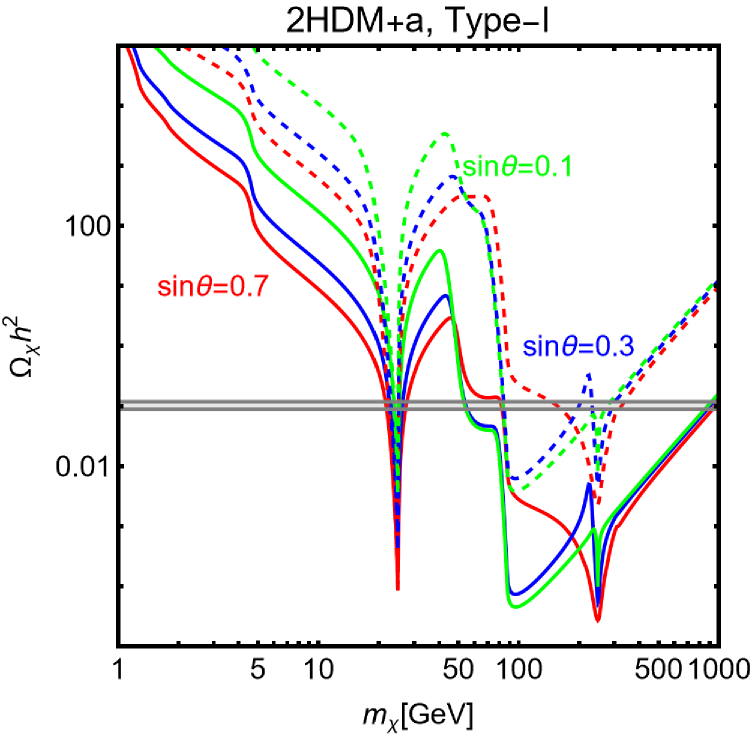}}
    \subfloat{\includegraphics[width=0.43\linewidth]{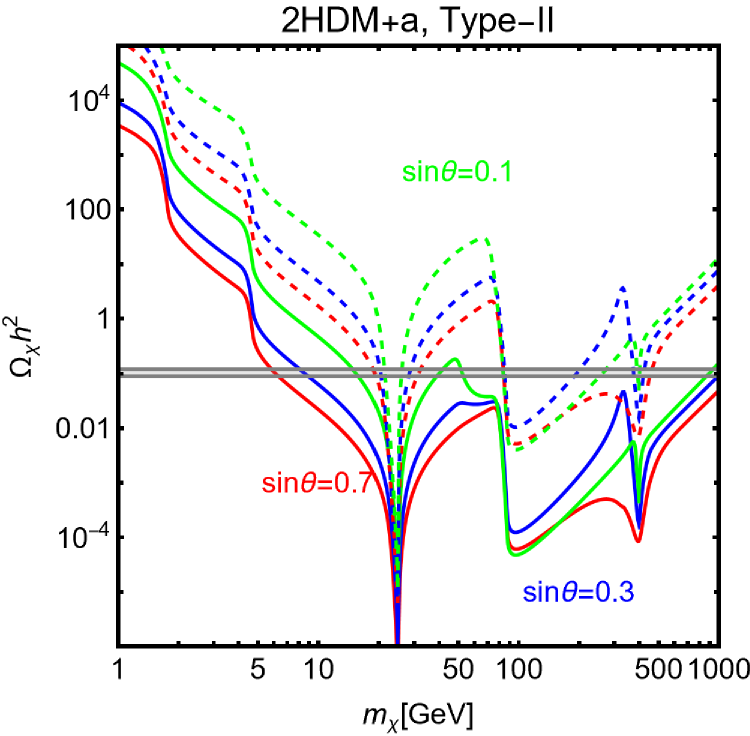}}
    \caption{Relic density $\Omega_\chi\, h^2$ as a function of the DM mass $m_\chi$ for the 2HDM+a model. The left panel refers to the Type-I configuration of the Yukawa couplings while the right one corresponds to Type-II. The different colours, namely, red, blue and green, correspond to different assignations of $\sin\theta$, $0.7,\,0.3,\,0.1$, respectively, as reported in the plots. Solid lines correspond to $y_\chi=1$ case while the dashed lines belong to $y_\chi=0.1$. In both cases, $m_a=50\,\mbox{GeV}$ while for what concerns $M_A=M_H=M_{H^{\pm}}$, a value of $500$ GeV has been considered for the Type-I scenario while a value of $800$ GeV has been considered for the Type-II case.}
    \label{fig:prelic2HDMa}
\end{figure*}

Additional annihilation channels into other combinations of the heavy BSM Higgs states might become relevant in the high DM mass regime. We do not show the corresponding cross-sections here for simplicity. An illustration of the behaviour of the DM relic density is provided in Fig.~\ref{fig:prelic2HDMa}.

The two panels of the Fig. \ref{fig:prelic2HDMa} show 
the behaviour of the DM relic density $\Omega_\chi h^2$ as a function of the DM mass $m_\chi$ for the Type-I (left) and Type-II (right) configurations. The solid (dashed) line corresponds
to $y_\chi=1 (0.1)$. The colours red, blue and green are used
to depict $\sin\theta=0.7,\, 0.3,\, 0.1$, respectively. A pair of grey lines presents the current limit on $\Omega_\chi h^2$. 
For what concerns the masses of the BSM Higgs bosons, the value $M_a=50\,\mbox{GeV}$ has been taken for the lightest pseudoscalar state while the other bosons have been assumed to be mass degenerate at $500$ GeV (left panel) and $800$ GeV (right panel). The value of $\tan\beta$ has been taken to be $5$ in both cases. By looking at the shape of the curves we can argue that for $m_\chi \lesssim 50\,\mbox{GeV}$,  the DM relic density is mostly accounted for, in both the Type-I and the Type-II scenarios, the DM annihilations into SM fermions with the sort of steps appearing at around $2$ and $5$ GeV, corresponding to the opening thresholds of the $\bar \tau \tau$ and $\bar b b$ final states respectively. For $m_\chi \sim M_a/2$, we have the first strong drop in the relic density due to the $s$-channel resonance of the DM annihilation cross-section. By comparing the two panels of the figure we see that the Type-II scenario corresponds to a lower relic density as a consequence of the $\tan\beta$ enhancement of the Yukawa couplings of the BSM Higgs bosons with leptons and $d$-type quarks. In the Type-I scenario, we notice another sharp drop of the relic density for $m_\chi \simeq 50\,\mbox{GeV}$, as the $\chi \chi \rightarrow aa$ process becomes kinematically accessible (this is less evident in the Type-II benchmark as annihilation into the SM fermion pairs is more efficient). Another sensitive decrease of the DM relic density, for both Type-I and Type-II, appears for $m_\chi \simeq 100\,\mbox{GeV}$. For our parameter assignation, this corresponds to the opening threshold of another very efficient annihilation process, i.e., into $ha$ final state. This annihilation channel, together possibly with the one into $hA$, mostly accounts for the relic density in the high DM mass regime, except the second "pole" of the annihilation cross-section into SM fermions, encountered for $m_\chi \sim M_A/2$.

\begin{figure*}
\vspace*{-5.7cm}
    \centerline{     \subfloat{\includegraphics[width=1.2\linewidth]{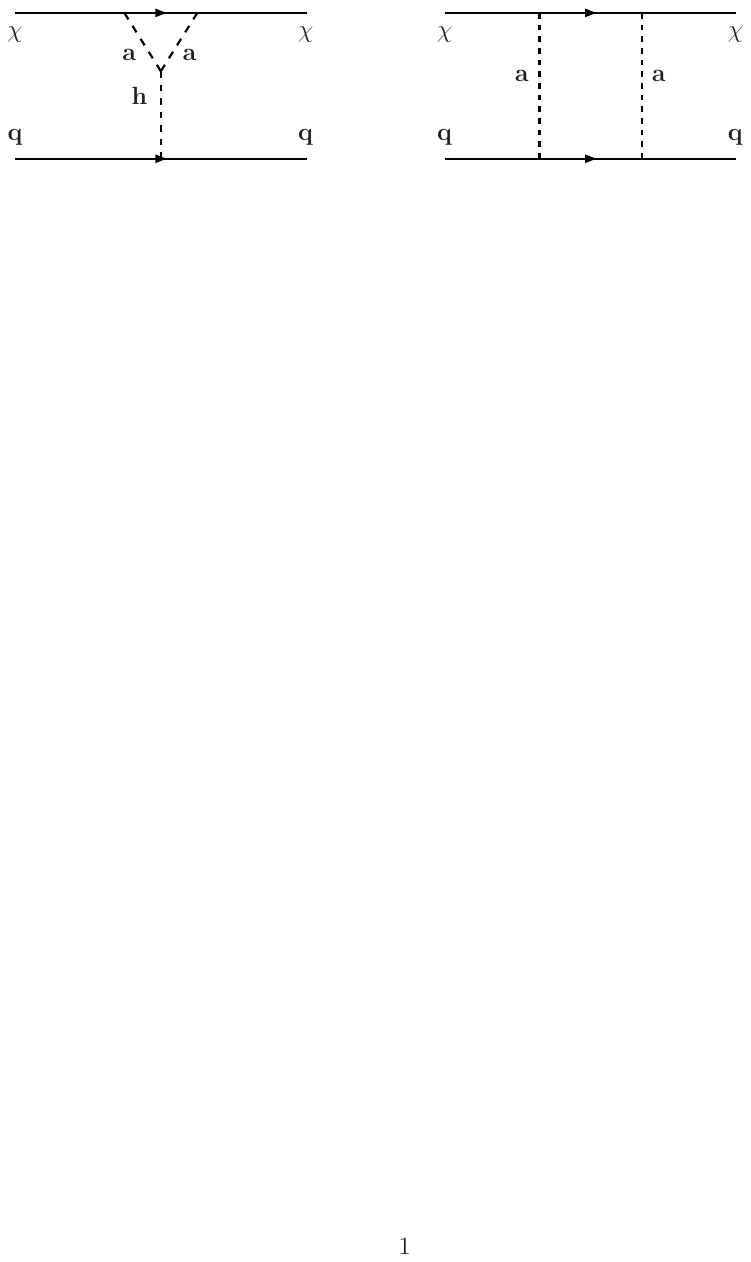}} }
\vspace*{-17.4cm}
    \caption{Generic Feynman diagrams for the loop induced scattering of the DM particle on quarks in the 2HD+a model. Analogous diagrams with $a$ replaced by $A$ and $h$ replaced by $H$ contribute to DD.}
    \label{fig:feynloop}
\vspace*{-2mm}
\end{figure*}

Moving to the DD, similar to its simplified counterpart, the most relevant contribution comes, for the 2HDM+a, at the one-loop level. An additional diagram topology, due to trilinear coupling between two CP-even and one CP-odd boson, is present though. The two Feymann's diagram topologies accounting for DM SI interactions at one loop are shown in Fig.\ref{fig:feynloop}. 

The Wilson coefficients contributing to the DM scattering cross-section are distributed in the following way \cite{Abe:2018emu}:
\begin{align}
    & C_q=C_q^{\rm tri}+C_q^{\rm box}, \nonumber\\
    & C_G=C_G^{\rm tri}+C_G^{\rm box}, \nonumber\\
    & C_q^{(1)}=C_q^{(1) \rm box},\nonumber\\
    & C_q^{(2)}=C_q^{(2) \rm box}.
\end{align}
with the label "tri" stemming from the triangle topology, proper of the 2HDM+a, while "box" refers to the box topology already present in the simplified model.
The contributions from the triangle loops are given by:
\begin{align}
    & C_q^{\rm tri}=-\sum_{\phi=h,H} \frac{\lambda_{\phi qq}}{m_\phi^2 v_h}C_{\phi \chi \chi}, ~~\mbox{where}\nonumber\\
    & C_{\phi \chi \chi}=-\frac{m_\chi y_\chi^2}{(4\pi)^2}\left \{ \lambda_{\phi aa} \cos^2 \theta \left[\frac{\partial}{\partial p^2}B_0 (p^2,M_a^2,m_\chi^2)\right]_{p^2=m_\chi^2}\right. \nonumber\\
    & \left. +\lambda_{\phi AA} \sin^2 \theta \left[\frac{\partial}{\partial p^2}B_0 (p^2,M_A^2,m_\chi^2)\right]_{p^2=m_\chi^2}\right.\nonumber\\
    & \left. + \frac{\lambda_{\phi aA} \sin 2 \theta}{M_A^2-M_a^2} \left[B_1 (m_\chi^2,M_A^2,m_\chi^2)-B_1 (m_\chi^2,M_a^2,m_\chi^2)\right]\right \},
\end{align}
with $\lambda_{\phi XY},\,\,X, Y=a,A$ are the trilinear couplings undefined between one CP-even and two CP-odd scalars.
and:
\begin{equation}
    C_G^{\rm tri}=\sum_{q=c,b,t}\frac{2}{27}C_q^{\rm tri},
\end{equation}
with $B_{0,1}$ being Passarino-Veltman functions \cite{tHooft:1978jhc,Passarino:1978jh}. The contributions from the box diagrams can be obtained just by adapting the results of the simplified model:
\begin{align}
    & C_q^{\rm box}=-\frac{m_\chi y_\chi^2}{(4\pi)^2}{\left(\frac{m_q}{v_h}\right)}^2\nonumber\\
    & \left \{\frac{\lambda_{Aqq}^2 \sin^2 \theta}{m_A^2}\left[G(m_\chi^2,0,M_A^2)-G(m_\chi^2,M_A^2,0)\right]\right. \nonumber\\
    & \left. +\frac{\lambda_{aqq}^2 \cos^2 \theta}{M_a^2}\left[G(m_\chi^2,0,M_a^2)-G(m_\chi^2,M_a^2,0)\right]\right. \nonumber\\
    &\left.+ \frac{\lambda_{aqq}\lambda_{Aqq} \sin \theta \cos \theta}{M_A^2-M_a^2}\left[G(m_\chi^2,0,M_A^2)-G(m_\chi^2,M_a^2,0)\right] \right \},
\end{align}
\begin{align}
    & C_q^{(1) \rm box}=-\frac{8 y_\chi^2}{(4\pi)^2}{\left(\frac{m_q}{v_h}\right)}^2\nonumber\\
    & \left \{\frac{\lambda_{Aqq}^2 \sin^2 \theta}{M_A^2}\left[X_{001}(p^2,m_\chi^2,0,M_A^2)-X_{001}(p^2,m_\chi^2,M_A^2,0)\right]\right. \nonumber\\
    & \left. +\frac{\lambda_{aqq}^2 \cos^2 \theta}{M_a^2}\left[X_{001}(p^2,m_\chi^2,0,M_a^2)-X_{001}(m_\chi^2,M_a^2,0)\right]\right. \nonumber\\
    &\left.+ \frac{\lambda_{aqq}\lambda_{Aqq} \sin \theta \cos \theta}{M_A^2-M_a^2}\left[X_{001}(p^2,m_\chi^2,0,M_A^2)-\right.\right. \nonumber\\
    &\left. \left. X_{001}(p^2,m_\chi^2,M_a^2,0)\right] \right \},
\end{align}
\begin{align}
    & C_q^{(2) \rm box}=-\frac{8 y_\chi^2}{(4\pi)^2}{\left(\frac{m_q}{v_h}\right)}^2\nonumber\\
    & \left \{\frac{\lambda_{Aqq}^2 \sin^2 \theta}{M_A^2}\left[X_{111}(p^2,m_\chi^2,0,M_A^2)-X_{111}(p^2,m_\chi^2,M_A^2,0)\right]\right. \nonumber\\
    & \left. +\frac{\lambda_{aqq}^2 \cos^2 \theta}{m_a^2}\left[X_{111}(p^2,m_\chi^2,0,M_a^2)-X_{111}(p^2,m_\chi^2,M_a^2,0)\right]\right. \nonumber\\
    &\left.+ \frac{\lambda_{aff}\lambda_{Aqq} \sin \theta \cos \theta}{M_A^2-M_a^2}\left[X_{111}(p^2,m_\chi^2,0,m_A^2)-\right.\right. \nonumber\\
    &\left. \left. X_{111}(p^2,m_\chi^2,M_a^2,0)\right] \right \},
\end{align}
\begin{align}
    & C_G^{\rm box}=\sum_{q=c,b,t}\frac{-m_\chi y_\chi^2}{432\pi^2}{\left(\frac{m_q}{v_h}\right)}^2\left[\lambda_{aqq}^2 \cos^2 \theta \frac{\partial F(M_a^2)}{\partial M_a^2}\right.\nonumber\\
    &\left. + \lambda_{Aqq}^2 \sin^2 \theta \frac{\partial F(M_A^2)}{\partial M_A^2} +\lambda_{Aqq}\lambda_{aqq} \sin 2\theta \frac{\left[F(M_A^2)-F(M_a^2)\right]}{M_A^2-M_a^2}\right],
\end{align}
The $G,\, X_{001},\,X_{111}$ and $F$ functions are defined by
Eqs. (\ref{eq:Gfn}), (\ref{eq:X001fn}), (\ref{eq:X111fn}) and (\ref{eq:Ffn}), respectively.

\begin{figure*}
\begin{center}
\subfloat{\includegraphics[width=0.33\linewidth]{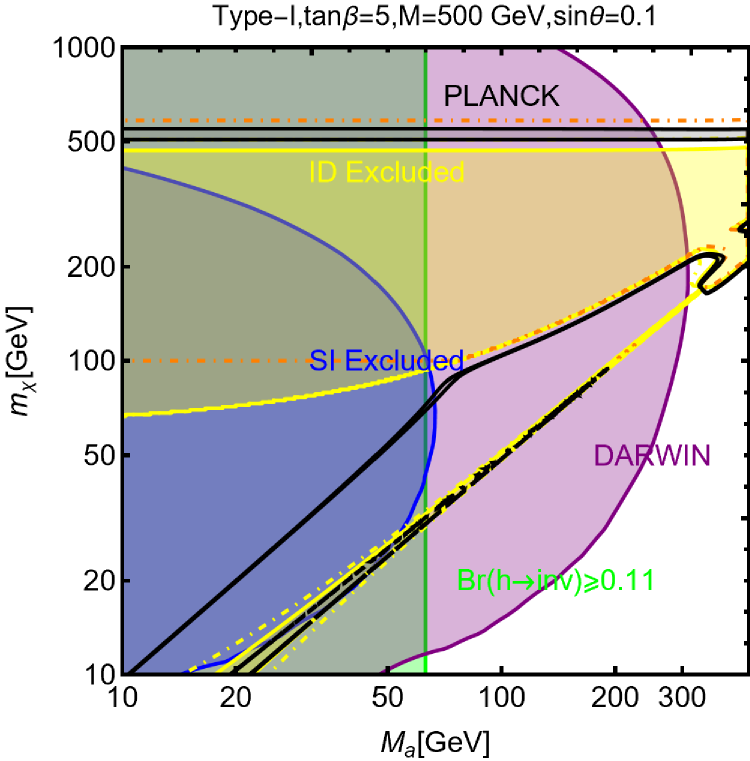}}
\subfloat{\includegraphics[width=0.33\linewidth]{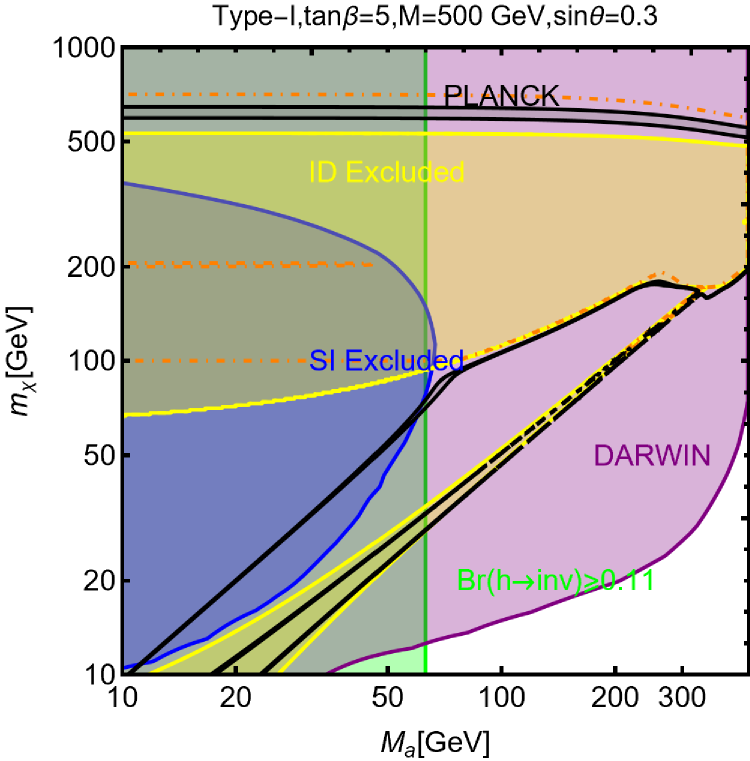}}
\subfloat{\includegraphics[width=0.33\linewidth]{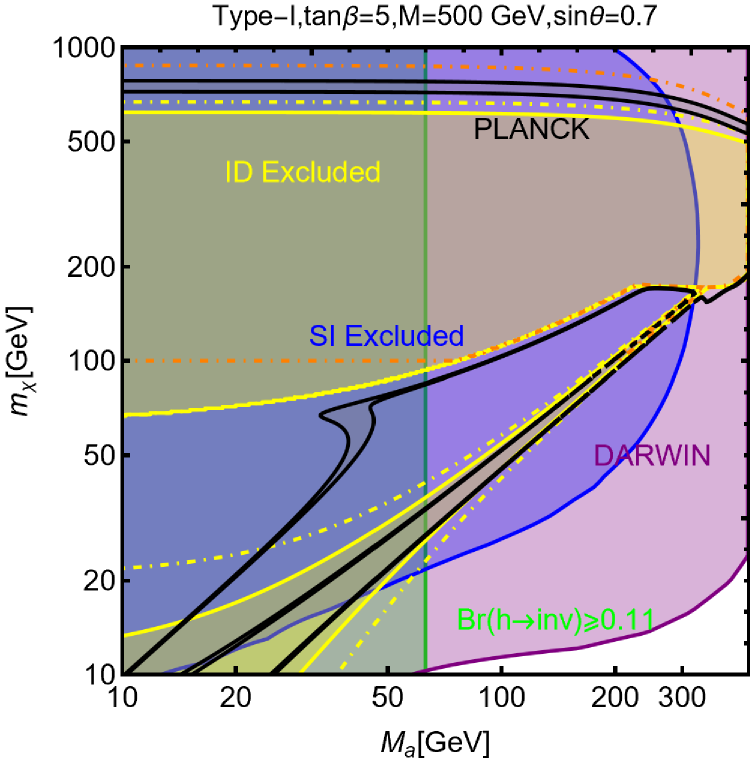}}\\
\subfloat{\includegraphics[width=0.33\linewidth]{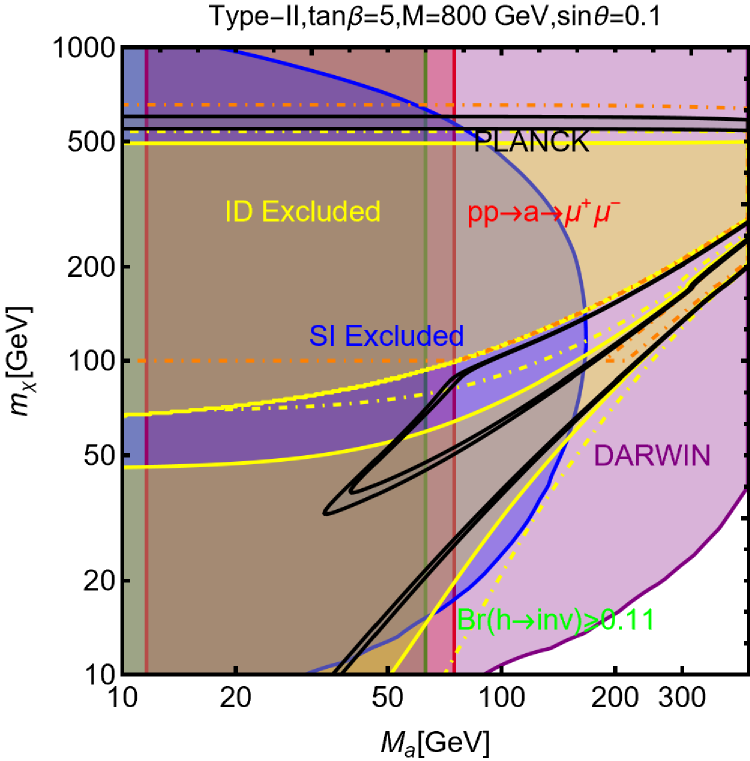}}
\subfloat{\includegraphics[width=0.33\linewidth]{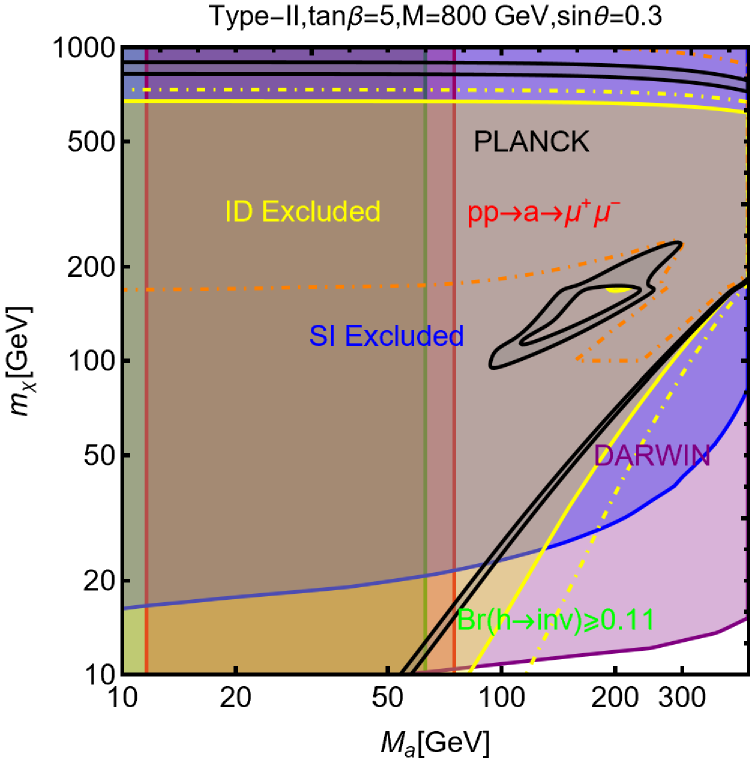}}
\subfloat{\includegraphics[width=0.33\linewidth]{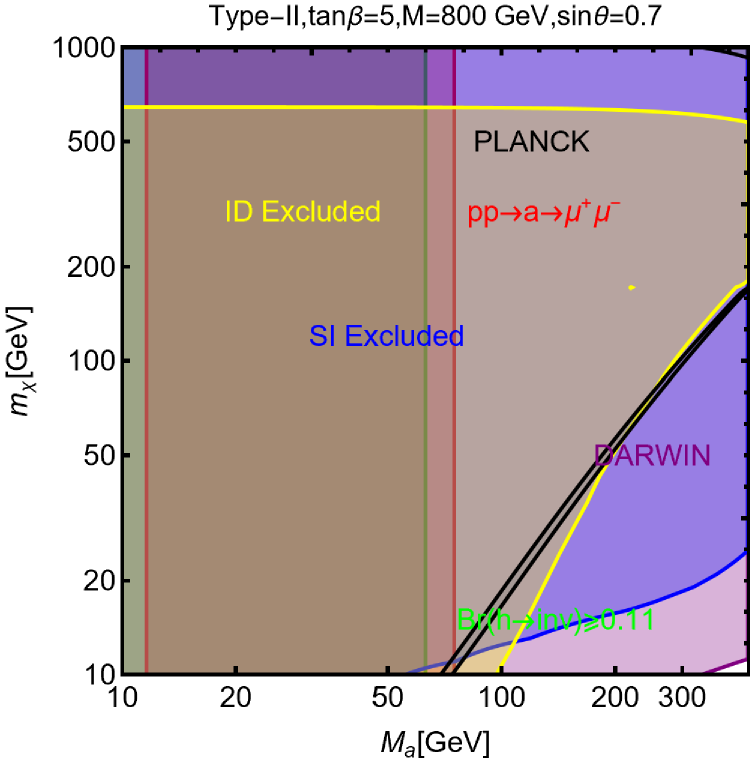}}
\end{center}
\caption{Summary of constraints in the $(m_\chi,m_a)$  plane for the 2HDM+a coupled to an SM gauge singlet fermionic DM. 
The top (bottom) row corresponds 
to the Type-I (Type-II) configuration of Yukawa couplings with $M=M_A=M_H=m_{H^\pm}=600~(800)$ GeV. $\tan\beta=5$ is fixed for all the plots while 
$\sin\theta=0.1$ (left column), 
$0.3$ (middle column) and $0.7$ (right column). The green (red) coloured region is excluded from
the bound on the invisible decay of the SM-like Higgs ($pp \to a\to \mu^+\mu^-$ cross-section). The remaining colour coding is the same as of Fig. \ref{fig:SD2HDMtypII}.}
\label{fig:p2HDMa}
\end{figure*}

The combination of the aforementioned constraints, in the $(m_\chi,M_a)$ bidimensional plane, is shown in Fig.~\ref{fig:p2HDMa}. Here, the top (bottom) row corresponds to the Type-I (Type-II) 2HDM configuration with $M=M_A=M_H=M_{H^\pm}=500$ $(800)$ GeV. $\tan\beta$ is fixed at
$5$ while three different assignations of $\sin\theta$
value, namely, $0.1$ (first column), $0.3$ (middle column)
and $0.7$ (last column) are chosen. According to the usual colour convention, these plots compare the requirement of the correct relic density (black coloured isocontours) with the present limits (future prospects) of the DD experiments using blue (purple) coloured regions) as well as from the ID (yellow coloured regions). Limits from the DM phenomenology have been complemented with the bound on the invisible decay of the SM-like Higgs (green coloured region) and the LHC searches of light resonances decaying into muon pairs (red coloured regions).

\begin{figure*}
    \centering
    \subfloat{\includegraphics[width=0.43\linewidth]{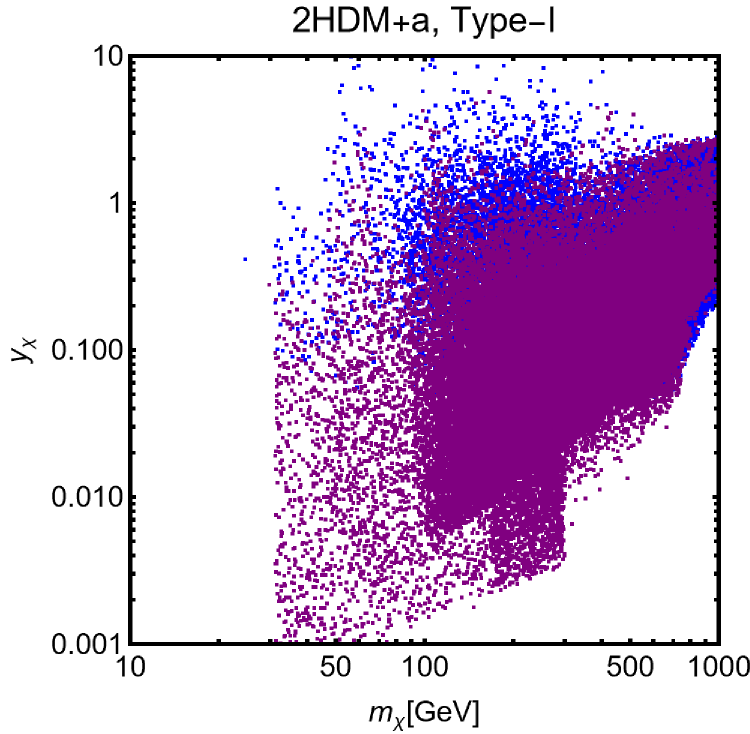}}
\subfloat{\includegraphics[width=0.41\linewidth]{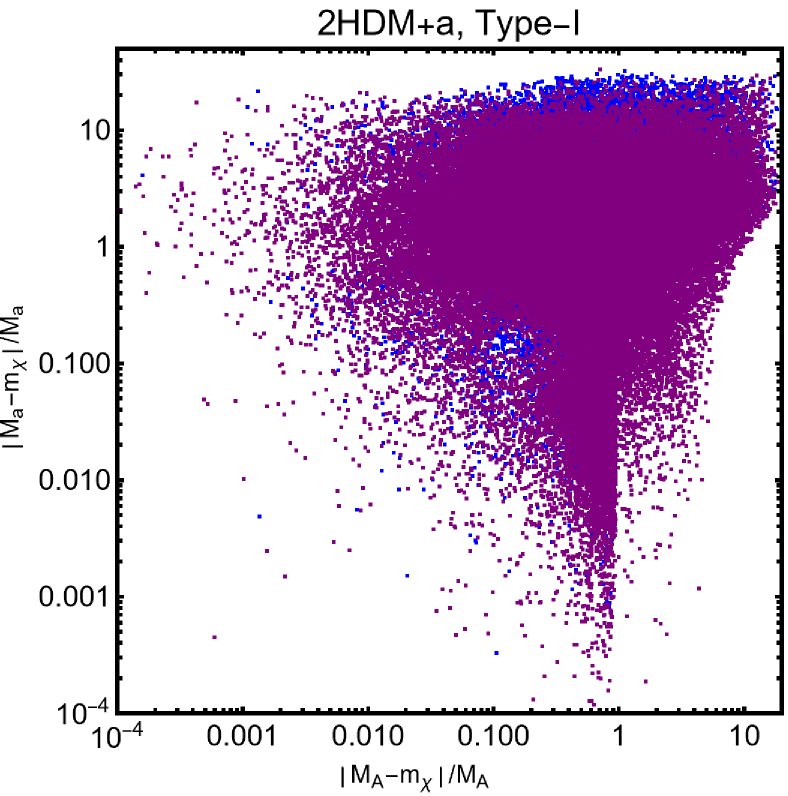}}\\    \subfloat{\includegraphics[width=0.43\linewidth]{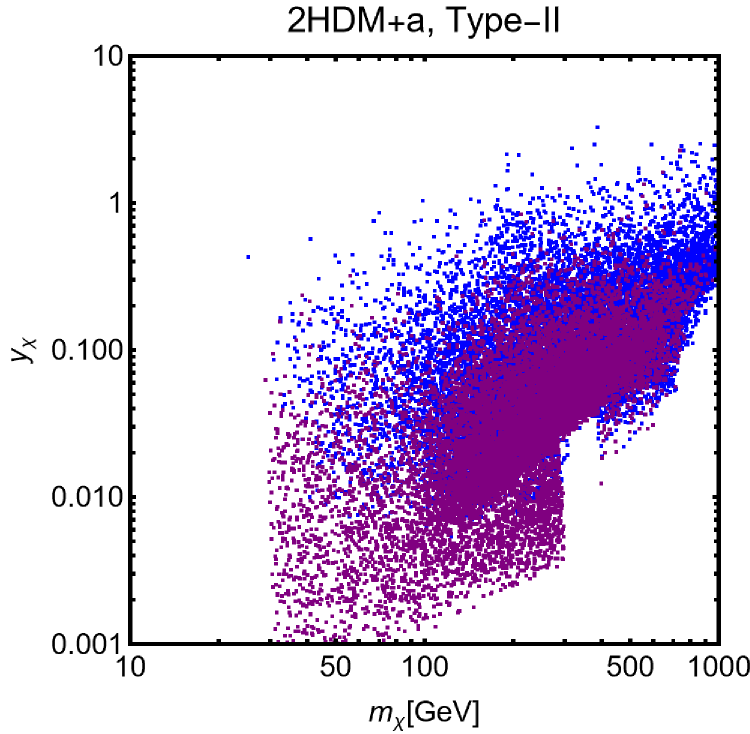}}
\subfloat{\includegraphics[width=0.41\linewidth]{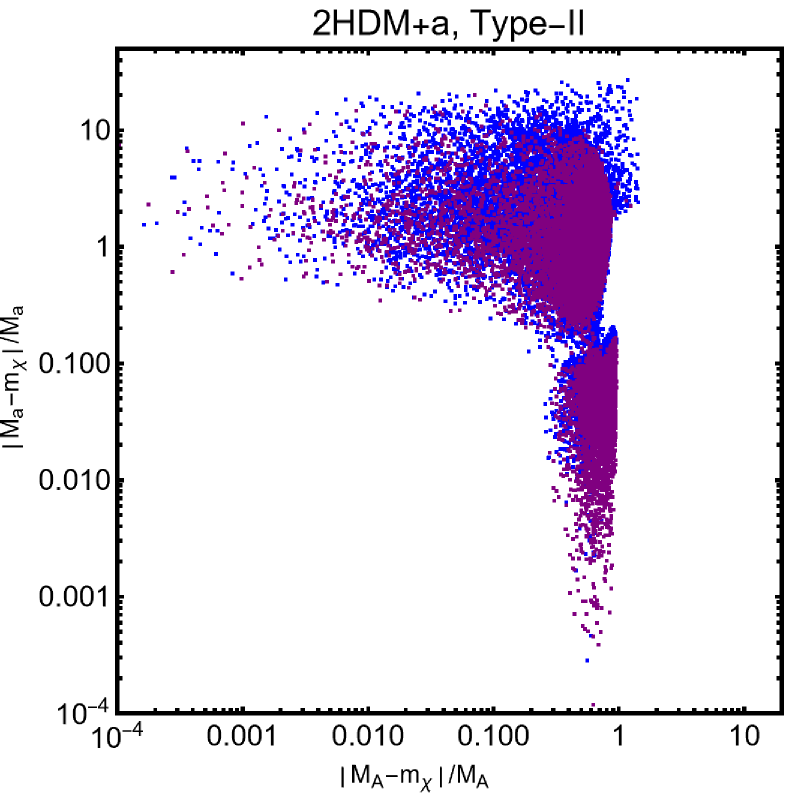}}
\caption{Parameter scan of the 2HDM+a coupled with an SM gauge singlet fermionic  DM (see main text for details). The colour convention is the same as Fig. \ref{fig:scanSU3}. The top (bottom) row refers to the Type-I (Type-II) configuration for Yukawa couplings.}
\label{fig:2HDMa_scan}
\end{figure*}

As the final step, we make a more systematic exploration of the parameter space of the model via a scan over the following parameters:
\begin{align}
    & m_\chi \in [1,1000]\,\mbox{GeV},\,\,\,\,\,M_a \in [10,600]\,\mbox{GeV},\nonumber\\
    & M_H=M_{H^{\pm}}=M_A \in [M_h,1500]\,\mbox{GeV},\,\,\,\,\,y_\chi \in \left[10^{-3},10\right],\nonumber\\
    & \sin\theta \in \left[-\frac{\pi}{4},\frac{\pi}{4}\right],\,\,\,\,\,\tan\beta \in [1,60].
\end{align}
The scan has been repeated for both Type-I and Type-II configurations of the Yukawa couplings. For the latter case, the minimal value for $M_{H,\, A, \, H^\pm}$ has been taken to be $800$ GeV to automatically comply with the constraints from B-physics.The scan results are shown in Fig. \ref{fig:2HDMa_scan}. As usual, only the model assignations (marked in blue colour) passing all the present constraints illustrated have been retained. By present constraints we intend: correct relic density, scattering cross-section compatible with DD, annihilation cross-section at present times below the exclusion limit by FERMI, constraints on the scalar potential as illustrated at the beginning of the subsection. EWPT are accounted for by assuming degenerate masses for the $H,A,H^\pm$ states while constraints on the Higgs signal strength are overcome via the alignment limit. In the case of the Type-II configuration for the Yukawa couplings between the BSM Higgs bosons and the SM fermions we have applied the lower bound $M_{H^{\pm}}\gtrsim 800\,\mbox{GeV}$. Regarding LHC, we have applied the bounds on the extra decays of the 125 Higgs and searches of light di-muon resonances, already evidenced in Fig.\ref{fig:p2HDMa}. Furthermore we have applied bounds from several LHC searches of heavy resonances, as listed in \cite{Arcadi:2022lpp} to which we refer for more details. General studies of the 2HDM+a, with focus on LHC signals, have been shown also in \cite{Robens:2021lov,Argyropoulos:2022ezr}.
Fig. \ref{fig:2HDMa_scan} shows also, as purple points, the model assignations compatible with the aforementioned constraints as well as with negative signals by DARWIN.

Before moving to next section we just mention that one could also consider extensions of two-Higgs doublet sector with complex scalar $SU(2)$ singlets. The interested read may refer, for example, to \cite{Muhlleitner:2016mzt,Biekotter:2021ovi,Biekotter:2022bxp}

\section{Models for spin-1 mediators}

In this section, we will discuss realistic completions of the simplified models where the DM coupled to an $s$-channel spin-$1$ mediator. These will allow to account for the direct coupling of the $Z'$ with the SM fermions, consistent with the gauge invariance and the perturbative unitarity. We illustrate below the general features common to the different DM realizations. For more details the interested reader might refer to \cite{Kahlhoefer:2015bea,Duerr:2016tmh}.
The $Z'$, again, is interpreted as the gauge boson of a new spontaneously broken $U(1)$ symmetry by the VEV of a scalar field $S$ which is explicitly introduced in the low energy Lagrangian which reads:
\begin{align}\label{eq:Lag-gauge-Higgs}
    & \mathcal{L}_{\rm gauge-higgs}={\left(D^\mu S\right)}^{\dagger}D_\mu S+\mu_S^2 S^\dagger S-\lambda_S {\left(S^\dagger S\right)}^2-\lambda_{HS}H^\dagger H S^\dagger S\nonumber\\
    & -g_X X_\mu \bar f \gamma^\mu \left(V_f - \gamma_5 A_f\right)f-\frac{1}{4}X^{\mu \nu}X_{\mu \nu}-\frac{1}{2}\sin \delta X^{\mu \nu}B_{\mu \nu}.
\end{align}
We denote by $X$ the gauge boson associated with the new gauge group and $Z^\prime$ represents the corresponding mass eigenstate after diagonalization of the mass matrix. To maintain full generality we have introduced the gauge and Lorentz invariant coupling between Higgs bilinears, $H^\dagger H |S|^2$, as well as the kinetic mixing operator \cite{Babu:1997st,Chun:2010ve}. The spontaneous breaking of the extra $U(1)$ symmetry, with associated gauge coupling $g_X$, dynamically generates a mass for the $X$ field:
\begin{equation}
\label{eq:mx}
    m_{X}=2 g_X v_S.
\end{equation}
In the case of a sizable value of the kinetic mixing parameter the mass of the new boson receives an additional contribution from the mixing with the Z-boson. Furthermore, as discussed in Ref.~\cite{Kahlhoefer:2015bea}, anomaly cancellation requires, in general, the SM $SU(2)_L$ doublets to be charged under the new symmetry. This induces a direct mass mixing $\delta m^2 X^\mu Z_\mu$:
\begin{equation}
    \delta m^2=\frac{1}{2}\frac{e^2 g_X q_H}{s_W c_W}v_h^2,
\end{equation}
with $q_H$ being the charge of the SM Higgs under the new symmetry. After the EWSB, 
it is possible to define three mass eigenstates, i.e., $A_\mu$, $Z_\mu$ and $Z'_\mu$, for the 
electrically neutral gauge bosons through the following two 
transformations~\cite{Babu:1997st,Chun:2010ve,Frandsen:2011cg,Mambrini:2011dw}: 
\begin{equation}
\label{eq:dtransformation}
\left(
\begin{array}{c}
B_\mu \\
W^3_\mu \\
X_\mu
\end{array}
\right)
=
\left(
\begin{array}{ccc}
1 & 0 & -t_\delta \\
0 & 1 & 0 \\
0 & 0 & 1/c_\delta
\end{array}
\right)
\left(
\begin{array}{ccc}
c_{\hat{W}} & -s_{\hat{W}} c_\xi & s_{\hat{W}} s_\xi \\
s_{\hat{W}} & c_{\hat{W}} c_\xi & -c_{\hat{W}} s_\xi \\
0 & s_\xi & c_\xi 
\end{array}
\right)
\left(
\begin{array}{c}
A_\mu \\
Z_\mu \\
Z'_\mu
\end{array}
\right),
\end{equation}
where $t_\delta,\,c_\delta = \tan\delta,\, \cos\delta$,
$c_{\hat W},\,s_{\hat W}=\cos\theta_{\hat W},\,\sin\theta_{\hat W}$, 
$c_{\xi},\,s_{\xi}=\cos\xi,\,\sin\xi$
and, the angle $\xi$ is defined by:
\begin{equation}
    \tan 2 \xi=\frac{-2 c_\delta (\delta m^2 +m_{Z_0}^2 s_W s_\delta)}{m_{X}^2-m_{Z_0}^2 c_\delta^2 +m_{Z_0}^2 s_W^2 s_\delta^2+2 \delta m^2 s_W s_\delta}.
\end{equation}
where $m_X$ is given by Eq. (\ref{eq:mx}) while $m_{Z_0}$ is the mass of the $Z$-boson as given in the SM in the absence of mixing with the $Z'$.  
The physical masses of the gauge bosons are given by:
\begin{align}
    & M_Z^2=m_{Z_0}^2 \left(1+\hat{s}_W \tan\xi \tan\delta\right),\nonumber\\
    & M_{Z'}^2=\frac{m_X^2+\delta m^2 \left(\hat{s}_W \sin\delta -\cos\delta \tan\xi\right)}{\cos^2 \delta \left(1+\hat{s}_W \tan\delta \tan \xi\right)},
\end{align}
Note that $\hat{s}_W,\hat{c}_W$ do not represent the experimentally measured Weinberg's angle. The latter, in the present setup, is defined through the relations:
\begin{align}
    & \hat{s}_W \hat{c}_W m_{Z_0}=s_W c_W M_Z,\nonumber\\
    & s_W^2 c_W^2 =\frac{\pi \alpha_{\rm em}(M_Z)}{\sqrt{2}G_F M_Z^2}.
\end{align}
Analogously, the invariance of the $W$-boson mass under the transformations 
of Eq.~(\ref{eq:dtransformation}) allows to relate 
the kinetic mixing parameter to the $\rho$ parameter as:
\begin{equation}
\rho=\frac{c_{\hat{W}}^2}{\left(1+s_{\hat{W}}\tan\delta \tan\xi\right)c_W^2},
\end{equation} 
which can be reformulated as:
\bea
\omega=s_W \tan\delta \tan\xi \simeq -\left(1-t_W^2\right) \Delta, 
\eea
where $\Delta=\rho-1~~{\rm and~~} t^2_W=\tan^2\theta_W$. In the physical basis, the couplings of the $Z/Z'$ with the SM states are given by:
\bea
 \mathcal{L}_{Z/Z',SM}&&=\ovl f \gamma^\mu \left(g_{f_L}^{Z}P_L+g_{f_R}^{Z}P_R\right) f Z_\mu \nonumber\\
 &&  +\ovl f \gamma^\mu \left(g_{f_L}^{Z'}P_L+g_{f_R}^{Z'}P_R\right) f Z'_\mu +g_W^Z [[W^+ W^-Z]] \nonumber\\
&& +g_W^{Z'} [[W^+ W^-Z']] +g_{hZZ} Z^\mu Z_\mu h \nonumber\\
&& + g_{hZZ'}  Z'_\mu Z^\mu h +g_{hZ'Z'}  Z'_\mu Z'^{\mu} h,
\eea
where:
\bea
\label{eq:Zcoup}
g_{f_L}^Z&&=-\frac{g}{c_W}\cos\xi \left \{T_3 \left(1+\frac{\omega}{2}\right)\right. \nonumber\\
&& \left. -Q \left[s_W^2
+\omega \left(\frac{2-t_W^2}{2(1-t_W^2)}\right)\right] \right \}+\frac{\sin\xi}{\cos \delta}g_X q_{f_L}, \nonumber\\
g_{f_R}^Z&&=\frac{g}{c_W}\cos\xi \left \{Q \left[s_W^2+\omega \left(\frac{2-t_W^2}{2(1-t_W^2)}\right)\right] \right \}\nonumber\\
&& +\frac{\sin\xi}{\cos \delta}g_X q_{f_R},
\eea
and,

\bea
\label{eq:Zprimecoup}
 g_{f_L}^{Z'}&&=-\frac{g}{c_W}\cos\xi \left (T_3 \left[s_W \tan\delta-\tan\xi\right.\right. \nonumber\\
&& \hspace*{1cm}\left.\left.
 +0.5 {\omega}\left(\tan\xi+{s_W t_W^2 \tan\delta (1-t_W^2)^{-1}}\right)\right]\right.\nonumber\\
&&\left.+ Q \left[s_W^2 \tan\xi-s_W \tan\delta \right.\right. \nonumber\\
&&\hspace*{1cm}\left. \left. +0.5 \, t_W^2 \omega (1-t_W^2)^{-1}
\left({\tan\xi-s_W \tan\delta}\right)\right] \right), \nonumber\\
&& -\frac{\cos\xi}{\cos \delta}g_X q_{f_L}, \nonumber\\
 g_{f_R}^{Z'}&&=-\frac{g}{c_W}\cos\xi \left \{Q \left[s_W^2 \tan\xi-s_W \tan\delta \right.\right.\nonumber\\
 && \left.\left. +0.5\,t_W^2 \omega (1-t_W^2)^{-1}\left({\tan\xi-s_W \tan\delta}\right)\right] \right\} \nonumber\\
 && -\frac{\cos\xi}{\cos \delta}g_X q_{f_L}.~~~
\eea
Besides,
\bea
g_W^Z&&=g c_W \cos\xi \left(1-\frac{\omega}{2 (c_W^2-s_W^2)}\right),\nonumber\\
g_W^{Z'}&&=-g c_W \sin\xi \left(1-\frac{\omega}{2 (c_W^2-s_W^2)}\right),
\eea
and,
\bea
 g_{hZZ}&&=\frac{m_{Z_0}^2}{v_h}\cos\xi^2 (1+\omega),\nonumber\\
 g_{hZZ'}&&=2\frac{m_{Z_0}^2}{v_h}\cos\xi^2\left[2 s_W \tan\delta-\tan\xi \right.\nonumber\\
 && \,\,\,\,\left.+\omega  \left(\tan\xi
 +{s_W t_W^2 \tan\delta (1-t^2_W)^{-1}}\right)\right],\nonumber\\
 g_{hZ'Z'}&&=\frac{m_{Z_0}^2}{v_h}\cos\xi^2\left[\tan^2\xi+s_W^2 \tan\xi \right.\nonumber\\
 && \,\,\,\,\left. - \omega\left(2+\tan^2 \xi-{s_W^2 t_W^2 \tan^2 \delta (1-t^2_W)^{-1}}\right)\right].~~
\eea
%
Here $T_3,\, Q$  are the isospin quantum number and electric charge of the associated SM fermions. $q_{f_{L,R}}$ are the charges of the left-/right-handed SM fermions under the new $U(1)$ symmetry. Unless differently stated we will work in the approximation which the $\omega=0$ in the definition of the previous couplings. In the presence of a non-zero $\lambda_{HS}$, mass mixing arises between the SM and the dark Higgs. The mass eigenstates can be defined as usual as:
\begin{align}
    & H_1=h \cos \theta+ s \sin \theta, \nonumber\\
    & H_2=-h \sin \theta + s \cos \theta, 
\end{align}
with $H_1$ identified as the $125$ GeV SM-like Higgs boson. It is useful to re-express the $H$ and $S$ self couplings, $\lambda_H$ and $\lambda_S$, as well as the portal coupling $\lambda_{HS}$, in terms of the masses of the physical states as:
\begin{align}
    & \lambda_h=\frac{1}{4v_h^2}\left[M_{H_1}^2+M_{H_2}^2+(M_{H_1}^2-M_{H_2}^2)\cos 2 \theta\right],\nonumber\\
    & \lambda_S=\frac{g_X^2}{m_{X}^2}\left[M_{H_1}^2+M_{H_2}^2+\left(M_{H_2}^2-M_{H_1}^2\right)\cos 2 \theta\right],\nonumber\\
    & \lambda_{HS}=\frac{g_X}{m_{X}v_h}\left(M_{H_1}^2-M_{H_2}^2\right)\sin 2 \theta.
\end{align}
Notice that the couplings $g_{hZZ},g_{hZ'Z'},g_{hZZ'}$ defined above will be modified in presence of the $H/S$ mixing. We do not report explicitly the corresponding expressions being them particularly lengthy.
Now let us discuss the general constraints on this setup besides the DM phenomenology. 
 A mass mixing between the $Z$ and the $Z'$ bosons induces a deviation in the EWPT with respect to the SM prediction. In the limit of small $\delta,\xi $ $\ll 1$, we can write the BSM contribution to the $S,T$ parameters as \cite{Frandsen:2011cg}: 
\begin{align}
    & \alpha_{\rm em} \Delta S=4 c_W^2 s_W \xi \left(\delta -s_W \xi\right),\nonumber\\
    & \alpha_{\rm em} \Delta T=\xi^2 \left(\frac{M_{Z'}}{M_Z}-2\right)+2 s_W \xi \delta,
\end{align}
while the deviation of the custodial symmetry parameter can be written as:
\begin{equation}
    \Delta \rho=\frac{c_W^2}{c_W^2-s_W^2}\xi^2 \left(\frac{M_{Z'}}{M_Z}-1\right).
\end{equation}
Another probe for the mixing between the $Z$ and the $Z'$ boson is represented by the atomic parity violation (APV). At the effective operator level, the APV is described by the following Lagrangian:
\begin{align}
    & \mathcal{L}_{\rm eff}=\frac{g^{Z}_{Ae}}{M_Z^2}\bar e \gamma^\mu \gamma_5 e \left(g^{Z}_{Ve} \bar u \gamma_\mu u+ g^{Z}_{Vd} \bar d \gamma_\mu d \right)\nonumber\\
   &  +\frac{g^{Z'}_{Ae}}{M_{Z'}^2}\bar e \gamma^\mu \gamma_5 e \left(g^{Z'}_{Vu} \bar u \gamma_\mu u+ g^{Z'}_{Vd} \bar d \gamma_\mu d \right).
\end{align}
where we have introduced the vector and axial couplings for the $Z,Z'$ bosons:
\begin{align}
    & g^Z_{Ve}=\frac{1}{2}\left(g^Z_{eL}+g^Z_{eR}\right),\,\,\,\,g^Z_{Ae}=\frac{1}{2}\left(g^Z_{eL}-g^Z_{eR}\right)\nonumber\\
    & g^{Z'}_{Ve}=\frac{1}{2}\left(g^{Z'}_{eL}+g^{Z'}_{eR}\right),\,\,\,\,g^{Z'}_{Ae}=\frac{1}{2}\left(g^{Z'}_{eL}-g^{Z'}_{eR}\right)
\end{align}
These microscopic interactions lead to the following parity violating Hamiltonian density for the electron field in the vicinity of the nucleus:
\begin{equation}
    \mathcal{H}_{\rm eff}=e^\dagger \left(\vec{r}\right)\gamma_5 e\left(\vec{r}\right) \frac{G_F}{2\sqrt{2}}Q_W^{\rm eff}(Z,N)\delta\left(\vec{r}\right),
\end{equation}
with $Q_W^{\rm eff}$ stemming for an effective weak nuclear charge which depends on the parameters of the underlying particle theory as well as on the number of protons $(Z)$ and neutrons $(N)$ for a given nucleus.  This should be compared with the SM value:
\begin{equation}
    Q_{W,SM}^{\rm eff}=Z(1-4 s_W^2)-N.
\end{equation}
The strongest constraint on the APV comes, at the present day, from the measurement of the weak charge of the Cesium \cite{Bennett:1999pd,Bennett:1999pd}:
\begin{equation}
    Q_{W,exp}^{eff}=-73.16(15),
\end{equation}
which is compatible with the SM prediction \cite{Marciano:1982mm}:
\begin{equation}
    Q_{W,SM}^{\rm eff}=-73.16(5).
\end{equation}
New interactions giving leading to further contribution to the APV should hence comply with the limit:
\begin{equation}
    \Delta Q=|Q_{W,exp}-Q_{W,SM}|<0.6.
\end{equation}
The most important probe in the regime $M_{Z'}>M_{Z}$ is represented by the collider searches. Indeed, spin-$1$ bosons can be efficiently produced in the proton collisions and can be searched via the resonance signals. The most effective ones are dijet, see e.g., Refs.~\cite{ATLAS:2019fgd,ATLAS:2019itm,CMS:2019emo}, and dileptons \cite{ATLAS:2019erb}.  In the case when the $Z'$ can decay into the DM pairs, monojet searches \cite{ATLAS:2017bfj} are a useful complement as well. 
Note that just like the LHC resonance searches, the LEP searches for deviations from the SM prediction to the dilepton production cross-section \cite{ALEPH:2006bhb} can also probe the case when the $Z'$ couples to the SM leptons.

\subsection{Pure kinetic mixing model}
We start by considering the case in which no SM state is charged under the new $U(1)$ symmetry, hence $\delta m^2=0$. We will also consider the mixing between two CP-even neutral scalars to be negligible,i.e. $\lambda_{HS} \ll 1$. In this setup, the model practically reduces to a $Z/Z'$ portal. We will consider a (complex) scalar $\chi$ and a Dirac fermionic DM  $\psi$ (we assume $\psi$ to be  vector-like  to avoid gauge anomalies). The relevant pieces of Lagrangian are:  
\begin{align}
\label{eq:KDMlagrangians}
& \mathcal{L}_{\rm DM, scalar}=(D^\mu \chi)^*D_\mu \chi-m_\chi^2 \chi^{*}\chi, \nonumber\\
& \mathcal{L}_{\rm DM. fermion}=\ovl \psi \gamma^\mu D_\mu \psi-m_\psi \ovl \psi \psi,
\end{align}
with $D_\mu=\partial_\mu-i g_X X_\mu$, the covariant derivative of the new gauge symmetry (for simplicity we have encoded the DM charge in the definition of the gauge coupling $g_X$). After the $Z-Z'$ mixing induced by the kinetic mixing, the DM portal Lagrangian is given by:
\bea
\mathcal{L}_\chi~&&=g_X \left(\chi^{*}\partial_\mu \chi-\chi \partial_\mu \chi^{*}\right)
\left(g_{\rm \chi}^X Z^\mu+g_{\rm \chi}^{Z'} Z'^\mu\right),\nonumber\\
\mathcal{L}_\psi~&&=g_X \ovl \psi \gamma_\mu \psi \left(g_{\rm \psi}^X Z^\mu+g_{\rm \psi}^{Z'} Z'^\mu\right),\nonumber\\
\eea
with $g_{\rm \chi,\psi}^{Z'}=\frac{\cos\xi}{\cos\delta}$ and $g_{\rm \chi,\psi}^{Z}=-\frac{\sin\xi}{\cos\delta}$. 

The mixing between the $Z$ and the $Z'$ bosons opens up final states accessible for the DM annihilations, which now include, besides the SM fermions, $WW$, $Z(Z')H_{1,2}$, $ZZ$, $Z'Z$ and $Z'Z'$ states. The corresponding cross sections read:
\begin{align}
    & \langle \sigma v \rangle (\chi^* \chi \rightarrow \bar f f)=\frac{4 n_f^c}{6\pi} m_\chi^2 \sqrt{1-\frac{m_f^2}{m_\chi^2}}\nonumber\\
    & \left[\left(G_{fL}^2+G_{fR}^2\right)\left(1-\frac{m_f^2}{4 m_\chi^2}\right)+\frac{3}{2}G_{fL}G_{fR}\frac{m_f^2}{m_\chi^2}\right] v^2, ~~\rm with\\
    & G_{fL}=\frac{g_\chi^Z g_{fL}^Z}{4m_\chi^2-M_Z^2}+\frac{g_\chi^{Z'} g_{fL}^{Z'}}{4m_\chi^2-M_{Z'}^2},\nonumber\\
     & G_{fR}=\frac{g_\chi^Z g_{fR}^Z}{4m_\chi^2-M_Z^2}+\frac{g_\chi^{Z'} g_{fR}^{Z'}}{4m_\chi^2-M_{Z'}^2}.
\end{align}

\begin{align}
    & \langle \sigma v \rangle (\chi^* \chi \rightarrow W^+ W^-)=\frac{2 m_\chi^2}{3\pi}G_W^2 {\left(1-\frac{M_W^2}{m_\chi^2}\right)}^{3/2}\nonumber\\
    & \times \left[\frac{m_\chi^4}{M_W^4}+5\frac{m_\chi^2}{M_W^2}+\frac{3}{4}\right]v^2,~~\rm with\nonumber\\
    & G_W=\frac{g_\chi^Z g_{W}^Z}{4m_\chi^2-M_Z^2}+\frac{g_\chi^{Z'} g_{W}^{Z'}}{4m_\chi^2-M_{Z'}^2}.
\end{align}

\begin{align}
    & \langle \sigma v \rangle (\chi^* \chi \rightarrow V X)=\frac{1}{8\pi}G^2_{VX}\sqrt{1-\frac{\overline{m}_V}{m_\chi^2}+{\left(\frac{\Delta m_{VX}^2}{4 m_\chi^2}\right)}^2}\nonumber\\
    & \left\{1+\frac{1}{2}\frac{m_\chi^2}{m_X^2}\left[1-\frac{\Delta m_{VX}^2}{2 m_\chi^2}+{\left(\frac{\Delta m_{VX}^2}{4 m_\chi^2}\right)}^2\right]\right\}, \rm ~~with \nonumber\\
    & G_{VX}=\frac{g_\chi^{Z} g_{XZV}}{4m_\chi^2-M_Z^2}+\frac{g_\chi^{Z'} g_{XZ'V}}{4m_\chi^2-M_{Z^{'}}^2},\nonumber\\
    & \overline{M}_V^2=\left(M_h^2+M_V^2\right)/2,\,\,\,\,\,\,\Delta M_{VX}^2=M_X^2-M_V^2,
\end{align}
where $V=Z,\,Z'$ and $X=H_1,\,H_2$.
\begin{align}
    & \langle \sigma v \rangle (\ovl\psi \psi \rightarrow \bar f f)=\frac{n_f^c}{2\pi}m_\psi^2 \sqrt{1-\frac{m_f^2}{m_\psi^2}}\left[\left(G_{fL}^2+G_{fR}^2\right)\right.\nonumber\\
    &\left. \times\left(1-\frac{m_f^2}{4 m_\psi^2}\right)+\frac{3}{2}G_{fL}G_{fR}\frac{m_f^2}{m_\psi^2}\right].
\end{align}

\begin{align}
    & \langle \sigma v \rangle (\ovl\psi \psi \rightarrow W^+ W^-)=\frac{m_\psi^2}{\pi}G_W^2 {\left(1-\frac{M_W^2}{m_\psi^2}\right)}^{3/2}\nonumber\\
    & \times \left[\frac{m_\psi^4}{M_W^4}+5\frac{m_\psi^2}{M_W^2}+\frac{3}{4}\right]\nonumber\\
\end{align}

\begin{align}
    & \langle \sigma v \rangle (\ovl\psi \psi \rightarrow V X )=\frac{1}{8\pi}G^2_{VX}\sqrt{1-\frac{\overline{M}_V}{m_\psi^2}+{\left(\frac{\Delta M_{XV}^2}{4 m_\psi^2}\right)}^2}\nonumber\\
    & \left\{1+\frac{1}{2}\frac{m_\psi^2}{M_Z^2}\left[1-\frac{\Delta M_{hV}^2}{2 m_\psi^2}+{\left(\frac{\Delta M_{XV}^2}{4 m_\psi^2}\right)}^2\right]\right\}.
\end{align}

$G_{f_{L, R}}, G_W, G_{VX}$ are the same as the scalar DM and are obtained just with the replacement $\chi \rightarrow \psi$.
The annihilation cross-sections into $ZZ,\,ZZ',\,Z'Z'$ final states have rather lengthy expressions and hence will be omitted for simplicity.

\begin{figure*}[!t]
\begin{center}
\subfloat{\includegraphics[width=0.33\linewidth]{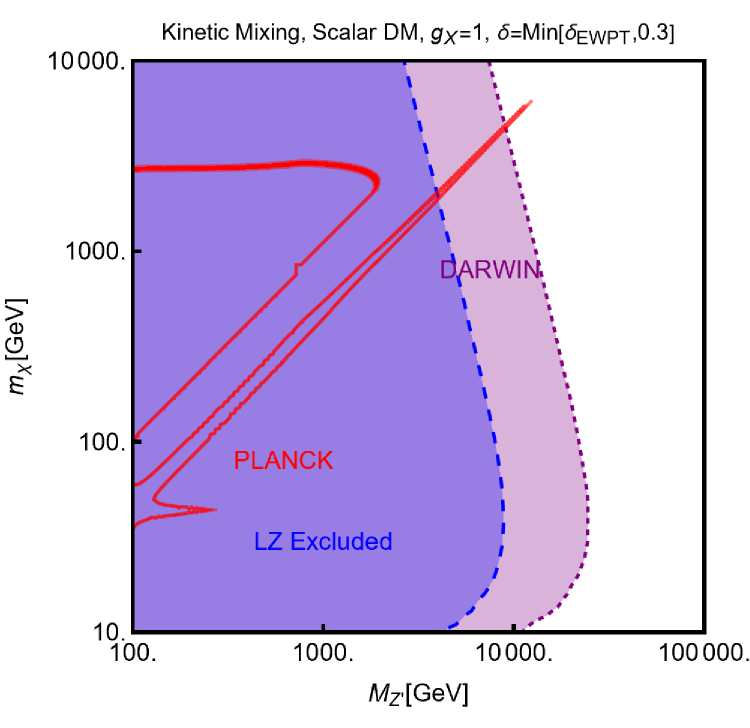}}
\subfloat{\includegraphics[width=0.33\linewidth]{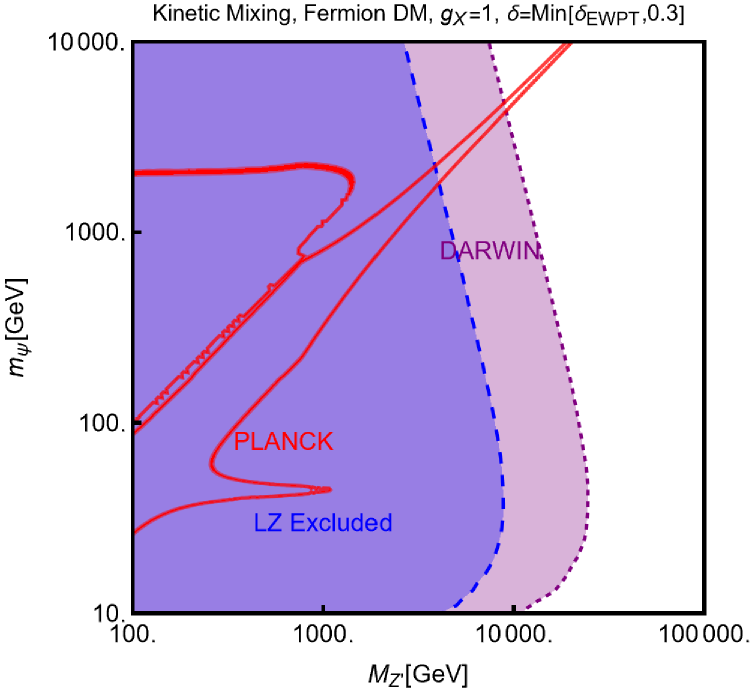}}\\
\subfloat{\includegraphics[width=0.33\linewidth]{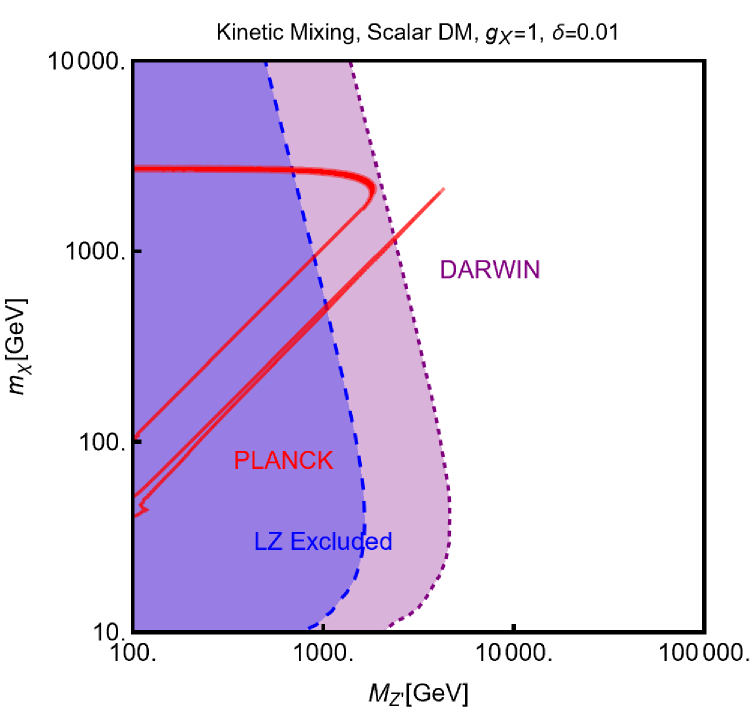}}
\subfloat{\includegraphics[width=0.33\linewidth]{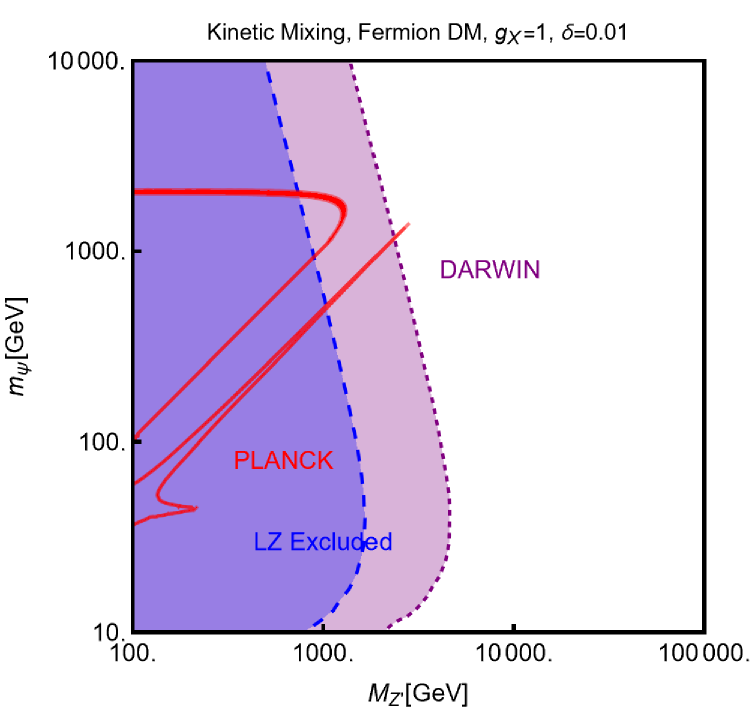}}
\end{center}
\caption{The combined DM constraints 
in the relevant bidimensional planes
$(m_{Z'},\,m_\chi)$ and $(m_{Z'},\,m_\psi)$ for the scalar (left column) and fermionic (right column) 
 DM, respectively,  interacting with a $Z'$, kinetically coupled to the SM 
$Z$ boson. In the top row, the kinetic mixing parameter $\delta$ has been set to the maximal value, 
as a function of $m_{Z'}$, consistent with the EWPT constraints while for the bottom row plots, $\delta$ 
has been set to a constant value of $0.01$. We set $g_X=1$ for all these plots. In these plots, the red coloured curve represents the isocontour of the correct DM relic density. The blue coloured region is excluded by 
the current constraints from LZ while the purple coloured regions correspond to the expected sensitivity reach of the DARWIN experiment.}
\label{fig:SKinetic_high}
\end{figure*}

Both the scalar and fermionic DM have the DD cross-section, induced by SI interactions, described by the following analytical expression: 
\bea
&& \sigma_{\rm DM \,p}^{\rm SI}=\frac{\mu_{\rm DM \,p}^2 }{\pi}{\left[b_p \frac{Z}{A}+b_n 
 \left(1-\frac{Z}{A}\right)\right]}^2,\nonumber\\
&& ~~{\rm where}~~ b_p=2 b_u+b_d,\,\,\,\,b_n=b_u+2 b_d, {\rm ~~with}\nonumber\\
&& ~~b_{f=u,d}=\frac{g_{DM}^Z\left(g_{f_L}^Z+g_{f_R}^{Z}\right)}{2 M_Z^2} +\frac{g_{DM}^{Z'}\left(g_{f_L}^{Z'}
 +g_{f_R}^{Z'}\right)}{2 M_{Z'}^2},
\eea
and ${\rm DM}=\chi,\,\psi$.
This setup has been already reviewed in \cite{Arcadi:2017kky}. We just provide and update of the results in Fig.~\ref{fig:SKinetic_high}, for two spin assignations of the DM, about the interplay of just DM relic density and DD.
%

\subsection{B-L Model with scalar DM}
A very popular application of the framework depicted above relies on the identification of the new gauge symmetry with the $B-L$ combination with $B$ and $L$ stemming for the baryon and lepton number respectively. The SM quarks and leptons are charged under this symmetry with quantum numbers being, respectively, equal to $1/3$ and $-1$. Anomaly cancellation requires to enlarge the spectrum of the theory further with three right-handed neutrinos (RHN) $N_{i=1,2,3}$ SM singlets but charged under the new gauge symmetry. We will consider in the next subsections the possibility that one of these RHN is the DM candidate. In this subsection, we will, instead, assume them to be cosmologically unstable. An intriguing possibility would be represented by the realisation of the see-saw mechanism for generating the SM neutrino masses. We will not investigate here such a case and rather assume the RHNs to be mass degenerate. The DM candidate is represented by a complex scalar field $\phi_{\rm DM}$ also charged under $B-L$ with the charge assignation suitably chosen to ensure its stability \cite{Rodejohann:2015lca}. The Lagrangian of the theory is hence given by: 
\begin{align}
    \mathcal{L}&=\mathcal{L}_{\rm SM}+\mathcal{L}_{\rm gauge-higgs}-\left(\frac{1}{2}\lambda_{N_i}S \ovl  {N^c_i} N_i+Y_{ij}\bar l_i H^\dagger N_i+\mbox{H.c.}\right)\nonumber\\
    & + (D_\mu \phi_{\rm DM})^\dagger (D^\mu \phi_{\rm DM})+\mu_{\rm DM}^2 \phi_{\rm DM}^\dagger \phi_{\rm DM}\nonumber\\
    & +\lambda_{H}\left(\phi_{\rm DM}^\dagger \phi_{\rm DM}\right) H^\dagger H +\lambda_{\rm DM}\left(\phi_{\rm DM}^\dagger \phi_{\rm DM}\right)|S|^2\nonumber\\
    & + \lambda_\phi \left(\phi_{\rm DM}^\dagger \phi_{\rm DM}\right)^2,
\end{align}
where $\mathcal{L}_{\rm gauge-higgs}$ is given by Eq.~(\ref{eq:Lag-gauge-Higgs}).
The scalar field $S$ has a charge equal to $2$ under the $B-L$ symmetry, so it generates Majorana mass terms for the RHNs once the $U(1)_{B-L}$ symmetry is spontaneously broken. The VEV of $S$, $v_S$ might also generate a contribution to the DM mass:
\begin{align}
    & m_{\phi}^2=\mu_{\phi_{\rm DM}}^2+\lambda_H\frac{v_h^2}{2}+\lambda_{\rm DM}\frac{v_S^2}{2},\nonumber\\
    & m_{N_i}=\frac{\lambda_i}{\sqrt{2}}v_S.
\end{align}
For simplicity, we will neglect here a kinetic mixing between the $U(1)$ bosons.

\begin{figure*}
    \centering
    \subfloat{\includegraphics[width=0.45\linewidth]{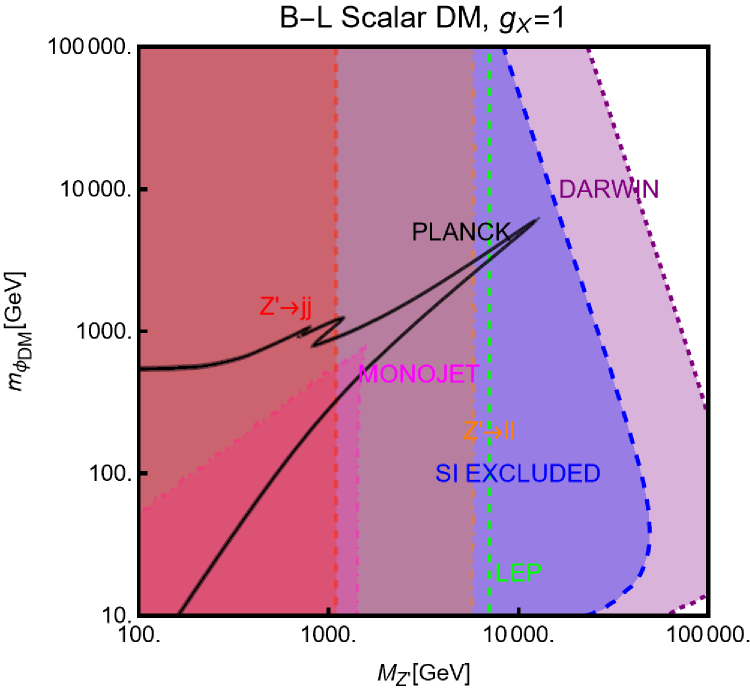}}
    \subfloat{\includegraphics[width=0.43\linewidth]{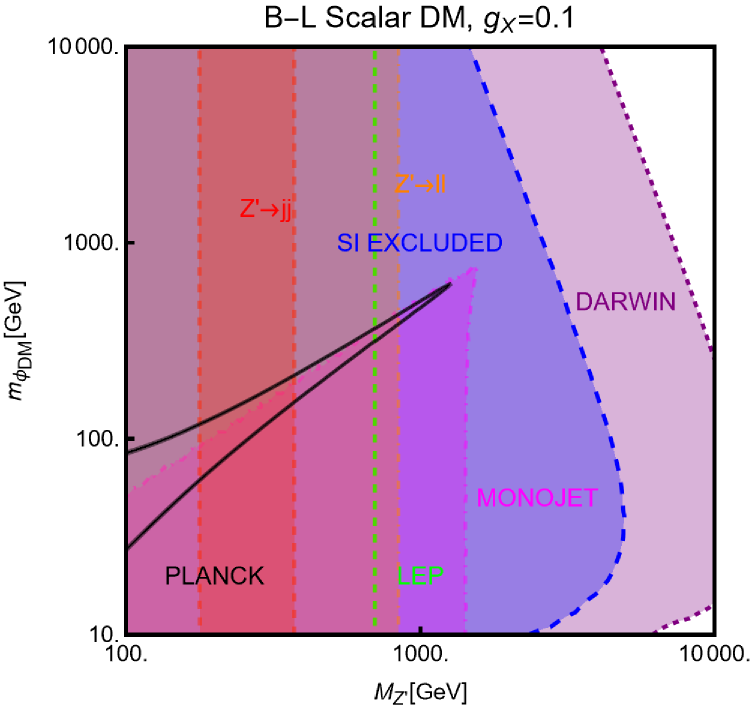}}
    \caption{Combination of constraints for $B-L$ model with a scalar DM $\phi_{\rm DM}$. The value of the $g_X$ is $1~(0.1)$ for the left (right) panel. The red, orange, pink and green coloured regions represent the exclusion limits from the LHC searches of dijets, di-leptons, monojet and the LEP experiment, respectively. The blue (purple) coloured region is based on the sensitivity reach of the current (next generation) DM detection experiments.}
    \label{fig:scalarBL}
\end{figure*}

Analytical expressions describing the relevant DM observables can be straightforwardly obtained by adapting the ones previously derived in this work, hence are not written here again. To provide a first illustration of the combined constraints, we will reduce the parameter space of the model by setting $\lambda_H=\lambda_{\rm DM}=\sin\theta=0$. In this setup, the DM phenomenology is determined by the gauge interactions of the scalar DM and strongly resembles the simplified model discussed at the beginning of this work.  The combined constraints are shown in Fig.~\ref{fig:scalarBL}. As already seen, the requirement of the correct relic density, due to the velocity suppression of the DM annihilation cross-section, cannot compete with the very strong exclusion bounds from the SI interactions. 

While not competitive, in the chosen setup, it is useful to show also collider constraints. In the fig. \ref{fig:scalarBL} are indeed present coloured regions corresponding to exclusion bounds from searches of dilepton resonances (orange), dijet resonances (red) and monojet events (magenta). Furthermore, the dashed green line represent the LEP bound from non resonant production of the $Z'$ which reads, for the B-L model \cite{Langacker:2008yv,ALEPH:2006bhb}:
\begin{equation}
    \frac{M_{Z'}}{g_{B-L}}> 7\,\mbox{TeV},
\end{equation}
with $g_{B-L}$ as the new gauge coupling.

\begin{figure*}
    \centering
    \subfloat{\includegraphics[width=0.43\linewidth]{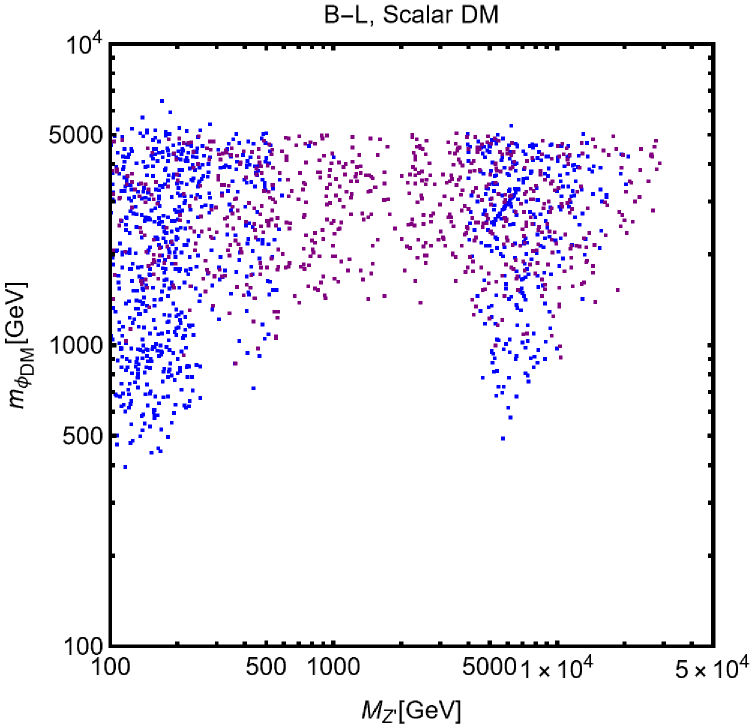}}
    \subfloat{\includegraphics[width=0.4\linewidth]{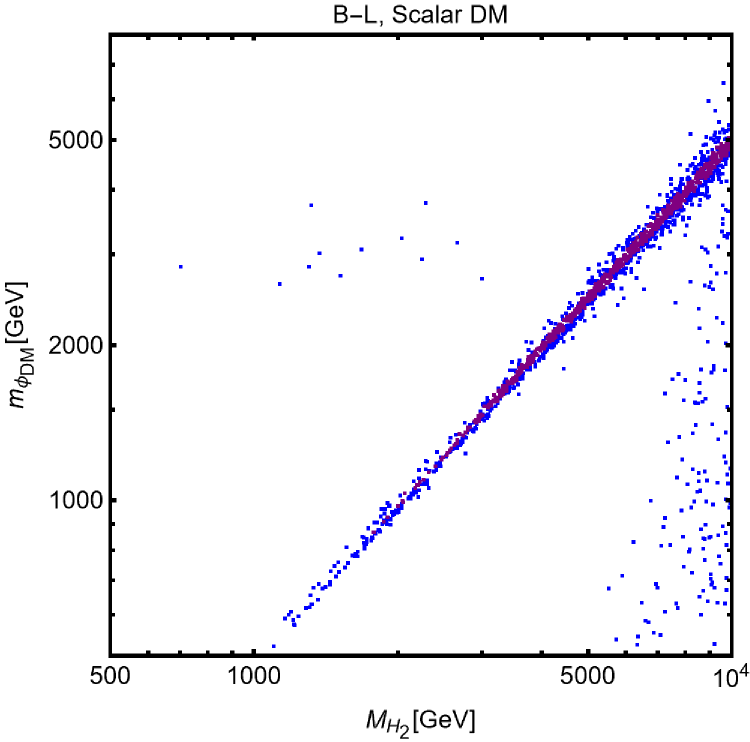}}
    \caption{Outcome of the parameter scan for the $B-L$ model with a scalar DM $\phi_{\rm DM}$. According to the usual colour coding, the blue coloured points correspond to the assignations of the model parameter compatible with the current experimental constraints. The purple coloured points will evade a negative signal by DARWIN. The left panel shows the $(M_{Z'},\,m_{\phi_{\rm DM}})$ plane while the right panel shows the $(M_{H_2},\,m_{\phi_{\rm DM}})$ one.}
    \label{fig:pscanBLscalar}
\end{figure*}

To account for the additional impact of the DM phenomenology on (1) the coupling of the DM and the dark Higgs and $(2)$ the mass mixing between the dark and the SM Higgs bosons, we have performed a parameter scan over the following parameters in certain ranges:
\begin{align}
    & \mu_{\phi_{\rm DM}} \in [1,10^4]\,\mbox{GeV},\,\,\,\,M_{Z'}\in [0.1,100]\,\mbox{TeV}, \nonumber\\
    & M_{H_2} \in [10,10^4]\,\mbox{GeV},\,\,\,\sin \theta \in [10^{-3},0.3],\nonumber\\
    & g_{B-L} \in \left[10^{-3},3\right],\,\,\,\lambda_{\rm DM}\in [10^{-3},10].
\end{align}
For simplicity we have assumed the DM charged under $B-L$ to be 1, so $g_{B-L}=g_X$. Furthermore, we have retained only values of the DM mass below $10\,\mbox{TeV}$. We have, moreover, assumed null kinetic mixing, at the tree level, between the electrically neutral gauge bosons. Direct mass mixing between the $Z$ and the $Z'$ boson is not present as the SM Higgs boson is uncharged under the $B-L$ symmetry. 
Following the usual convention, the parameter assignations passing all current constraints from the relic density, the DD and LHC searches have been shown, in Fig. \ref{fig:pscanBLscalar}, in the $(M_{Z'},m_{\phi_{\rm DM}})$ and $(M_{H_2},m_{\phi_{\rm DM}})$ bidimensional planes. 
Purple coloured points are, instead, the points which would comply with a negative signal from the DARWIN experiment.
In agreement with the findings of Ref.~\cite{Rodejohann:2015lca}, the $B-L$ scalar DM model is very constrained as it allows viable DM only for masses above $500$ GeV. The right panel of the Fig. \ref{fig:pscanBLscalar} also provides a clear indication that the vast majority of the allowed parameter space corresponds to the $m_{\phi_{\rm DM}} \sim M_{H_2}/2$ pole.  

\subsection{$B-L$ Model with a fermionic DM}

As pointed out in the previous subsection, one of the three RHNs, typically the lightest one, might be adopted as the DM candidate $N_1$. To obtain this result an ad-hoc $Z_2$ symmetry is introduced, to forbid Yukawa couplings of the concerned state with the SM leptons. In this setup, the DM interacts axially with the $Z$ and $Z'$ bosons. A further double $H_{1,2}$ portal is present in case of a sizeable $\sin\theta$. Analytical expressions of the DM annihilation cross-sections, responsible for the DD, can be obtained from the ones written previously in this work hence they will not be repeated here. 

Concerning the DD we have mainly SI interactions which can be described via the following cross-section:
\begin{align}
    & \sigma_{N_1 p}^{\rm SI}=\frac{4 \mu_{N_1 p}^2}{\pi}\left \{\frac{y_{N_1} m_p}{v_h}\sin \theta \cos \theta \left(\frac{1}{M_{H_1}^2}-\frac{1}{M_{H_2}^2}\right) \right.\nonumber\\
    &\left. \times \ \left[\sum_{q=u,d,s}f_q^p+\frac{2}{27}f_{TG}\right]\right.\nonumber\\
    & \left. + m_p \sum_{q=u,d,s} f_q^p f_q+\sum_{q=u,d,s,c,b} \frac{3}{4}m_p\left(q(2)+ \bar q(2)\right) \left (g_q^{(1)}+g_q^{(2)}\right)\right.\nonumber\\
    &\left. -\frac{8\pi}{9\alpha_s}f_{TG}f_G\right \},
\end{align}
with
\begin{align}
    f_q&=\frac{g_{N_1}^{Z'\,2}}{64\pi^2}\left( \frac{g_{H_1 Z' Z'}}{M_{H_1}^2}+\frac{g_{H_2 Z' Z'}}{M_{H_2}^2}\right)g_H\left(\frac{M_{Z^{'}}^2}{m_{N_1}^2}\right)\nonumber\\
    & +\frac{g_{N_1}^{Z^{'}\,2}}{M_{Z'}^3}\left(g_{Z'q}^{V^2}-g_{Z'q}^{A^2}\right)g_S\left(\frac{M_{Z^{'}}^2}{m_{N_1}^2}\right),\nonumber\\
    g_q^{(1)}&=\frac{2 g_{N_1}^{Z^{'}\,2}}{M_{Z'}^3}\left(g_{Z'q}^{V^2}+g_{Z'q}^{A^2}\right) g_{T1}\left(\frac{M_{Z^{'}}^2}{m_{N_1}^2},\right)\nonumber\\
    g_q^{(2)}&=\frac{2 g_{N_1}^{Z^{'}\,2}}{M_{Z'}^3}\left(g_{Z'q}^{V^2}+g_{Z'q}^{A^2}\right) g_{T2}\left(\frac{M_{Z^{'}}^2}{m_{N_1}^2}\right),
\end{align}
with $g_{H_{1,2}Z'Z'}$ being the couplings between the Higgs bosons and two Z'.Here, the loop functions $g_S,g_{T1},g_{T2}$ are the same given once discussing the simplified spin-$1$ portals while:
\begin{align}
    g_H(x)&=-\frac{2}{b_x}(2+2x-x^2){\tan}^{-1} \left(\frac{2b_x}{\sqrt{x}}\right)\nonumber\\
    & +2 \sqrt{x}(2-x\log(x)),
\end{align}
where we remind that $b_x=\sqrt{1-x/4}$.
The scattering cross-section is the combination of a tree-level contribution, associated with the $t$-channel exchange of the $H_{1,2}$ bosons, and a loop-level contribution originated by the coupling of the DM with the spin-$1$ electrically neutral bosons. This kind of contribution might become relevant in the case of very small values of $\sin\theta$. For simplicity, we have assumed negligible $\sin\xi$ so that $f_q,g_q^{(1)},g_q^{(2)}$ are accounted for only loops involving the $Z'$.

\begin{figure*}
    \centering
    \subfloat{\includegraphics[width=0.33\linewidth]{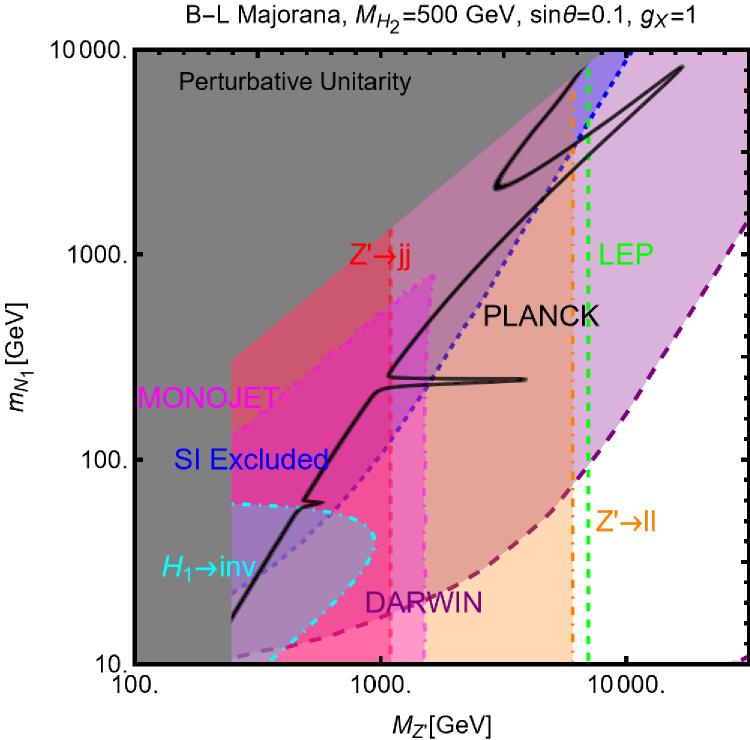}}
    \subfloat{\includegraphics[width=0.33\linewidth]{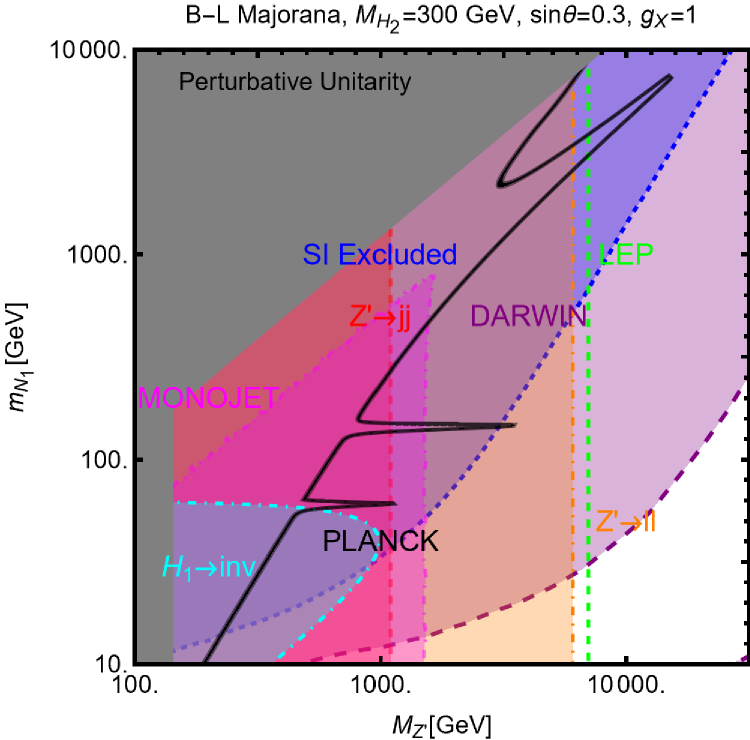}}
    \subfloat{\includegraphics[width=0.33\linewidth]{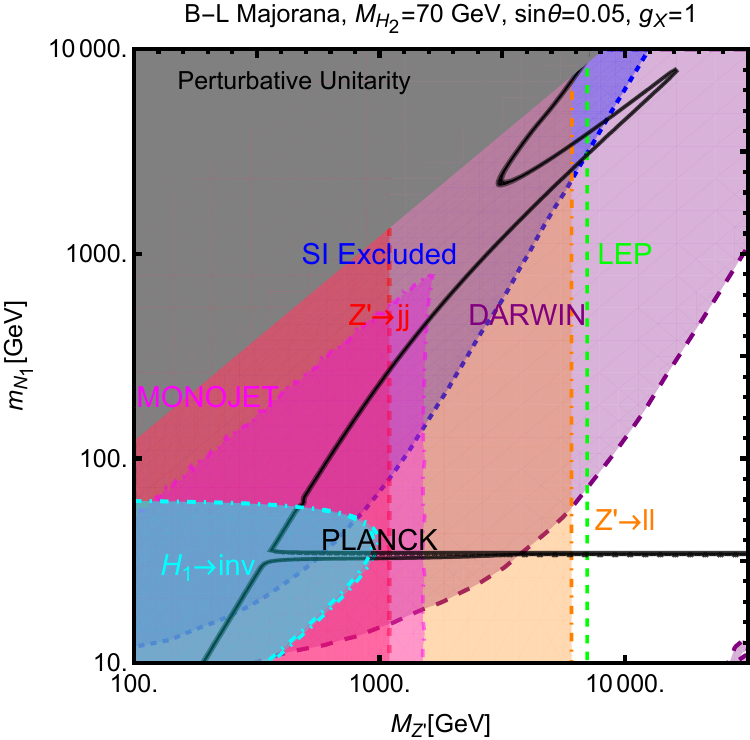}}
    \caption{Summary of constraints in the $(m_{N_1},m_{Z'})$ plane for three benchmark parameter assignations ($M_{H_2},\,\sin\theta,\,g_X$) of the $B-L$ model with a Majorana Neutrino DM $N_1$. The parameter $g_X$ is kept fixed at $1$ while $M_{H_2},\,\sin\theta$ values are $(300~{\rm GeV},\,0.1)$ (left), $(500~{\rm GeV},\,0.3)$ (middle) and $(70~{\rm GeV},\,0.05)$ (right), respectively. The cyan (black) coloured region is excluded from the bound on the invisible decay of the SM-like Higgs (perturbative unitarity bound on the model parameters). The remaining colour coding is the same as in Fig. \ref{fig:scalarBL}. }
    \label{fig:pBLmajo}
\end{figure*}

Following our conventional procedure we first provide, in Fig.~\ref{fig:pBLmajo}, an illustration of the combined constraints on the model in the $(M_{Z'},m_{N_1})$ bidimensional plane. The three panels consider three benchmark assignations of $(M_{H_2},\sin\theta)$, namely, $(500,0.1)$ (left panel), $(300,0.3)$ (middle panel) and $(70,0.05)$ 
(right panel), respectively, setting $g_X=1$ for all three.  The shape of the correct relic density contours (black coloured) strongly resembles the ones of the spin-$1$ mediator for simplified models discussed in the first part of this paper (see fig. \ref{fig:Zpportal}). The main difference is the appearance of "spikes" corresponding to the $m_{N_1}\sim M_{H_{1,2}}/2$ poles. The DD proves to be very effective in constraining this scenario with an eventual negative signal by DARWIN possibly ruling out the entire parameter space, except for very narrow $s$-channel resonances. Sticking on the present constraints, the strongest one is actually coming from resonant and non-resonant collider searches of the dilepton signals.

\begin{figure*}
    \centering
    \subfloat{\includegraphics[width=0.35\linewidth]{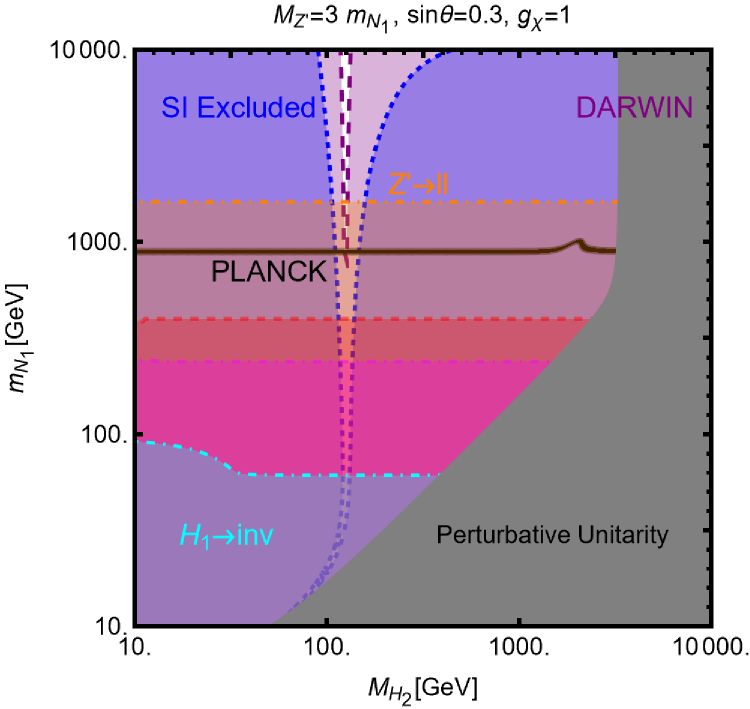}}
    \subfloat{\includegraphics[width=0.35\linewidth]{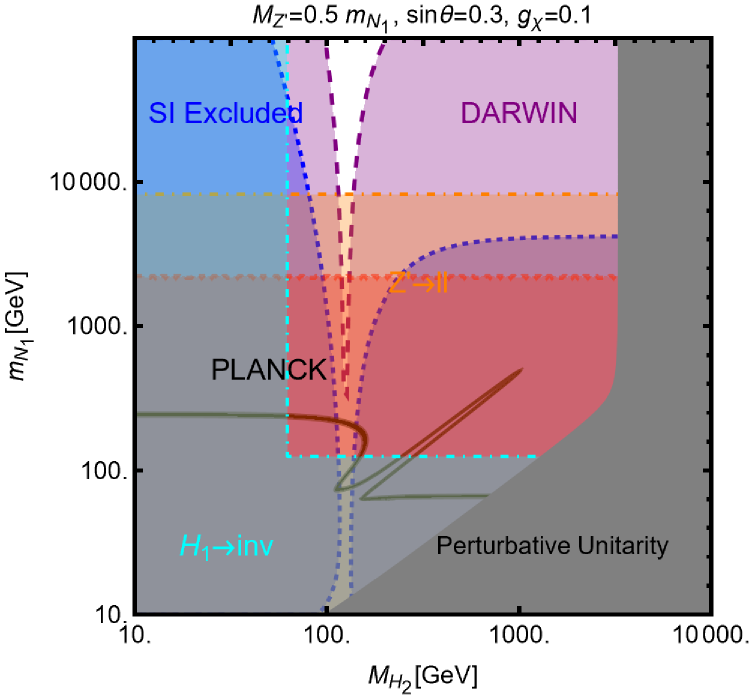}}\\
   \subfloat{\includegraphics[width=0.35\linewidth]{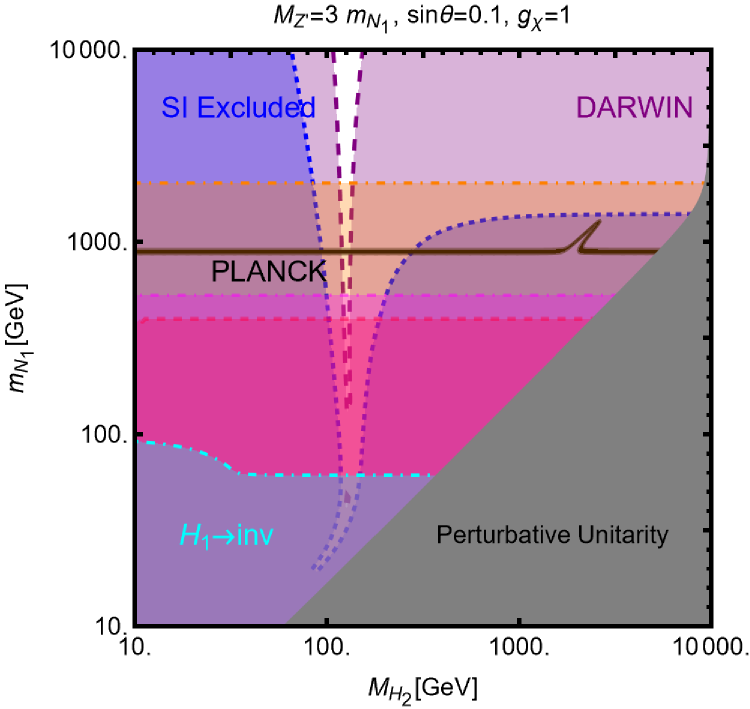}}
    \subfloat{\includegraphics[width=0.35\linewidth]{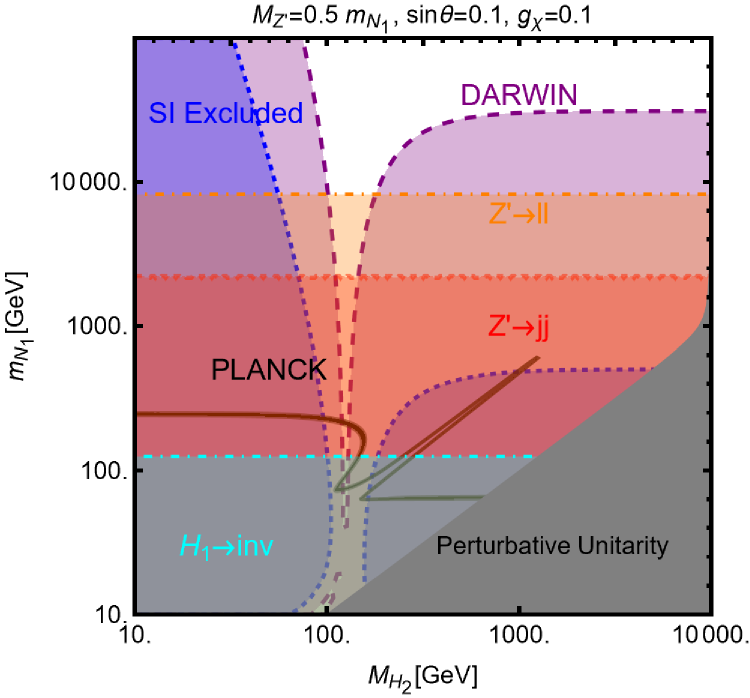}}\\
    \subfloat{\includegraphics[width=0.35\linewidth]{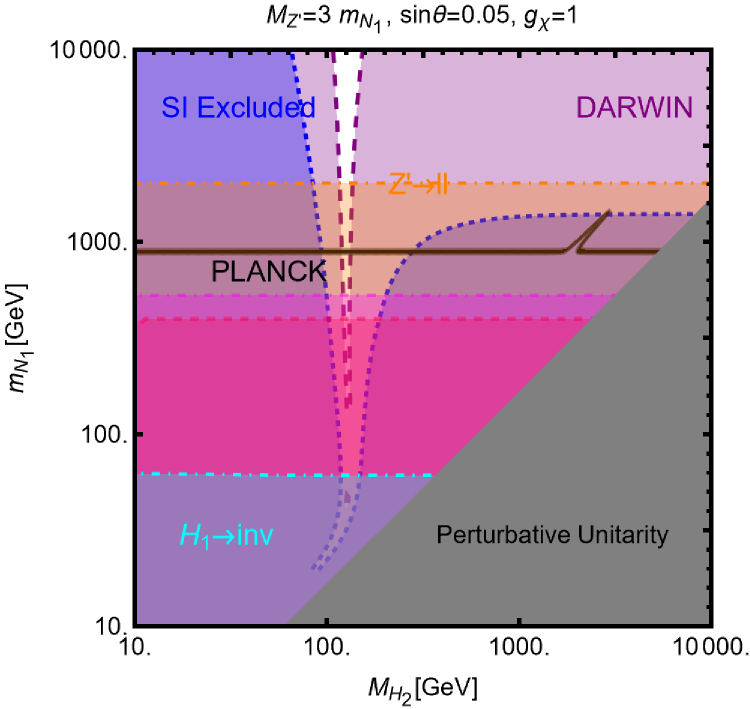}}
    \subfloat{\includegraphics[width=0.35\linewidth]{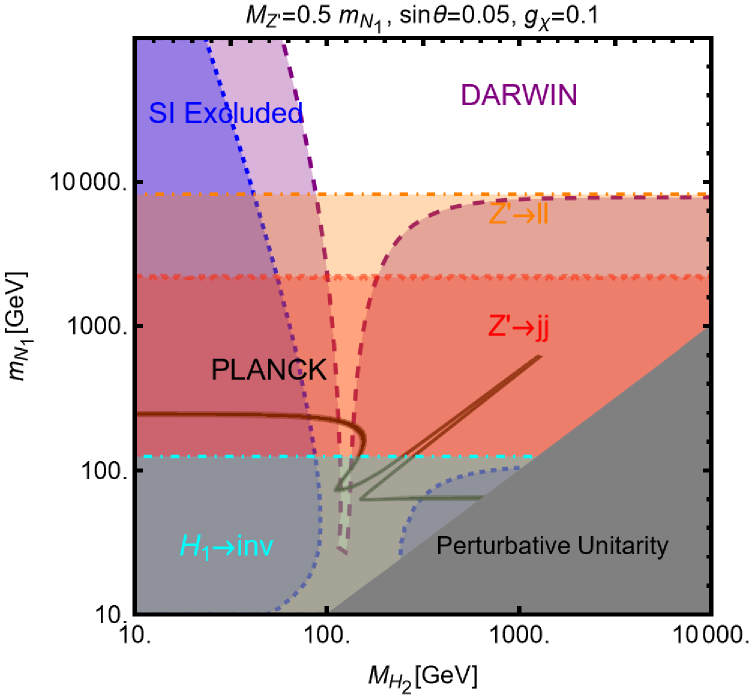}}
    \caption{Same as Fig. \ref{fig:pBLmajo} but considering the $(M_{H_2},m_{N_1})$ plane. The three rows correspond to the three assignations of $\sin\theta$, namely, $0.3$ (top), $0.1$ (middle), and $0.05$ (bottom), respectively. The left column corresponds to the assignation $M_{Z'}/m_{N_1}=3$ and $g_X=1$, while the right column to $M_{Z'}/m_{N_1}=0.5$ and $g_X=1$.}
    \label{fig:pBLmajo_bis}
\end{figure*}

While Fig. \ref{fig:pBLmajo} focused mostly on the interplay between the DM and $Z'$ masses, we consider in Fig. \ref{fig:pBLmajo_bis} the correlation between $m_{N_1}$ and $M_{H_2}$. Again we have considered three possible assignations of $\sin \theta$,
$0.3$ (top row), $0.1$ (middle row)
and $0.05$ (bottom row), respectively.
The left (right) column corresponds
to $M_{Z'}/m_{N_1}, g_X$ values
$3,1~(0.5,\,1)$. Looking at the left column, one notices that the relic density contour is mostly a horizontal line 

\begin{figure*}
    \centering
    \subfloat{\includegraphics[width=0.36\linewidth]{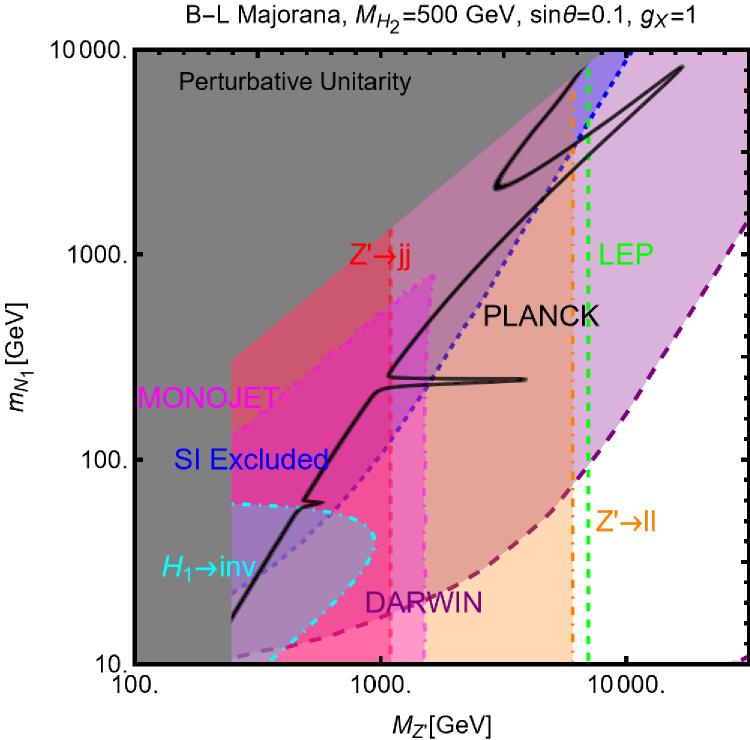}}
    \subfloat{\includegraphics[width=0.34\linewidth]{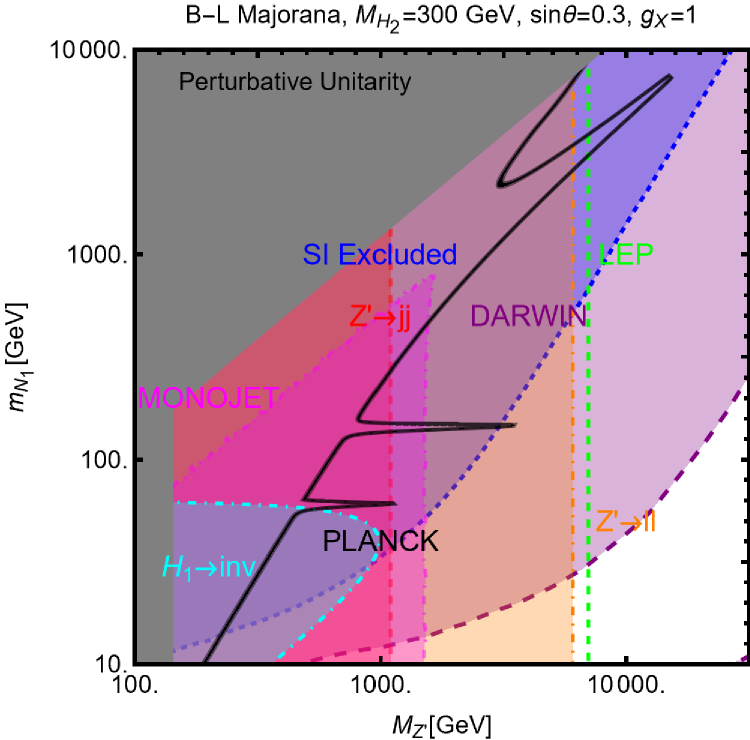}}
    \subfloat{\includegraphics[width=0.35\linewidth]{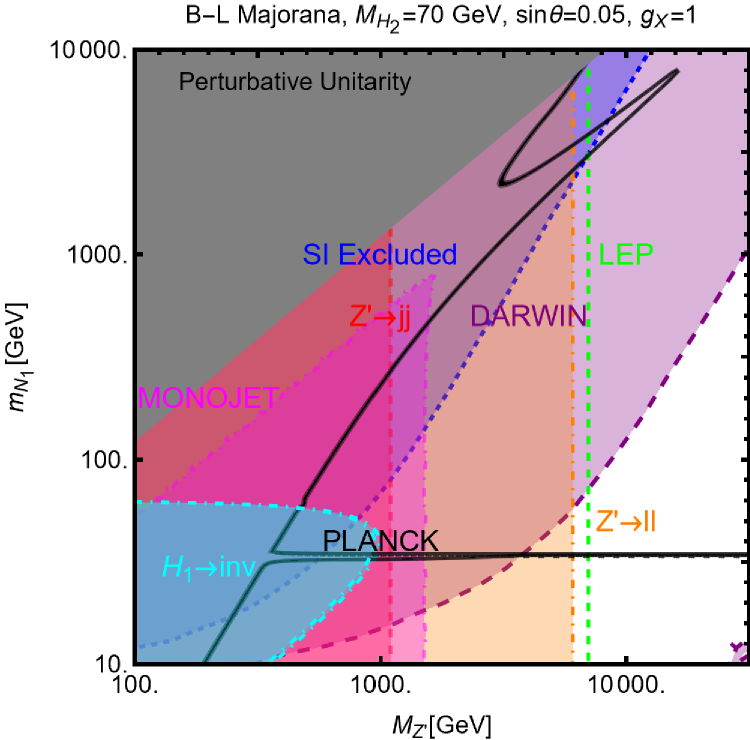}}
    \caption{Outcome of the parameter scan of the $B-L$ model with a Majorana neutrino DM $N_1$. Using the conventional colour coding, as of Fig. \ref{fig:pscanBLscalar}, the model assignations complying with the current and near future constraints are shown in the 
    $(M_{Z'},m_{N_1})$ plane (left panel), $(m_{N_1},g_X)$ plane (middle panel) and $(M_{H_2}, \sin\theta)$ plane (right panel), respectively.}
    \label{fig:pscanBLmajo}
\end{figure*}
To provide a final broader overview of the concerned model we have considered the following parameter scan:
\begin{align}
    & m_{N_1} \in \left[10,10^4\right]\,\mbox{GeV},\,\,\,\,M_{Z'}\in \left[0.1,100\right 
  ]\,\mbox{TeV},\nonumber\\
  & M_{H_2} \in \left[10,10^4\right]\,\mbox{GeV}, \nonumber\\
  & g_{X}\in \left[10^{-2},3\right]\,\,\,\sin \theta \in \left[10^{-3},0.3\right].
\end{align}
The outcome of the parameter scan is shown, according to the usual convention, in Fig. \ref{fig:pscanBLmajo} in the $(M_{Z'},m_{N_1})$, $(m_{N_1},g_X)$ (middle panel) and $(M_{H_2},\sin\theta)$ (right panel) planes. Contrary to the case of a scalar DM, previously discussed, in the viable parameter regions, the DM relic density is mostly accounted for the $Z'$ portal contributions, around the $m_{N_1} \sim M_{Z'}/2$ pole configuration. The strong constraints on the model, already highlighted by the specific benchmarks illustrated above, are passed only for the DM masses above the TeV scale. While not relevant for the relic density, the mixing angle $\theta$ is nevertheless constrained by the DD. The right panel of fig. \ref{fig:pscanBLmajo} evidences that current constraints give the upper bound of $\sin \theta \lesssim 0.1$. A negative detection from DARWIN will strengthen the upper bound to around $0.05$.

An alternative realisation of a fermionic DM consists of a Dirac vector-like fermion, charged under $B-L$. In such a case its interaction Lagrangian is given by:
\begin{align}
    & \mathcal{L}=i \ovl \psi \slashed{\partial}\psi-g_X \ovl \psi \gamma^\mu \psi X_\mu.
\end{align}
Being a vector-like fermion, its mass does not necessarily originate from the breaking of the $U(1)$-symmetry, hence it is a completely independent parameter from the gauge coupling $g_X$. For the case of a scalar DM, cosmologically unstable RHNs should nevertheless be present to ensure anomaly cancellation.

\begin{figure*}
    \centering
    \subfloat{\includegraphics[width=0.5\linewidth]{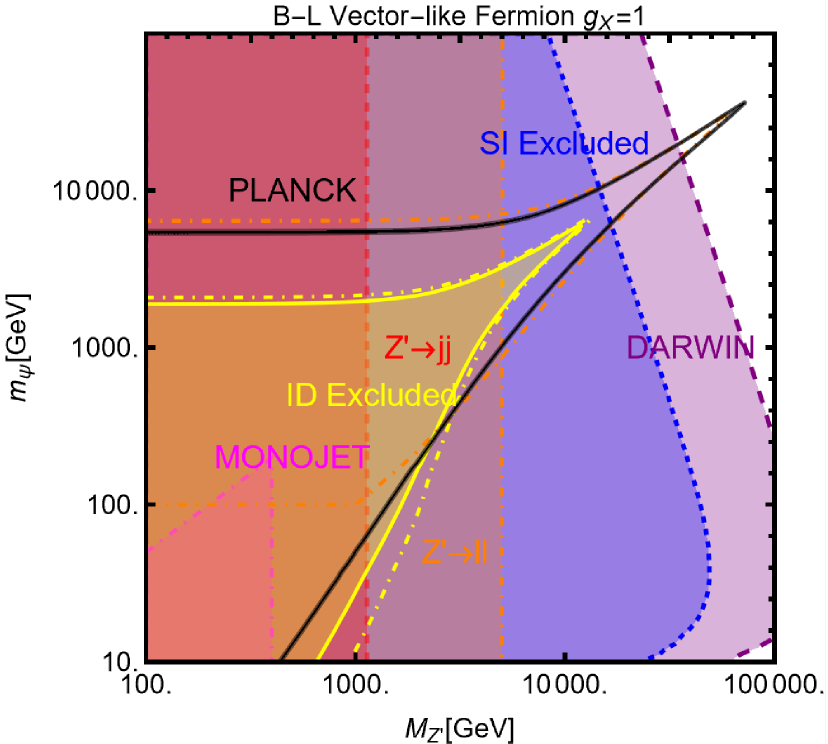}}
    \subfloat{\includegraphics[width=0.46\linewidth]{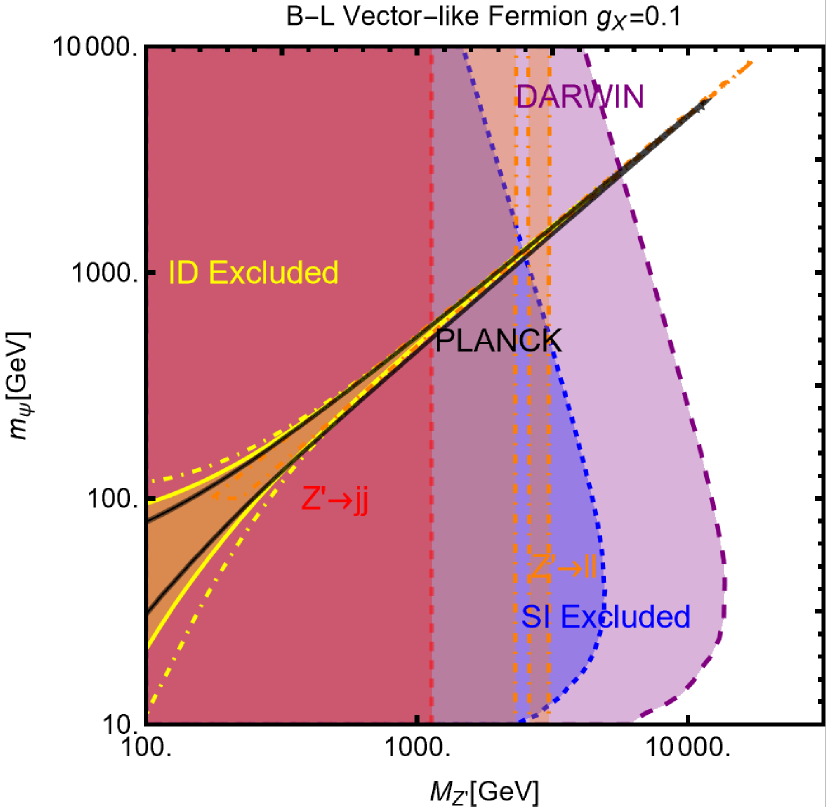}}
    \caption{Same as Fig.~\ref{fig:scalarBL} but for a vector-like Dirac fermion DM $\psi$, charged under $B-L$, including exclusion region (yellow coloured) from the ID limits of the DM. The value of $g_X$ is $1~(0.1)$ for the right (left) plot, as written on the top.}
    \label{fig:plotVLLBL}
\end{figure*}

We show in Fig. \ref{fig:plotVLLBL} the combined constraints on the concerned model in the $(M_{Z'},m_\psi)$ plane for the two assignations of the DM coupling, $1$ (left) and $0.1$ (right), respectively, setting the tree-level $ZZ'$ kinetic mixing to be zero.  The DM being a Dirac fermion, the most competitive constraint is the one associated with the DM DD which, in the case $g_X=1$, excludes masses of the $Z'$ above $10$ TeV. Contrary to the very similar scenario shown in Fig. \ref{fig:scalarBL} for a scalar DM, the more efficient annihilation processes of a Dirac fermion DM, still allow for a viable parameter region in correspondence with the $m_\psi \sim M_{Z'}/2$ pole. Since the DM annihilation cross-section is s-wave dominated, ID constraints apply, corresponding to the yellow region in the plot (dot-dashed yellow and orange contours represent near future experimental sensitivity by FERMI and CTA, respectively). As evident, they are competitive with DD and LHC constraints only around the pole region. Similar to the previous scenarios, one might consider the impact of the coupling of the DM with the dark Higgs. We will neglect here this possibility since this would require the introduction of a further free parameter, the DM Yukawa since the DM mass is not strictly related to the breaking of the $B-L$ symmetry.

\subsection{Model with a Majorana DM and only axial couplings}

Finally, let us consider a scenario in which the charge assignment of the SM fermions with respect to the new $U(1)$ group imposes the $Z'$ boson to interact with them only via the axial couplings. The DM $\chi$ is assumed to be a Majorana fermion with a mass dynamically generated by the breaking of the $U(1)$ symmetry. As shown in Ref.~\cite{Kahlhoefer:2015bea}, if no further exotic fermions, besides the DM, are added to the theory, anomaly cancellation favours flavour universal coupling for the $Z'$. Importantly, this imposes the SM doublet to be charged under the new $U(1)$ symmetry and the existence of a direct mass mixing between the $U(1)$ gauge bosons. 

\begin{figure*}
    \centering
    \subfloat{\includegraphics[width=0.35\linewidth]{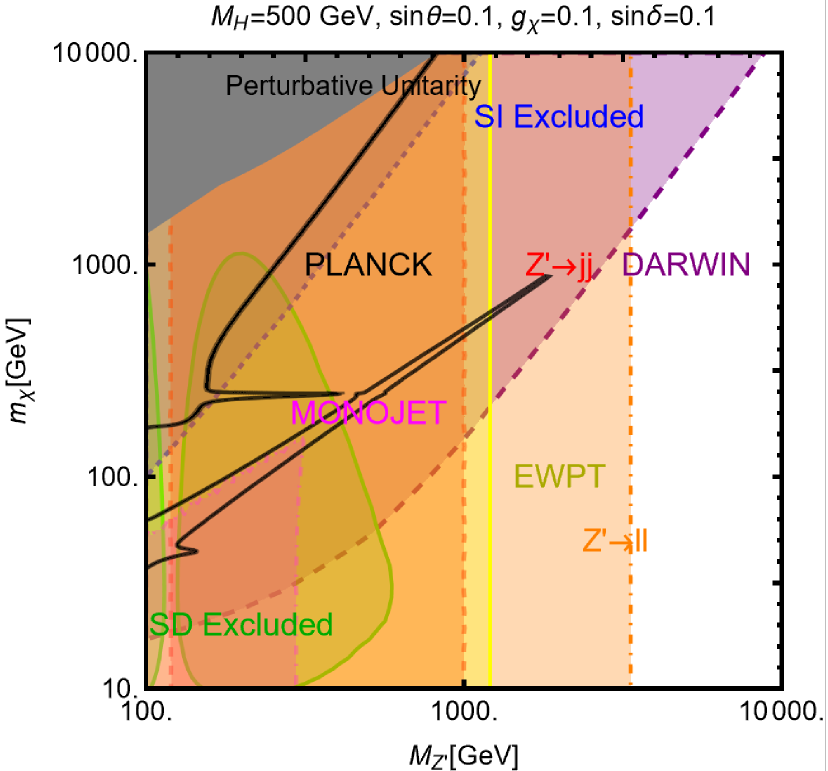}}
    \subfloat{\includegraphics[width=0.35\linewidth]{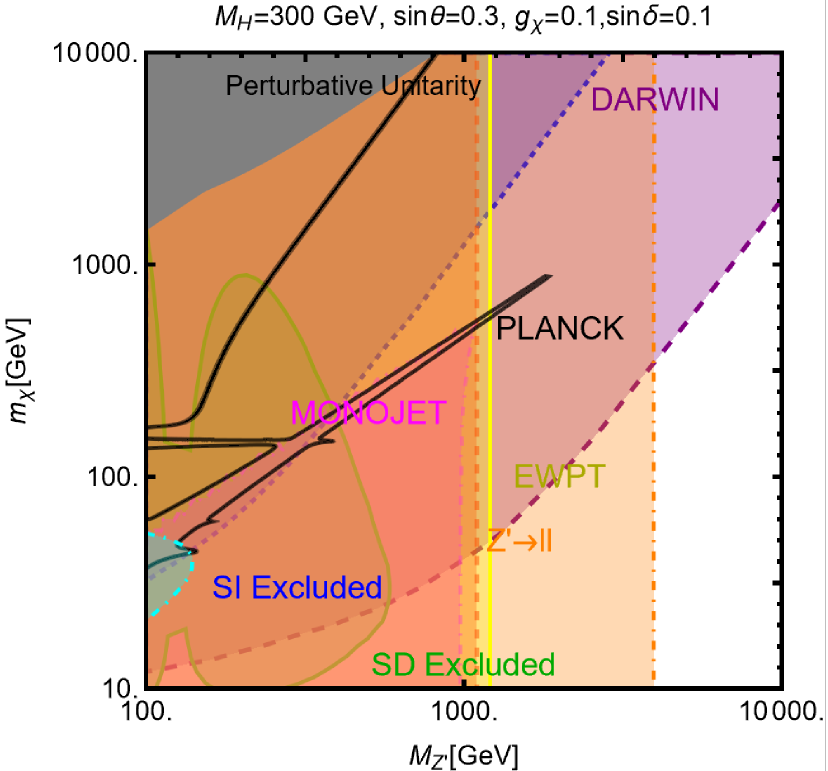}}
    \subfloat{\includegraphics[width=0.35\linewidth]{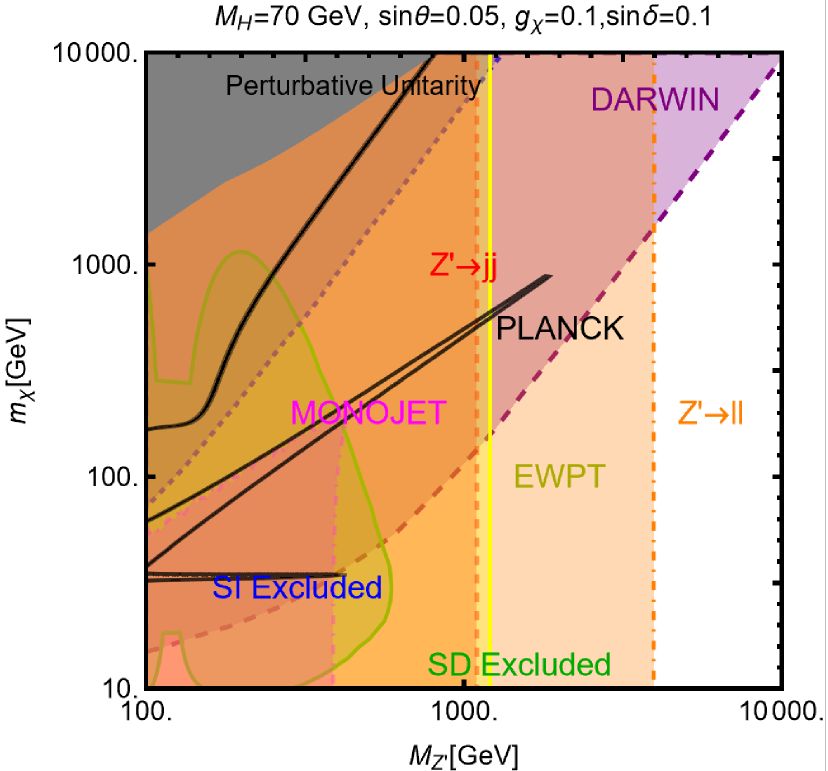}}
    \caption{Similar to Fig.~\ref{fig:pBLmajo} but for a model with a Majorana fermion DM $\chi$ and a $Z'$,having only axial couplings with the SM fermions, including green and yellow coloured regions which are excluded from the SD interactions and the EWPT, respectively. Unlike Fig.~\ref{fig:pBLmajo}, here we considered $g_X=0.1$ and $\sin\delta=0.1$ although panel-wise
    assignations of $M_{H_2},\,\sin\theta$ remain the same.}
    \label{fig:pAA}
\end{figure*}

The viable parameter region for this setup is shown in the $(M_{Z'},m_\chi)$ bidimensional plane,
for the three conventional assignations of the $(M_{H_2},\sin\theta)$ pair, in Fig.~\ref{fig:pAA}, keeping $g_X=0.1$ this time. For generality, we have also considered the presence of a non-zero kinetic mixing, $\sin\delta=0.1$. The shape of the relic density contours is interpreted with the fact that, given the velocity suppression of the DM annihilation cross-section into the SM fermions, the thermally favoured value is achieved mostly around the $m_\chi \sim M_{H_2}/2,\, M_{Z'}/2$ poles or when the $Z'Z'$ and/or $Z' H_2$ final states are kinematically accessible. This last possibility is however limited by the bound on the perturbative unitarity of the axial coupling of the DM with the $Z/Z'$ bosons. For what the DD is concerned, one notices exclusion bounds both from the SI interactions, due to the DM coupling with the Higgs, and the SD interactions, arising from the axial couplings of the $Z/Z'$ bosons with the DM and the SM fermions. The LHC searches, especially the ones from dilepton searches, provide at the moment more competitive constraints. Contrary to other $Z'$ models considered here, the presence of a direct $Z/Z'$ mixing increases the importance of constraints from the EWPT, which appear to be comparable with the ones from the DD and the LHC. Considering all together, a strongly disfavoured scenario seems to emerge.

\subsection{$U(1)_X$ and 2HDM}

In this subsection we consider to incorporate the $Z'$ portal, originated from a spontaneously broken $U(1)_X$ symmetry, in a 2HDM model. This option is very intersting as the $U(1)_X$ symmetry can be used, in spite of ad-hoc $Z_2$ symmetries, to impose only specific coupling assignations of the Higgs doublets to the SM fermions and, hence, define the Type-I, Type-II, Type-X and Type-Y models.
In turn, the extended Higgs sector provides new couplings, enriching the DM phenomenology.
The scalar sector of the theory can be characterised via the following potential:
\begin{align}\label{eq:2HDMU1X}
    V&=V_{\rm 2HDM}+m_S^2 S^\dagger S+\frac{\lambda_s}{2}{\left( S^\dagger S\right)}^2\nonumber\\
    & +\mu_1 S^\dagger S \Phi_1^\dagger \Phi_1+\mu_2 S^\dagger S \Phi_2^\dagger \Phi_2\nonumber\\
    & +\left(\mu \Phi_1^\dagger \Phi_2 S+\mbox{H.c.}\right),
\end{align}
with $V_{\rm 2HDM}$ already defined in Eq.~\eqref{eq:2HDM_potential} while $S$ is the field responsible for breaking the $U(1)_X$ symmetry.
Assuming, as usual, that CP is preserved in the scalar sector, the physical mass spectrum in the scalar sector will resemble one of the 2HDM+s models. Contrary to this, we will assume negligible mass mixing between the $SU(2)$ Higgs doublet and singlets so that the physical states will be the 125 GeV SM-like Higgs $h$, a doublet-like state $H$, and a singlet like state $h_s$ with mass:
\begin{equation}
    M_{h_s}^2 \simeq \lambda_s v_S^2.
\end{equation}
To ensure the smallness of the singlet-doublet mixing, the $\mu$ parameter of Eq.~(\ref{eq:2HDMU1X}) should be small enough. However, it also controls the mass of the charged Higgs:
\begin{equation}
    M_{H^\pm}^2 \simeq \frac{\left(\sqrt{2}\mu v_S -\lambda_4 v_1 v_2\right)}{2 v_1 v_2}v_h^2,
\end{equation}
consequently, it should satisfy:
\begin{equation}
    \mu > \frac{\lambda_4 v_1 v_2}{\sqrt{2}v_S}.
\end{equation}
For simplicity, we will consider, in this setup, just the case in which the $U(1)_X$ symmetry is, again, $B-L$, and the DM candidate is the lightest RH Majorana neutrino. Its Yukawa Lagrangian will hence be of the form:
\begin{equation}
    \mathcal{L}_{\rm yuk}= Y^{N_1} \overline{N_1^c}S N_1+\mbox{H.c.},
\end{equation}
with $m_{N_1}=y^{N_1} v_S/(2 \sqrt{2})$. One can also incorporate, via the heavier RHNs $N_{2,3}$, a Type-I see-saw mechanism for the generation of neutrino masses \cite{Camargo:2019ukv,Arcadi:2020aot}. We will not discuss here this possibility.

The mass Lagrangian for the gauge bosons is given by:
\begin{align}
    \mathcal{L}_{\rm gauge}&=(D_\mu \Phi_1)^\dagger (D^\mu \Phi_1)+(D_\mu \Phi_2)^\dagger (D^\mu \Phi_2)\nonumber\\
    & +(D_\mu S)^\dagger (D^\mu S)+\frac{1}{4}g_2^2 v_h^2 W^{- \mu}W^+_\mu \nonumber\\
    &+\frac{1}{8}g_1^{2}v_h^2 Z^{\mu}Z_{\mu}-\frac{1}{4}g_1\left(G_{X_1}v_1^2+G_{X_2}v_2^2\right)Z^{\mu}X_\mu \nonumber\\
    & +\frac{1}{8}\left(v_1^2 G_{X_1}^2+v_2^2 G_{X_2}^2 +v_S^2 G_{X_S}^2 g_X^2\right)X_\mu X^\mu,
\end{align}
where the covariant derivative is defined as:
\begin{equation}
D_\mu=\partial_\mu+ig_2T^a W^a_\mu+ig_1\frac{Q_Y}{2}B_\mu+ig_X \frac{Q_X}{2}X_\mu,
\end{equation}
while the couplings of the electrically neutral gauge bosons are defined as:
\begin{equation}
    G_{X_i}=\frac{g_1 \delta Q_{Y_i}}{c_W}+g_X Q_{X_i},
\end{equation}
with $\delta$ representing a small kinetic mixing parameter $\sin\delta \simeq \delta$.
Again we have a mass mixing (direct and kinetic) between the $Z$ and the $X$ bosons analogous to the one presented previously in this work. For this case, the mixing angle $\xi$ is given by:
\begin{equation}
    \tan 2 \xi=\frac{2 g_1 \left(G_{X_1}v_1^2+G_{X_2}v_2^2\right)}{m^2_{Z_0}-m_X^2}.
\end{equation}
Assuming for simplicity a small mixing regime, we can write, in good approximation:
\begin{align}
    & M_Z^2 \simeq m_{Z^0}^2=\frac{1}{4}g_1^{2}v_h^2,\nonumber\\
    & M_{Z'}^2=\frac{v_S^2}{4} g_X^2 q_{X_S}^2+g_X^2 \sin^2 2 \beta v_h^2 (q_{X_1}-q_{X_2})^2, \nonumber\\
    & \sin \xi \simeq \frac{G_{X_1}v_1^2+G_{X_2}v_2^2}{M_{Z'}^2}.
\end{align}
Finally, the Lagrangian containing the "portal" interactions mediated by the $Z/Z'$:
\begin{align}
    \mathcal{L}_{\rm NC}&= \left[\left(g^Z_{f_L}+g^Z_{f_R}\right)\bar f \gamma^\mu f \right. \nonumber\\
    & \left. +\left(g^Z_{f_R}-g^Z_{f_L}\right)\bar f \gamma^\mu \gamma_5 f\right]Z_\mu \nonumber\\
    & +\left[\left(g^{Z'}_{f_R}+g^{Z'}_{f_R}\right)\bar f \gamma^\mu f +\left(g^{Z'}_{f_R}-g^{Z'}_{f_L}\right)\bar f \gamma^\mu \gamma_5 f\right]Z^{'}_\mu \nonumber\\
    & -\frac{1}{4} g_X \cos \xi N_1 \gamma^\mu \gamma_5 N_1 Z^{'}_\mu\nonumber\\
    &+\frac{1}{4} g_X \sin \xi N_1 \gamma^\mu \gamma_5 N_1 Z_\mu.
\end{align}

with $g_{f_{L,R}}^{Z,Z'}$ given by Eqs. (\ref{eq:Zcoup}) and (\ref{eq:Zprimecoup}) in the limit of negligible kinetic mixing.

Concerning the DM phenomenology, the main difference between this setup and the previously considered models is the presence of the additional $Z' W^+ W^-$, $Z^{'}W^\pm H^\mp$, $HZ'Z'$ and $HZZ'$ vertices, which can open new annihilation channels in the heavy DM regime. For what the DD is concerned, the relevant analytical expressions substantially coincide with the ones given for the $B-L$ model of the previous subsection, besides a redefinition of the couplings.

\begin{figure*}
    \centering
    \subfloat{\includegraphics[width=0.45\linewidth]{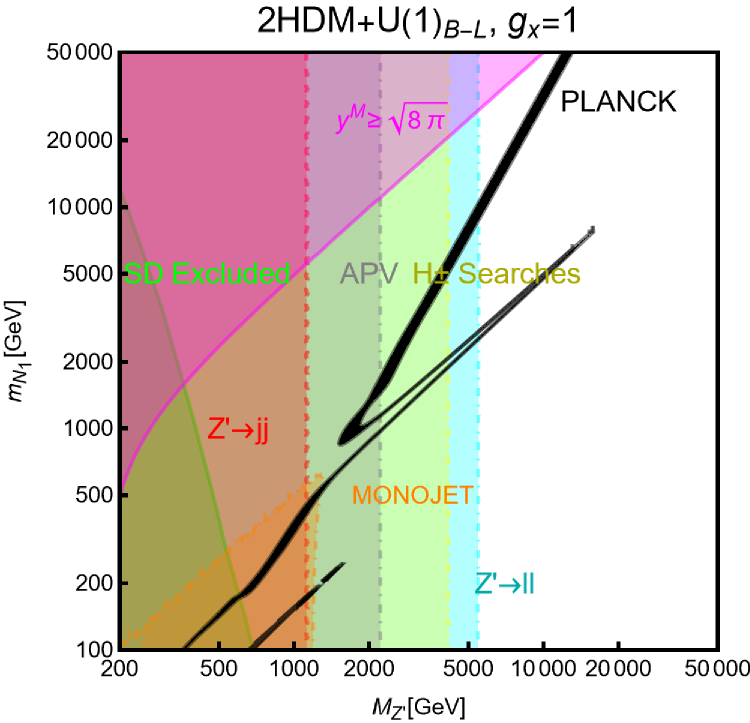}}
    \subfloat{\includegraphics[width=0.45\linewidth]{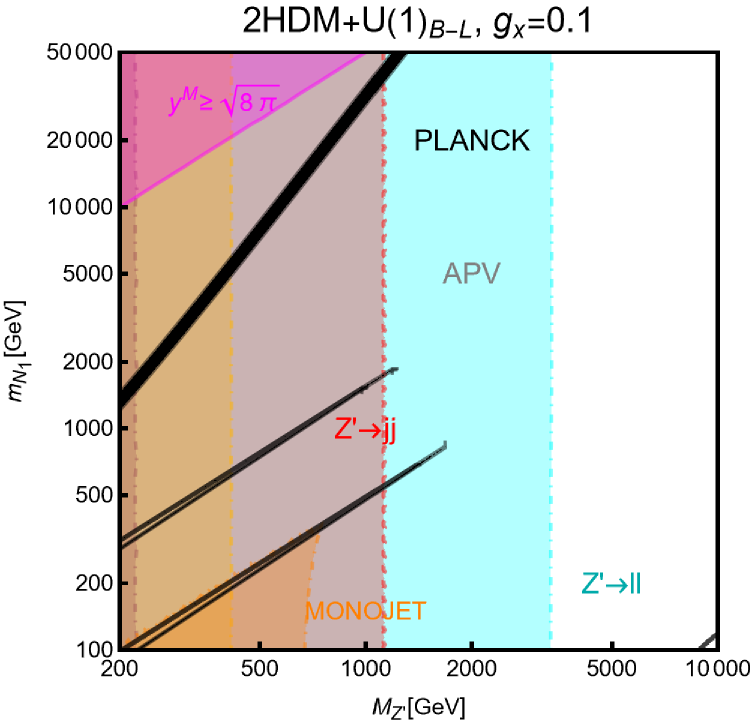}}
    \caption{Summary of constraints for the $2HDM+U(1)_X$ model with a Majorana neutrino DM $N_1$ and $X=B-L$. The results are illustrated in the $(M_{Z'},m_{N_1})$ bidimensional plane for two values of $g_X$, namely, $1$ (left) and $0.1$ (right). Following the usual conventions, the relic density isocontours (black coloured) are compared with the various excluded regions. The red, orange, cyan, purple and, grey coloured regions are excluded from the LHC searches of the dijet, monojet, dilepton, the perturbative unitarity bound on $Y^{N_1}$ and, the APV, respectively. For the right panel, additional exclusions appear from the SD interactions (green coloured)
    and $H^\pm$ searches at the collider (light green coloured).}
    \label{fig:p2HDMU1}
\end{figure*}

To illustrate the DM phenomenology of the model we limit ourselves just to the usual combination of the constraints in the $(M_{Z'},m_{N_1})$ bidimensional plane. The latter combination is shown in Fig. \ref{fig:p2HDMU1} for the two assignations $1$ (left) and $0.1$ (right) of the DM gauge coupling $g_X$. As evident, the shapes of the relic density contours show two cusps, one corresponding to $m_{N_1} \sim M_{Z'}/2$ and the other to $m_{N_1}\sim M_{h_s}/2$. For the parameter assignation considered in the figure, one always has $M_{h_s}< M_{Z'}$. Having neglected the mixing between the $h_s$ state and the ones coming from the doublet, the $h_s$ state mediates mostly the DM annihilation into a pair of $ZZ$ while, on the contrary, the $Z'$ mediates annihilations over a broader variety of the final states. For such reasons, for $g_X=1$, the cusp associated with the $h_s$ resonance is less pronounced. For $g_X=0.1$, the $h_s$ and $Z'$ cusps are of comparable size due to the suppression of the $Z'$ mediated annihilations. In the absence of mixing between the singlet and the doublet Higgs, the loop induced SI cross-section lies below the present and future experimental sensitivity, hence no exclusion region is visible in the plot. The most relevant constraint for the model comes from the LHC, in the form of searches of the dilepton resonance. The only viable region of the parameter space appears to be the $m_{N_1} \sim M_{Z'}/2$, for DM masses above the TeV scale, for $g_X=1$.


\section{Conclusions}
\label{sec:Conclusions}
In light of the current and upcoming direct and indirect detection data, we have reviewed the thermal production of dark matter in many simplified models and some UV complete theories. We have systematically explored the mechanisms of dark matter production, including s-channel processes via spin-0 mediators (both CP-even and CP-odd) and spin-1 mediators, t-channel production, and have delved into widely discussed dark matter portals such as the Higgs portal and Z portal. Our investigation also extends to models incorporating scalar, fermion, and vector dark matter candidates to assess the implications of different spin properties. In particular, we outlined the region of parameter space in which a vector dark matter under Abelian and non-Abelian gauge groups reproduces the correct relic density in agreement with current and future data. Moreover, we outline the parameter space in which Z' and dark photon mediators, in the presence or absence of kinetic mixing, offer a plausible road to thermal DM. Advancing beyond these preliminary models, our discourse has also covered thermal dark matter production within Two Higgs Doublet Models (Type-I and Type-II), highlighting scenarios where a scalar, pseudoscalar, or vector boson facilitates the interaction between the dark and visible sectors. Moreover, we have revisited thermal dark matter production scenarios under the well-regarded B-L gauge symmetry, examining both straightforward B-L extensions with right-handed neutrinos and those embedded within a Two Higgs Doublet Model framework.

Regarding constraints on these models, our observations underline significant bounds arising from collider data, notably from the invisible decay of the Higgs, monojet, dijet, and dilepton events. In the domain of indirect detection, we have leveraged data from the Fermi-LAT collaboration, alongside potential prospects offered by CTA. In the context of direct detection, our analysis incorporates exclusion limits from the XENONnT and LZ collaborations, as well as the projected sensitivity of the DARWIN experiment. 

Our findings reveal a profound synergy among direct and indirect detection methodologies and collider experiments, emphatically supporting the allure of the Weakly Interacting Massive Particle (WIMP) paradigm. This synergy underscores the complementary nature of varied dark matter search strategies. By mapping the constraints onto specific parameter spaces of dark matter models, we observe comprehensive coverage across distinct regions, with some overlaps, illustrating the inherently multifaceted approach required in the pursuit of dark matter within the WIMP framework.

We have found that the most simple dark sectors are severely constrained by direct detection experiments, especially those featuring a DM-nucleon spin-independent scattering cross-section. With upcoming data, those portals will be fully or nearly excluded for DM masses below $1$TeV. One needs to either to push the dark matter mass towards multi-TeV scale or invoke non-standard cosmology to shift the relic curve and allow DM masses below $1$TeV, while keeping the DM complementarity search ongoing. The exclusion from the DD constraints can be relaxed somehow in the next-to-minimal scenarios, featuring multiple mediators or new states lighter than the DM (we have reviewed the example of a light pseudoscalar), etc.

In conclusion, we have combined different experimental datasets and theoretical models, computed the DM relic density, and the direct, indirect, and collider observables to paint a clear picture of where the WIMP paradigm stands and the near future prospects. The most simple WIMP models will be fully probed with upcoming data, strengthening the need for the next generation of experiments. 

Moreover, our work shows that some DM constructions will survive the null results from the collider, direct and indirect experiments, suggesting that a further step in sensitivity reach is needed to falsify the WIMP paradigm. Our conclusion is based on the thermal production of dark matter. If one invokes a new production mechanism, new regions of parameter space will be viable and the current excluded regions may open up.

\begin{acknowledgements}
The authors also thank Francesco D'Eramo for his valuable comments. JPN warmly thanks the Galileo Galilei Institute for Theoretical Physics for the hospitality for partial support during the completion of this work. D.C.A. acknowledges funding from the Spanish MCIN/AEI/10.13039/501100011033 through grant
CNS2022-135262 funded by the “European Union NextGenerationEU/PRTR”, P. G. acknowledges the IITD SEED grant support 
IITD/Plg/budget/2018-2019/21924, continued as \\
IITD/Plg/budget/2019-2020/173965, IITD Equipment Matching Grant IITD/IRD/MI02120/208794, and Start-up Research Grant (SRG) support SRG/2019/000064 from the Science and Engineering Research Board (SERB), Department of Science and Technology, Government of India. M.~D. acknowledges support by an appointment to the NASA Postdoctoral Program at the NASA Goddard Space Flight Center, administered by Oak Ridge Associated Universities through a contract with NASA. S.~P.'s work was partly supported by the U.S. Department of Energy grant number DE-SC0010107. 
This work is also supported by the Spanish MICINN's Consolider-Ingenio 2010 Programme under grant Multi-Dark 
{\bf CSD2009 - 00064}, the contract {\bf FPA2010 - 17747}, the France-US PICS no. 06482 and the LIA-TCAP of CNRS. M.P. acknowledges support by the Deutsche Forschungsgemeinschaft (DFG, German Research Foundation) under Germany's Excellence Strategy – EXC 2121 “Quantum Universe” – 390833306.  Y.~M. acknowledges partial support the ERC advanced grants Higgs@LHC and MassTeV. This research was also supported in part by the Research Executive Agency (REA) of the European Union under the Grant Agreement {\bf PITN-GA2012-316704} (``HiggsTools''). JPN acknowledges support from
Coordenação de Aperfeiçoamento de Pessoal de Nível Superior (CAPES) under Grant No. 88887.712383/2022-0 and the PRINT-UFRN program under the Grant No. 88887.912033/2023-00. FSQ is supported by Simons Foundation (Award Number:1023171-RC), FAPESP Grant 2018/25225-9, 2021/01089-1, 2023/01197-4, ICTP-SAIFR FAPESP Grants 2021/14335-0, CNPq Grants 307130/2021-5, and ANID-Millennium Science Initiative Program ICN2019\textunderscore044. 
\end{acknowledgements}

\appendix



\bibliography{bibfilenew2}{}

\end{document}